\documentclass[12pt]{article}
\usepackage{tikz}
\usepackage{tikz-cd}

\usetikzlibrary{decorations.pathmorphing}

\usepackage{jheppub}

\usepackage{pgfplots}
\usepackage{pgfplotstable}
\usepackage{pgfkeys}
\allowdisplaybreaks
\usepackage{ytableau}
\usepackage{comment}
\usepackage[vcentermath]{youngtab}
\usepackage{bm}
\usepackage{ulem}
\usepackage{xcolor}
\definecolor{nosaka}{rgb}{0.0, 0.7, 0.0}

\newcommand\be{\begin{equation}}
\newcommand\ee{\end{equation}}

\newcommand\Tr{\mathrm{Tr}}

\newcommand\bt{\mathfrak{t}}
\usepackage{braket}
\usepackage{subfigure}
\usepackage{mathrsfs}

\title{Wilson line defect indices for M2-brane SCFTs}
\abstract{
We investigate line defect Coulomb and Higgs indices in 3d $\mathcal{N}\ge 4$ supersymmetric gauge theories.
The normalized line defect indices defined by ratios of the decorated defect indices to the ordinary indices yield  well-defined integer values,
the line defect Witten indices in the unit R-charge fugacity limit.
We propose general formulae for these Witten indices in both Coulomb and Higgs limits for the theories without Chern-Simons terms, and we show that both count the states invariant under the residual gauge group left unbroken at a generic point of the branch, which we read off from the vacuum equations.
Furthermore, we find powerful algebraic recursion relations for Wilson line defect Coulomb indices
in the 3d SCFTs describing M2-branes, including ADHM-type theories and ABJ(M) theories.
These relations imply that higher rank line defect indices are recursively determined by lower rank data
across $U(N)$, $USp(2N)$, $SO(2N)$ and $SO(2N+1)$ ADHM-type theories and ABJ(M) theories.
}

\author[a]{Hirotaka Hayashi,}
\emailAdd{h.hayashi@tokai.ac.jp}
\affiliation[a]{Department of Physics, School of Science, Tokai University,\\
4-1-1 Kitakaname, Hiratsuka-shi, Kanagawa 259-1292, Japan}

\author[b,c]{Tomoki Nosaka}
\emailAdd{nosaka@simis.cn}
\affiliation[b]{
Fudan Center for Mathematics and Interdisciplinary Study, Fudan University, Shanghai 200433, China
}
\affiliation[c]{
Shanghai Institute for Mathematics and Interdisciplinary Sciences,\\
Block A, International Innovation Plaza, No.~657 Songhu Road, Yangpu District, Shanghai, China
}
\author[d]{and Tadashi Okazaki}
\emailAdd{tokazaki@seu.edu.cn}
\affiliation[d]{
School of Physics and Shing-Tung Yau Center, Southeast University,\\
Yifu Architecture Building, No.2 Sipailou, Xuanwu district, Nanjing, Jiangsu, 210096, China}

\begin{document}
\maketitle

\ytableausetup{boxsize=1.5mm}

\section{Introduction and summary}
Line defects, such as Wilson lines, are fundamental non-local observables
that probe the phase structure, spectrum, and dualities of quantum field theories.
Specifically, endpoints or junctions of BPS line defects host the BPS local operators.
The line defect index \cite{Dimofte:2011py,Gang:2012yr,Drukker:2012sr,Cordova:2016uwk}
that is the supersymmetric index decorated by an insertion of the BPS line operators counts the BPS local operators sitting at the junction of line operators,
providing a highly refined version of the ordinary supersymmetric index \cite{Kinney:2005ej,Romelsberger:2005eg,Bhattacharya:2008zy}.
This refinement allows us to isolate non-trivial BPS sectors that are usually hidden in the absence of the line operators.
Understanding this refined spectral data is crucial for performing precision tests of three-dimensional dualities and the AdS/CFT correspondence,
where these defects map to fundamental strings or heavy probe branes in the bulk geometry.
In contrast to 4d $\mathcal{N}=4$ supersymmetric gauge theories,
line operator dualities and their holographic realization for 3d SCFTs remain comparatively unexplored
though the line defect indices have already provided non-trivial tests of dualities of line operators with Abelian gauge groups \cite{Drukker:2012sr,Hayashi:2025guk}.

In this paper we study the line defect indices in three-dimensional $\mathcal{N}=4$ supersymmetric gauge theories.
In particular, we investigate their specific fugacity limits,
i.e.~the Coulomb and Higgs limits \cite{Razamat:2014pta}, which are responsible for counting the BPS operators on the Coulomb and Higgs branches respectively.
To extract the intrinsic data of the one-dimensional effective defect theory,
we introduce the \textit{normalized line defect index}:
\begin{align}
\langle\mathcal{W}_{\mathcal{R}_{1}}\cdots\mathcal{W}_{\mathcal{R}_{k}}\rangle^{G(C/H)}(\mathfrak{t})
=\frac{\langle W_{\mathcal{R}_{1}}\cdots W_{\mathcal{R}_{k}}\rangle^{G(C/H)}(\mathfrak{t})}
{\mathcal{I}^{G(C/H)}(\mathfrak{t})}.
\end{align}
Here $\langle W_{\mathcal{R}_{1}}\cdots W_{\mathcal{R}_{k}}\rangle^{G(C/H)}(\mathfrak{t})$
is the line defect Coulomb/Higgs index of the Wilson lines in representation $\mathcal{R}_j$, $j=1,\cdots, k$,
and $\mathcal{I}^{G(C/H)}(\mathfrak{t})$ is the Coulomb/Higgs index,
where $\mathfrak{t}$ is the R-charge fugacity.

While the Coulomb/Higgs indices can be identified with the Hilbert series for the Coulomb/Higgs branch \cite{Razamat:2014pta} and typically exhibit non-negative coefficients,
the line defect indices lead to the frequent appearance of negative coefficients in the Taylor expansion.
Deriving the exact closed-form rational functions for these indices e.g.~via residue calculations are technically daunting and generally lead to prohibitive complexity.

Our first result is the following conjectural formula for the $\mathfrak{t}\rightarrow 1$ limit of the normalized index,
defining the \textit{line defect Witten index}.
The conjecture concerns theories without Chern-Simons terms and is supported by extensive exact computations across a broad class of 3d $\mathcal{N}\ge 4$ supersymmetric gauge theories.
\begin{itemize}
    \item \textbf{Coulomb branch index:}
    The Coulomb line defect index at $\mathfrak{t} \rightarrow 1$ reduces to a contour integral over the maximal torus of the gauge group $G$:
    \begin{align}
        \langle\mathcal{W}_{\mathcal{R}_{1}}\cdots\mathcal{W}_{\mathcal{R}_{k}}\rangle^{G(C)}(1)
        = \oint \prod_{i=1}^{r}\frac{ds_{i}}{2\pi is_{i}} \prod_{j=1}^{k}\chi_{\mathcal{R}_{j}}(s).
    \end{align}
    Physically, this implies that the low energy dynamics on the Coulomb branch projects the defect states onto the gauge invariant sector
    under the unbroken maximal torus $T \subset G$.
    For $k=2$ and $\mathcal{R}_2=\overline{\mathcal{R}_1}$ with $\mathcal{R}_1$ a minuscule representation, this integral elegantly saturates the lower bound given by the dimension of the representation.

    \item \textbf{Higgs branch index:} In the Higgs limit, the Witten index simplifies drastically to the product of the dimensions of the representations:
    \begin{align}
        \langle\mathcal{W}_{\mathcal{R}_{1}}\cdots\mathcal{W}_{\mathcal{R}_{k}}\rangle^{G(H)}(1) = \prod_{j=1}^{k}\text{dim}\,\mathcal{R}_{j}.
    \end{align}
    This is consistent with the physical picture where the gauge group $G$ is completely broken
    by the vacuum expectation values of matter fields, freezing the holonomy to the identity element $e\in G$.

\end{itemize}

Our second main result is the discovery of a universal class of \textit{algebraic recursion relations} for Wilson line defect Coulomb indices
associated with minuscule representations in 3d SCFTs on M2-branes.
Extending the $U(N)$ ADHM case \cite{Hayashi:2024jof}, we establish analogous exact recursive structures for $USp(2N)$ and $SO(2N)$ gauge theories,
and remarkably, for ABJ(M) theories \cite{Aharony:2008ug,Aharony:2008gk}.
Microscopically, the gauge group rank $N$ counts the number of coincident M2-branes.
The exact factorization of the line defect index into products of lower rank indices thus admits a compelling physical interpretation,
i.e.~the geometric decoupling of a sub-stack of branes.
A Wilson line in a minuscule representation, such as the rank-$k$ antisymmetric of $U(N)$ or the fundamental of $USp(2N)$ and $SO(2N)$, acts as a rigid probe.
Mathematically, its weights belong to a single Weyl orbit with unit multiplicity, which induces a precisely uniform shift in the matrix integral.
Physically, this uniform shift due to the line defect peels $k$ (or $1$) M2-branes away from the remaining $N-k$ (or $N-1$) branes in the stack.
Furthermore, for the $SO(2N+1)$ gauge theories, whose fundamental representation fails to be minuscule due to the zero weight,
we discover that the obstruction to factorization is due to the identity contribution
and upon its subtraction the two-point function satisfies an exact recursion relation with the rank-one $SO(3)$ theory.
Crucially, these recursion relations allow us to systematically extract the large $N$ limit,
where the defect observable crystallizes into a simple closed-form expression.
Furthermore, this factorization guarantees that the defect index precisely inherits the palindromic symmetry
($\mathfrak{t} \rightarrow \mathfrak{t}^{-1}$) from the bulk indices,
a hallmark property that inherently breaks down for generic non-minuscule representations.

\subsection{Structure}
The remainder of this paper is organized as follows.
In section \ref{sec_lineind} we introduce the line defect Witten index for 3d $\mathcal{N}=4$ supersymmetric gauge theories,
propose general formulae for its Coulomb and Higgs limits, examine the theories to which neither applies, and explain their physical interpretation.
In section \ref{sec_recursion} we propose recursion relations for line defect Coulomb indices.
In section \ref{sec_largeN} we discuss the large $N$ limit of the line defect Coulomb indices.
Explicit expressions and computational techniques of the line defect indices are deferred to appendices \ref{app_algorithms} and \ref{app_Higgs} to maintain a concise physics oriented presentation in the main text.

\subsection{Future works}

\begin{itemize}

\item While the normalized line defect index is generically formulated as a complicated rational function of fugacity,
our finding ensures that the index consistently evaluates to an integer upon turning off the fugacity.
This structural property strongly suggests that
the seemingly complex closed-form expressions of the normalized indices could potentially be repackaged into elegant, compact formulations.
In fact, such formulae have been found in 4d $\mathcal{N}=4$ super Yang-Mills theory,
where line defect indices and half-indices elegantly reduce to $q$-binomial coefficients or the principal specialization of Schur polynomials \cite{Hatsuda:2025zvi}.
Exploring such expressions for the 3d line defect indices remains an interesting open problem.

\item Mathematically proving our proposed formulae \eqref{Witten_indC_conj_int} and \eqref{Witten_indH_conj} for the line defect Witten indices, and the recursive relations,
directly from the rigorous evaluation of the index formulae, poses an important and highly non-trivial challenge.
In section \ref{sec_globalandCS} we find that averaging the inserted characters over the residual gauge group reproduces the Witten indices of a number of theories with Chern-Simons terms. A general formula is not known, as we discuss in section \ref{sec_validity}, and looking for one, and understanding the physical mechanism behind it, would be an interesting problem.

\item Although in this paper we focus on the factorized recursion relation of the Coulomb line defect indices in the minuscule representations, except the recursion relation for the fundamental representation in $SO(2N+1)$, it would be interesting to investigate the recursion relation for more general setups.
Indeed, factorized recursion relation seems to hold for the one-point function of the rank $2$ (anti)symmetric representation for the $SO(2N+1)$ gauge theory (see appendix \ref{app_SOoddC_moregeneral}) and $USp(2N)$ gauge theory (see appendix \ref{app_USpC_moregeneral}) as well.
In the antisymmetric case of the $USp(2N)$ gauge theory the relation factorizes only after the identity contribution is added, exactly as for the fundamental Wilson lines in the $SO(2N+1)$ gauge theory.
Moreover, as we see in \eqref{R_uN_C_Wsym2} the algebraic recursion relations, although not factorized, also exist at least for the symmetric Wilson lines in $U(N)$ ADHM theories.
Again, as observed in appendices \ref{app_SOoddC_moregeneral} and \ref{app_USpC_moregeneral}, recursion relations of the same structure might hold also for the one-point function of the rank $4$ symmetric representation and the two-point function of the rank $2$ (anti)symmetric representation in the $SO(2N+1)$ gauge theory and the $USp(2N)$ gauge theory.
As found in \cite{Hayashi:2024jof} the recursion relations also exist for the Higgs line defect indices of the antisymmetric Wilson lines in the $U(N)$ ADHM theory with one flavor.
Furthermore, since the Higgs (line defect) indices of the 3d ADHM theories can be related to the instanton partition function of 5d super Yang-Mills theories with some additional hypermultiplets (see e.g.~\cite{Hayashi:2024jof}), it would also be possible to write down the recursion relation for these line defect indices from the Nakajima-Yoshioka blowup relation for the 5d instanton partition functions \cite{Nakajima:2005fg,Kim:2019uqw}.
It would be interesting to investigate such new recursion relations for the Higgs line defect indices and argue their interpretations in the M2-brane SCFTs.

\item It would also be interesting to generalize our analysis of the Coulomb line defect indices of the $U(N)_k\times U(N)_{-k}$ ABJM theory and of the $SO(2N+\gamma)_{2k}\times USp(2N)_{-k}$ ABJ theory ($\gamma=0,1$) to those theories with rank deformation, $U(N_1)_k\times U(N_2)_{-k}$ and $SO(2N_1+\gamma)_{2k}\times USp(2N_2)_{-k}$.
When the rank difference is at most $1$, the monopoles surviving in the Coulomb limit can be classified via the ``clustering pattern'' as explained in appendix \ref{app_OSp_SO2Nchi-1}.
For more general rank difference, however, bare monopole operators of negative dimension appear. The Coulomb limit is well defined only after taking into account that the bare monopoles are not gauge invariant, which requires a new method of analysis.
Some first steps in this direction are taken in this paper.
In appendix \ref{app_ABJrank} the $U(1)_k\times U(1+M)_{-k}$ theories are treated flux sector by flux sector, and in section \ref{sec_rankABJ} the orthosymplectic index \eqref{OSpABJ_index2} is generalized to unequal ranks.
For $M=1$ every bare monopole operator has non-negative dimension and the character average reproduces the Coulomb limit. For $M\ge2$, where the residual gauge group acquires a non-Abelian factor, monopole operators of negative dimension contribute to the line defect indices once the representation is large compared with the Chern-Simons level. Whether and when the character average reproduces the Coulomb limit in that case is an open question.
For the rank deformed orthosymplectic theories we have not evaluated the Coulomb line defect indices, since monopole operators of negative dimension occur there once the ranks differ. When the leftover gauge factor carries massless charged matter, as happens for $\gamma=1$ with $N_2>N_1$, that matter may have to be included in the character average, as for the corresponding interfaces of four dimensional gauge theories.
It may also be significant that the Coulomb index without line defects came out independent of the rank difference in every case we computed for the unitary theories, and for the orthosymplectic ones with $N_2>N_1$.
What is still missing is an analytic treatment for general ranks, in which the grand canonical sum factorizes and closed form expressions are available, as in appendices \ref{ABJMalgorithm} and \ref{app_OSpgammachinot0-1}.
The building blocks of that sum admit a free fermion form not only for the unitary theories but also for the orthosymplectic theories without rank deformation.
This structure also holds for theories with the difference of the effective ranks being $1$ as we demonstrated in $SO(2N)^{(-\zeta)}_{2k}\times USp(2N)_{-k}$ ABJ theory in appendix \ref{app_OSp_SO2Nchi-1}, which is what one would want to generalize further.
We expect the recursion relations to be the natural organizing principle here.
The factorized relations \eqref{R_uNuNABJM_W} and \eqref{SO2Nchi+recreln} for the ABJM and ABJ theories should admit counterparts for the rank deformed theories, in which the rank difference enters as an additional parameter, and understanding these relations would be the most efficient route to the closed form expressions and to the $\mathfrak{t}\rightarrow1$ limits of these theories.

\item It is a compelling future problem to investigate the giant graviton expansion for the M2-brane line defect indices.
Recently, the giant graviton expansion \cite{Arai:2020qaj,Gaiotto:2021xce} has been successfully extended to line defect indices
and half-indices in four-dimensional $\mathcal{N}=4$ super Yang-Mills theory \cite{Imamura:2024lkw,Imamura:2024pgp,Hatsuda:2024uwt,Beccaria:2024lbt,Beccaria:2024oif,Imamura:2024zvw,Imamura:2025fqa}.
The algebraic factorizations that allow us to find the closed-form expressions in our work provide crucial clues for synthesizing these issues.
Establishing the giant graviton expansion for our M2-brane defect indices
would provide profound insights into the finite $N$ holography involving wrapped branes in M-theory.

\item Mirror symmetry \cite{Intriligator:1996ex} exchanges the Coulomb and Higgs branches
and maps the Wilson lines to the vortex line defects \cite{Assel:2015oxa,Dimofte:2019zzj}.
At the level of the line defect indices, the mirror map acts as $t\rightarrow t^{-1}$ together with the exchange of the flavor and topological fugacities
and therefore interchanges the Coulomb and Higgs limits.
Consequently, our two conjectures (\ref{Witten_indC_conj_int}) and (\ref{Witten_indH_conj})
are mirror images of a pair of predictions for the disorder type line defects.
The \textit{Higgs} limit of the vortex line defect Witten index of a theory $T$ is computed by the character integral of the mirror gauge group of $T^{\vee}$,
while its \textit{Coulomb} limit is governed by the dimension formula.
For Abelian gauge theories these statements can be verified exactly.
The dualities of the Wilson, vortex, and mixed vortex-Wilson lines for the SQED, ADHM and Abelian quiver CS theories were established
in \cite{Hayashi:2025guk} through matching pairs of line defect indices, where the identities were proven in the Coulomb and Higgs limits.
Extending these statements to the non-Abelian vortex lines, whose mirror duals
of non-Abelian Wilson lines generically involve couplings to one-dimensional
$\mathcal{N}=4$ quantum mechanics or sums of vortex lines \cite{Assel:2015oxa,Dimofte:2019zzj},
is an interesting problem which we leave for future work.

\end{itemize}

\section{Line defect Witten indices}
\label{sec_lineind}
The supersymmetric index can be enriched by line defect operators.
According to the supersymmetric localization \cite{Kim:2009wb,Imamura:2011su,Kapustin:2011jm},
the index of 3d $\mathcal{N}=4$ gauge theory with gauge group $G$ and hypermultiplets
transforming in a representation $\mathbf{R}$ can be expressed as
a discrete sum over GNO magnetic fluxes combined with a continuous matrix integral over the gauge holonomies.
When we further introduce the half-BPS Wilson lines $W_{\mathcal{R}_j}$, $j=1,\cdots, k$ wrapping the temporal circle,
the line defect index takes the form\footnote{We use the definition in \cite{Okazaki:2019ony}. Also see \cite{Hayashi:2022ldo,Hayashi:2025guk}. }
\begin{align}
&
\langle W_{\mathcal{R}_1}\cdots W_{\mathcal{R}_k}\rangle^{G}(t, \mathbf{x}, \mathbf{z}; q)
\nonumber\\
&= \sum_{\mathbf{m} \in \Gamma_{\mathfrak{g}}}
\frac{1}{|\mathscr{W}|}
\oint_{T} \prod_{i=1}^{\text{rk}(G)} \frac{ds_i}{2\pi i s_i}
\prod_{\alpha \in \Delta_{\mathfrak{g}}}
\left(1-q^{\frac{|\alpha(\mathbf{m})|}{2}} s^\alpha\right)
\frac{\left(q^{\frac{1}{2}+\frac{|\alpha(\mathbf{m})|}{2}} t^2 s^{\alpha}; q\right)_{\infty}}
{\left(q^{\frac{1}{2}+\frac{|\alpha(\mathbf{m})|}{2}} t^{-2} s^\alpha; q\right)_{\infty}}
\nonumber\\
&\quad \times \prod_{\rho \in \mathbf{R}}
\prod_{w \in \mathbf{R}_F}
\frac{\left(q^{\frac{3}{4}+\frac{|\rho(\mathbf{m})|}{2}} t^{-1} s^{-\rho} \mathbf{x}^{-w}; q\right)_{\infty}}
{\left(q^{\frac{1}{4}+\frac{|\rho(\mathbf{m})|}{2}} t s^\rho \mathbf{x}^w; q\right)_{\infty}}
q^{\frac{\Delta(\mathbf{m})}{4}}t^{-\Delta(\mathbf{m})}
\mathbf{z}^{\mathbf{m}}
\prod_{j=1}^{k}
\chi_{\mathcal{R}_j}(s),
\end{align}
where the summation is taken over the GNO magnetic fluxes $\mathbf{m}$ in the cocharacter lattice $\Gamma_{\mathfrak{g}}$ of $G$ and $\mathscr{W}$ denotes the Weyl group.
The integration is performed over the gauge fugacities $s = (s_1, \dots, s_{\text{rk}(G)})$ parameterizing the maximal torus $T \subset G$.
The term $q^{\frac{\Delta(\mathbf{m})}{4}}t^{-\Delta(\mathbf{m})} \mathbf{z}^{\mathbf{m}}$ captures the contributions from the bare monopoles.
The classical insertion of the Wilson lines simply amounts to the product of characters $\chi_{\mathcal{R}_j}(s)$ evaluated on the maximal torus.
Furthermore $\Delta_{\mathfrak{g}}$ represents the set of roots of $G$ comprising the vector multiplet contributions,
while $\rho$ and $w$ denote the weights of the hypermultiplet representation under the gauge group $G$ and the flavor symmetry group $G_F$
with fugacities $\mathbf{x}$ respectively.
The ordinary full supersymmetric index $\mathcal{I}^G$ is smoothly recovered
by taking the trivial representations for all $\mathcal{R}_j$, yielding the empty defect insertion $\prod \chi = 1$.

To isolate the spectrum of BPS local operators bound to the line defects,
the full supersymmetric index can be specialized to the Coulomb and Higgs limits \cite{Razamat:2014pta}.
In these limits, the full index in the absence of the line operators
exactly reduces to the Hilbert series enumerating the half-BPS operators in the respective chiral rings.
We define the Coulomb and Higgs limits of the Wilson line defect index as
\begin{align}
\langle W_{\mathcal{R}_1} \cdots W_{\mathcal{R}_k}\rangle^{G(C)}(z;\mathfrak{t})
&:=
\lim_{\substack{q\rightarrow 0 \\ \mathfrak{t}=q^{\frac{1}{4}}t^{-1}:\,\text{fixed}}}
\langle W_{\mathcal{R}_1}\cdots W_{\mathcal{R}_k}\rangle^{G}(t, \mathbf{x}, \mathbf{z}; q),
\\
\langle W_{\mathcal{R}_1} \cdots W_{\mathcal{R}_k}\rangle^{G(H)}(z;\mathfrak{t})
&:=
\lim_{\substack{q\rightarrow 0 \\ \mathfrak{t}=q^{\frac{1}{4}}t:\,\text{fixed}}}
\langle W_{\mathcal{R}_1}\cdots W_{\mathcal{R}_k}\rangle^{G}(t, \mathbf{x}, \mathbf{z}; q).\label{Higgs}
\end{align}
These specialized correlators serve as the precise Hilbert series for the 1d line operator algebra, vastly generalizing the well-studied bulk Hilbert series \cite{Benvenuti:2006qr,Benvenuti:2010pq,Hanany:2011db,Cremonesi:2013lqa,Mekareeya:2015bla,Hayashi:2022ldo} to the line defect sector
(see e.g.~\cite{Dimofte:2019zzj,Crew:2020psc,Nawata:2021nse,Crew:2024fom,Hayashi:2024jof,Hayashi:2025guk}).
In the following, for notational simplicity, we suppress the flavor fugacity $z$
whenever it is set to unity or absent, and regard the index as a function solely of the R-charge fugacity $\mathfrak{t}$.

The index that counts the genuine one-dimensional degrees of freedom should be identified with the normalized line defect index,
namely the ratio of the index in the presence of the line insertion to the index without the line defect.
We define the normalized Coulomb/Higgs line defect indices by
\begin{align}
\langle \mathcal{W}_{\mathcal{R}_1}\cdots \mathcal{W}_{\mathcal{R}_k}\rangle^{G(C/H)}(\mathfrak{t})
&=\frac{\langle W_{\mathcal{R}_1}\cdots W_{\mathcal{R}_k}\rangle^{G(C/H)} (\mathfrak{t})}
{\mathcal{I}^{G(C/H)}(\mathfrak{t})},
\end{align}
where $\langle W_{\mathcal{R}_1}\cdots W_{\mathcal{R}_k}\rangle^{G(C/H)}$ is the Coulomb/Higgs line defect index of the
Wilson lines $W_{\mathcal{R}_j}$ transforming in the representation $\mathcal{R}_{j}$ of $G$ gauge theory
and $\mathcal{I}^{G(C/H)}$ is the Coulomb/Higgs index of $G$ gauge theory.

The ordinary Coulomb/Higgs indices as well as the line defect indices generally exhibit divergences as the fugacity $\mathfrak{t}$ goes to unity.
However, the normalized line defect indices admit a finite and well-defined limit.
Interpreted from the perspective of the effective one-dimensional defect theory,
this quantity naturally plays the role of the Witten index, and we will call it the \textit{line defect Witten index}.
The ordinary line defect index is expressed directly as a matrix integral.
By contrast, the line defect Witten index arises from the $\mathfrak{t}\rightarrow 1$ limit of a ratio of such integrals,
with cancellations of the singularities occurring at $\mathfrak{t}=1$.
Consequently, deriving exact closed-form formulae appears considerably more non-trivial from the viewpoint of matrix integrals
while one naturally expects these quantities to take integer values from the physical perspective.
Motivated by explicit evaluations of Coulomb/Higgs indices and line defect indices,
we propose several conjectures for the closed-form formulae of the line defect Witten indices in the following subsections.
We mainly consider rather general results in this section and the explicit computational results of the Coulomb/Higgs line defect Witten indices for some examples are relegated to appendix \ref{app_algorithms} and \ref{app_Higgs} respectively.
The examples cover ADHM-type theories, $\mathcal{N}=4$ SQCD and Chern-Simons matter theories, so that the conclusions below are not tied to one class of matter content but hold across a broad range of 3d $\mathcal{N}\ge4$ theories.

\subsection{Coulomb indices}
\label{sec_Coulombindices}
For the Coulomb line defect Witten index of the Wilson lines $W_{\mathcal{R}_j}$, $j=1,\cdots, k$,
transforming in the representation $\mathcal{R}_j$ of gauge group $G$,
we propose the following conjecture:
\begin{align}
\label{Witten_indC_conj_int}
\langle \mathcal{W}_{\mathcal{R}_1}\cdots \mathcal{W}_{\mathcal{R}_k}\rangle^{G (C)}(1)
&=\oint \prod_{i=1}^{r} \frac{ds_i}{2\pi is_i}
\prod_{j=1}^{k} \chi_{\mathcal{R}_j} (s),
\end{align}
motivated by the explicit results in appendix \ref{app_algorithms} (as well as other unlisted results).
Here $\chi_{\mathcal{R}}(s)$ is the character of the representation $\mathcal{R}$ of $G$ and $r$ is the rank of $G$.
From the viewpoint of the matrix model expression, the simplicity of our conjecture \eqref{Witten_indC_conj_int} is highly non-trivial.
Generically, the normalized line defect index is defined as a ratio of intricate matrix integrals.
The physical motivation behind this character integral formula deeply originates from the structure of the residual gauge group on the Coulomb branch.
A generic point on the Coulomb branch preserves the maximal torus $T \subset G$.
The zero modes parameterizing the maximal torus $T$ on the Coulomb branch result in the integration over the torus coordinates $s_i$, which acts as a projection onto the gauge invariant sector of the low-energy effective theory.

The argument applies to any good (in the sense of Gaiotto-Witten \cite{Gaiotto:2008ak}) three dimensional $\mathcal{N}=4$ gauge theory without Chern-Simons terms, whatever its matter content.
Indeed, in the Coulomb limit the hypermultiplets drop out of the integrand and survive only through the dimensions of the bare monopole operators, so that the line defect index becomes the monopole sum \eqref{monopoleformula}, in which the magnetic flux $\mathbf{m}$ enters through an overall power of $\mathfrak{t}$ and through the subgroup $H_{\mathbf{m}}\subseteq G$ that it leaves unbroken.
As $\mathfrak{t}\rightarrow1$ the sum is dominated by the generic fluxes, for which the unbroken subgroup $H_{\mathbf{m}}\subseteq G$ of \eqref{monopoleformula} is the maximal torus $T$ and the integrand reduces to $(1-\mathfrak{t}^2)^{-r}\prod_j\chi_{\mathcal{R}_j}(s)$, so that the divergent prefactor cancels in the normalized index and \eqref{Witten_indC_conj_int} follows.
The matter content controls which fluxes dominate but not the surviving integrand, which is why \eqref{Witten_indC_conj_int} does not refer to it.
We have verified this both for the ADHM-type theories of appendices \ref{app_SO2N+1algorithm}-\ref{app_USp}, which always contain a hypermultiplet in the adjoint or in the (anti)symmetric representation, and for $\mathcal{N}=4$ SQCD, whose only matter consists of fundamental hypermultiplets, in appendix \ref{app_C_SQCD}.

For a finite dimensional representation $\mathcal{R}$ the character $\chi_{\mathcal{R}}(s)$ can be written as
\begin{align}
\label{ch_decomp}
\chi_{\mathcal{R}}(s)
&=\sum_{\mu=(\mu_1,\cdots,\mu_r) \in \Lambda}m_{\mathcal{R}}(\mu)s_1^{\mu_1}\cdots s_r^{\mu_r},
\end{align}
where $\Lambda$ is the weight lattice of $G$ and $\mu$ runs over the weights in $\mathcal{R}$.
The expansion coefficient $m_{\mathcal{R}}(\mu)$ $\in \mathbb{Z}_{\ge0}$ is the multiplicity of the weight $\mu$.
Plugging the weight decomposition (\ref{ch_decomp}) of the character into the integral (\ref{Witten_indC_conj_int}), we obtain
\begin{align}
\langle \mathcal{W}_{\mathcal{R}_1}\cdots \mathcal{W}_{\mathcal{R}_k}\rangle^{G (C)}(1)
&=\sum_{\mu^{(1)}_i+\cdots+\mu^{(k)}_i=0}\prod_{j=1}^{k}m_{\mathcal{R}_j}(\mu^{(j)}).
\end{align}
In particular, the line defect Witten index derived from the two-point functions of Wilson lines
transforming in a pair of conjugate representations $\mathcal{R}$ and $\overline{\mathcal{R}}$ is
\begin{align}
\label{Witten_indC_2pt}
\langle \mathcal{W}_{\mathcal{R}} \mathcal{W}_{\overline{\mathcal{R}}}\rangle^{G (C)}(1)
&=\sum_{\mu}m_{\mathcal{R}}(\mu)^2.
\end{align}
This formula admits a natural combinatorial interpretation.
For $G=U(N)$, the weight multiplicity $m_{\mathcal{R}_\lambda}(\mu)$ of the irreducible
representation labeled by a partition $\lambda$ coincides with the Kostka number $K_{\lambda\mu}$,
i.e.~the character (Schur polynomial) admits the weight expansion
\begin{align}
\chi_\lambda(s) = s_\lambda(s_1,\dots,s_N) = \sum_{\mu} K_{\lambda\mu}\, s^{\mu},
\end{align}
where $K_{\lambda\mu}$ counts the number of semistandard Young tableaux of shape $\lambda$
and content $\mu$. Consequently, the two-point line defect Witten index~\eqref{Witten_indC_2pt}
for $G=U(N)$ is written combinatorially as
\begin{align}
\label{Witten_indC_2ptuN}
\langle \mathcal{W}_{\mathcal{R}_\lambda} \mathcal{W}_{\overline{\mathcal{R}_\lambda}}\rangle^{U(N)(C)}(1)
&=\sum_{\mu} K_{\lambda\mu}^2.
\end{align}

This combinatorial structure generalizes naturally to the other classical gauge groups
$G=USp(2N)$, $SO(2N+1)$, and $SO(2N)$, although the precise combinatorial rules differ among the three cases.
For these groups the characters $\chi_\lambda(s)$ are no longer
Schur polynomials but their symplectic and orthogonal analogues, given by the Weyl character
formula for type $C_N$, $B_N$, and $D_N$ respectively. Their expansion in the weight basis
defines three \textit{a priori distinct} families of generalized Kostka numbers,
\begin{align}
\label{BCDch_exp}
\chi_\lambda^{\mathfrak{usp}(2N)}(s) = \sum_{\mu} K^{C}_{\lambda\mu}\, s^{\mu}, \qquad
\chi_\lambda^{\mathfrak{so}(2N+1)}(s) = \sum_{\mu} K^{B}_{\lambda\mu}\, s^{\mu}, \qquad
\chi_\lambda^{\mathfrak{so}(2N)}(s) = \sum_{\mu} K^{D}_{\lambda\mu}\, s^{\mu},
\end{align}
introduced by King \cite{MR480895,MR552445,MR725604}.
Each of $K^{C}_{\lambda\mu}$, $K^{B}_{\lambda\mu}$, and $K^{D}_{\lambda\mu}$ admits a
combinatorial characterization in terms of semistandard tableaux of shape $\lambda$
and content $\mu$, but the underlying alphabets and filling rules are genuinely different:
\begin{itemize}
\item[$C_N$:] entries are drawn from the ordered alphabet
$1<\bar1<2<\bar2<\cdots<N<\bar N$, with the extra condition that every entry in row $i$
is $\ge i$ (symplectic tableaux);
\item[$B_N$:] entries are drawn from the same alphabet with an additional middle letter
$0$ inserted, $1<\bar1<\cdots<N<\bar N<0$ (or an equivalent ordering depending on convention),
subject to a parallel row-bound condition (odd orthogonal tableaux);
\item[$D_N$:] entries are drawn from the alphabet
$1<\bar1<2<\bar2<\cdots<N<\bar N$ without the middle letter $0$, but subject to a
\textit{different} filling rule along the last column, and, when $\lambda_N\neq0$,
an extra label distinguishing the two inequivalent spin representations related by the outer automorphism of $D_N$ (even orthogonal tableaux).
\end{itemize}

Accordingly, for $G=USp(2N)$, $SO(2N+1)$ and $SO(2N)$ we have
\begin{subequations}
\begin{align}
\label{Witten_indC_2pt_BCD}
\langle \mathcal{W}_{\mathcal{R}_\lambda} \mathcal{W}_{\mathcal{R}_\lambda}\rangle^{USp(2N)(C)}(1)
&=\sum_{\mu} \big(K^{C}_{\lambda\mu}\big)^2, \\
\langle \mathcal{W}_{\mathcal{R}_\lambda} \mathcal{W}_{\mathcal{R}_\lambda}\rangle^{SO(2N+1)(C)}(1)
&=\sum_{\mu} \big(K^{B}_{\lambda\mu}\big)^2, \\
\langle \mathcal{W}_{\mathcal{R}_\lambda} \mathcal{W}_{\mathcal{R}_\lambda}\rangle^{SO(2N)(C)}(1)
&=\sum_{\mu} \big(K^{D}_{\lambda\mu}\big)^2.
\end{align}
\end{subequations}
using self-conjugacy $\overline{\mathcal{R}_\lambda}=\mathcal{R}_\lambda$ for $USp(2N)$,
$SO(2N+1)$, and for $SO(2N)$ whenever $\lambda_N=0$.
This is a direct consequence of (\ref{Witten_indC_2pt}) together with the weight expansions (\ref{BCDch_exp})
of the symplectic/orthogonal character, exactly as in the $U(N)$ derivation.

On the other hand, the dimension of the representation $\mathcal{R}$ is
\begin{align}
\dim \mathcal{R}&=\sum_{\mu} m_{\mathcal{R}}(\mu).
\end{align}
Since weight multiplicities $m_{\mathcal{R}}(\mu)$  are non-negative integers,
their squares $m_{\mathcal{R}}(\mu)^2$ are naturally greater than or equal to their linear values.
This simple algebraic fact dictates that the Coulomb index is always bounded from below by the representation's dimension
\begin{align}
\langle \mathcal{W}_{\mathcal{R}} \mathcal{W}_{\overline{\mathcal{R}}}\rangle^{G (C)}(1)
&\ge \dim \mathcal{R}.
\end{align}
Recall that a dominant weight $\lambda_m$ is called \textit{minuscule} \cite{MR2109105}
if the Weyl group $\mathscr{W}$ acts transitively on the weights of the irreducible representation $\mathcal{R}_{\lambda_m}$ with highest weight $\lambda_m$,
or equivalently if $\langle \lambda_m, \alpha^{\vee}\rangle \in \{0,\pm1\}$ for every root $\alpha$.
In this case all weights lie in the single Weyl orbit $\mathscr{W}\cdot\lambda_m$ and occur with multiplicity one.
When the representation $\mathcal{R}$ is associated with the minuscule weight $\lambda_{m}$,
which we denote by $\mathcal{R}_{\lambda_m}$,
the set of weights form a single Weyl orbit and the character can be written as
\begin{align}
\chi_{\mathcal{R}_{\lambda_m}}(s)&=\sum_{\mu\in \mathscr{W}\cdot \lambda_m}s^{\mu},
\end{align}
where every weight occurs exactly with multiplicity one.
Accordingly, the Witten index for the two-point function coincides with the dimension of the representation $\mathcal{R}_{\lambda_m}$
\begin{align}
\langle \mathcal{W}_{\mathcal{R}_{\lambda_m}} \mathcal{W}_{\overline{\mathcal{R}_{\lambda_m}}}\rangle^{G (C)}(1)
&=\dim \mathcal{R}_{\lambda_m}.
\end{align}
One important remark is that this relation will always hold for minuscule representations $\mathcal{R}_{\lambda_m}$,
but the converse is not true.
Namely, even if the two-point function associated with the representation $\mathcal{R}_{\lambda}$ labeled by $\lambda$
coincides with $\dim \mathcal{R}_{\lambda}$,
this does not necessarily imply that the representation $\mathcal{R}_{\lambda}$ is minuscule.

\paragraph{Symmetric representations}
The case of a single-row Young diagram $\lambda=(m)$ gives a particularly transparent
illustration of the line defect Witten indices and the generalized Kostka numbers,
since the relevant characters admit simple closed forms in terms of the $j$-th complete homogeneous symmetric function
\begin{align}
h_j(x_1,x_1^{-1},\dots,x_N,x_N^{-1})
&=\sum_{a_i,b_i\ge0,\,\sum(a_i+b_i)=j}
\prod_i x_i^{a_i-b_i}.
\end{align}
For $USp(2N)$ the symmetric power $\mathrm{Sym}^m(V)$ of the fundamental $V$ is already irreducible.
There is no symplectic-trace to subtract since the invariant form on $V$ is antisymmetric.
Hence
\begin{align}
\chi^{\mathfrak{usp}(2N)}_{(m)}(s) = h_m(s_1,s_1^{-1},\dots,s_N,s_N^{-1}).
\end{align}
For $SO(2N+1)$ and $SO(2N)$, by contrast, $\mathrm{Sym}^m(V)$ decomposes into traceless
pieces, $\mathrm{Sym}^m(V)=\bigoplus_{k=0}^{\lfloor m/2\rfloor}\mathcal{R}_{(m-2k)}$, so that
the one-row representation is obtained by subtracting one trace shell,
\begin{subequations}
\begin{align}
\chi^{\mathfrak{so}(2N+1)}_{(m)}(s) &= h_m(s_1,s_1^{-1},\dots,s_N,s_N^{-1},1) - h_{m-2}(s_1,s_1^{-1},\dots,s_N,s_N^{-1},1), \\
\chi^{\mathfrak{so}(2N)}_{(m)}(s) &= h_m(s_1,s_1^{-1},\dots,s_N,s_N^{-1}) - h_{m-2}(s_1,s_1^{-1},\dots,s_N,s_N^{-1}),
\end{align}
\end{subequations}
where the extra variable $1$ for $SO(2N+1)$ accounts for the additional zero weight
in the $(2N+1)$-dimensional fundamental representation.
Writing $a_i+b_i=|\mu_i|+2c_i$ and summing over the resulting simplex yields the closed-form weight multiplicities,
valid for $m\ge|\mu|:=\sum_i|\mu_i|$,
\begin{subequations}
\begin{align}
\label{onerow_mult}
K^{C}_{(m),\mu} &= \binom{N-1+\tfrac{m-|\mu|}{2}}{N-1}, &
K^{D}_{(m),\mu} &= \binom{N-2+\tfrac{m-|\mu|}{2}}{N-2}, &
(m\equiv|\mu|\!\!\!\!\mod 2), \\
K^{B}_{(m),\mu} &= \binom{N-1+\big\lfloor\tfrac{m-|\mu|}{2}\big\rfloor}{N-1}, & & &
\end{align}
\end{subequations}
and zero otherwise. Unlike the $U(N)$ case, where $K_{(m),\mu}=1$ for every weight of
$\mathrm{Sym}^m(\mathbb{C}^N)$ (the Schur polynomial $h_m$ being multiplicity-free),
these generalized Kostka numbers grow with $N$ and generically exceed one already at $m=2$.

As an explicit check, consider $USp(4)$ ($N=2$) with $m=2$. The weight $\mu=(0,0)$ has
$K^{C}_{\ydiagram{2},(0,0)}=\binom{2}{1}=2$, while each of the eight weights with $|\mu|=2$ has
multiplicity $K^{C}_{(2),\mu}=\binom{1}{1}=1$. Summing over all ten weights reproduces
$\dim\mathcal{R}_{\ydiagram{2}}=10$, while
\begin{align}
\langle \mathcal{W}_{\ydiagram{2}} \mathcal{W}_{\ydiagram{2}}\rangle^{USp(4)(C)}(1)
&= \sum_\mu \big(K^C_{\ydiagram{2},\mu}\big)^2
\nonumber\\
&= 2^2 + 8\times1^2
= 12>10 = \dim\mathcal{R}_{\ydiagram{2}},
\end{align}
strictly exceeding the dimension, in contrast to the minuscule fundamental representation discussed above.

The same computation for $SO(5)$ ($N=2$, $B_2$) gives $\dim\mathcal{R}_{\ydiagram{2}}=14$ and
\begin{align}
\langle \mathcal{W}_{\ydiagram{2}}\mathcal{W}_{\ydiagram{2}}\rangle^{SO(5)(C)}(1)=2^2+4\times1^2+8\times1^2
=16>14=\dim\mathcal{R}_{\ydiagram{2}}.
\end{align}
For $SO(4)$ ($N=2$, $D_2$) the multiplicities are all equal to one since $K^{D}_{(m),\mu}=\binom{d}{0}=1$ identically at rank $N=2$, giving
$\dim\mathcal{R}_{\ydiagram{2}}=9$ and
\begin{align}
\langle \mathcal{W}_{\ydiagram{2}}\mathcal{W}_{\ydiagram{2}}\rangle^{SO(4)(C)}(1)=9=\dim\mathcal{R}_{\ydiagram{2}},
\end{align}
saturating the bound exactly. This saturation is special to $D_2\cong A_1\times A_1$, and
generic multiplicities greater than one reappear for $D_N$ with $N\ge3$.

\paragraph{Antisymmetric representations}
Antisymmetric representations exhibit a role reversal complementary to the symmetric case above,
reflecting the fact that the invariant bilinear form is antisymmetric for $USp(2N)$ but symmetric for $SO(N)$:

\begin{itemize}
\item For $USp(2N)$ the symplectic form $\omega\in\Lambda^2 V^*$ provides
a contraction $\Lambda^m V\to\Lambda^{m-2}V$ so that $\Lambda^m V$ is \textit{not} irreducible.
It decomposes as $\Lambda^m V = \mathcal{R}_{(1^m)}\oplus \mathcal{R}_{(1^{m-2})}\oplus\cdots$,
and the one-column representation requires subtracting one trace shell,
\begin{align}
\chi^{\mathfrak{usp}(2N)}_{(1^m)}(s) = e_m(s_1,s_1^{-1},\dots,s_N,s_N^{-1}) - e_{m-2}(s_1,s_1^{-1},\dots,s_N,s_N^{-1}),
\end{align}
where $e_j$ is the degree $j$ elementary symmetric polynomial.

\item For $SO(2N+1)$ and $SO(2N)$ the metric $g\in\mathrm{Sym}^2 V^*$
cannot contract an antisymmetric tensor, so $\Lambda^m V$ is already irreducible, with
\begin{align}
\chi^{\mathfrak{so}(2N+1)}_{(1^m)}(s) = e_m(s_1,s_1^{-1},\dots,s_N,s_N^{-1},1), \qquad
\chi^{\mathfrak{so}(2N)}_{(1^m)}(s) = e_m(s_1,s_1^{-1},\dots,s_N,s_N^{-1}),
\end{align}
with no trace subtraction required. This irreducibility holds for $1\le m\le N-1$ in both cases,
and extends to $m=N$ for $SO(2N+1)$, where $\Lambda^N V$ is irreducible with highest weight $2\omega_N$ (distinct from the spin representation $\omega_N$ itself).
For $SO(2N)$, by contrast, $m=N$ is exceptional, since $\Lambda^N V$ splits into a pair of self-dual/anti-self-dual pieces
related by the outer automorphism, in parallel with the spin representations, and requires separate treatment.
\end{itemize}

This is the mirror image of the symmetric case, where $SO(N)$ required trace subtraction while
$USp(2N)$ did not. Here $USp(2N)$ requires trace subtraction while $SO(N)$ does not. Indeed,
at $m=2$ this dichotomy is precisely the statement that the adjoint representation is
$\mathrm{Sym}^2 V$ for $USp(2N)$ but $\Lambda^2 V$ for $SO(N)$,
so that the $B_N$/$D_N$ adjoint computation carried out above is simultaneously the $m=2$ special case of the present antisymmetric power discussion,
with no independent computation needed.

For $USp(2N)$ and $SO(2N)$, whose characters are built from the same $2N$ paired variables
$s_i^{\pm1}$ (with no unpaired variable), expanding $e_j$ as a sum over degree $j$ square free monomials gives the weight multiplicity,
for a weight $\mu$ with entries in $\{-1,0,1\}$ and support size $s=\sum_i|\mu_i|$,
\begin{align}
[e_j]^{C,D}_\mu = \binom{N-s}{(j-s)/2}, \qquad j\equiv s \!\!\!\mod 2,
\end{align}
and zero when $j\not\equiv s\!\!\mod2$, so that, e.g.~for $USp(2N)$,
\begin{align}
K^{C}_{(1^m),\mu} = \binom{N-s}{\tfrac{m-s}{2}} - \binom{N-s}{\tfrac{m-s}{2}-1}.
\end{align}
For $SO(2N+1)$, the extra unpaired variable $1$ in $\chi^{\mathfrak{so}(2N+1)}_{(1^m)}(s)$ removes the parity constraint.
Writing the generating function per index as $(1+x^2)$ for the $N-s$ vanishing
weights and $(1+x)$ for the extra slot, one finds instead
\begin{align}
K^{B}_{(1^m),\mu} = [e_m]^{B}_\mu = \binom{N-s}{\big\lfloor \tfrac{m-s}{2}\big\rfloor},
\end{align}
valid for any parity of $m-s$, with no trace subtraction needed since $\Lambda^m V$ is already
irreducible for $SO(2N+1)$.

As an explicit illustration, let us take $m=2$ and the rank-$2$ antisymmetric representation.
For $USp(4)$ with $N=2$, the weight $\mu=(0,0)$ has multiplicity $K^C_{\ydiagram{1,1},(0,0)}=\binom{2}{1}-\binom{2}{0}=1$,
while the four weights $(\pm1,\pm1)$ each have multiplicity $\binom{0}{0}=1$.
Summing gives
$\dim\mathcal{R}_{\ydiagram{1,1}}=5$ and
\begin{align}
\langle \mathcal{W}_{\ydiagram{1,1}}\mathcal{W}_{\ydiagram{1,1}}\rangle^{USp(4)(C)}(1)
&= 4\times1^2+1^2 = 5 = \dim\mathcal{R}_{\ydiagram{1,1}},
\end{align}
saturating the bound.
This saturation, however, is special to $N=2$.
For $USp(6)$ with $N=3$ the zero weight instead has multiplicity $K^{C}_{\ydiagram{1,1},(0,0,0)}=\binom{3}{1}-\binom{3}{0}=2$,
while the twelve weights of support two each retain multiplicity one, giving
$\dim\mathcal{R}_{\ydiagram{1,1}}=12+2=14$ but
\begin{align}
\langle \mathcal{W}_{\ydiagram{1,1}}\mathcal{W}_{\ydiagram{1,1}}\rangle^{USp(6)(C)}(1)
&=12\times1^2+2^2 = 16 > 14 = \dim\mathcal{R}_{\ydiagram{1,1}},
\end{align}
so that, unlike the $N=2$ case, the bound is no longer saturated and the inequality is already strict at rank $N=3$.

The orthogonal groups exhibit the same qualitative growth of multiplicities despite the absence of trace subtraction.
Note that irreducibility of the representation does \textit{not} imply that its weight multiplicities are trivial.
Already at $m=2$ (the adjoint), the zero weight carries multiplicity $N$, as computed above
\begin{align}
\sum_\mu(K^B_{\ydiagram{1,1},\mu})^2=3N^2
\end{align}
for $B_N$ and
\begin{align}
\sum_\mu(K^D_{\ydiagram{1,1},\mu})^2=3N^2-2N
\end{align}
for $D_N$.
We illustrate the growth at higher $m$ with two further examples.

For $B_3=\mathfrak{so}(7)$ ($N=3$) and $m=3$, using the floor formula $K^B_{\ydiagram{1,1,1},\mu}=\binom{3-s}{\lfloor(3-s)/2\rfloor}$ gives
\begin{align}
K^{B}_{\ydiagram{1,1,1},\mu} =
\begin{cases}
1, & \text{support}(\mu)=3 \quad (8\text{ weights}), \\
1, & \text{support}(\mu)=2 \quad (12\text{ weights}), \\
2, & \text{support}(\mu)=1 \quad (6\text{ weights}), \\
3, & \mu=0 \quad (1\text{ weight}),
\end{cases}
\end{align}
which correctly reproduces $\dim\mathcal{R}_{\ydiagram{1,1,1}}=8+12+12+3=35=\binom{7}{3}$.
The line defect Witten index is
\begin{align}
\langle \mathcal{W}_{\ydiagram{1,1,1}}\mathcal{W}_{\ydiagram{1,1,1}}\rangle^{SO(7)(C)}(1)
&= 8\times1^2+12\times1^2+6\times2^2+1\times3^2
\nonumber\\
&= 8+12+24+9=53 >35=\dim\mathcal{R}_{\ydiagram{1,1,1}},
\end{align}
a substantially stronger violation of saturation than at $m=2$.

For $D_5=\mathfrak{so}(10)$ ($N=5$) and $m=3<N$, the parity-constrained formula $[e_j]^{C,D}_\mu$ gives
weight multiplicities $K^{D}_{\ydiagram{1,1,1},\mu}=1$ for the $80$ weights of support $3$,
and $K^{D}_{\ydiagram{1,1,1},\mu}=4$ for the $10$ weights of support $1$, giving
$\dim\mathcal{R}_{\ydiagram{1,1,1}}=80+40=120=\binom{10}{3}$
(as expected, since $\Lambda^3$ of the vector representation has the same dimension as for $GL(10)$ when $m<N$),
while
\begin{align}
\langle \mathcal{W}_{\ydiagram{1,1,1}}\mathcal{W}_{\ydiagram{1,1,1}}\rangle^{SO(10)(C)}(1)
= 80\times1^2+10\times4^2 = 80+160=240 = 2\times120,
\end{align}
exactly twice the dimension.

Finally, for $U(N)$ the antisymmetric representation is irreducible and minuscule with character
\begin{align}
\chi_{(1^m)}(s)=e_m(s_1,\dots,s_N)
\end{align}
and every weight multiplicity equal to one.
By the general result established around (\ref{Witten_indC_2ptuN}), this immediately gives
\begin{align}
\langle \mathcal{W}_{(1^m)}\mathcal{W}_{\overline{(1^m)}}\rangle^{U(N)(C)}(1)=\dim\Lambda^m(\mathbb{C}^N)=\binom{N}{m},
\end{align}
saturating the bound exactly for every $N$ and $m$, which is in sharp contrast to the symplectic and orthogonal cases above,
where the excess over $\dim\mathcal{R}$ grows rapidly with $m$ regardless of whether trace subtraction is required.

\subsection{Higgs indices}
\label{sec_Higgsindices}
In contrast to the Coulomb branch,
where the gauge group left unbroken at a generic point is the maximal torus $T\subset G$,
the Higgs branch of a 3d $\mathcal{N}=4$ gauge theory with sufficiently many hypermultiplets is realized as a hyperk\"{a}hler quotient
at which the gauge group $G$ is generically completely broken.
Let us consider a Wilson line operator $W_\mathcal{R}$ in a representation $\mathcal{R}$ of the gauge group $G$.
At a generic point on the Higgs branch, the complete breaking of $G$ triggers the standard Higgs mechanism.
The gauge fields $A$ acquire massive gaps proportional to the matter vacuum expectation values and decouple from the low-energy effective theory.
From a physical perspective, the gauge field ceases to be a fluctuating degree of freedom in the infrared.
The holonomy is effectively frozen to the identity element $e\in G$.
Therefore, evaluating the Wilson line in the Higgs branch reduces precisely to evaluating the trace of the identity operator within the representation space:
\begin{align}
W_{\mathcal{R}}
&={\Tr}_{\mathcal{R}}\left[\mathcal{P}\exp\left(i\oint A\right)\right]
\xrightarrow{\text{Higgsing}}{\Tr}_{\mathcal{R}}(\mathbf{1})
=\dim \mathcal{R}.
\end{align}
It is therefore natural to expect that
a Wilson line inserted in a representation $\mathcal{R}$ of $G$ probes the character of $\mathcal{R}$ only at the identity element, i.e.~its dimension,
once the $\mathfrak{t}\rightarrow 1$ limit projects onto the genuine one-dimensional degrees of freedom.

Motivated by the explicit results (see appendix \ref{app_Higgs}), we propose the following conjecture for the Higgs line defect Witten indices:
for Wilson lines $W_{\mathcal{R}_1},\dots,W_{\mathcal{R}_k}$ transforming in representations $\mathcal{R}_1,\dots, \mathcal{R}_k$ of $G$,
\begin{align}
\label{Witten_indH_conj}
\langle \mathcal{W}_{\mathcal{R}_1}\cdots \mathcal{W}_{\mathcal{R}_k}\rangle^{G(H)}(1)
= \prod_{j=1}^k \dim \mathcal{R}_j\,.
\end{align}
Equivalently, writing $\dim \mathcal{R}_j=\chi_{\mathcal{R}_j}(e)$ with $e\in G$ the identity element,
\begin{align}
\label{Witten_indH_char}
\langle \mathcal{W}_{\mathcal{R}_1}\cdots \mathcal{W}_{\mathcal{R}_k}\rangle^{G(H)}(1)
= \prod_{j=1}^k \chi_{\mathcal{R}_j}(e)\,,
\end{align}
in parallel with the formula (\ref{Witten_indC_conj_int}) for the Coulomb limit,
in which the analogous quantity is expressed as an integral of characters over the maximal torus
rather than an evaluation at a single group element.
In particular, for a Wilson line and its conjugate,
\begin{align}
\label{Witten_indH_2pt}
\langle \mathcal{W}_\mathcal{R} \mathcal{W}_{\overline{\mathcal{R}}}\rangle^{G(H)}(1) = (\dim \mathcal{R})^2,
\end{align}
to be compared with the Coulomb line defect two-point function (\ref{Witten_indC_2pt}).

Let us emphasize that \eqref{Witten_indH_conj} is not tied to the ADHM-type quivers,
and that it can in fact be understood in general as the Higgs counterpart of the character integral \eqref{Witten_indC_conj_int}.
For any good theory the Higgs limit receives contributions only from the zero magnetic flux sector, and the line defect Higgs index collapses to a finite dimensional integral over the maximal torus $T\subset G$,
\begin{align}
\label{Higgs_general_int}
&\langle W_{\mathcal{R}_1}\cdots W_{\mathcal{R}_k}\rangle^{G(H)}(\mathfrak{t})\nonumber\\
&\qquad=\frac{(1-\mathfrak{t}^2)^{r}}{|\mathscr{W}|}\oint_{T}\prod_{i=1}^{r}\frac{ds_i}{2\pi is_i}\,
\prod_{j=1}^{k}\chi_{\mathcal{R}_j}(s)\,
\frac{\prod_{\alpha\in\Delta_{\mathfrak{g}}}(1-s^{\alpha})(1-\mathfrak{t}^2s^{\alpha})}
{\prod_{\rho\in\Omega}(1-\mathfrak{t}\,s^{\rho})}\,,
\end{align}
where $r=\text{rank}\,G$, $\mathscr{W}$ is the Weyl group of $G$, $\Delta_{\mathfrak{g}}$ is the set of roots of $\mathfrak{g}$ and $\Omega$ is the multiset of weights of the hypermultiplet representation $\mathcal{H}\oplus\overline{\mathcal{H}}$, so that $|\Omega|=2\dim_{\mathbb{C}}\mathcal{H}$.
Here the flavor fugacities are suppressed for brevity, and restoring them simply replaces $s^{\rho}$ by $s^{\rho}y^{\rho}$.
The behavior of \eqref{Higgs_general_int} as $\mathfrak{t}\rightarrow 1$ is controlled by the neighborhood of the identity $s=e$.
Writing $\mathfrak{t}=e^{-\varepsilon}$ and $s_i=e^{\mathrm{i}\theta_i}$ and rescaling $\theta_i=\varepsilon\hat{\theta}_i$,
the measure together with the vector multiplet numerator scales as $\varepsilon^{2r+2|\Delta|}=\varepsilon^{2\dim G}$
while the hypermultiplet denominator scales as $\varepsilon^{|\Omega|}=\varepsilon^{2\dim_{\mathbb{C}}\mathcal{H}}$,
and the rescaled integrand approaches a finite $\varepsilon$-independent function of $\hat{\theta}$.
Hence the integral is dominated by the shrinking region $|\theta|=\mathcal{O}(\varepsilon)$ and diverges as
\begin{align}
\langle W_{\mathcal{R}_1}\cdots W_{\mathcal{R}_k}\rangle^{G(H)}(\mathfrak{t})
\;\sim\;\varepsilon^{-2\dim_{\mathbb{H}}\mathcal{M}_{H}},
\qquad
\dim_{\mathbb{H}}\mathcal{M}_{H}=\dim_{\mathbb{C}}\mathcal{H}-\dim G\,.
\end{align}
Since the characters are continuous and $\chi_{\mathcal{R}_j}(s)\rightarrow\chi_{\mathcal{R}_j}(e)=\dim\mathcal{R}_j$ uniformly on this region,
they simply factor out of the leading divergence and cancel against the same divergence of $\mathcal{I}^{G(H)}$, which is exactly \eqref{Witten_indH_conj}.
This is the analytic counterpart of the physical statement that the holonomy is frozen to the identity,
and it makes clear that \eqref{Witten_indH_conj} requires complete Higgs breaking, $\dim_{\mathbb{C}}\mathcal{H}>\dim G$,
in the same way that \eqref{Witten_indC_conj_int} requires the theory to be good.
Beyond the ADHM-type theories, we have verified \eqref{Witten_indH_conj} explicitly for $\mathcal{N}=4$ SQCD, i.e.~for theories whose only matter consists of fundamental hypermultiplets,
with unitary, symplectic and orthogonal gauge groups of ranks up to three, and for Wilson lines in a variety of representations.
See appendix \ref{app_Higgs_SQCD}.
The argument above also makes clear what can go wrong. If the gauge group is enlarged so that a generic point of the Higgs branch is left invariant by a non-trivial subgroup, the holonomy is no longer frozen to the identity alone and \eqref{Witten_indH_conj} is modified.
This does happen already for $O(N_c)$ SQCD, as we discuss in section \ref{sec_orthSQCD}.

\subsection{Other cases}
\label{sec_globalandCS}
The two formulae \eqref{Witten_indC_conj_int} and \eqref{Witten_indH_conj} were read off from two different physical pictures. On the Coulomb branch the gauge group is broken to its maximal torus, whereas on a Higgs branch with enough matter it is broken completely.
Neither picture applies to every theory. A torus survives on the Higgs branch of $\mathcal{N}=8$ super Yang-Mills theory and on the moduli space of a Chern-Simons matter theory, a reflection survives in orthogonal SQCD at the lower end of the good range, and a non-Abelian factor survives in the rank deformed ABJ theories.
In all of these the $\mathfrak{t}\rightarrow1$ limit is given by neither \eqref{Witten_indC_conj_int} nor \eqref{Witten_indH_conj}.

In this subsection we consider the average of the inserted characters over the gauge group $H_0$ that survives at a generic point of the branch one expands around, and we collect the models in which we have checked that this average gives the Witten index.
What follows is a list of examples rather than a general statement.
Section \ref{sec_H0} defines $H_0$ and explains how to compute it, section \ref{sec_generalconj} writes down that average, sections \ref{sec_orthSQCD} to \ref{sec_rankABJ} go through the models one by one, and section \ref{sec_validity} collects the open problems.
In practice the two limits remain most conveniently used in the separate forms \eqref{Witten_indC_conj_int} and \eqref{Witten_indH_conj}, and this is how we shall state our results in the rest of the paper.

\subsubsection{The residual gauge group}
\label{sec_H0}
The formula (\ref{Witten_indC_conj_int}) for the Coulomb limit crucially relied on the fact that,
in an ordinary $\mathcal{N}=4$ gauge theory without Chern-Simons terms,
a generic point of the Coulomb branch is parameterized by an independent set of
adjoint scalar zero modes $\sigma$ together with the dual photons,
under which the gauge group $G$ is broken to its maximal torus $T\subset G$
while the matter fields remain massless only along the Higgs branch directions.
Likewise, the formula (\ref{Witten_indH_conj}) relied on the fact that
a generic point of the Higgs branch, realized as a hyperk\"{a}hler quotient by the matter fields,
completely breaks $G$ and freezes the holonomy of the gauge field to the identity.
In this sense, the two conjectures directly reflect the residual symmetry breaking pattern associated to each branch.
It is therefore natural to take that residual symmetry itself as the basic input.

Fix the branch of the moduli space that one is expanding around, which is the Coulomb branch when the Coulomb limit of the index is taken and the Higgs branch when the Higgs limit is taken, and pick a vacuum $p$ on it which is generic, in the sense that it lies on the smooth locus of the branch and, within it, in the dense open subset on which the unbroken gauge group is as small as possible.
The scalar vacuum expectation values at $p$ break $G$ down to its stabilizer at $p$,
\begin{align}
\label{H0def}
H_0=\bigl\{\,h\in G\;\bigm|\;h\cdot p=p\,\bigr\}\;\subseteq\; G\,,
\end{align}
where $p$ stands for the collection of scalar vacuum expectation values at that vacuum, the adjoint scalars of the vector multiplet together with the hypermultiplet scalars, and $h\cdot p$ for the action on them of the constant gauge transformation $h$.
This $H_0$ is the gauge group left unbroken at $p$, which in a theory without Chern-Simons terms is the gauge group of the low energy effective theory there. We call it the stabilizer or the residual gauge group interchangeably.

In an ordinary $\mathcal{N}=4$ gauge theory $H_0$ is computed by the Higgs mechanism.
Write $X\cdot p$ for the infinitesimal version of the action appearing in \eqref{H0def},
$X\cdot p=\frac{d}{d\epsilon}\bigl(e^{\epsilon X}\cdot p\bigr)\big|_{\epsilon=0}$ with $X$ a generator of $\mathfrak{g}$,
that is for the gauge variation of the scalar vacuum expectation values.
On a scalar $\Phi$ in a representation $R$ of $G$ this is simply the action of $X$ in $R$, which for the matter appearing in this paper reads
$[X,\Phi]$ in the adjoint representation,
$X\Phi$ in the fundamental or vector one,
$X\Phi+\Phi X^{T}$ in the rank two symmetric or antisymmetric one,
and $X_a\Phi-\Phi X_b$ on a bifundamental of two gauge nodes $a$ and $b$, whether the hypermultiplet is twisted or not.
The covariant kinetic term of the scalars contains $|X\cdot p|^2$, so that the gauge fields along the directions with $X\cdot p\neq0$ acquire a mass proportional to the vacuum expectation values, while those along
\begin{align}
\label{H0lin}
X\cdot p=0\,,\qquad X\in\mathfrak{g}\,,
\end{align}
stay massless.
The solutions of this linear equation form the Lie algebra $\mathfrak{h}_0$ of $H_0$, and $\dim G-\dim H_0$ is the number of vector multiplets that become massive and are eaten.
Solving instead the finite equation in \eqref{H0def} gives $H_0$ itself, including any disconnected part, which \eqref{H0lin} cannot see.
How \eqref{H0def} and \eqref{H0lin} are solved in practice depends on the theory, and we work this out case by case below, where the answers turn out to be $T$, the trivial group, a disconnected group, a torus other than $T$, and a group with a non-Abelian factor.

Note that \eqref{H0lin} is only a convenient way of computing $H_0$.
In a Chern-Simons matter theory the Chern-Simons term makes every gauge field topologically massive, so that there is no massless vector multiplet to count and \eqref{H0lin} loses its usual meaning, while the structure of the moduli space is different as well, as we recall in section \ref{sec_CSmatter}.
The finite equation in \eqref{H0def} nevertheless continues to make sense, and it is this classical stabilizer that we use throughout.
What we actually verify is a statement about the value of the limit rather than about massless gauge bosons.
In the theories collected below, the $\mathfrak{t}\rightarrow1$ limit of the normalized index equals the average of the inserted characters over $H_0$ as defined in \eqref{H0def}, taken with the Haar measure of $H_0$.
The reading in terms of light gauge fields enforcing their Gauss law, which we give in section \ref{sec_generalconj}, is an interpretation of this result rather than a derivation of it.
We return to the identification of $H_0$, and to the situations in which this equality does not apply, in section \ref{sec_validity}.

Two different generic points of the same branch have stabilizers that are conjugate to each other inside $G$, and since everything below is built out of characters, which are class functions, none of our formulae depends on which generic $p$ is chosen.\footnote{
In mathematical terminology $H_0$ is the principal isotropy subgroup of the $G$ action on the branch.}
Two features of \eqref{H0def} are worth noting.
First, $H_0$ need not be connected. Its identity component is the one determined by \eqref{H0lin}, but a disconnected piece $\pi_0(H_0)$ can survive as well, and it enters the average on the same footing as the identity component.
This is what makes the answer sensitive to the global form of the gauge group, as we shall see when passing from $SO(N_c)$ to $O(N_c)$ in section \ref{sec_orthSQCD}.
Second, $H_0$ is determined by the branch, so that the Coulomb and the Higgs limits differ only through which $H_0$ they produce.

It remains to fix the notation for the group average.
We write $dh$ for the Haar measure of $H_0$ normalized to unit total volume, $\int_{H_0}dh=1$.
For a finite group this means $\int_{H_0}dh\,f(h)=\frac{1}{|H_0|}\sum_{h\in H_0}f(h)$,
and for a torus $H_0=U(1)^{r_0}$ of rank $r_0$ it is the contour integral $\oint\prod_{i=1}^{r_0}\frac{ds_i}{2\pi is_i}$ already appearing in \eqref{Witten_indC_conj_int}.

\subsubsection{Average of characters}
\label{sec_generalconj}
Given $H_0$, the natural quantity to compare the Witten index with is the average of the inserted characters over it,
\begin{align}
\label{Witten_ind_unified}
\lim_{\mathfrak{t}\rightarrow 1}\langle \mathcal{W}_{\mathcal{R}_1}\cdots \mathcal{W}_{\mathcal{R}_k}\rangle^{G(C/H)}(\mathfrak{t})
=\int_{H_0}dh\,\prod_{j=1}^k\chi_{\mathcal{R}_j}(h)\,.
\end{align}
The right hand side has a simple meaning.
Let $V$ be a finite dimensional representation of a compact group $H_0$ and denote by
\begin{align}
\label{invariants_def}
V^{H_0}=\bigl\{\,v\in V\;\bigm|\;h\cdot v=v\;\;\text{for all }h\in H_0\,\bigr\}\;\subseteq\;V
\end{align}
the subspace of $H_0$ invariant vectors.
Averaging the character of $V$ over the group is the projector onto that subspace, so that its trace gives
$\int_{H_0}dh\,\chi_{V}(h)=\dim V^{H_0}$.
Applying this to $V=\mathcal{R}_1\otimes\cdots\otimes\mathcal{R}_k$, which is a representation of $H_0$ by restriction from $G$ and whose character is $\prod_{j}\chi_{\mathcal{R}_j}$, we may equivalently write \eqref{Witten_ind_unified} as
\begin{align}
\label{Witten_ind_unified_inv}
\lim_{\mathfrak{t}\rightarrow 1}\langle \mathcal{W}_{\mathcal{R}_1}\cdots \mathcal{W}_{\mathcal{R}_k}\rangle^{G(C/H)}(\mathfrak{t})
=\dim\Bigl[\bigl(\mathcal{R}_1\otimes\cdots\otimes\mathcal{R}_k\bigr)^{H_0}\Bigr]\,.
\end{align}
This has a simple physical reading.
A Wilson line in the representation $\mathcal{R}$ can be traded for a one dimensional quantum mechanics living on the line, coupled to the bulk gauge field and with the representation space $\mathcal{R}$ itself as its Hilbert space, the trace in $\mathrm{Tr}_{\mathcal{R}}\mathcal{P}\exp\bigl(i\oint A\bigr)$ being the trace over that Hilbert space.
Inserting $k$ Wilson lines therefore adds the finite dimensional space $\mathcal{R}_1\otimes\cdots\otimes\mathcal{R}_k$ to the states of the theory.
At a generic point of the branch the gauge fields outside $H_0$ are massive and their holonomies are frozen to the identity, so they impose no constraint on this space, whereas those inside $H_0$ remain light and their Gauss law still has to be imposed.
Only the $H_0$ invariant part of $\mathcal{R}_1\otimes\cdots\otimes\mathcal{R}_k$ therefore survives, and averaging the characters over $H_0$ is exactly the projection onto it.
The $\mathfrak{t}\rightarrow1$ limit counts these states.
Written as \eqref{Witten_ind_unified_inv} the answer is a non-negative integer by construction, which is a non-trivial check on our results. For example, the values $2,5,14,3,13,8$ obtained for $O(3)$ with $N_f=2$ in appendix \ref{app_Higgs_SQCD} are each one half of a sum of two integers, and yet all of them come out integral.

Two cases are immediate. On the Coulomb branch of an ordinary $\mathcal{N}=4$ gauge theory the hypermultiplet scalars vanish and the only vacuum expectation value is the adjoint scalar, which can be brought into the Cartan subalgebra by a gauge transformation. For $G=U(N)$ the vacuum is $p=\sigma=\mathrm{diag}(\sigma_1,\cdots,\sigma_N)$, so that \eqref{H0lin} reads $(X\cdot p)_{ij}=[X,\sigma]_{ij}=X_{ij}(\sigma_i-\sigma_j)=0$, which at a generic point forces $X$ to be diagonal. Hence $H_0=T$, the subspace $(\mathcal{R}_1\otimes\cdots\otimes\mathcal{R}_k)^{T}$ is the zero weight subspace, and \eqref{Witten_ind_unified} is the character integral \eqref{Witten_indC_conj_int}.
On a Higgs branch the adjoint scalar vanishes instead. For $\mathcal{N}=4$ SQCD with gauge group $U(N_c)$ and $N_f$ fundamental hypermultiplets, whose scalars $q_\alpha$ are vectors of $\mathbb{C}^{N_c}$, equation \eqref{H0lin} reads $X\cdot p=(Xq_1,\cdots,Xq_{N_f})=0$, and for $N_f\ge N_c$ the $q_\alpha$ generically span $\mathbb{C}^{N_c}$, so that $X=0$ is the only solution. Hence $H_0=\{e\}$, every state is invariant, and \eqref{Witten_ind_unified} reduces to $\prod_j\dim\mathcal{R}_j$, which is \eqref{Witten_indH_conj}.

The remaining cases are the ones for which neither \eqref{Witten_indC_conj_int} nor \eqref{Witten_indH_conj} applies, and we take them up model by model in the three subsections that follow.

\subsubsection{Orthogonal SQCD}
\label{sec_orthSQCD}
Consider $\mathcal{N}=4$ SQCD with $N_f$ hypermultiplets in the fundamental representation and gauge group $SO(N_c)$ or $O(N_c)=SO(N_c)\rtimes\mathbb{Z}^{\mathcal{C}}_2$, which is good for $N_f\ge N_c-1$.
This is again a Higgs branch statement, so the vacuum is the one used above for unitary SQCD with the scalars now real, but this time the finite equation \eqref{H0def} has to be used, since the reflections it produces are invisible to \eqref{H0lin}.
The vacuum expectation values of the $N_f$ hypermultiplets generically span a $\min(N_f,N_c)$ dimensional subspace of $\mathbb{R}^{N_c}$, so that
\begin{align}
\label{H0discrete}
H_0=\bigl\{h\in O(N_c)\;\bigm|\;hq_\alpha=q_\alpha,\ \alpha=1,\cdots,N_f\bigr\}=O(N_c-N_f)\,,
\end{align}
acting on the orthogonal complement of the span.
Inside $SO(N_c)$ this is $SO(N_c-N_f)$, which is trivial for every $N_f\ge N_c-1$, so that \eqref{Witten_indH_conj} holds for the $SO(N_c)$ theory throughout the good range.
For the $O(N_c)$ theory, however, the balanced case $N_f=N_c-1$ leaves the reflection
\begin{align}
g_0=\mathrm{diag}(\underbrace{1,\cdots,1}_{N_c-1},-1)\in O(N_c)
\end{align}
unbroken, so that $H_0=O(1)=\{e,g_0\}$.
The Haar measure of a two element group is $\int_{H_0}dh\,f(h)=\frac12\bigl(f(e)+f(g_0)\bigr)$, so \eqref{Witten_ind_unified} becomes
\begin{align}
\label{Witten_indH_O}
\lim_{\mathfrak{t}\rightarrow 1}\langle \mathcal{W}_{\mathcal{R}_1}\cdots \mathcal{W}_{\mathcal{R}_k}\rangle^{O(N_c)\text{-}[N_c-1](H)}(\mathfrak{t})
=\frac{1}{2}\Bigl(\prod_{j=1}^k\dim\mathcal{R}_j+\prod_{j=1}^k\chi_{\mathcal{R}_j}^{\mathfrak{so}(N_c)}(g_0)\Bigr),
\end{align}
whereas for $N_f\ge N_c$ the stabilizer is trivial and \eqref{Witten_indH_conj} is recovered.
Since $g_0^2=e$, any representation $\mathcal{R}$ of $O(N_c)$ splits into the eigenspaces $\mathcal{R}^{\pm}$ of $g_0$ and $\chi_{\mathcal{R}}^{\mathfrak{so}(N_c)}(g_0)=\dim\mathcal{R}^{+}-\dim\mathcal{R}^{-}$, so that for a single Wilson line \eqref{Witten_indH_O} simply counts the $g_0$ even states, $\frac12(\dim\mathcal{R}+\chi_{\mathcal{R}}^{\mathfrak{so}(N_c)}(g_0))=\dim\mathcal{R}^{+}$, as it must by \eqref{Witten_ind_unified_inv}.
The same conclusion follows from the index directly.
Repeating the scaling argument of section \ref{sec_Higgsindices} in the sector twisted by $g_0$, the integral localizes around $s=g_0$ instead of $s=e$ and diverges as $\varepsilon^{-2(\dim_{\mathbb{C}}\mathcal{H}^{g_0}-\dim G^{g_0})}$, where the superscript denotes the $g_0$ invariant part. For $SO(N_c)$ with $N_f$ fundamental hypermultiplets the two exponents differ precisely by $N_f-(N_c-1)$.
The twisted and untwisted sectors therefore diverge at the same rate exactly at $N_f=N_c-1$, and the characters are evaluated at $e$ and at $g_0$ with equal weight, which is \eqref{Witten_indH_O}.
We have verified \eqref{Witten_indH_O} for $O(3)$ with $N_f=2$ and for $O(5)$ with $N_f=4$, and \eqref{Witten_indH_conj} for $O(3)$ with $N_f=3,4$ and $O(5)$ with $N_f=5$ (see appendix \ref{app_Higgs_SQCD}).
Whether the reflection survives is a property of the matter content and not of the global form alone, as the orthogonal ADHM-type theories relevant for M2-branes show.
There the fundamental hypermultiplets are accompanied by one in the rank two symmetric or antisymmetric representation, which breaks the reflection as well, so that $H_0=\{e\}$ and \eqref{Witten_indH_conj} holds for the $O(N_c)$ theory just as it does for $SO(N_c)$.
This is what we find in appendix \ref{app_Higgs_SON}, where the Higgs limits of the $O(3)$ and $SO(3)$ theories with a rank two tensor hypermultiplet and four fundamental hypermultiplets agree with each other and with $\prod_j\dim\mathcal{R}_j$.

\subsubsection{Chern-Simons matter theories}
\label{sec_CSmatter}
For theories with two gauge nodes we write $\mathcal{W}_{\mathcal{R}_1;\mathcal{R}_2}$ for the Wilson line in the representation $\mathcal{R}_1$ of the first gauge node and $\mathcal{R}_2$ of the second, so that its insertion in the matrix model is the product of characters $\chi_{\mathcal{R}_1}(s)\chi_{\mathcal{R}_2}(\sigma)$ evaluated on the two holonomies, and $\mathbf{1}$ for the trivial representation of the node in question.

\paragraph{ABJM theories}
In 3d $\mathcal{N}>4$ Chern-Simons matter theories, such as the ABJM theory \cite{Aharony:2008ug} and its generalizations,
the structure recalled in section \ref{sec_H0} is qualitatively different, as anticipated there.
The equation of motion of the auxiliary field sourced by the Chern-Simons coupling relates the adjoint scalar $\sigma$ to the matter field bilinears,
schematically $\sigma \sim \mu(\Phi)$, with $\mu$ the moment map of the matter fields,
so that $\sigma$ no longer parameterizes an independent direction of the moduli space once the matter fields are turned on.
Moreover, since the Chern-Simons term renders the gauge field topologically massive,
there is no propagating dual photon and the corresponding continuous Coulomb branch direction is absent.
What survives is only a discrete identification of order $k$, the Chern-Simons level, implemented by monopole operators
carrying non-trivial gauge charge under the residual $U(1)$'s.
Consequently, a generic point of the moduli space of Chern-Simons matter theories is realized as a single mixed branch
generated by gauge invariant combinations of monopole operators dressed by matter fields,
e.g.~$\mathbb{C}^4/\mathbb{Z}_k$ for the $U(1)_k\times U(1)_{-k}$ ABJM theory.
In particular a generic point does not completely break $G$, since the matter vacuum expectation values leave unbroken a diagonal torus $U(1)^N \subset G$,
in direct analogy with the residual maximal torus $T$ that characterizes the Coulomb branch of an ordinary $\mathcal{N}=4$ gauge theory.
Both limits are therefore governed by \eqref{Witten_ind_unified} with $H_0=U(1)^N_{\text{diag}}$, i.e.~by a torus character integral of the Coulomb type
rather than by the dimension formula \eqref{Witten_indH_conj} associated to complete breaking.

We solve \eqref{H0def} explicitly here, since the same computation is repeated with a rank deformation below.
Write $A^{a}$, $a=1,2$, for the bifundamental scalars of the $U(N)_k\times U(N)_{-k}$ ABJM theory in $(\mathbf{N},\overline{\mathbf{N}})$ and $B_a$ for those in $(\overline{\mathbf{N}},\mathbf{N})$, transforming as $A^{a}\rightarrow h_1A^{a}h_2^{-1}$ and $B_a\rightarrow h_2B_ah_1^{-1}$ under $h=(h_1,h_2)$.
The Chern-Simons coupling makes the equation of motion of the auxiliary field algebraic,
\begin{align}
\label{ABJvac1}
\frac{k}{2\pi}\sigma_1=\sum_{a}\bigl(A^{a}A^{a\dagger}-B_a^{\dagger}B_a\bigr)\,,\qquad
-\frac{k}{2\pi}\sigma_2=\sum_{a}\bigl(B_aB_a^{\dagger}-A^{a\dagger}A^{a}\bigr)\,,
\end{align}
which is the precise form of the relation $\sigma\sim\mu(\Phi)$ quoted above, the right hand sides being the moment maps of the two gauge groups and the left hand sides carrying the two Chern-Simons levels $k$ and $-k$. The scalar potential contains in addition
\begin{align}
\label{ABJvac2}
\sigma_1A^{a}-A^{a}\sigma_2=0\,,\qquad
\sigma_2B_a-B_a\sigma_1=0\,.
\end{align}
A generic solution, describing $N$ M2-branes sitting at distinct points, has $A^{a}=\alpha^{a}$ and $B_a=\beta_a$ diagonal and invertible, and then \eqref{ABJvac1} gives $\sigma_1=\sigma_2=\frac{2\pi}{k}\sum_a\bigl(|\alpha^{a}|^2-|\beta_a|^2\bigr)$, consistently with \eqref{ABJvac2}.
Imposing \eqref{H0lin} on it with $X=(X_1,X_2)$ gives $X_1\alpha^{a}-\alpha^{a}X_2=0$ and $X_2\beta_a-\beta_aX_1=0$, that is $X_2=(\alpha^{a})^{-1}X_1\alpha^{a}=\beta_aX_1\beta_a^{-1}$, which at a generic point forces $X_1$ to be diagonal and $X_2=X_1$. Hence
\begin{align}
\label{H0ABJM}
H_0=U(1)^N_{\text{diag}}\subset U(N)\times U(N)\,,
\end{align}
the relative phases being broken by the bifundamental vacuum expectation values while the diagonal ones, under which these are neutral, survive.

The two limits reach this same $H_0$ from opposite directions, and this is why the Coulomb and Higgs limits of these theories look alike.
In the Coulomb limit the ordinary bifundamental hypermultiplets drop out of the localization integrand and the surviving operators are the monopoles dressed by the twisted hypermultiplets, while in the Higgs limit the roles are exchanged and the monopoles are dressed by the ordinary ones.
The ABJ(M) theories are self-mirror, and the mirror map exchanges the hypermultiplets with the twisted hypermultiplets, so the two arguments are the same argument and must give the same $H_0$, which is \eqref{H0ABJM}.
This is why \eqref{OSpABJt1limits} and its unitary counterpart show the Coulomb and Higgs limits agreeing, and it is the reason a torus rather than the trivial group appears even in the Higgs limit.

This expectation can in fact be verified explicitly for the $U(N)_k\times U(N)_{-k}$ ABJM theory.
We have confirmed that
the Wilson line $\mathcal{W}_{\lambda;\bar\lambda}$ of the ABJM theory,
labeled by a representation $\lambda$ of $U(N)_k$ and its conjugate $\bar\lambda$ of $U(N)_{-k}$,
reduces on the residual diagonal torus $U(1)^N \subset U(N)_k\times U(N)_{-k}$
to the product of characters $\chi_\lambda(s)\chi_\lambda(s^{-1})$ evaluated at the common torus coordinate $s=(s_1,\dots,s_N)$ in the Coulomb/Higgs limit.
See appendix \ref{ABJMalgorithm}.
Expanding the Schur polynomial $\chi_\lambda(s)=\sum_\mu K_{\lambda\mu}s^\mu$
in terms of Kostka numbers $K_{\lambda\mu}$, which coincide precisely with the $U(N)$ weight
multiplicities $m_{\mathcal{R}_\lambda}(\mu)$, the torus integral over $U(1)^N$ projects onto the diagonal term and reproduces
\begin{align}
\langle \mathcal{W}_{\lambda;\bar\lambda}\rangle^{U(N)_k\times U(N)_{-k}\text{ ABJM}(C)}(1)
&=\oint \prod_{i=1}^N \frac{ds_i}{2\pi is_i}\,\chi_\lambda(s)\chi_\lambda(s^{-1})
=\sum_\mu K_{\lambda\mu}^2,
\end{align}
which is precisely the two-point function formula (\ref{Witten_indC_2ptuN})
for the line defect Coulomb index of the $U(N)$ gauge theory with $\mathcal{R}=\mathcal{R}_\lambda$.
This confirms that, despite the different physical origin of the residual torus,
the ABJM line defect Witten index is governed by exactly the same character integral structure as (\ref{Witten_indC_conj_int}),
once restricted to the surviving diagonal $U(N)$ subgroup.

\paragraph{Orthosymplectic ABJ theories}
The orthosymplectic ABJ theories $SO(2N+\gamma)_{2k}\times USp(2N)_{-k}$ with $\gamma=0,1$ work in exactly the same way. The bifundamental vacuum expectation values pair the $2N$ directions of the vector representation with the $2N$ symplectic colors two by two, so that the diagonal torus survives, embedded as $N$ rotation planes inside the vector representation of the orthogonal factor and as the maximal torus of the symplectic one,
\begin{align}
\label{H0OSpABJ}
H_0=U(1)^N_{\text{diag}}\;\subset\; SO(2N+\gamma)_{2k}\times USp(2N)_{-k}\,,
\end{align}
independently of $\gamma$, the unpaired direction of the vector representation present for $\gamma=1$ being a singlet of the diagonal torus.
Writing $u_i=s_i+s_i^{-1}$, the two fundamental characters restrict to $\chi^{\mathfrak{so}(2N+\gamma)}_{\ydiagram{1}}=\gamma+\sum_{i=1}^{N}u_i$ and $\chi^{\mathfrak{usp}(2N)}_{\ydiagram{1}}=\sum_{i=1}^{N}u_i$, the term $\gamma$ being the direction of the vector representation that the branes do not occupy, so that \eqref{Witten_ind_unified} gives
\begin{align}
\label{OSpABJunified}
\int_{H_0}dh\,\chi_{\ydiagram{1};\ydiagram{1}}(h)
=\oint\prod_{i=1}^{N}\frac{ds_i}{2\pi is_i}\Bigl(\gamma+\sum_{i=1}^{N}u_i\Bigr)\Bigl(\sum_{j=1}^{N}u_j\Bigr)=2N\,,
\end{align}
independently of $\gamma$, in agreement with \eqref{OSpABJt1limits}.
Here the interplay with the global form can be seen once more. For $\gamma=1$ the unoccupied direction is also left invariant by the reflection $g_0=\mathrm{diag}(1,\cdots,1,-1)$, so that the $O(2N+1)_{2k}\times USp(2N)_{-k}$ theory has the disconnected $H_0=U(1)^N_{\text{diag}}\times\mathbb{Z}_2$.
The extra average over the two elements does not change \eqref{OSpABJunified}, because $\chi^{\mathfrak{so}(2N+1)}_{\ydiagram{1}}(g_0)=\pm1$ enters as a torus singlet multiplying $\chi^{\mathfrak{usp}(2N)}_{\ydiagram{1}}$, whose torus integral vanishes.
In agreement with this, the index of $SO(2N+1)^{(\chi\zeta)}_{2k}\times USp(2N)_{-k}$, where the superscripts are the signs $\chi=\pm1$ and $\zeta=\pm1$ of the background fugacities for the charge conjugation $\mathbb{Z}^{\mathcal{C}}_2$ and for the magnetic $\mathbb{Z}^{\mathcal{M}}_2$ of the orthogonal node, does not depend on $\chi$ at all, that is the two $\mathbb{Z}^{\mathcal{C}}_2$ sectors contribute equally, which is what one expects when the $\mathbb{Z}_2$ is unbroken.
For $\gamma=0$ there is no such direction, the reflection is broken, and correspondingly the $\chi=-1$ sector of $SO(2N)_{2k}\times USp(2N)_{-k}$ carries a strictly weaker divergence than the $\chi=+1$ one.
The $O(2N)_{2k}\times USp(2N)_{-k}$ theory therefore has $H_0=U(1)^N_{\text{diag}}$ as well, and \eqref{OSpABJunified} again gives $2N$.
We have checked all of this on the exact expressions, finding the value $2N$ for $SO(2N+1)^{(\chi+)}$, $SO(2N)^{(++)}$, $O(2N)^{(+)}$ and $O(2N+1)^{(+)}$, and confirming that the poles of $\mathcal{I}^{SO(2N)^{(\pm+)}_{2k}\times USp(2N)_{-k}\text{ABJ}(C)}$ at $\mathfrak{t}=1$ are indeed of different order.
Beyond the bifundamental Wilson line, we have also evaluated the Coulomb limit for Wilson lines in the fundamental and symmetric representations of either node and in bi-symmetric combinations, always finding agreement with \eqref{Witten_ind_unified}, as reported in appendix \ref{app_OSp_otherreps}.

\subsubsection{Rank deformed ABJ theories}
\label{sec_rankABJ}
\paragraph{Unitary rank deformations}
The rank deformed ABJ theories $U(N)_k\times U(N+M)_{-k}$ are the case in which $H_0$ acquires a non-Abelian factor. We go through them in detail, since here \eqref{H0def} is hardest to guess.
The matter content is the one of section \ref{sec_CSmatter}, with $A^{a}$ now an $N\times(N+M)$ matrix and $B_a$ an $(N+M)\times N$ one, and the vacuum equations are again \eqref{ABJvac1} and \eqref{ABJvac2}.
A generic solution, describing $N$ M2-branes sitting at distinct points, is
\begin{align}
\label{ABJvev}
&A^{a}=\bigl(\alpha^{a}\,\big|\,0_{N\times M}\bigr)\,,\qquad
B_a=\begin{pmatrix}\beta_a\\ 0_{M\times N}\end{pmatrix}\,,\nonumber\\
&\sigma_1=\mathrm{diag}\bigl(\sigma^{(1)},\cdots,\sigma^{(N)}\bigr)\,,\qquad
\sigma_2=\mathrm{diag}\bigl(\sigma^{(1)},\cdots,\sigma^{(N)},0,\cdots,0\bigr)\,,
\end{align}
with $\alpha^{a}$ and $\beta_a$ diagonal and invertible and $\frac{k}{2\pi}\sigma^{(i)}=\sum_a(|\alpha^{a}_i|^2-|\beta_{a,i}|^2)$.
The last $M$ colors of the second node are not used by the vacuum expectation values.
Now impose \eqref{H0lin}. Writing $X=(X_1,X_2)$ and decomposing $X_2$ into blocks of sizes $N$ and $M$,
$X_2=\left(\begin{smallmatrix}Y&Z\\ W&V\end{smallmatrix}\right)$, the two conditions become
\begin{align}
\label{ABJstab}
&X_1A^{a}-A^{a}X_2=\bigl(X_1\alpha^{a}-\alpha^{a}Y\,\big|\,-\alpha^{a}Z\bigr)=0\,,\nonumber\\
&X_2B_a-B_aX_1=\begin{pmatrix}Y\beta_a-\beta_aX_1\\ W\beta_a\end{pmatrix}=0\,.
\end{align}
Since $\alpha^{a}$ and $\beta_a$ are invertible the off-diagonal blocks must vanish, $Z=0$ and $W=0$, and what is left is $Y=(\alpha^{a})^{-1}X_1\alpha^{a}=\beta_aX_1\beta_a^{-1}$.
For diagonal $\alpha^{a}$ and $\beta_a$ these read $(X_1)_{ij}\bigl(\alpha^{a}_j/\alpha^{a}_i-\beta_{a,i}/\beta_{a,j}\bigr)=0$, so that at a generic point $X_1$ is diagonal and then $Y=X_1$.
The block $V$ is not constrained at all, since it acts only on the colors that the vacuum expectation values do not use.
Hence $\mathfrak{h}_0$ consists of the pairs $\bigl(X_1,\mathrm{diag}(X_1,V)\bigr)$ with $X_1$ diagonal and $V\in\mathfrak{u}(M)$ arbitrary, that is
\begin{align}
\label{H0ABJrank}
H_0=U(1)^N_{\text{diag}}\times U(M)\,.
\end{align}
Setting $M=0$ recovers \eqref{H0ABJM}, and what survives at low energy on the extra factor is a topological $U(M)_k$ Chern-Simons sector.
The same steps with the orthogonal and symplectic blocks give the stabilizers quoted below for the orthosymplectic theories.
Each flux sector of the index is labeled by a bare monopole operator, and once the ranks differ these operators need not all have non-negative dimension.
For $N=1$ and $M=1$, where $H_0=U(1)_{\text{diag}}\times U(1)$ is a torus, every bare monopole operator has non-negative dimension, and we find that \eqref{Witten_ind_unified} reproduces the Coulomb limit at every level we examined, see appendix \ref{app_ABJrank}.
For $M\ge2$ the leftover $U(M)$ is non-Abelian, but bare monopole operators of negative dimension contribute to the line defect indices there, and we leave these cases for future work.

\paragraph{Orthosymplectic rank deformations}
The orthosymplectic ABJ theories can be rank deformed as well, and the same block computation applies with the unitary blocks replaced by orthogonal and symplectic ones.
For $SO(2N_1+\gamma)_{2k}\times USp(2N_2)_{-k}$ the bifundamental vacuum expectation values pair $2\min(N_1,N_2)$ directions of the vector representation with as many symplectic colors, and whichever node is larger keeps an untouched factor, so that
\begin{align}
\label{H0OSpABJrank}
H_0=\begin{cases}
U(1)^{N_1}_{\text{diag}}\times USp\bigl(2(N_2-N_1)\bigr)\,, & N_2>N_1\,,\\[2pt]
U(1)^{N_2}_{\text{diag}}\times SO\bigl(2(N_1-N_2)+\gamma\bigr)\,, & N_1>N_2\,.
\end{cases}
\end{align}
Setting $N_1=N_2$ recovers \eqref{H0OSpABJ}.
We have generalized \eqref{OSpABJ_index2} to unequal ranks, and for $N_2>N_1$ the Coulomb index without line defects coincides, in the cases we have checked, with that of the equal rank theory with $N=N_1$.
We do not quote line defect results for these theories. Once the ranks differ, bare monopole operators of negative dimension occur in general and can contribute to the line defect indices, and we leave the rank deformed orthosymplectic theories for future work.

Both families are instances of a statement that follows from the vacuum equations alone, and that therefore fixes $H_0$ for a general $\mathcal{N}=4$ Chern-Simons quiver.
Take a circular quiver with gauge group $\prod_{a}U(N_a)_{k_a}$, the edges carrying bifundamental hypermultiplets and twisted hypermultiplets and only two of the Chern-Simons levels being non-zero.
The equation of motion of the auxiliary field ties the adjoint scalars to the matter fields, so that the vacuum is specified by the bifundamental vacuum expectation values alone, and at a generic point these have the common rank
$N=\min_a N_a$, the number of M2-branes that can move away.
The argument below uses only the pattern of gauge charges, which is the same for the two types of bifundamental matter, so it does not depend on how the two are distributed over the edges.
The stabilizer condition $h_aA_ah_{a+1}^{-1}=A_a$ along each edge, with $A_a$ the bifundamental vacuum expectation value on that edge, then identifies the phases of the $N$ occupied colors at consecutive nodes and leaves the unoccupied ones free, so that
\begin{align}
\label{H0quiver}
H_0=U(1)^N_{\text{diag}}\times\prod_{a}U(N_a-N)\,,\qquad N=\min_a N_a\,.
\end{align}
The diagonal torus is the gauge group on the world-volumes of the mobile branes and the factors $U(N_a-N)$ are those of the colors that stay behind at the node $a$.
The ABJM theory and the rank deformed ABJ theories are the two node cases of \eqref{H0quiver}, and the orthosymplectic theories obey the same statement with the leftover unitary factor replaced by the leftover orthogonal or symplectic one.
We have not tested \eqref{H0quiver} beyond two nodes.
For a three node theory such as $U(1)_k\times U(1)_0\times U(1+M)_{-k}$, a rank deformation of the $\mathcal{N}=4$ circular quiver Chern-Simons theory $U(N)_k\times U(N)_0\times U(N)_{-k}$ allowed by the bound on the number of fractional M2-branes, \eqref{H0quiver} gives $H_0=U(1)_{\text{diag}}\times U(M)$, with a non-Abelian factor at the deformed node.
Evaluating the Coulomb limit there requires control over the contributions of bare monopole operators of negative dimension, as in the two node theories above, which we have not attempted, so we leave this case for future work.

The reading of section \ref{sec_generalconj} assumes that the leftover factors, namely those acting on the colors that the vacuum expectation values do not use, carry no massless charged matter, so that at low energy they support a pure Chern-Simons sector.
This holds in the unitary quivers, where the bifundamental components joining an occupied color to a leftover one are given a mass by the vacuum expectation values, but not in the orthosymplectic theories with $\gamma=1$ and $N_2>N_1$, where the unpaired direction of the vector representation leaves a half hypermultiplet charged under the leftover symplectic factor massless.

\subsubsection{Open problems}
\label{sec_validity}
Beyond the cases collected above we do not know a formula for the $\mathfrak{t}\rightarrow1$ limit of the line defect index. The character average \eqref{Witten_ind_unified} is not expected to apply in general, and we do not know a criterion that decides when it does.
Much of the landscape is also unexplored, in particular Chern-Simons quivers with more than two nodes, the rank deformed ABJ theories with $M\ge2$ and their orthosymplectic counterparts, and representations larger than the ones we have been able to evaluate.
We record here the situations in which we know that \eqref{Witten_ind_unified} does not apply.

First, the limit may not exist at all.
For the Coulomb limit this requires the theory to be good or ugly in the sense of \cite{Gaiotto:2008ak}, so that all BPS monopole operators carry strictly positive (resp.~non-negative) Coulomb branch scaling dimensions, the flux sum converges and the Coulomb index is a well-defined formal power series in $\mathfrak{t}$.
For the Higgs limit the analogous requirement is that the branch exists and is of the expected dimension, $\dim_{\mathbb{C}}\mathcal{H}>\dim G-\dim H_0$, which for $H_0=\{e\}$ is the condition $\dim_{\mathbb{C}}\mathcal{H}>\dim G$ of section \ref{sec_Higgsindices}.

Second, the limit may fail to count states with unit weight.
By \eqref{Witten_ind_unified_inv} the right hand side of \eqref{Witten_ind_unified} is the dimension of a vector space, so the left hand side can only be a plain count of the defect ground states.
If a background fugacity $\zeta=\pm1$ or $\chi=\pm1$ for a discrete global symmetry is switched on, every state is weighted by its charge and the trace becomes a difference of two counts rather than a single one.
Such graded quantities lie outside the scope of \eqref{Witten_ind_unified} and are not integers in general, as one sees in \eqref{SO2N+1fundzeta-1t1limit}, \eqref{SO2Nchi+fundzeta-1t1limit} and \eqref{SO2Nchi-fundzeta-1t1limit}.
They can be brought back within its scope by gauging the discrete symmetry, which is the passage from $SO(N_c)$ to $O(N_c)$ of section \ref{sec_orthSQCD} and from $SO(2N+1)$ to $Spin(2N+1)$ in appendix \ref{app_algorithms}, since the sum over sectors is again a plain count, now for a larger $H_0$.

Third, the inserted representations may be charged under the center of $H_0$.
All the Chern-Simons matter checks reported above involve products $\mathcal{R}\otimes\overline{\mathcal{R}}$, which carry no such charge.
For an insertion that does carry one the answer can differ, because the factor of $H_0$ left over at a deformed node supports a topological Chern-Simons sector at level $k$, in which a Wilson line whose charge is a multiple of $k$ can be screened by a monopole operator and is not an independent line operator.
In the rank deformed theory $U(1)_k\times U(2)_{-k}$, for instance, we find for $k=2,3$ and $m\le3$
\begin{align}
\label{ABJonepoint}
\lim_{\mathfrak{t}\rightarrow1}\langle\mathcal{W}_{\mathbf{1};(m)}\rangle=
\begin{cases}1\,, & m\equiv0 \bmod k\,,\\ 0\,, & \text{otherwise}\,,\end{cases}
\end{align}
whereas $\dim[((m))^{H_0}]$ vanishes for every $m\ge1$.

Finally, even in the Chern-Simons matter theories where the average does reproduce the Witten index we do not understand why. Solving the vacuum equations gives $H_0$ there just as it does in ordinary theories, but the argument of section \ref{sec_generalconj}, that the heavy gauge fields impose no constraint while the light ones impose their Gauss law, is not available there because every gauge field is massive.

\section{Recursion relations}
\label{sec_recursion}
In this section we examine the Coulomb line defect indices of Wilson lines for the 3d $\mathcal{N} \ge 4$ SCFTs describing a stack of M2-branes and turn to the second main result of this paper.
Holographically, these M2-brane SCFTs admit descriptions in either M-theory or Type II string theory
depending on the specific coupling regimes and singularity structures.
Under this AdS/CFT correspondence, the supersymmetric Wilson line defects are holographically addressed as fundamental strings
or macroscopic heavy probe branes that extend along the $AdS_4$ boundary and wrap appropriate cycles within the internal compact manifold.
We propose algebraic recursion relations for the line defect Coulomb indices for the M2-brane SCFTs.

\subsection{$U(N)$ ADHM theories}
\label{sec_recursion_A}
While the evaluation of general line defect indices yields complicated rational functions,
the Wilson lines transforming in the antisymmetric representations corresponding to the minuscule weights of $U(N)$ exhibit remarkably elegant structural properties.
In \cite{Hayashi:2024jof} the following factorized recursion relation
for the line defect Coulomb index of the Wilson line $W_{(1^n)}$ in the rank-$n$ antisymmetric representations in the $U(N)$ ADHM theory was found:
\begin{align}
\label{R_uN_C_Wasym}
\langle \mathcal{W}_{(1^n)} \mathcal{W}_{\overline{(1^n)}}\rangle^{\textrm{$U(N)$ ADHM-$[l]$} (C)}(z;\mathfrak{t})
&=\frac{\mathcal{I}^{\textrm{$U(N-n)$ ADHM-$[l]$} (C)}(z;\mathfrak{t}) \mathcal{I}^{\textrm{$U(n)$ ADHM-$[l]$} (C)}(z;\mathfrak{t})}
{\mathcal{I}^{\textrm{$U(N)$ ADHM-$[l]$} (C)}(z;\mathfrak{t})}.
\end{align}

As is evident from the structure of the recurrence relation (\ref{R_uN_C_Wasym}),
the two-point functions of the antisymmetric Wilson line defects inherit the palindromic property of the Coulomb indices,
leading to the following relations:
\begin{align}
\langle \mathcal{W}_{(1^k)} \mathcal{W}_{\overline{(1^k)}}\rangle^{\textrm{$U(N)$ ADHM-$[l]$} (C)}(\mathfrak{t}^{-1})
&=\langle \mathcal{W}_{(1^k)} \mathcal{W}_{\overline{(1^k)}}\rangle^{\textrm{$U(N)$ ADHM-$[l]$} (C)}(\mathfrak{t}).
\end{align}

Note that the recursion relation, although not of a simple factorized form like \eqref{R_uN_C_Wasym}, also exists for the Coulomb line defect indices of symmetric Wilson lines.
For example, for the rank $2$ symmetric representation we find
\begin{align}
&\langle W_{\ydiagram{2}}W_{\overline{\ydiagram{2}}}\rangle^{U(N)\text{ ADHM-}[l](C)}(z;\mathfrak{t})\nonumber \\
&
=
{\cal I}^{U(1)\text{ ADHM-}[l](C)}(z;\mathfrak{t})
{\cal I}^{U(N-1)\text{ ADHM-}[l](C)}(z;\mathfrak{t})\nonumber \\
&\quad
+\Bigl[-(1+2\mathfrak{t}^2){\cal I}^{U(2)\text{ ADHM-}[l](C)}(z;\mathfrak{t})\nonumber \\
&\quad\quad
+(1+\mathfrak{t}^2)({\cal I}^{U(1)\text{ ADHM-}[l](C)}(z;\mathfrak{t}))^2
\Bigr]
{\cal I}^{U(N-2)\text{ ADHM-}[l](C)}(z;\mathfrak{t}),
\label{R_uN_C_Wsym2}
\end{align}
which follows from equation (3.69) in \cite{Hayashi:2024jof}.

\subsection{$USp(2N)$ gauge theories}
\label{sec_recursion_C}
In the case of the 3d $\mathcal{N}=4$ $USp(2N)$ gauge theories, which describe M2-branes probing a D-type singularity,
a similar recursive structure emerges for the fundamental Wilson line.
Unlike the $U(N)$ theories, it is the fundamental representation that constitutes the minuscule weight for the $USp(2N)$ gauge group.
We conjecture that the normalized two-point function of the fundamental Wilson lines in the $USp(2N)$ theory
coupled to an antisymmetric (or symmetric) hypermultiplet and $2l$ fundamental half-hypermultiplets is governed by the following recurrence relation:\footnote{
As written in \eqref{USprecrelnappendix}, the same recurrence relation holds also for the $USp(2N)$ theory with $2l+1$ fundamental half-hypermultiplets.
}
\begin{align}
\label{R_usp_C_W1}
&
\langle \mathcal{W}_{\ydiagram{1}} \mathcal{W}_{\ydiagram{1}}\rangle^{\textrm{$USp(2N)+$(A/S)-$[2l]$} (C)}(\mathfrak{t})
\nonumber\\
&=\frac{\mathcal{I}^{\textrm{$USp(2N-2)+$(A/S)-$[2l]$} (C)}(\mathfrak{t})
\mathcal{I}^{\textrm{$USp(2)+$(A/S)-$[2l]$} (C)}(\mathfrak{t})
}
{\mathcal{I}^{\textrm{$USp(2N)+$(A/S)-$[2l]$} (C)}(\mathfrak{t})}
\langle \mathcal{W}_{\ydiagram{1}} \mathcal{W}_{\ydiagram{1}} \rangle^{\textrm{$USp(2)+$(A/S)-$[2l]$} (C)}(\mathfrak{t}).
\end{align}
See appendix \ref{app_USp_fund} for the evidence for the recurrence relation.
Here the Coulomb index $\mathcal{I}^{\textrm{$USp(2N)+$(A/S)-$[2l]$} (C)}(\mathfrak{t})$ coincides with the Hilbert series of
$\mathrm{Sym}^N (\mathbb{C}^2/\widehat{D}_{l-2+2\epsilon})$,
where $\epsilon $ is $+1$ (resp.~$0$) for the theory with a symmetric (resp.~antisymmetric) hypermultiplet.
Here $\widehat{D}_{l-2+2\epsilon}$ is the binary dihedral group of order $4(l-2+2\epsilon)$.
$\langle \mathcal{W}_{\ydiagram{1}} \mathcal{W}_{\ydiagram{1}} \rangle^{\textrm{$USp(2)+$(A/S)-$[2l]$} (C)}(\mathfrak{t})$
is the two-point function of the fundamental Wilson line in the $USp(2)$ gauge theory
coupled to an antisymmetric (resp.~symmetric) hypermultiplet and $2l$ fundamental half-hypermultiplets.
Note that the latter only depends on the flavor data through the specific combination
\begin{align}
\label{uspDelta}
\Delta&=l-2+2\epsilon.
\end{align}
In other words, the two-point function for the $USp(2)$ theory with a symmetric hypermultiplet and $2l$ half-hypers
coincides exactly with that for the theory with an antisymmetric hypermultiplet and $2(l+2)$ half-hypers.
The two-point function of the fundamental Wilson line in the $USp(2)$ gauge theory is exactly computed as
\begin{align}
\label{usp2_C_W1}
\langle \mathcal{W}_{\ydiagram{1}} \mathcal{W}_{\ydiagram{1}}\rangle^{\textrm{$USp(2)+$(A/S)-$[2l]$} (C)}(\mathfrak{t})
&=\frac{(1+\mathfrak{t}^2)(1+\mathfrak{t}^{2\Delta})} {1+\mathfrak{t}^{2\Delta+2}}.
\end{align}
According to the recurrence relation (\ref{R_usp_C_W1}),
the two-point functions for higher rank $USp(2N)$ gauge theories also depend on the effective flavor parameter (\ref{uspDelta}).

The relation (\ref{R_usp_C_W1}) highlights that
the higher rank line defect index is completely determined by the bulk Coulomb indices and the exact two-point function of the minimal rank $USp(2)$ theory.

For example, for the $USp(4)$ gauge theory, we obtain
\begin{align}
&
\langle \mathcal{W}_{\ydiagram{1}} \mathcal{W}_{\ydiagram{1}}\rangle^{\textrm{$USp(4)+$(A/S)-$[2l]$} (C)}(\mathfrak{t})
\nonumber\\
&=\frac{(1 + \mathfrak{t}^2)(1 + \mathfrak{t}^4)(1 + \mathfrak{t}^{2\Delta})^2 (1 + \mathfrak{t}^{2 + 2\Delta})}
{1 + \mathfrak{t}^{2\Delta+2} + \mathfrak{t}^{2\Delta+4} + \mathfrak{t}^{2\Delta+6}
+ \mathfrak{t}^{4\Delta+2} + \mathfrak{t}^{4\Delta+4} + \mathfrak{t}^{4\Delta+6} + \mathfrak{t}^{6\Delta+8}}.
\end{align}
For the $USp(6)$ gauge theory, we find
\begin{align}
&
\langle \mathcal{W}_{\ydiagram{1}} \mathcal{W}_{\ydiagram{1}}\rangle^{\textrm{$USp(6)+$(A/S)-$[2l]$} (C)}(\mathfrak{t})
\nonumber\\
&=\frac{1}{P_{USp(6),\Delta}(\mathfrak{t})}
(1 + \mathfrak{t}^2) (1 + \mathfrak{t}^4 + \mathfrak{t}^8) (1 + \mathfrak{t}^{2\Delta}) (1 + \mathfrak{t}^{2\Delta} + \mathfrak{t}^{4\Delta})\nonumber \\
&\times (1 + \mathfrak{t}^{2\Delta+2} + \mathfrak{t}^{2\Delta+4} + \mathfrak{t}^{2\Delta+6} + \mathfrak{t}^{4\Delta+2} + \mathfrak{t}^{4\Delta+4} + \mathfrak{t}^{4\Delta+6} + \mathfrak{t}^{6\Delta+8}),
\end{align}
where
\begin{align}
&
P_{USp(6),\Delta}(\mathfrak{t})
\nonumber\\
&=1 + \mathfrak{t}^{2\Delta+2} + \mathfrak{t}^{2\Delta+4} + \mathfrak{t}^{2\Delta+6} + \mathfrak{t}^{2\Delta+8}
+ \mathfrak{t}^{2\Delta+10} + \mathfrak{t}^{4\Delta+2} + 2\mathfrak{t}^{4\Delta+4} + 2\mathfrak{t}^{4\Delta+6} \nonumber\\
& + 2\mathfrak{t}^{4\Delta+8} + 2\mathfrak{t}^{4\Delta+10} + \mathfrak{t}^{4\Delta+12} + \mathfrak{t}^{4\Delta+14}
+ \mathfrak{t}^{6\Delta+2} + \mathfrak{t}^{6\Delta+4} + 3\mathfrak{t}^{6\Delta+6} + 2\mathfrak{t}^{6\Delta+8} \nonumber\\
&+ 2\mathfrak{t}^{6\Delta+10} + 3\mathfrak{t}^{6\Delta+12} + \mathfrak{t}^{6\Delta+14} + \mathfrak{t}^{6\Delta+16}
+ \mathfrak{t}^{8\Delta+4} + \mathfrak{t}^{8\Delta+6} + 2\mathfrak{t}^{8\Delta+8} + 2\mathfrak{t}^{8\Delta+10} \nonumber\\
&+ 2\mathfrak{t}^{8\Delta+12} + 2\mathfrak{t}^{8\Delta+14} + \mathfrak{t}^{8\Delta+16}
+ \mathfrak{t}^{10\Delta+8} + \mathfrak{t}^{10\Delta+10} + \mathfrak{t}^{10\Delta+12}
+ \mathfrak{t}^{10\Delta+14} + \mathfrak{t}^{10\Delta+16} + \mathfrak{t}^{12\Delta+18}.
\end{align}

It follows from the properties of the Coulomb indices together with the recursion relation (\ref{R_usp_C_W1})
that the fundamental Wilson line two-point function is invariant under the transformation $\mathfrak{t}\rightarrow \mathfrak{t}^{-1}$
\begin{align}
\langle \mathcal{W}_{\ydiagram{1}} \mathcal{W}_{\ydiagram{1}}\rangle^{\textrm{$USp(2N)+$(A/S)-$[2l]$} (C)}(\mathfrak{t}^{-1})
&=\langle \mathcal{W}_{\ydiagram{1}} \mathcal{W}_{\ydiagram{1}}\rangle^{\textrm{$USp(2N)+$(A/S)-$[2l]$} (C)}(\mathfrak{t}).
\end{align}

\subsection{$SO(2N)$ gauge theories}
\label{sec_recursion_D}
We further propose that an analogous recursive relation holds for the line defect indices associated with the fundamental Wilson line in the $SO(2N)$ gauge theory.
This generalizes the structures previously observed for the antisymmetric Wilson line in the $U(N)$ ADHM theory and the fundamental Wilson line in the $USp(2N)$ gauge theory.
We conjecture that the normalized line defect indices satisfy recursion relations of the form\footnote{
As written in \eqref{SO2Nchi+recrelnappendix} and \eqref{SO2Nchi-recrelnappendix}, the same recursion relations hold also for the theories with $\mathbb{Z}^{{\cal M}}_2$ fugacity $\zeta=-1$.
}
\begin{subequations}
\begin{align}
\label{R_so2N_C_W1}
\langle \mathcal{W}_{\ydiagram{1}} \mathcal{W}_{\ydiagram{1}} \rangle^{\textrm{$SO(2N)^{(++)}+$(A/S)-$[2l]$} (C)}(\mathfrak{t})
&=\frac{
\mathcal{I}^{\textrm{$SO(2N-2)^{(++)}+$(A/S)-$[2l]$} (C)}(\mathfrak{t})
\mathcal{I}^{\textrm{$SO(2)^{(++)}+$(A/S)-$[2l]$} (C)}(\mathfrak{t})
}
{\mathcal{I}^{\textrm{$SO(2N)^{(++)}+$(A/S)-$[2l]$} (C)}(\mathfrak{t})}, \\
\langle \mathcal{W}_{\ydiagram{1}} \mathcal{W}_{\ydiagram{1}} \rangle^{\textrm{$SO(2N)^{(-+)}+$(A/S)-$[2l]$} (C)}(\mathfrak{t})
&=\frac{
\mathcal{I}^{\textrm{$SO(2N-2)^{(-+)}+$(A/S)-$[2l]$} (C)}(\mathfrak{t})
\mathcal{I}^{\textrm{$SO(2)^{(++)}+$(A/S)-$[2l]$} (C)}(\mathfrak{t})
}
{\mathcal{I}^{\textrm{$SO(2N)^{(-+)}+$(A/S)-$[2l]$} (C)}(\mathfrak{t})},
\end{align}
\end{subequations}
for $N>1$.
See appendices \ref{app_SO2Nchi+fund} and \ref{app_SOevenchim1fund} for the evidence for the recursion relations.
Remarkably, all these quantities are completely determined by the Coulomb indices
and their structure closely parallels those for the $U(N)$ ADHM theory.
The Coulomb index $\mathcal{I}^{\textrm{$SO(2N)^{(\chi \zeta)}+$(A/S)-$[2l]$} (C)}(\mathfrak{t})$ for the $SO(2N)$ gauge theory
depends on the discrete fugacities $\chi$ and $\zeta$ for the $\mathbb{Z}_2^{\mathcal{C}}$ charge conjugation symmetry
and the $\mathbb{Z}_2^{\mathcal{M}}$ magnetic symmetry respectively \cite{Aharony:2013kma}.
We can construct the Coulomb index for the $O(2N)^+$ gauge theory by gauging the $\mathbb{Z}_2^{\mathcal{C}}$ symmetry
\begin{align}
\mathcal{I}^{\textrm{$O(2N)^{(+)}+$(A/S)-$[2l]$} (C)}(\mathfrak{t})
&=\frac12
\left(
\mathcal{I}^{\textrm{$SO(2N)^{(++)}+$(A/S)-$[2l]$} (C)}(\mathfrak{t})
+
\mathcal{I}^{\textrm{$SO(2N)^{(-+)}+$(A/S)-$[2l]$} (C)}(\mathfrak{t})
\right),
\end{align}
which is identified with the Hilbert series of $\mathrm{Sym}^N (\mathbb{C}^2/\widehat{D}_{l+2\epsilon})$, with $\epsilon=+1$ (resp.~$0$) corresponding to the presence of a symmetric (resp.~antisymmetric) hypermultiplet.
The recursive structure encoded in (\ref{R_so2N_C_W1}) extends to the $O(2N)$ gauge theories,
implying that the two-point function of the fundamental Wilson line satisfies the same recursion relation
\begin{align}
\langle \mathcal{W}_{\ydiagram{1}} \mathcal{W}_{\ydiagram{1}} \rangle^{\textrm{$O(2N)^{(+)}+$(A/S)-$[2l]$} (C)}(\mathfrak{t})
&=\frac{
\mathcal{I}^{\textrm{$O(2N-2)^{(+)}+$(A/S)-$[2l]$} (C)}(\mathfrak{t})
\mathcal{I}^{\textrm{$SO(2)^{(++)}+$(A/S)-$[2l]$} (C)}(\mathfrak{t})
}
{\mathcal{I}^{\textrm{$O(2N)^{(+)}+$(A/S)-$[2l]$} (C)}(\mathfrak{t})}.
\end{align}
The Coulomb line defect indices in the $SO(2N)$ gauge theories depend on the flavor data only through
\begin{align}
\Delta&=l+2\epsilon.
\end{align}
As a consequence, one finds an equivalence between
the two-point functions of the theory with a symmetric hypermultiplet and $2l$ fundamental half-hypermultiplets, and that with an antisymmetric hypermultiplet and $2(l+2)$ of them.
The $SO(2)$ gauge theory Coulomb indices, appearing as a multiplicative factor in all cases, admit an exact evaluation given by
\begin{align}
\mathcal{I}^{\textrm{$SO(2)^{(++)}+$(A/S)-$[2l]$} (C)}(\mathfrak{t})
&=\frac{1 + \mathfrak{t}^{2\Delta}}{(1 - \mathfrak{t}^2)(1 - \mathfrak{t}^{2\Delta})}.
\end{align}

For example, in the case of the $SO(4)$ gauge theories,
the two-point functions of the fundamental Wilson lines can be computed exactly and take the following forms:
\begin{subequations}
\begin{align}
\langle \mathcal{W}_{\ydiagram{1}} \mathcal{W}_{\ydiagram{1}} \rangle^{\textrm{$SO(4)^{(++)}+$(A/S)-$[2l]$} (C)}(\mathfrak{t})
&=\frac{(1 + \mathfrak{t}^2)^2 (1 + \mathfrak{t}^{2\Delta})^3}{1 + \mathfrak{t}^{4\Delta} + 2\mathfrak{t}^{2 + 2\Delta} + \mathfrak{t}^{4 + 2\Delta} + 2\mathfrak{t}^{2 + 4\Delta} + \mathfrak{t}^{4 + 6\Delta}}, \\
\langle \mathcal{W}_{\ydiagram{1}} \mathcal{W}_{\ydiagram{1}} \rangle^{\textrm{$SO(4)^{(-+)}+$(A/S)-$[2l]$} (C)}(\mathfrak{t})
&=\frac{(1 + \mathfrak{t}^4)(1 + \mathfrak{t}^{2\Delta})}{1 + \mathfrak{t}^{4 + 2\Delta}}.
\end{align}
\end{subequations}
Gauging the $\mathbb{Z}_2^{\mathcal{C}}$ charge conjugation symmetry yields the $O(4)$ gauge theory two-point function
\begin{align}
\langle \mathcal{W}_{\ydiagram{1}} \mathcal{W}_{\ydiagram{1}} \rangle^{\textrm{$O(4)^{(+)}+$(A/S)-$[2l]$} (C)}(\mathfrak{t})
=
\langle {\cal W}_{\ydiagram{1}}{\cal W}_{\ydiagram{1}}\rangle^{USp(4)+\text{(A/S)-}[2l+4](C)}(\mathfrak{t}).
\end{align}
Let us turn to the $SO(6)$ gauge theories.
Although the computation becomes somewhat more involved, we obtain
\begin{align}
\langle \mathcal{W}_{\ydiagram{1}} \mathcal{W}_{\ydiagram{1}} \rangle^{\textrm{$SO(6)^{(++)}+$(A/S)-$[2l]$} (C)}(\mathfrak{t})
&=\frac{1}{P_{SO(6)^{++},\Delta}(\mathfrak{t})}\Bigl(
(1 + \mathfrak{t}^4)(1 + \mathfrak{t}^2 + \mathfrak{t}^4)(1 + \mathfrak{t}^{2\Delta} + \mathfrak{t}^{4\Delta})
\nonumber\\
&\times (1 + \mathfrak{t}^{4\Delta} + 2\mathfrak{t}^{2 + 2\Delta} + \mathfrak{t}^{4 + 2\Delta} + 2\mathfrak{t}^{2 + 4\Delta} + \mathfrak{t}^{4 + 6\Delta})
\Bigr)
\end{align}
for $(\chi\zeta)$ $=$ $(++)$
and
\begin{align}
&
\langle \mathcal{W}_{\ydiagram{1}} \mathcal{W}_{\ydiagram{1}} \rangle^{\textrm{$SO(6)^{(-+)}+$(A/S)-$[2l]$} (C)}(\mathfrak{t})
\nonumber\\
&=\frac{(1 + \mathfrak{t}^2)^2 (1 - \mathfrak{t}^2 + \mathfrak{t}^4) (1 + \mathfrak{t}^{2\Delta})^2 (1 + \mathfrak{t}^{4 + 2\Delta})}{1 + \mathfrak{t}^{2 + 2\Delta} + \mathfrak{t}^{6 + 2\Delta} + \mathfrak{t}^{8 + 2\Delta} + \mathfrak{t}^{4 + 4\Delta} + \mathfrak{t}^{6 + 4\Delta} + \mathfrak{t}^{10 + 4\Delta} + \mathfrak{t}^{12 + 6\Delta}}
\end{align}
for $(\chi\zeta)$ $=$ $(-+)$.
Here
\begin{align}
\label{P_so6++}
P_{SO(6)^{++},\Delta}(\mathfrak{t})&=
1- \mathfrak{t}^{2\Delta} + \mathfrak{t}^{4\Delta} + \mathfrak{t}^{2+2\Delta} + 2\mathfrak{t}^{4+2\Delta}
+ \mathfrak{t}^{6+2\Delta} + \mathfrak{t}^{8+2\Delta} + \mathfrak{t}^{2+4\Delta} + \mathfrak{t}^{4+4\Delta} \nonumber\\
&+ 2\mathfrak{t}^{6+4\Delta} + \mathfrak{t}^{8+4\Delta} + \mathfrak{t}^{10+4\Delta} + \mathfrak{t}^{2+6\Delta} + \mathfrak{t}^{4+6\Delta} + 2\mathfrak{t}^{6+6\Delta} + \mathfrak{t}^{8+6\Delta} \nonumber \\
&+ \mathfrak{t}^{10+6\Delta} + \mathfrak{t}^{12+6\Delta} + \mathfrak{t}^{4+8\Delta} + \mathfrak{t}^{6+8\Delta} + 2\mathfrak{t}^{8+8\Delta} + \mathfrak{t}^{10+8\Delta} - \mathfrak{t}^{12+8\Delta} + \mathfrak{t}^{12+10\Delta}.
\end{align}
Gauging the $\mathbb{Z}_2^{\mathcal{C}}$ symmetry again leads to the normalized two-point function for the $O(6)$ gauge theory
\begin{align}
\langle \mathcal{W}_{\ydiagram{1}} \mathcal{W}_{\ydiagram{1}} \rangle^{\textrm{$O(6)^{(+)}+$(A/S)-$[2l]$} (C)}(\mathfrak{t})
=
\langle {\cal W}_{\ydiagram{1}}{\cal W}_{\ydiagram{1}}\rangle^{USp(6)+\text{(A/S)-}[2l+4](C)}(\mathfrak{t}).
\end{align}

According to the recursion relations (\ref{R_so2N_C_W1})
the two-point functions of the fundamental Wilson lines are invariant under the transformation $\mathfrak{t}\rightarrow \mathfrak{t}^{-1}$
\begin{subequations}
\begin{align}
\langle \mathcal{W}_{\ydiagram{1}} \mathcal{W}_{\ydiagram{1}} \rangle^{\textrm{$SO(2N)^{(++)}+$(A/S)-$[2l]$} (C)}(\mathfrak{t}^{-1})
&=\langle \mathcal{W}_{\ydiagram{1}} \mathcal{W}_{\ydiagram{1}} \rangle^{\textrm{$SO(2N)^{(++)}+$(A/S)-$[2l]$} (C)}(\mathfrak{t}), \\
\langle \mathcal{W}_{\ydiagram{1}} \mathcal{W}_{\ydiagram{1}} \rangle^{\textrm{$SO(2N)^{(-+)}+$(A/S)-$[2l]$} (C)}(\mathfrak{t}^{-1})
&=\langle \mathcal{W}_{\ydiagram{1}} \mathcal{W}_{\ydiagram{1}} \rangle^{\textrm{$SO(2N)^{(-+)}+$(A/S)-$[2l]$} (C)}(\mathfrak{t}), \\
\langle \mathcal{W}_{\ydiagram{1}} \mathcal{W}_{\ydiagram{1}} \rangle^{\textrm{$O(2N)^{(+)}+$(A/S)-$[2l]$} (C)}(\mathfrak{t}^{-1})
&=\langle \mathcal{W}_{\ydiagram{1}} \mathcal{W}_{\ydiagram{1}} \rangle^{\textrm{$O(2N)^{(+)}+$(A/S)-$[2l]$} (C)}(\mathfrak{t}).
\end{align}
\end{subequations}

\subsection{$SO(2N+1)$ gauge theories}
\label{sec_recursion_B}
We next turn to the $SO(2N+1)$ gauge theories coupled to a symmetric or antisymmetric hypermultiplet and $l$ hypermultiplets in the fundamental representation.
In contrast to the $U(N)$, $USp(2N)$ and $SO(2N)$ gauge theories, the fundamental representation of $SO(2N+1)$ is not minuscule.
Its weight system consists of the single Weyl orbit of the highest weight together with the zero weight.
Correspondingly, we find that
the two-point function of the fundamental Wilson lines does \textit{not} obey a factorized recursion relation of the form
(\ref{R_uN_C_Wasym}), (\ref{R_usp_C_W1}) or (\ref{R_so2N_C_W1}).

Remarkably, however, the failure of the factorization turns out to be recovered by subtracting the contribution of the identity.
We conjecture that the two-point function satisfies the exact recursion relation
\begin{align}
\label{R_so2N+1_C_W1}
&
\langle \mathcal{W}_{\ydiagram{1}} \mathcal{W}_{\ydiagram{1}}\rangle^{\textrm{$SO(2N+1)^{(\chi+)}+$(A/S)-$[2l]$} (C)}(\mathfrak{t}) - 1
\nonumber\\
&=\frac{\mathcal{I}^{\textrm{$SO(2N-1)^{(\chi+)}+$(A/S)-$[2l]$} (C)}(\mathfrak{t})\,
\mathcal{I}^{\textrm{$SO(3)^{(\chi+)}+$(A/S)-$[2l]$} (C)}(\mathfrak{t})}
{\mathcal{I}^{\textrm{$SO(2N+1)^{(\chi+)}+$(A/S)-$[2l]$} (C)}(\mathfrak{t})}
\left(
\langle \mathcal{W}_{\ydiagram{1}} \mathcal{W}_{\ydiagram{1}}\rangle^{\textrm{$SO(3)^{(\chi+)}+$(A/S)-$[2l]$} (C)}(\mathfrak{t}) - 1
\right)
\end{align}
for $N>1$.
See appendix \ref{app_SO2N+1fund} for the evidence for the recursion relation.
Equivalently, in terms of the unnormalized correlators, the recursion relation takes the simple form
\begin{align}
\label{R_so2N+1_C_W1_unnorm}
&
\langle W_{\ydiagram{1}} W_{\ydiagram{1}}\rangle^{SO(2N+1)^{(\chi+)}}(\mathfrak{t})
- \mathcal{I}^{SO(2N+1)^{(\chi+)}}(\mathfrak{t})
\nonumber\\
&=\mathcal{I}^{SO(2N-1)^{(\chi+)}}(\mathfrak{t})
\left(
\langle W_{\ydiagram{1}} W_{\ydiagram{1}}\rangle^{SO(3)^{(\chi+)}}(\mathfrak{t})
- \mathcal{I}^{SO(3)^{(\chi+)}}(\mathfrak{t})
\right),
\end{align}
where we have suppressed the matter content.
The identity subtracted two-point function for the $SO(2N+1)$ theory factorizes into
the Coulomb index of the $SO(2N-1)$ theory times the subtracted two-point function of the $SO(3)$ theory.
Upon regarding the subtracted two-point function as the two-point function,
the recursion relation takes the same structure as the recursion relation (\ref{R_usp_C_W1}) for the two-point function of the $USp(2N)$ theory.
The recursion relation (\ref{R_so2N+1_C_W1}) extends straightforwardly to $O(2N+1)$ gauge theories by gauging the $\mathbb{Z}_2^{\mathcal{C}}$ charge conjugation symmetry
\begin{align}
\label{R_o2N+1_C_W1}
&
\langle \mathcal{W}_{\ydiagram{1}} \mathcal{W}_{\ydiagram{1}}\rangle^{\textrm{$O(2N+1)^{(+)}+$(A/S)-$[2l]$} (C)}(\mathfrak{t}) - 1
\nonumber\\
&=\frac{\mathcal{I}^{\textrm{$O(2N-1)^{(+)}+$(A/S)-$[2l]$} (C)}(\mathfrak{t})\,
\mathcal{I}^{\textrm{$O(3)^{(+)}+$(A/S)-$[2l]$} (C)}(\mathfrak{t})}
{\mathcal{I}^{\textrm{$O(2N+1)^{(+)}+$(A/S)-$[2l]$} (C)}(\mathfrak{t})}
\left(
\langle \mathcal{W}_{\ydiagram{1}} \mathcal{W}_{\ydiagram{1}}\rangle^{\textrm{$O(3)^{(+)}+$(A/S)-$[2l]$} (C)}(\mathfrak{t}) - 1
\right).
\end{align}
Note that for the odd orthogonal groups the charge conjugation is an inner automorphism,
$O(2N+1)=SO(2N+1)\times\mathbb{Z}_2^{\mathcal{C}}$
with the $\mathbb{Z}_2^{\mathcal{C}}$ factor generated by the central element $-\mathbf{I}$
so that the Coulomb indices and the line defect indices are independent of the discrete fugacity $\chi$ and coincide for $SO(2N+1)^{(\pm+)}$ and $O(2N+1)^{(+)}$, in contrast to the even orthogonal case.

The Coulomb indices and the line defect indices again depend on
the flavor data only through the combination
\begin{align}
\label{soDelta_B}
\Delta&=l+2\epsilon,
\end{align}
where $\epsilon=1$ (resp.~$\epsilon=0$) for the theory with a symmetric
(resp.~antisymmetric) hypermultiplet.
The seed of the recursion, i.e.~the $SO(3)$ theory admits the exact closed-forms
\begin{align}
\label{so3_exact}
\mathcal{I}^{\textrm{$SO(3)^{(\chi+)}+$(A/S)-$[2l]$}(C)}(\mathfrak{t})
&=\frac{1+\mathfrak{t}^{2\Delta+2}}{(1-\mathfrak{t}^4)(1-\mathfrak{t}^{2\Delta})},
\nonumber\\
\langle \mathcal{W}_{\ydiagram{1}} \mathcal{W}_{\ydiagram{1}}\rangle^{\textrm{$SO(3)^{(\chi+)}+$(A/S)-$[2l]$} (C)}(\mathfrak{t}) - 1
&=\frac{(1+\mathfrak{t}^2)\left(\mathfrak{t}^2+2\mathfrak{t}^{2\Delta}-\mathfrak{t}^{2\Delta+2}\right)}{1+\mathfrak{t}^{2\Delta+2}}.
\end{align}
The Coulomb index in (\ref{so3_exact}) is the Hilbert series of $\mathbb{C}^2/\widehat{D}_{l+2\epsilon}$
and more generally we find that the $SO(2N+1)$ Coulomb indices agree with
the Hilbert series of $\mathrm{Sym}^N(\mathbb{C}^2/\widehat{D}_{l+2\epsilon})$.

For example, in the case of the $SO(5)$ gauge theory,
the Coulomb index and the normalized two-point function of the fundamental Wilson lines can be computed exactly.
We obtain
\begin{align}
\label{so5_exact}
&
\langle \mathcal{W}_{\ydiagram{1}} \mathcal{W}_{\ydiagram{1}}\rangle^{\textrm{$SO(5)^{(\chi+)}+$(A/S)-$[2l]$} (C)}(\mathfrak{t})
\nonumber\\
&=\frac{1}
{1+\left(\mathfrak{t}^{2}+\mathfrak{t}^{4}+\mathfrak{t}^{6}\right)
\left(\mathfrak{t}^{2\Delta}+\mathfrak{t}^{4\Delta}\right)
+\mathfrak{t}^{8+6\Delta}}
\nonumber\\
&\times
\Bigl(
1 + \mathfrak{t}^2 + \mathfrak{t}^4 + \mathfrak{t}^6 + \mathfrak{t}^8
+ \left(2 + 3\mathfrak{t}^2 + 4\mathfrak{t}^4 + 4\mathfrak{t}^6 + \mathfrak{t}^8+\mathfrak{t}^{10}\right)\mathfrak{t}^{2\Delta}
\nonumber\\
&
+ \left(2 + 4\mathfrak{t}^2 + 4\mathfrak{t}^4 + 4\mathfrak{t}^6 + \mathfrak{t}^8\right)\mathfrak{t}^{4\Delta}
+ \left(2\mathfrak{t}^2 + \mathfrak{t}^4 + \mathfrak{t}^6 + 2\mathfrak{t}^8 - \mathfrak{t}^{10}\right)\mathfrak{t}^{6\Delta}
\Bigr).
\end{align}
Similarly, for the $SO(7)$ gauge theory we obtain the exact expressions
\begin{align}
\label{so7_exact}
\langle \mathcal{W}_{\ydiagram{1}} \mathcal{W}_{\ydiagram{1}}\rangle^{\textrm{$SO(7)^{(\chi+)}+$(A/S)-$[2l]$} (C)}(\mathfrak{t})
&=\frac{Q_{SO(7),\Delta}(\mathfrak{t})}{P_{SO(7),\Delta}(\mathfrak{t})},
\end{align}
with
\begin{align}
\label{so7_N}
P_{SO(7),\Delta}(\mathfrak{t})
&=1
+\left(\mathfrak{t}^{2}+\mathfrak{t}^{4}+\mathfrak{t}^{6}+\mathfrak{t}^{8}+\mathfrak{t}^{10}\right)\mathfrak{t}^{2\Delta}
+\left(\mathfrak{t}^{2}+2\mathfrak{t}^{4}+2\mathfrak{t}^{6}+2\mathfrak{t}^{8}+2\mathfrak{t}^{10}+\mathfrak{t}^{12}+\mathfrak{t}^{14}\right)\mathfrak{t}^{4\Delta}
\nonumber\\
&
+\left(\mathfrak{t}^{2}+\mathfrak{t}^{4}+3\mathfrak{t}^{6}+2\mathfrak{t}^{8}+2\mathfrak{t}^{10}+3\mathfrak{t}^{12}+\mathfrak{t}^{14}+\mathfrak{t}^{16}\right)\mathfrak{t}^{6\Delta}
\nonumber\\
&
+\left(\mathfrak{t}^{4}+\mathfrak{t}^{6}+2\mathfrak{t}^{8}+2\mathfrak{t}^{10}+2\mathfrak{t}^{12}+2\mathfrak{t}^{14}+\mathfrak{t}^{16}\right)\mathfrak{t}^{8\Delta}
\nonumber\\
&
+\left(\mathfrak{t}^{8}+\mathfrak{t}^{10}+\mathfrak{t}^{12}+\mathfrak{t}^{14}+\mathfrak{t}^{16}\right)\mathfrak{t}^{10\Delta}
+\mathfrak{t}^{18+12\Delta}
\end{align}
and
\begin{align}
\label{so7_Q}
Q_{SO(7),\Delta}(\mathfrak{t})
&=\sum_{i=0}^{6}\mathfrak{t}^{2i}
+\left(2 + 3\mathfrak{t}^2 + 4\mathfrak{t}^4 + 5\mathfrak{t}^6 + 6\mathfrak{t}^8 + 6\mathfrak{t}^{10}
+ 3\mathfrak{t}^{12} + 3\mathfrak{t}^{14} + 2\mathfrak{t}^{16} + \mathfrak{t}^{18}\right)\mathfrak{t}^{2\Delta}
\nonumber\\
&
+\left(2 + 5\mathfrak{t}^2 + 9\mathfrak{t}^4 + 12\mathfrak{t}^6 + 13\mathfrak{t}^8 + 13\mathfrak{t}^{10}
+ 10\mathfrak{t}^{12} + 8\mathfrak{t}^{14} + 4\mathfrak{t}^{16} + \mathfrak{t}^{18}\right)\mathfrak{t}^{4\Delta}
\nonumber\\
&
+\left(2 + 6\mathfrak{t}^2 + 10\mathfrak{t}^4 + 16\mathfrak{t}^6 + 15\mathfrak{t}^8 + 16\mathfrak{t}^{10}
+ 15\mathfrak{t}^{12} + 10\mathfrak{t}^{14} + 6\mathfrak{t}^{16} + \mathfrak{t}^{18} + \mathfrak{t}^{20}\right)\mathfrak{t}^{6\Delta}
\nonumber\\
&
+\left(4\mathfrak{t}^2 + 8\mathfrak{t}^4 + 11\mathfrak{t}^6 + 13\mathfrak{t}^8 + 13\mathfrak{t}^{10}
+ 13\mathfrak{t}^{12} + 9\mathfrak{t}^{14} + 5\mathfrak{t}^{16} + \mathfrak{t}^{18}\right)\mathfrak{t}^{8\Delta}
\nonumber\\
&
+\left(2\mathfrak{t}^2 + 3\mathfrak{t}^4 + 4\mathfrak{t}^6 + 6\mathfrak{t}^8 + 6\mathfrak{t}^{10}
+ 6\mathfrak{t}^{12} + 4\mathfrak{t}^{14} + 3\mathfrak{t}^{16} + \mathfrak{t}^{18}\right)\mathfrak{t}^{10\Delta}
\nonumber\\
&
+\left(2\mathfrak{t}^{8} + \mathfrak{t}^{10} + \mathfrak{t}^{12} + \mathfrak{t}^{14} + \mathfrak{t}^{16}
+ 2\mathfrak{t}^{18} - \mathfrak{t}^{20}\right)\mathfrak{t}^{12\Delta}.
\end{align}

The structure of (\ref{R_so2N+1_C_W1}) differs from the recursion relations of the previous subsections
as the subtraction of the identity is required.
Writing the character at $\chi=+1$ as $\chi^{\mathfrak{so}(2N+1)}_{\ydiagram{1}}(s,+1)=\widetilde{\chi}(s)+1$,
where $\widetilde{\chi}(s)=\sum_{i=1}^{N}(s_i+s_i^{-1})$ is the sum over
the Weyl orbit of the highest weight and $1$ is the contribution of
the zero weight, the two-point insertion decomposes into three channels,
\begin{align}
\label{Bchannel}
\chi^{\mathfrak{so}(2N+1)}_{\ydiagram{1}}(s,+1)^2
&=\widetilde{\chi}(s)^2+2\,\widetilde{\chi}(s)+1.
\end{align}
The last term shifts no magnetic flux and simply reproduces the Coulomb index.
This is precisely the identity subtracted in (\ref{R_so2N+1_C_W1}).
For $\chi=-1$ the zero weight contributes $\chi$ rather than $1$, so that the middle channel is multiplied by $\chi$ while the last one is unchanged, and the subtraction of the identity is the same.
We emphasize that
neither the full character nor the pure Weyl-orbit part $\widetilde{\chi}$ alone satisfies a factorized recursion relation.
The physical interpretation of this structure in terms of a frozen fractional brane will be discussed in section \ref{sec_factorization}.

Finally, unlike the $USp(2N)$ and $SO(2N)$ cases,
the recursion relation (\ref{R_so2N+1_C_W1}) does \textit{not} imply the invariance of the two-point function under $\mathfrak{t}\rightarrow\mathfrak{t}^{-1}$.
The subtraction of the identity is not covariant under this transformation and indeed the exact seed (\ref{so3_exact}) is not palindromic.
This realizes the breakdown of the palindromic property for non-minuscule representations.

\subsection{$U(N)_k \times U(N)_{-k}$ ABJM theories}
The exact recursive property is not limited to the 3d $\mathcal{N}=4$ ADHM-type descriptions
but extends beautifully to the 3d $\mathcal{N}=6$ $U(N)_k \times U(N)_{-k}$ ABJM theories.
Here, the appropriate operators are the Wilson lines transforming in the bi-antisymmetric representations,
which naturally correspond to the minuscule weights of the product gauge group.
Specifically, the line defect index for the Wilson line $W_{(1^n);\overline{(1^n)}}$
transforming in the rank-$n$ antisymmetric representation of $U(N)_k$ and the conjugate rank-$n$ antisymmetric representation of $U(N)_{-k}$
satisfies the following relation:
\begin{align}
\label{R_uNuNABJM_W}
&
\langle \mathcal{W}_{(1^n);\overline{(1^n)}}\rangle^{U(N)_k\times U(N)_{-k}}(\mathfrak{t})
\nonumber\\
&=\frac{
\mathcal{I}^{U(N-n)_k\times U(N-n)_{-k}}(\mathfrak{t})
\mathcal{I}^{U(n)_k\times U(n)_{-k}}(\mathfrak{t})
}
{\mathcal{I}^{U(N)_k\times U(N)_{-k}}(\mathfrak{t})}
\langle \mathcal{W}_{(1^n);\overline{(1^n)}}\rangle^{U(n)_k\times U(n)_{-k}}(\mathfrak{t}).
\end{align}
This demonstrates that the hierarchical decoupling of minuscule line defects is a highly robust and universal feature of M2-brane SCFTs,
consistently bridging across theories whether they are dual to M-theory on $\mathbb{C}^4/\mathbb{Z}_k$.

For example, the two-point function of the Wilson lines in the bifundamental representation of the higher-rank ABJM theory
can be computed from the following general formula for the line defect index for the Wilson line of charge $(+1,-1)$
in the $U(1)_k\times U(1)_{-k}$ ABJM theory:
\begin{align}
\label{u1u1_W1-1}
\langle \mathcal{W}_{+1;-1}\rangle^{U(1)_k \times U(1)_{-k}}(\mathfrak{t})
&=\frac{\mathfrak{t}(1+\mathfrak{t}^{k-2})}{1+\mathfrak{t}^{k}}.
\end{align}

The biantisymmetric Wilson line defect index exhibits an exact $\mathfrak{t} \leftrightarrow \mathfrak{t}^{-1}$ symmetry,
as dictated by the recursion relation (\ref{R_uNuNABJM_W}) and the properties of the Coulomb indices
\begin{align}
\langle \mathcal{W}_{(1^n);\overline{(1^n)}}\rangle^{U(N)_k\times U(N)_{-k}}(\mathfrak{t}^{-1})
&=\langle \mathcal{W}_{(1^n);\overline{(1^n)}}\rangle^{U(N)_k\times U(N)_{-k}}(\mathfrak{t}).
\end{align}

\subsection{$SO(2N+\gamma)_{2k}\times USp(2N)_{-k}$ ABJ theories}

We also find the similar recursion relation for the 3d ${\cal N}=5$ $SO(2N)^{(\chi+)}_{2k}\times USp(2N)_{-k}$ ABJ theory and the 3d ${\cal N}=5$ $SO(2N+1)^{(\chi+)}_{2k}\times USp(2N)_{-k}$ ABJ theory.
For the Wilson line transforming in the fundamental representation in both $SO(2N)$ (or $SO(2N+1)$) and $USp(2N)$, we find
\begin{subequations}
\label{SO2Nchi+recreln}
\begin{align}
&\langle {\cal W}_{\ydiagram{1};\ydiagram{1}}\rangle^{SO(2N)^{(\chi+)}_{2k}\times USp(2N)_{-k}\text{ABJ}(C)}(\mathfrak{t})\nonumber \\
&=\frac{
{\cal I}^{SO(2N-2)^{(\chi+)}_{2k}\times USp(2N-2)_{-k}\text{ABJ}(C)}(\mathfrak{t}){\cal I}^{SO(3)^{(\chi+)}_{2k}\times USp(2)_{-k}\text{ABJ}(C)}(\mathfrak{t})
}{
{\cal I}^{SO(2N)^{(\chi+)}_{2k}\times USp(2N)_{-k}\text{ABJ}(C)}(\mathfrak{t})
}\nonumber \\
&\quad\times
\langle {\cal W}_{\ydiagram{1};\ydiagram{1}}\rangle^{SO(3)^{(\chi+)}_{2k}\times USp(2)_{-k}\text{ABJ}(C)}(\mathfrak{t}),\\
&\langle {\cal W}_{\ydiagram{1};\ydiagram{1}}\rangle^{SO(2N+1)^{(\chi+)}_{2k}\times USp(2N)_{-k}\text{ABJ}(C)}(\mathfrak{t})\nonumber \\
&=
\frac{
{\cal I}^{SO(2N-1)^{(\chi+)}_{2k}\times USp(2N-2)_{-k}\text{ABJ}(C)}(\mathfrak{t})
{\cal I}^{SO(3)^{(\chi+)}_{2k}\times USp(2)_{-k}\text{ABJ}(C)}(\mathfrak{t})
}{
{\cal I}^{SO(2N+1)^{(\chi+)}_{2k}\times USp(2N)_{-k}\text{ABJ}(C)}(\mathfrak{t})
}\nonumber \\
&\quad\times
\langle {\cal W}_{\ydiagram{1};\ydiagram{1}}\rangle^{SO(3)^{(\chi+)}_{2k}\times USp(2)_{-k}\text{ABJ}(C)}(\mathfrak{t}),
\end{align}
\end{subequations}
with $N\ge 2$ and
\begin{subequations}
\begin{align}
&{\cal I}^{SO(3)^{(\chi+)}_{2k}\times USp(2)_{-k}\text{ABJ}(C)}(\mathfrak{t})
=
\frac{
1 + \mathfrak{t}^{2k+2}
}{
(1 - \mathfrak{t}^4) (1 - \mathfrak{t}^{2 k})
},\\
&\langle {\cal W}_{\ydiagram{1};\ydiagram{1}}\rangle^{SO(3)^{(\chi\zeta)}_{2k}\times USp(2)_{-k}\text{ABJ}(C)}(\mathfrak{t})
=\frac{\mathfrak{t}(1 + \mathfrak{t}^2) (1 + \mathfrak{t}^{2 k-2})}{1 + \mathfrak{t}^{2k+2}}.
\end{align}
\end{subequations}
The recursion relations extend to the theories with the $\mathbb{Z}^{{\cal C}}_2$ symmetry gauged as
\begin{subequations}
\begin{align}
&\langle {\cal W}_{\ydiagram{1};\ydiagram{1}}\rangle^{O(2N)^{(+)}_{2k}\times USp(2N)_{-k}\text{ABJ}(C)}(\mathfrak{t})\nonumber \\
&=\frac{
{\cal I}^{O(2N-2)^{(+)}_{2k}\times USp(2N-2)_{-k}\text{ABJ}(C)}(\mathfrak{t}){\cal I}^{SO(3)^{(\chi+)}_{2k}\times USp(2)_{-k}\text{ABJ}(C)}(\mathfrak{t})
}{
{\cal I}^{O(2N)^{(+)}_{2k}\times USp(2N)_{-k}\text{ABJ}(C)}(\mathfrak{t})
}\nonumber \\
&\quad\times
\langle {\cal W}_{\ydiagram{1};\ydiagram{1}}\rangle^{SO(3)^{(\chi+)}_{2k}\times USp(2)_{-k}\text{ABJ}(C)}(\mathfrak{t}),\\
&\langle {\cal W}_{\ydiagram{1};\ydiagram{1}}\rangle^{O(2N+1)^{(+)}_{2k}\times USp(2N)_{-k}\text{ABJ}(C)}(\mathfrak{t})\nonumber \\
&=
\frac{
{\cal I}^{O(2N-1)^{(+)}_{2k}\times USp(2N-2)_{-k}\text{ABJ}(C)}(\mathfrak{t})
{\cal I}^{SO(3)^{(\chi+)}_{2k}\times USp(2)_{-k}\text{ABJ}(C)}(\mathfrak{t})
}{
{\cal I}^{O(2N+1)^{(+)}_{2k}\times USp(2N)_{-k}\text{ABJ}(C)}(\mathfrak{t})
}\nonumber \\
&\quad\times
\langle {\cal W}_{\ydiagram{1};\ydiagram{1}}\rangle^{SO(3)^{(\chi+)}_{2k}\times USp(2)_{-k}\text{ABJ}(C)}(\mathfrak{t}).
\end{align}
\end{subequations}
See
appendices \ref{app_OSp2Nchi1} and \ref{app_OSp2Nchim1} for the evidence for the recursion relations.

From the recursion relation it also follows that the bifundamental Wilson line defect indices are invariant under the exchange of $\mathfrak{t}\leftrightarrow \mathfrak{t}^{-1}$
\begin{subequations}
\begin{align}
&\langle {\cal W}_{\ydiagram{1};\ydiagram{1}}\rangle^{SO(2N+1)^{(\chi+)}_{2k}\times USp(2N)_{-k}\text{ABJ}(C)}(\mathfrak{t}^{-1})
=\langle {\cal W}_{\ydiagram{1};\ydiagram{1}}\rangle^{SO(2N+1)^{(\chi+)}_{2k}\times USp(2N)_{-k}\text{ABJ}(C)}(\mathfrak{t}),\\
&\langle {\cal W}_{\ydiagram{1};\ydiagram{1}}\rangle^{SO(2N)^{(\chi+)}_{2k}\times USp(2N)_{-k}\text{ABJ}(C)}(\mathfrak{t}^{-1})
=\langle {\cal W}_{\ydiagram{1};\ydiagram{1}}\rangle^{SO(2N)^{(\chi+)}_{2k}\times USp(2N)_{-k}\text{ABJ}(C)}(\mathfrak{t}),\\
&\langle {\cal W}_{\ydiagram{1};\ydiagram{1}}\rangle^{O(2N+1)^{(+)}_{2k}\times USp(2N)_{-k}\text{ABJ}(C)}(\mathfrak{t}^{-1})
=\langle {\cal W}_{\ydiagram{1};\ydiagram{1}}\rangle^{O(2N+1)^{(+)}_{2k}\times USp(2N)_{-k}\text{ABJ}(C)}(\mathfrak{t}),\\
&\langle {\cal W}_{\ydiagram{1};\ydiagram{1}}\rangle^{O(2N)^{(+)}_{2k}\times USp(2N)_{-k}\text{ABJ}(C)}(\mathfrak{t}^{-1})
=\langle {\cal W}_{\ydiagram{1};\ydiagram{1}}\rangle^{O(2N)^{(+)}_{2k}\times USp(2N)_{-k}\text{ABJ}(C)}(\mathfrak{t}).
\end{align}
\end{subequations}

\subsection{Factorization and brane decoupling}
\label{sec_factorization}
The exact factorization into lower rank indices without any subtraction is realized when $\mathcal{R}$ is a minuscule representation,
such as the rank-$n$ antisymmetric representation of $U(N)$ or the fundamental representations of $USp(2N)$ and $O(2N)$.\footnote{
Although we do not discuss in this paper, the recursion relations for the theories with a connected orthogonal gauge group
might also have a brane interpretation by an appropriate boundary condition on the discrete background gauge field.
}
By definition the weights of a minuscule representation form a single Weyl orbit with unit multiplicity, $m_{\mathcal{R}}(\mu)=1$.
Each such weight shifts a single component of the magnetic flux by one unit, leaving the remaining components and the residual Weyl symmetry intact
so that the shifted flux sectors reassemble into those of a decoupled product of lower rank theories.
This algebraic structure admits a simple interpretation in terms of branes.
In the Type IIA realization, the gauge theory arises on a stack of $N$ D2-branes and the supersymmetric Wilson line is engineered by
fundamental strings ending on the stack \cite{Assel:2015oxa}.
More specifically, the Wilson line in the rank-$n$ antisymmetric representation of $U(N)$ is realized by $n$ fundamental strings each of which ends on different $n$ D2-branes out of the $N$ D2-branes. The Wilson line in the fundamental representation of $USp(2N)$ or $O(2N)$ is given by a fundamental string ending on a D2-brane out of the $N$ D2-branes.
Upon lifting to M-theory, the D2-branes become $N$ coincident M2-branes, while the fundamental string, wrapping the M-theory circle,
becomes a probe M2-brane ending on the stack.
The exact factorization then demonstrates that a minuscule defect acts as a structurally pure probe which cleanly partitions the bulk BPS states,
splitting the stack into two decoupled sub-stacks of $n$ (or $1$) and $N-n$ (or $N-1$) M2-branes.
The seed of the recursion is the minimal rank theory, $U(n)$, $USp(2)$ or $SO(2)$, describing the peeled branes in isolation.

This uniform shift structure and the brane decoupling picture are put to a sharp test by the $O(2N+1)$ gauge theories,
for which, as shown in section \ref{sec_recursion_B},
the factorization holds only after the identity subtraction and involves the $O(3)$ theory as its seed.
Among the O2-plane variants realizing the orthosymplectic gauge theories on D2-branes \cite{Sethi:1998zk,Hanany:2000fq,Bergman:2020ifi},
the $\widetilde{\textrm{O2}}^{-}$-plane realizing $O(2N+1)$ is distinguished by a half D2-brane frozen at the fixed plane,
which extends the Chan-Paton space of the D2-branes from $2N$ to $2N+1$ dimensions and thereby
produces both the odd orthogonal gauge group and the zero weight of its vector representation.
The frozen half-brane is part of the definition of the orientifold variant rather than a removable object.
An O2-plane lifts to a pair of OM2-planes at the two fixed points of the M-theory circle
and the $\widetilde{\textrm{O2}}^{-}$-plane lifts to a pair of $\textrm{OM2}^{+}$-planes,
each carrying $+\frac{3}{16}$ units of M2-brane charge in place of the $-\frac{1}{16}$ units of the $\textrm{OM2}^{-}$-planes
arising from the $\textrm{O2}^{-}$-plane \cite{Sethi:1998zk,Hanany:2000fq}.
The half unit of D2-brane charge is precisely accounted for by this difference, $2\times\left(\frac{3}{16}+\frac{1}{16}\right)=\frac{1}{2}$.
The frozen half-brane does not lift to a separate stuck membrane but dissolves entirely into the discrete type of the fixed planes,
with no position moduli and no localized object left behind.
This is compatible with the above result that no modification of the inserted character can eliminate its effect.

Let us comment on the spinor Wilson line of $Spin(2N+1)$.
While the spinor representation is minuscule,
its weights $\frac{1}{2}(\pm e_1\pm e_2\cdots\pm e_N)$ shift all $N$ components of the magnetic flux simultaneously.
Such a line defect globally perturbs the entire eigenvalue configuration and leaves no unshifted stack
so that no decomposition into a peeled brane and a residual stack would be available.
It is also unclear whether theories with spinor gauge groups and Wilson lines in spinor representations admit an interpretation
as effective descriptions of specific M2-brane configurations.
If so, it would be highly desirable to identify the corresponding brane realization.
Although we do not pursue this question here, it represents an interesting direction for future investigation.

We close this section with a remark on the scope of these factorizations.
All of the recursion relations above, subtracted or not, hold for the \textit{Coulomb index} but not for the full supersymmetric index.
In the absence of line operators, the Coulomb index captures the monopole operators and the vector multiplet scalars,
in agreement with the Hilbert series of the geometry probed by the M2-branes.
In the full index, splitting the M2-brane stack activates light hypermultiplets representing open strings stretched
between the sub-stacks, whose residual interactions obstruct a clean tensor product factorization of the underlying Hilbert space.
The Coulomb limit would filter out these interpolating excitations so that the separated sub-stacks remain non-interacting, thereby enabling the exact factorizations.

\section{Large $N$ limit}
\label{sec_largeN}
In the large $N$ limit the recursion relations of section
\ref{sec_recursion} trivialize. The ratio of the bulk Coulomb indices
tends to one, and the normalized line defect index reduces to the
contribution of the peeled branes alone, i.e.~to the unnormalized
correlator of the seed theory.
In every case the resulting expression coincides with the Hilbert series
of the moduli space of the peeled M2-branes, providing a geometric
confirmation of the brane decoupling picture of section
\ref{sec_factorization}.
Note that the large $N$ expressions diverge as
$\mathfrak{t}\rightarrow1$, in accordance with the line defect Witten
indices of section \ref{sec_lineind} growing with the rank.
The limits $N\rightarrow\infty$ and $\mathfrak{t}\rightarrow1$ do not
commute.

\subsection{$U(\infty)$}
An immediate consequence of the recursion relation (\ref{R_uN_C_Wasym}) is that the large $N$ two-point function in the Coulomb limit reduces to the Coulomb index itself.
In other words, we have \cite{Hayashi:2024jof}
\begin{align}
\langle \mathcal{W}_{(1^k)} \mathcal{W}_{\overline{(1^k)}} \rangle^{\textrm{$U(\infty)$ ADHM-$[l]$} (C)}(z;\mathfrak{t})
&=\mathcal{I}^{\textrm{$U(k)$ ADHM-$[l]$} (C)}(z;\mathfrak{t}).
\end{align}
This is precisely the Coulomb index for the theory of $k$ M2-branes probing the $A_{l-1}$ singularity.
At infinite $N$ the normalized defect index retains no memory of the residual stack
so that the $k$ branes pulled off by the antisymmetric Wilson line behave as an isolated stack on $\mathbb{C}^2/\mathbb{Z}_l$.
For example, for the fundamental Wilson line with $k=1$ we have
\begin{align}
\label{largeN_uN_PE}
\langle \mathcal{W}_{\ydiagram{1}} \mathcal{W}_{\overline{\ydiagram{1}}} \rangle^{\textrm{$U(\infty)$ ADHM-$[l]$} (C)}(z;\mathfrak{t})
&=\frac{1-\mathfrak{t}^{2l}}{(1-\mathfrak{t}^2) (1-\mathfrak{t}^{l}z^{l}) (1-\mathfrak{t}^lz^{-l})}
\nonumber\\
&=\mathrm{PE}\left[
\mathfrak{t}^2 + \mathfrak{t}^l z^l + \mathfrak{t}^l z^{-l} - \mathfrak{t}^{2l}
\right],
\end{align}
where the single-particle index enumerates the generators and the relation of the chiral ring of $\mathbb{C}^2/\mathbb{Z}_l$.
The adjoint scalar $\varphi$ contributes $\mathfrak{t}^2$, the minimal monopole operators $v_{\pm}$ contribute $\mathfrak{t}^l z^{\pm l}$,
and the chiral ring relation $v_+ v_- = \varphi^l$ is encoded in the fermionic letter $-\mathfrak{t}^{2l}$.

\subsection{$USp(2\infty)$, $SO(2\infty)$, $SO(2\infty+1)$}
\label{sec_largeN_BCD}
We now derive new large $N$ results for gauge theories of type BCD.
For the orthosymplectic gauge theories the large $N$ two-point functions
of the fundamental Wilson lines take a single universal form.
From the recursion relations (\ref{R_usp_C_W1}) and (\ref{R_so2N_C_W1}) together with the seed expressions, we obtain
\begin{align}
\label{largeN_universal}
\langle \mathcal{W}_{\ydiagram{1}} \mathcal{W}_{\ydiagram{1}}\rangle^{G(\infty) (C)}(\mathfrak{t})
&=\frac{1+\mathfrak{t}^{2\Delta}}{(1-\mathfrak{t}^2) (1-\mathfrak{t}^{2\Delta})}
\end{align}
for $G(\infty)=USp(2\infty)$, $SO(2\infty)^{(\pm+)}$ and $O(2\infty)^{(+)}$,
where $\Delta$ is given by (\ref{uspDelta}) for $USp$ and by $\Delta=l+2\epsilon$ for the orthogonal series.

The odd orthogonal gauge theories join the same universal form despite the identity subtraction at finite $N$.
The recursion relation (\ref{R_so2N+1_C_W1}) then yields
\begin{align}
\langle \mathcal{W}_{\ydiagram{1}} \mathcal{W}_{\ydiagram{1}}\rangle^{SO(2\infty+1)(C)}(\mathfrak{t})
&=1+\mathcal{I}^{SO(3)(C)}(\mathfrak{t})
\left(\langle \mathcal{W}_{\ydiagram{1}} \mathcal{W}_{\ydiagram{1}}\rangle^{SO(3)(C)}(\mathfrak{t})-1\right)
\nonumber\\
&=1+\frac{(1+\mathfrak{t}^2)\left(\mathfrak{t}^2+2\mathfrak{t}^{2\Delta}-\mathfrak{t}^{2\Delta+2}\right)}{(1-\mathfrak{t}^4)(1-\mathfrak{t}^{2\Delta})}
=\frac{1+\mathfrak{t}^{2\Delta}}{(1-\mathfrak{t}^2)(1-\mathfrak{t}^{2\Delta})},
\end{align}
which coincides exactly with the universal expression (\ref{largeN_universal}).
The effect of the subtracted identity and of the frozen half-brane, being associated with $O(1)$ degrees of freedom,
becomes invisible at large $N$, and the flavor data enter only through the single combination $\Delta$, uniformly across all the orthosymplectic series and their discrete projections.

\subsection{Branes on the double cover}
The universal expression (\ref{largeN_universal}) admits a suggestive geometric interpretation.
Its plethystic decomposition
\begin{align}
\frac{1+\mathfrak{t}^{2\Delta}}{(1-\mathfrak{t}^2)(1-\mathfrak{t}^{2\Delta})}
&=\mathrm{PE}\left[\mathfrak{t}^2+2\mathfrak{t}^{2\Delta}-\mathfrak{t}^{4\Delta}\right]
\end{align}
is the Hilbert series of the $A_{2\Delta-1}$ singularity $\mathbb{C}^2/\mathbb{Z}_{2\Delta}$,
with the three letters corresponding to the generators $\varphi$, $V_{\pm}$ and the relation $V_+V_-=\varphi^{2\Delta}$ of its coordinate ring,
realized in the seed theory as the vector multiplet scalar and the minimal monopole operators.
The bulk Coulomb branch, on the other hand, is the symmetric product of the $D$-type singularity $\mathbb{C}^2/\widehat{D}_{\Delta}$.
Since the maximal cyclic subgroup
$\mathbb{Z}_{2\Delta}\subset\widehat{D}_{\Delta}$ has index two,
\begin{align}
\mathbb{C}^2/\widehat{D}_{\Delta}
=\left(\mathbb{C}^2/\mathbb{Z}_{2\Delta}\right)/\mathbb{Z}_2,
\end{align}
the single M2-brane peeled off by the pair of the fundamental Wilson lines explores
the $\mathbb{Z}_2$ double cover of the $D$-type singularity probed by the bulk M2-branes.
This is consistent with the appearance of the $SO(2)^{(++)}$ seed, rather than the $O(2)$ seed,
in the recursion relations (\ref{R_so2N_C_W1}) of section \ref{sec_recursion}.

\subsection{$U(\infty)_{k}\times U(\infty)_{-k}$}
In the large $N$ limit, the normalized Wilson line defect Coulomb index for the rank-$l$ bi-antisymmetric representation
becomes identical to the corresponding unnormalized index in the $U(l)_{k}\times U(l)_{-k}$ ABJM theory
\begin{align}
\langle \mathcal{W}_{(1^l);\overline{(1^l)}}\rangle^{U(\infty)_k\times U(\infty)_{-k}}(\mathfrak{t})
&=\langle W_{(1^l);\overline{(1^l)}}\rangle^{U(l)_k\times U(l)_{-k}}(\mathfrak{t}),
\end{align}
which follows from the recursion relation (\ref{R_uNuNABJM_W}).
The $l$ M2-branes pulled off by the bi-antisymmetric Wilson line behave as an isolated stack of $l$ M2-branes probing $\mathbb{C}^4/\mathbb{Z}_k$.

\subsection{$SO(2\infty+\gamma)_{2k}\times USp(2\infty)_{-k}$}
From the recursion relations \eqref{SO2Nchi+recreln} it follows that the normalized Wilson line defect indices for the bifundamental representation in the orthosymplectic ABJ theories become, both for $SO(2N+1)_{2k}^{(\chi+)}$ and $SO(2N)^{(\chi+)}_{2k}$, the unnormalized Wilson line defect index of $SO(3)^{(\chi+)}_{2k}\times USp(2)_{-k}$ ABJ theory
\begin{align}
&\langle {\cal W}_{\ydiagram{1};\ydiagram{1}}\rangle^{SO(2\infty)^{(\chi+)}_{2k}\times USp(2\infty)_{-k}\text{ABJ}(C)}=
\langle {\cal W}_{(\ydiagram{1};\ydiagram{1})}\rangle^{SO(2\infty+1)^{(\chi+)}_{2k}\times USp(2\infty)_{-k}\text{ABJ}(C)}\nonumber \\
&=
\frac{
\mathfrak{t} + \mathfrak{t}^{2 k - 1}
}{(1 - \mathfrak{t}^2) (1 - \mathfrak{t}^{2 k})
}.
\end{align}

\acknowledgments

The work of H.H.~is supported in part by JSPS KAKENHI Grant Number JP23K03396.
The work of T.N.~is supported by the Startup Funding no.~2302-SRFP-2024-0012 of Shanghai Institute for Mathematics and Interdisciplinary Sciences.
The work of T.O.~was supported by the Startup Funding no.~4007012317 of the Southeast University.

\appendix

\section{Factorized integrals in Coulomb limit}
\label{app_algorithms}

As we utilized in \cite{Hayashi:2024jof}, the $N$ dimensional integral over the gauge holonomy in the line defect indices in the 3d ADHM theory with $U(N)$ gauge group factorizes in the Coulomb limit to the lower dimensional integrations over the holonomies associated with the same values of the monopole charges.
In this appendix we explain the algorithm to calculate the Coulomb line defect indices via similar approach in the theories with $SO$ or $USp$ gauge group, as well as the ABJ(M) theories with $U(N)_k\times U(N)_{-k}$ or $SO(2N+\gamma)_{2k}\times USp(2N)_{-k}$ gauge group ($\gamma=0,1$).

\subsection{$SO(2N+1)^{(\chi\zeta)}$ and $SO(2N)^{(+\zeta)}$}
\label{app_SO2N+1algorithm}

First let us consider the theory with $SO(2N+\gamma)$ gauge group ($\gamma=0,1$) with one hypermultiplet in symmetric or antisymmetric representation and $l$ fundamental hypermultiplets, that is $2l$ fundamental half-hypermultiplets, which is what the label $[2l]$ counts.
The superconformal index with the insertion of single variable functions $\prod_{i=1}^Nf(s_i)$ is given by \cite{Hayashi:2022ldo}
\begin{align}
&\biggl\langle \prod_{i=1}^Nf(s_i)\biggr\rangle^{SO(2N+\gamma)^{(\chi\zeta)}+\text{(A/S)-}[2l]}(t,x,y_{\alpha};q)\nonumber \\
&
=\frac{1}{2^{N+\gamma-1}N!}\frac{(q^{\frac12}t^2;q)_{\infty}^N}
{(q^{\frac12}t^{-2};q)_{\infty}^N}
\sum_{m_1,\cdots, m_{N}\in \mathbb{Z}}\oint \prod_{i=1}^N \frac{ds_i}{2\pi is_i}
f(s_i)
\prod_{i=1}^N
\prod_\pm
(1-\chi q^{\frac{|m_i|}{2}}s_i^{\pm 1})^{\gamma}\nonumber \\
&\quad\times
\prod_{i<j}
\prod_\pm
(1-q^{\frac{|m_i -m_j|}{2}} s_i^{\pm 1}s_j^{\mp 1})
(1-q^{\frac{|m_i +m_j|}{2}} s_i^{\pm 1}s_j^{\pm 1})
\left[\prod_{i=1}^N
\prod_\pm
\frac{(\chi q^{\frac12+\frac{|m_i|}{2}}t^2 s_i^{\mp 1};q)_{\infty}}
{(\chi q^{\frac12+\frac{|m_i|}{2}}t^{-2} s_i^{\pm 1};q)_{\infty}}
\right]^{\gamma}\nonumber \\
&\quad\times
\prod_{i<j}
\prod_\pm
\frac{(q^{\frac12+\frac{|m_i-m_j|}{2}}t^2 s_i^{\mp 1}s_j^{\pm 1};q)_{\infty}}
{(q^{\frac12+\frac{|m_i-m_j|}{2}}t^{-2} s_i^{\pm 1}s_j^{\mp 1};q)_{\infty}}
\frac{(q^{\frac12+\frac{|m_i+m_j|}{2}}t^2 s_i^{\mp 1}s_j^{\mp 1};q)_{\infty}}
{(q^{\frac12+\frac{|m_i+m_j|}{2}}t^{-2} s_i^{\pm 1}s_j^{\pm 1};q)_{\infty}}
\prod_\pm
\frac{(q^{\frac34}t^{-1}x^{\mp 1};q)_{\infty}^{N+\gamma\epsilon}}
{(q^{\frac14}tx^{\pm 1};q)_{\infty}^{N+\gamma\epsilon}}\nonumber \\
&\quad\times
\prod_{i<j}
\prod_\pm
\frac{(q^{\frac34+\frac{|m_i-m_j|}{2}} t^{-1}s_i^{\mp 1} s_j^{\pm 1}x^{\mp 1};q)_{\infty}}
{(q^{\frac14+\frac{|m_i-m_j|}{2}} ts_i^{\pm 1} s_j^{\mp 1}x^{\pm 1};q)_{\infty}}
\frac{(q^{\frac34+\frac{|m_i-m_j|}{2}} t^{-1}s_i^{\mp 1} s_j^{\pm 1}x^{\pm 1};q)_{\infty}}
{(q^{\frac14+\frac{|m_i-m_j|}{2}} ts_i^{\pm 1} s_j^{\mp 1}x^{\mp 1};q)_{\infty}}
\nonumber\\
&\quad\times
\prod_{i<j}
\prod_\pm
\frac{(q^{\frac34+\frac{|m_i+m_j|}{2}} t^{-1}s_i^{\mp 1} s_j^{\mp 1}x^{\mp 1};q)_{\infty}}
{(q^{\frac14+\frac{|m_i+m_j|}{2}} ts_i^{\pm 1} s_j^{\pm 1}x^{\pm 1};q)_{\infty}}
\frac{(q^{\frac34+\frac{|m_i+m_j|}{2}} t^{-1}s_i^{\mp 1} s_j^{\mp 1}x^{\pm 1};q)_{\infty}}
{(q^{\frac14+\frac{|m_i+m_j|}{2}} ts_i^{\pm 1} s_j^{\pm 1}x^{\mp 1};q)_{\infty}}
\nonumber\\
&\quad\times
\left[
\prod_{i=1}^N
\prod_\pm
\frac{(\chi q^{\frac34+\frac{|m_i|}{2}} t^{-1}s_i^{\mp 1}x^{\mp 1};q)_{\infty}}
{(\chi q^{\frac14+\frac{|m_i|}{2}} ts_i^{\pm 1}x^{\pm 1};q)_{\infty}}
\frac{(\chi q^{\frac34+\frac{|m_i|}{2}} t^{-1}s_i^{\mp 1}x^{\pm 1};q)_{\infty}}
{(\chi q^{\frac14+\frac{|m_i|}{2}} ts_i^{\pm 1}x^{\mp 1};q)_{\infty}}
\right]^{\gamma}
\nonumber\\
&\quad\times
\left[
\prod_{i=1}^{N}
\prod_\pm
\frac{(q^{\frac34+|m_i|}t^{-1}s_i^{\mp 2}x^{\mp 1};q)_{\infty}}
{(q^{\frac14+|m_i|}t s_i^{\pm 2}x^{\pm 1};q)_{\infty}}
\frac{(q^{\frac34+|m_i|}t^{-1}s_i^{\mp 2}x^{\pm 1};q)_{\infty}}
{(q^{{\frac14}+|m_i|}t s_i^{\pm 2}x^{\mp 1};q)_{\infty}}
\right]^{\epsilon}
\nonumber\\
&\quad\times
\prod_{i=1}^N
\prod_{\alpha=1}^{2l}
\prod_\pm
\frac{(q^{\frac34+\frac{|m_i|}{2}} t^{-1}s_i^{\mp 1}y_{\alpha}^{-1};q)_{\infty}}
{(q^{\frac14+\frac{|m_i|}{2}} ts_i^{\pm 1}y_{\alpha};q)_{\infty}}
\left[
\prod_{\alpha=1}^{2l}
\frac{(\chi q^{\frac34}t^{-1}y_{\alpha}^{-1};q)_{\infty}}
{(\chi q^{\frac14}ty_{\alpha};q)_{\infty}}
\right]^{\gamma}
\nonumber\\
&\quad\times
q^{\frac{\Delta}{2}\sum_{i=1}^{N}|m_i|}
t^{-2\Delta \sum_{i=1}^{N}|m_i|}
\zeta^{\sum_{i=1}^N m_i}
\label{SOoddandevenwithchi1fullindex}
\end{align}
with $\epsilon=0$ (resp.~$\epsilon=1$) for the hypermultiplet with fugacity $x$ being in antisymmetric (resp.~symmetric) representation, and the flavor fugacities $y_\alpha$ satisfying $y_{l+\alpha}=y_\alpha^{-1}$.
Here $\zeta=\pm 1$ and $\chi=\pm 1$ are discrete fugacities for $\mathbb{Z}^{{\cal M}}_2$ and $\mathbb{Z}^{{\cal C}}_2$.
In this subsection we consider only the cases except $(\gamma,\chi)=(0,-1)$, which is treated separately in section \ref{app_SOevenchim1}.
Here $\Delta$ is defined by
\begin{align}
\Delta=l+2\epsilon.
\label{DeltaSO}
\end{align}
Here we consider the insertion of the product of single variable function $\prod_i f(s_i)$ rather than the characters $\chi_\lambda(s)$, which directly give the Wilson line defect indices, for the purpose of stating our algorithm in a simple manner.

First of all, let us notice that the integrand of \eqref{SOoddandevenwithchi1fullindex} for each fixed monopole charge is invariant under the replacement $(m_i,s_i)\rightarrow (-m_i,s_i^{-1})$ for any single $i$ if $f(s)$ satisfies $f(s)=f(s^{-1})$, which is always the case for our purpose for calculating the line defect indices through the insertion of $\prod_if(s_i)$.
Therefore, we can replace the summation over $m_i\in\mathbb{Z}$ to $m_i\ge 0$ as
\begin{align}
\sum_{m_i\in\mathbb{Z}}(\cdots)\rightarrow
\sum_{m_i=0}(\cdots)+2\sum_{m_i=1}^\infty(\cdots)
=
\sum_{m_i=0}^\infty 2^{1-\delta_{m_i,0}}(\cdots).
\label{Z2trick}
\end{align}
We can use the same trick for the case $(\gamma,\chi)=(0,-1)$ and in the theory with $USp(2N)$ gauge group.

In the Coulomb limit $q,t\rightarrow 0$ with $\mathfrak{t}=q^{\frac{1}{4}}t^{-1}$ fixed, the index reduces to
\begin{align}
&\biggl\langle \prod_{i=1}^Nf(s_i)\biggr\rangle^{SO(2N+\gamma)^{(\chi\zeta)}+\text{(A/S)-}[2l](C)}(\mathfrak{t})\nonumber \\
&
=\frac{1}{2^{N+\gamma-1}N!}
\frac{1
}{
(1-\mathfrak{t}^2)^N
}
\sum_{m_1,\cdots,m_N\ge 0}\prod_{i=1}^N 2^{1-\delta_{m_i,0}}\oint \prod_{i=1}^N \frac{ds_i}{2\pi is_i}
f(s_i)
\prod_{\substack{i=1\\ (m_i=0)}}^N
\prod_\pm
\left[
\frac{
1-\chi s_i^{\pm 1}
}{
1-\chi\mathfrak{t}^2s_i^{\pm 1}
}
\right]^{\gamma}\nonumber \\
&\quad\times \prod_{\substack{i<j\\ (m_i=m_j)}}^N
\prod_\pm
\frac{
1-s_i^{\pm 1}s_j^{\mp 1}
}{
1-\mathfrak{t}^2 s_i^{\pm 1}s_j^{\mp 1}
}
\prod_{\substack{i<j\\ (m_i=-m_j)}}^N
\prod_\pm
\frac{
1-s_i^{\pm 1}s_j^{\pm 1}
}{
1-\mathfrak{t}^2 s_i^{\pm 1}s_j^{\pm 1}
}
\mathfrak{t}^{2\Delta \sum_{i=1}^N|m_i|}
\zeta^{\sum_{i=1}^N m_i}.
\label{SOoddC1}
\end{align}
Note that the condition $m_i=-m_j$ is satisfied only when $m_i=m_j=0$ due to our restriction $m_i\ge 0$.
Namely, for each $\{m_i\}$, the $N$-dimensional integral factorizes into the lower dimensional integrals each of which contains only the integration variables associated with $m_i=0$, or those associated with $m_i=m\ge 1$.
By using the multiplicities $\nu_m=\#\{i|m_i=m\}$ to label the monopole charges, and taking into account that there are $\frac{N!}{\prod_m \nu_m!}$ different choices of the monopole charges which correspond to the same set of multiplicities, we can rewrite the Coulomb line defect index \eqref{SOoddC1} as
\begin{align}
&\biggl\langle \prod_{i=1}^Nf(s_i)\biggr\rangle^{SO(2N+\gamma)^{(\chi\zeta)}+\text{(A/S)-}[2l](C)}(\mathfrak{t})\nonumber \\
&=\frac{1}{2^{\gamma-1}}\sum_{\substack{\nu_0,\nu_m=0\\ (\nu_0+\sum_{m=1}^\infty \nu_m=N)}}^\infty (\zeta\mathfrak{t}^{2\Delta})^{\sum_{m=1}^\infty m \nu_m}
\tilde{\cal I}^{(0)}_{B,\gamma,\chi}[f;\nu_0]\prod_{m=1}^\infty\tilde{\cal I}^{(1)}[f;\nu_m],
\label{SOoddC2}
\end{align}
with
\begin{subequations}
\begin{align}
&\tilde{\cal I}_{B,\gamma,\chi}^{(0)}[f;\nu]\nonumber \\
&=\frac{1}{2^\nu(1-\mathfrak{t}^2)^\nu\nu!}\oint\prod_{i=1}^\nu\frac{d\sigma_i}{2\pi i\sigma_i}f(\sigma_i)\prod_\pm\left[\frac{1-\chi\sigma_i^{\pm 1}}{1-\mathfrak{t}^2\chi\sigma_i^{\pm 1}}\right]^\gamma\nonumber \\
&\quad\times
\prod_{i<j}^\nu\prod_\pm\frac{
(1-\sigma_i^{\pm 1}\sigma_j^{\mp 1})(1-\sigma_i^{\pm 1}\sigma_j^{\pm 1})
}{
(1-\mathfrak{t}^2\sigma_i^{\pm 1}\sigma_j^{\mp 1})(1-\mathfrak{t}^2\sigma_i^{\pm 1}\sigma_j^{\pm 1})
},\label{calI0SOodd}\\
&\tilde{\cal I}^{(1)}[f;\nu]
=\frac{1}{(1-\mathfrak{t}^2)^{\nu} \nu !}\oint
\prod_{i=1}^{\nu}\frac{d\sigma_i}{2\pi i\sigma_i}f(\sigma_i)
\prod_{i<j}^{\nu}\prod_\pm
\frac{
1-\sigma_i^{\pm 1} \sigma_j^{\mp 1}
}{
1-\mathfrak{t}^2 \sigma_i^{\pm 1} \sigma_j^{\mp 1}
}.
\label{calI1SOodd}
\end{align}
\end{subequations}
Note that $\tilde{\cal I}^{(1)}[f;\nu]$ can be evaluated systematically through the Fermi gas formalism \cite{Hayashi:2024jof}.

The distribution of the values of $\nu_m$ can be classified by the integer partition $\lambda$ with $|\lambda|\le N$ as
\begin{align}
&\lambda=(\underbrace{x_1,\cdots,x_1}_{\kappa_1},\underbrace{x_2,\cdots,x_2}_{\kappa_2},\cdots,\underbrace{x_n,\cdots,x_n}_{\kappa_n})\nonumber \\
&\leftrightarrow
\nu_0=N-|\lambda|,\quad
\nu_{m_1^{(1)}}=\nu_{m_2^{(1)}}= \cdots=\nu_{m_{\kappa_1}^{(1)}}=x_1,\quad
\nu_{m_1^{(2)}}=\nu_{m_2^{(2)}}= \cdots=\nu_{m_{\kappa_2}^{(2)}}=x_2,\nonumber \\
&\quad \cdots,\quad
\nu_{m_1^{(n)}}=\nu_{m_2^{(n)}}= \cdots=\nu_{m_{\kappa_n}^{(n)}}=x_n,
\end{align}
with $m^{(a)}_i\ge 1$, $m^{(a)}_i\neq m^{(b)}_{j}$ for $(a,i)\neq (b,j)$, and $m^{(a)}_i<m^{(a)}_{i+1}$.
For each $\lambda$, the summation over $\nu_m$'s in \eqref{SOoddC2} reduces to the summation over $m^{(a)}_i$'s, of which the factor $\tilde{\cal I}^{(0)}[f;\nu_0]\prod_{m=1}^\infty\tilde{\cal I}^{(1)}[f;\nu_m]$ is independent.
As a result, we obtain
\begin{align}
&\biggl\langle \prod_{i=1}^Nf(s_i)\biggr\rangle^{SO(2N+\gamma)^{(\chi\zeta)}+\text{(A/S)-}[2l](C)}\nonumber(\mathfrak{t}) \\
&=
\frac{1}{2^{\gamma-1}}
\sum_{\kappa=0}^N
\tilde{\cal I}_{\gamma,\chi}^{(0)}[f;N-\kappa]
\sum_{\substack{\lambda\\ (|\lambda|=\kappa)}}
G_\lambda(\zeta,\mathfrak{t})
\prod_{i\ge 1}
\frac{
\tilde{\cal I}^{(1)}[f;i]^{m_i(\lambda)}
}{
m_i(\lambda)!},
\label{SOoddCfiniteN}
\end{align}
where $m_i(\lambda)=\#\{j|\lambda_j=i\}$.
Here $G_\lambda(\zeta,\mathfrak{t})$ is the form factor which arises from the summation over $m^{(a)}_i$'s
\begin{align}
\frac{G_\lambda(\zeta,\mathfrak{t})}{\prod_im_i(\lambda)!}=\sum_{\substack{
1\le m^{(1)}_1<\cdots<m^{(1)}_{\kappa_1},\cdots,\\
1\le m^{(n)}_1<\cdots<m^{(n)}_{\kappa_n}\\
(m^{(a)}_i\neq m^{(b)}_j\text{ if }(a,i)\neq (b,j))}}
(\zeta \mathfrak{t}^{2\Delta})^{\sum_{a=1}^n\sum_{i=1}^{\kappa_a} m^{(a)}_i},
\label{formfactorG}
\end{align}
which can be calculated systematically through the recursion relations obtained by reorganizing the nested sums \cite{Hayashi:2024jof}
\begin{subequations}
\begin{align}
&G_{(\lambda_1)}(\zeta,\mathfrak{t})=\sum_{m=1}^\infty (\zeta\mathfrak{t}^{2\Delta})^{m \lambda_1}=\frac{(\zeta\mathfrak{t}^{2\Delta})^{\lambda_1}}{1-(\zeta\mathfrak{t}^{2\Delta})^{\lambda_1}},\\
&G_{(\lambda_1,\cdots,\lambda_\ell)}(\zeta,\mathfrak{t})=G_{(\lambda_1)}(\zeta,\mathfrak{t})G_{(\lambda_2,\cdots,\lambda_{\ell})}(\zeta,\mathfrak{t})-\sum_{i=2}^{\ell}G_{(\lambda_i+\lambda_1,\lambda_2,\cdots,\hat{\lambda}_i,\cdots,\lambda_\ell)}(\zeta,\mathfrak{t}).
\end{align}
\end{subequations}

We also note that the decomposition of the index with the insertion of $\prod_if(s_i)$ \eqref{SOoddC2} implies that the grand canonical sum with respect to the rank $N$
\begin{align}
\Xi_{B,\gamma,\chi}[f;u]=\sum_{N=0}^\infty u^N\biggl\langle \prod_{i=1}^Nf(s_i)\biggr\rangle^{SO(2N+\gamma)^{(\chi\zeta)}+\text{(A/S)-}[2l](C)}(\mathfrak{t})
\end{align}
factorizes into the contributions from different values of the monopole charges $m$ as
\begin{align}
\Xi_{B,\gamma,\chi}[f;u]=\frac{1}{2^{\gamma-1}}\tilde{\Xi}^{(0)}_{B,\gamma,\chi}[f;u]\prod_{m=1}^\infty\tilde{\Xi}^{(1)}[f;\zeta^m\mathfrak{t}^{2m\Delta} u],
\label{XiBfactorized}
\end{align}
where
\begin{align}
\tilde{\Xi}_{B,\gamma,\chi}^{(0)}[f;u]=\sum_{\nu=0}^\infty u^\nu \tilde{\cal I}^{(0)}_{B,\gamma,\chi}[f;\nu],\quad
\tilde{\Xi}^{(1)}[f;u]=\sum_{\nu=0}^\infty u^\nu \tilde{\cal I}^{(1)}[f;\nu].\label{Xitilde(1)}
\end{align}
As is the case for the ADHM theory with $U(N)$ gauge group \cite{Hayashi:2022ldo,Hayashi:2024jof}, once we figure out $\tilde{\cal I}^{(0)}_{B,\gamma,\chi}[f;\nu]$ and $\tilde{\cal I}^{(1)}[f;\nu]$ for general $\nu$, the grand canonical sum is useful for obtaining the large $N$ limit of the line defect indices as well as their finite $N$ corrections, which albeit we do not analyze in this article.

\subsubsection{Fundamental Wilson lines in $SO(2N+1)^{(\chi\zeta)}$}
\label{app_SO2N+1fund}

Let us apply the method explained above to the one- and two-point functions of the fundamental Wilson line in the $SO(2N+1)$ theory.
These line defect indices are calculated by the insertion of the characters $\chi^{\mathfrak{so}(2N+1)}_{\ydiagram{1}}(s,\chi)=\chi+\sum_i(s_i+s_i^{-1})$ and $\chi^{\mathfrak{so}(2N+1)}_{\ydiagram{1}}(s,\chi)^2$, where $\chi=\pm 1$ is the $\mathbb{Z}_2^{{\cal C}}$ fugacity, which can be extracted by inserting $\prod_if(s_i)$ with $f(s)=(1+a(s+s^{-1}))(1+b(s+s^{-1}))$ as
\begin{subequations}
\label{SOADHMfunnd1ptand2ptfromgeneratingfcn}
\begin{align}
\langle W_{\ydiagram{1}}\rangle^{SO(2N+1)^{(\chi\zeta)}+\text{(A/S)-}[2l](C)}(\mathfrak{t})
&=
\chi\biggl\langle \prod_if(s_i)\biggr\rangle^{SO(2N+1)^{(\chi\zeta)}+\text{(A/S)-}[2l](C)}(\mathfrak{t})\Bigr|_{a^0b^0}\nonumber \\
&\quad
+\biggl\langle \prod_if(s_i)\biggr\rangle^{SO(2N+1)^{(\chi\zeta)}+\text{(A/S)-}[2l](C)}(\mathfrak{t})\Bigr|_{a^1b^0},\\
\langle W_{\ydiagram{1}} W_{\ydiagram{1}}\rangle^{SO(2N+1)^{(\chi\zeta)}+\text{(A/S)-}[2l](C)}(\mathfrak{t})
&=
\biggl\langle \prod_if(s_i)\biggr\rangle^{SO(2N+1)^{(\chi\zeta)}+\text{(A/S)-}[2l](C)}(\mathfrak{t})\Bigr|_{a^0b^0}\nonumber \\
&\quad
+2\chi\biggl\langle \prod_if(s_i)\biggr\rangle^{SO(2N+1)^{(\chi\zeta)}+\text{(A/S)-}[2l](C)}(\mathfrak{t})\Bigr|_{a^1b^0}\nonumber \\
&\quad
+\biggl\langle \prod_if(s_i)\biggr\rangle^{SO(2N+1)^{(\chi\zeta)}+\text{(A/S)-}[2l](C)}(\mathfrak{t})\Bigr|_{a^1b^1}.
\end{align}
\end{subequations}
Here $(\cdots)|_{a^ib^j}$ stands for the coefficient of $a^ib^j$ in the small $a,b$-expansion of the expression $(\cdots)$.

For this choice of $f(s)$, we propose the following expressions for $\tilde{\cal I}^{(0)}_{B,1,\chi}[f;\nu]$ \eqref{calI0SOodd} and $\tilde{\cal I}^{(1)}[f;\nu]$ \eqref{calI1SOodd} for general $\nu$ and to the order ${\cal O}(a^1,b^1)$
\begin{subequations}
\begin{align}
&\tilde{\cal I}^{(0)}_{B,1,\chi}[(1+a(s+s^{-1}))(1+b(s+s^{-1}));\nu]\nonumber \\
&=
\frac{1}{(\mathfrak{t}^4;\mathfrak{t}^4)_\nu}\biggl[1-\chi(a+b)(1-\mathfrak{t}^{2\nu})+ab\Bigl(2(1-\mathfrak{t}^{2\nu})+\frac{\mathfrak{t}^2(1-\mathfrak{t}^{4\nu})}{1-\mathfrak{t}^2}\Bigr)\biggr]+{\cal O}(a^2,b^2),\label{SOoddcalI0fundgeneralnu} \\
&\tilde{\cal I}^{(1)}[(1+a(s+s^{-1}))(1+b(s+s^{-1}));\nu]=\frac{1}{(\mathfrak{t}^2;\mathfrak{t}^2)_\nu}\Bigl(1+\frac{2ab(1-\mathfrak{t}^{2\nu})}{1-\mathfrak{t}^2}\Bigr)+{\cal O}(a^2,b^2),
\label{calI1fundgeneralnu}
\end{align}
\end{subequations}
which we have confirmed for $\nu\le 4$.
We can evaluate the grand canonical sum of the indices \eqref{XiBfactorized} by plugging \eqref{SOoddcalI0fundgeneralnu} and \eqref{calI1fundgeneralnu} into \eqref{Xitilde(1)}, which read
\begin{subequations}
\begin{align}
&\tilde{\Xi}^{(0)}_{B,1,\chi}[(1+a(s+s^{-1}))(1+b(s+s^{-1}));u]\nonumber \\
&=\frac{1}{(u;\mathfrak{t}^4)_\infty}\biggl[1+\chi(a+b)\Bigl(-1+\frac{(u;\mathfrak{t}^4)_\infty}{(u\mathfrak{t}^2;\mathfrak{t}^4)_\infty}\Bigr)+ab\Bigl(2-\frac{2(u;\mathfrak{t}^4)_\infty}{(u\mathfrak{t}^2;\mathfrak{t}^4)_\infty}+\frac{u\mathfrak{t}^2}{1-\mathfrak{t}^2}\Bigr)+{\cal O}(a^2,b^2)\biggr],\\
&\tilde{\Xi}^{(1)}[(1+a(s+s^{-1}))(1+b(s+s^{-1}));u]=\frac{1}{(u;\mathfrak{t}^2)_\infty}\biggl[1+\frac{2abu}{1-\mathfrak{t}^2}+{\cal O}(a^2,b^2)\biggr],
\end{align}
\end{subequations}
as
\begin{align}
&\Xi_{B,1,\chi}[(1+a(s+s^{-1}))(1+b(s+s^{-1}));u]\nonumber \\
&=
\tilde{\Xi}^{(0)}_{B,1,\chi}[(1+a(s+s^{-1}))(1+b(s+s^{-1}));u]\nonumber \\
&\quad\times
\prod_{m=1}^\infty \tilde{\Xi}^{(1)}[(1+a(s+s^{-1}))(1+b(s+s^{-1}));\zeta^m\mathfrak{t}^{2m\Delta} u]\nonumber \\
&=
\Xi_{B,1,\chi}[1;u]
\biggl[1
+\chi(a+b)
\Bigl(-1
+\frac{(u;\mathfrak{t}^4)_\infty}{(u\mathfrak{t}^2;\mathfrak{t}^4)_\infty}
\Bigr)\nonumber \\
&\quad
+ab
\Bigl(
2
-\frac{2(u;\mathfrak{t}^4)_\infty}{(u\mathfrak{t}^2;\mathfrak{t}^4)_\infty}
+\frac{u\mathfrak{t}^2}{1-\mathfrak{t}^2}
+
\frac{2u\zeta\mathfrak{t}^{2\Delta}}{(1-\mathfrak{t}^2)(1-\zeta\mathfrak{t}^{2\Delta})}
\Bigr)
+{\cal O}(a^2,b^2)\biggr].
\label{SO2N+1fundGC}
\end{align}
Here
\begin{align}
\Xi_{B,1,\chi}[1;u]
=
\sum_{N=0}^\infty u^N{\cal I}^{SO(2N+1)^{(\chi\zeta)}+\text{(A/S)-}[2l](C)}(\mathfrak{t})
=\frac{1}{(u;\mathfrak{t}^4)_\infty}
\prod_{m=1}^\infty \frac{1}{(\zeta^m \mathfrak{t}^{2m\Delta}u;\mathfrak{t}^2)_\infty}
\label{ISOoddGC}
\end{align}
is the grand canonical sum of the Coulomb indices without Wilson lines.

Combining the coefficients of $a^0b^0$, $a^1b^0$ and $a^1b^1$ in \eqref{SO2N+1fundGC} according to \eqref{SOADHMfunnd1ptand2ptfromgeneratingfcn}, we find the grand canonical sum of the one- and the two-point function is given as
\begin{align}
&\sum_{N=0}^\infty u^N \langle W_{\ydiagram{1}} \rangle^{SO(2N+1)^{(\chi\zeta)}+\text{(A/S)-}[2l](C)}(\mathfrak{t})=\frac{\chi}{(u\mathfrak{t}^2;\mathfrak{t}^4)_\infty}\prod_{m=1}^\infty\frac{1}{(\zeta^m\mathfrak{t}^{2m\Delta}u;\mathfrak{t}^2)_\infty},\\
&\sum_{N=0}^\infty u^N \langle W_{\ydiagram{1}} W_{\ydiagram{1}}\rangle^{SO(2N+1)^{(\chi\zeta)}+\text{(A/S)-}[2l](C)}(\mathfrak{t})=
\Xi_{B,1,\chi}[f=1;u]\Bigl(1+\frac{u(\mathfrak{t}^2+\zeta \mathfrak{t}^{2\Delta} (2-\mathfrak{t}^2))}{(1-\mathfrak{t}^2)(1-\zeta \mathfrak{t}^{2\Delta})}
\Bigr).
\end{align}
The second formula implies that the two-point function satisfies the recursion relation as claimed in \eqref{R_so2N+1_C_W1_unnorm}
\begin{align}
&\langle W_{\ydiagram{1}} W_{\ydiagram{1}}\rangle^{SO(2N+1)^{(\chi\zeta)}+\text{(A/S)-}[2l](C)}(\mathfrak{t})-{\cal I}^{SO(2N+1)^{(\chi\zeta)}+\text{(A/S)-}[2l](C)}(\mathfrak{t})\nonumber \\
&=
{\cal I}^{SO(2N-1)^{(\chi\zeta)}+\text{(A/S)-}[2l](C)}(\mathfrak{t})(
\langle W_{\ydiagram{1}} W_{\ydiagram{1}}\rangle^{SO(3)^{(\chi\zeta)}+\text{(A/S)-}[2l](C)}(\mathfrak{t})-{\cal I}^{SO(3)^{(\chi\zeta)}+\text{(A/S)-}[2l](C)}(\mathfrak{t})
),
\label{SO2N+1recrelnappendix}
\end{align}
with
\begin{subequations}
\begin{align}
&{\cal I}^{SO(3)^{(\chi\zeta)}+\text{(A/S)-}[2l](C)}(\mathfrak{t})=
\frac{1+\zeta \mathfrak{t}^{2\Delta+2}}{(1-\mathfrak{t}^4)(1-\zeta\mathfrak{t}^{2\Delta})},\\
&\langle W_{\ydiagram{1}} W_{\ydiagram{1}}\rangle^{SO(3)^{(\chi\zeta)}+\text{(A/S)-}[2l](C)}(\mathfrak{t})=
\frac{1+\mathfrak{t}^2+\mathfrak{t}^4+\zeta\mathfrak{t}^{2\Delta}(2+2\mathfrak{t}^2-\mathfrak{t}^4)}{(1-\mathfrak{t}^4)(1-\zeta\mathfrak{t}^{2\Delta})}.
\end{align}
\end{subequations}

In particular, for $\zeta=1$ or the theory with $\mathbb{Z}^{{\cal M}}_2$ gauged, we find
\begin{subequations}
\begin{align}
&\lim_{\mathfrak{t}\rightarrow 1}
\langle {\cal W}_{\ydiagram{1}}\rangle^{SO(2N+1)^{(\chi+)}+\text{(A/S)-}[2l](C)}(\mathfrak{t})
=\lim_{\mathfrak{t}\rightarrow 1}
\langle {\cal W}_{\ydiagram{1}}\rangle^{Spin(2N+1)^{(\chi)}+\text{(A/S)-}[2l](C)}(\mathfrak{t})
=
\chi,\\
&\lim_{\mathfrak{t}\rightarrow 1}
\langle {\cal W}_{\ydiagram{1}} {\cal W}_{\ydiagram{1}}\rangle^{SO(2N+1)^{(\chi+)}+\text{(A/S)-}[2l](C)}(\mathfrak{t})
=
\lim_{\mathfrak{t}\rightarrow 1}
\langle {\cal W}_{\ydiagram{1}} {\cal W}_{\ydiagram{1}}\rangle^{Spin(2N+1)^{(\chi)}+\text{(A/S)-}[2l](C)}(\mathfrak{t})\nonumber \\
&
=2N+1,
\end{align}
\end{subequations}
which are consistent with our conjecture \eqref{Witten_indC_conj_int}.
Here we have defined
\begin{align}
\langle {\cal W}_{{\cal R}_1} \cdots {\cal W}_{{\cal R}_k}\rangle^{Spin(2N+\gamma)^{(\chi)}+\text{(A/S)-}[2l](C)}(\mathfrak{t})
=
\frac{
\langle W_{{\cal R}_1} \cdots W_{{\cal R}_k}\rangle^{Spin(2N+\gamma)^{(\chi)}+\text{(A/S)-}[2l](C)}(\mathfrak{t})
}{
{\cal I}^{Spin(2N+\gamma)^{(\chi)}+\text{(A/S)-}[2l](C)}(\mathfrak{t})
}
\end{align}
with
\begin{subequations}
\begin{align}
&{\cal I}^{Spin(2N+\gamma)^{(\chi)}+\text{(A/S)-}[2l](C)}(\mathfrak{t})=\frac{1}{2}\sum_{\zeta=\pm 1}{\cal I}^{SO(2N+\gamma)^{(\chi\zeta)}+\text{(A/S)-}[2l](C)}(\mathfrak{t}),\\
&\langle W_{{\cal R}_1}\cdots W_{{\cal R}_k}\rangle^{Spin(2N+\gamma)^{(\chi)}+\text{(A/S)-}[2l](C)}(\mathfrak{t})=\frac{1}{2}\sum_{\zeta=\pm 1} \langle W_{{\cal R}_1}\cdots W_{{\cal R}_k}\rangle^{SO(2N+\gamma)^{(\chi\zeta)}+\text{(A/S)-}[2l](C)}(\mathfrak{t}).
\end{align}
\end{subequations}

On the other hand, for $\zeta=-1$ we find
\begin{subequations}
\label{SO2N+1fundzeta-1t1limit}
\begin{align}
&\lim_{\mathfrak{t}\rightarrow 1}\langle {\cal W}_{\ydiagram{1}}\rangle^{SO(2N+1)^{(\chi-)}+\text{(A/S)-}[2l](C)}(\mathfrak{t})=\begin{cases}
\displaystyle \chi&\quad (N\text{: even}),\vspace{0.2cm}\\
\displaystyle \frac{\chi(\Delta-1)}{\Delta+1}&\quad (N\text{: odd})
\end{cases},\\
&\lim_{\mathfrak{t}\rightarrow 1}\langle {\cal W}_{\ydiagram{1}}{\cal W}_{\ydiagram{1}}\rangle^{SO(2N+1)^{(\chi-)}+\text{(A/S)-}[2l](C)}(\mathfrak{t})=\begin{cases}
\displaystyle 1&\quad (N\text{: even}),\vspace{0.2cm}\\
\displaystyle \frac{3(\Delta-1)}{\Delta+1}&\quad (N\text{: odd})
\end{cases}.
\end{align}
\end{subequations}
We also find
\begin{align}
&\langle {\cal W}_{\ydiagram{1}}\rangle^{SO(2N+1)^{(\chi\zeta)}+\text{(A/S)-}[2l](C)}(\mathfrak{t}^{-1})
=
\langle {\cal W}_{\ydiagram{1}}\rangle^{SO(2N+1)^{(\chi\zeta)}+\text{(A/S)-}[2l](C)}(\mathfrak{t}).
\end{align}

\subsubsection{Spinor Wilson lines in $Spin(2N+1)^{(\chi)}$}
Next let us consider the two-point function of the Wilson lines in the spinor representation of the $Spin(2N+1)$ gauge group.
Since the character of the spinor representation is $\chi^{\mathfrak{so}(2N+1)}_{\text{sp}}(s)=\prod_{i=1}^N(s_i^{\frac{1}{2}}+s_i^{-\frac{1}{2}})$, the two-point function can be calculated
by choosing the single variable function $f(s)=(s^{\frac{1}{2}}+s^{-\frac{1}{2}})^2=s+s^{-1}+2$.

For this choice of $f(s)$, we find
\begin{subequations}
\label{tildecalISOoddspinor}
\begin{align}
&\tilde{\cal I}_{B,1,1}^{(0)}[s+s^{-1}+2;\nu]=
\frac{1}{(\mathfrak{t}^2;\mathfrak{t}^2)_\nu},\quad
\tilde{\cal I}_{B,1,-1}^{(0)}[s+s^{-1}+2;1]=
\frac{3 - \mathfrak{t}^2}{1 - \mathfrak{t}^4},\\
&\tilde{\cal I}_{B,1,-1}^{(0)}[s+s^{-1}+2;2]=
\frac{5 - 4 \mathfrak{t}^2 + \mathfrak{t}^4}{(1 - \mathfrak{t}^2) (1 - \mathfrak{t}^8)},\\
&\tilde{\cal I}_{B,1,-1}^{(0)}[s+s^{-1}+2;3]=
\frac{7 - 4 \mathfrak{t}^2 + 3 \mathfrak{t}^4 - 5 \mathfrak{t}^6 + 4 \mathfrak{t}^8 - \mathfrak{t}^{10}}{(1 - \mathfrak{t}^2) (1 - \mathfrak{t}^8) (1 - \mathfrak{t}^{12})},\\
&\tilde{\cal I}_{B,1,-1}^{(0)}[s+s^{-1}+2;4]=
\frac{9 - 4 \mathfrak{t}^{2} + 5 \mathfrak{t}^{4} - 3 \mathfrak{t}^{6} - 3 \mathfrak{t}^{8} + 5 \mathfrak{t}^{10} - 3 \mathfrak{t}^{12} + 5 \mathfrak{t}^{14} - 4 \mathfrak{t}^{16} + \mathfrak{t}^{18}}{(1 - \mathfrak{t}^{2}) (1 - \mathfrak{t}^{8}) (1 - \mathfrak{t}^{12}) (1 - \mathfrak{t}^{16})},\\
&\tilde{\cal I}^{(1)}[s+s^{-1}+2;1]=
\frac{2}{1 - \mathfrak{t}^2},\quad
\tilde{\cal I}^{(1)}[s+s^{-1}+2;2]=
\frac{3 + \mathfrak{t}^2}{(1 - \mathfrak{t}^2) (1 - \mathfrak{t}^4)},\\
&\tilde{\cal I}^{(1)}[s+s^{-1}+2;3]=
\frac{2 (2 + \mathfrak{t}^2 + \mathfrak{t}^4)}{(1 - \mathfrak{t}^2) (1 - \mathfrak{t}^4) (1 - \mathfrak{t}^6)},\\
&\tilde{\cal I}^{(1)}[s+s^{-1}+2;4]=
\frac{5 + 3 \mathfrak{t}^2 + 4 \mathfrak{t}^4 + 3 \mathfrak{t}^6 + \mathfrak{t}^8}{(1 - \mathfrak{t}^2) (1 - \mathfrak{t}^4) (1 - \mathfrak{t}^6) (1 - \mathfrak{t}^8)},
\end{align}
\end{subequations}
with which we find, after gauging $\mathbb{Z}^{{\cal M}}_2$ symmetry,
{\fontsize{9pt}{1pt}\selectfont
\begin{subequations}
\begin{align}
&\langle W_{\text{sp}}W_{\text{sp}} \rangle^{Spin(3)^{(+)}+\text{(A/S)-}[2l](C)}(\mathfrak{t})
=\frac{1 + \mathfrak{t}^{4 \Delta}}{(1 - \mathfrak{t}^2) (1 - \mathfrak{t}^{4 \Delta})},\\
&\langle W_{\text{sp}}W_{\text{sp}} \rangle^{Spin(5)^{(+)}+\text{(A/S)-}[2l](C)}(\mathfrak{t})
=\frac{1 + \mathfrak{t}^{4 \Delta} (3 + 3 \mathfrak{t}^2) + \mathfrak{t}^{ 8 \Delta + 2}}{(1 - \mathfrak{t}^2) (1 - \mathfrak{t}^4) (1 - \mathfrak{t}^{4 \Delta})^2},\\
&\langle W_{\text{sp}}W_{\text{sp}} \rangle^{Spin(7)^{(+)}+\text{(A/S)-}[2l](C)}(\mathfrak{t})
=\frac{1}{(1 - \mathfrak{t}^2) (1 - \mathfrak{t}^4) (1 - \mathfrak{t}^6) (1 - \mathfrak{t}^{4 \Delta})^2 (1 - \mathfrak{t}^{12 \Delta})}
(1
+ \mathfrak{t}^{4 \Delta} (3 + 6 \mathfrak{t}^2 + 6 \mathfrak{t}^4 + \mathfrak{t}^6)\nonumber \\
&\quad
+ \mathfrak{t}^{8 \Delta} (6 + 10 \mathfrak{t}^2 + 10 \mathfrak{t}^4 + 5 \mathfrak{t}^6)
+ \mathfrak{t}^{12 \Delta} (5 + 10 \mathfrak{t}^2 + 10 \mathfrak{t}^4 + 6 \mathfrak{t}^6)
+ \mathfrak{t}^{16\Delta} (1 + 6 \mathfrak{t}^2 + 6 \mathfrak{t}^4 + 3 \mathfrak{t}^6)
+ \mathfrak{t}^{20 \Delta + 6}
),\\
&\langle W_{\text{sp}}W_{\text{sp}} \rangle^{Spin(9)^{(+)}+\text{(A/S)-}[2l](C)}(\mathfrak{t})\nonumber \\
&=
\frac{
1}{(1 - \mathfrak{t}^2) (1 - \mathfrak{t}^4) (1 - \mathfrak{t}^6) (1 - \mathfrak{t}^8) (1 - \mathfrak{t}^{4 \Delta})^2 (1 - \mathfrak{t}^{8 \Delta}) (1 - \mathfrak{t}^{12 \Delta})}
(1
+ \mathfrak{t}^{4 \Delta} (3 + 6 \mathfrak{t}^2 + 9 \mathfrak{t}^4 + 7 \mathfrak{t}^6 + 4 \mathfrak{t}^8 + \mathfrak{t}^{10})\nonumber \\
&\quad
+ \mathfrak{t}^{8 \Delta} (10 + 19 \mathfrak{t}^2 + 29 \mathfrak{t}^4 + 32 \mathfrak{t}^6 + 23 \mathfrak{t}^8 + 13 \mathfrak{t}^{10} + 2 \mathfrak{t}^{12})
+ \mathfrak{t}^{12 \Delta} (9 + 31 \mathfrak{t}^2 + 50 \mathfrak{t}^4 + 57 \mathfrak{t}^6 + 45 \mathfrak{t}^8 + 26 \mathfrak{t}^{10} + 7 \mathfrak{t}^{12})\nonumber \\
&\quad
+ \mathfrak{t}^{16 \Delta} (7 + 26 \mathfrak{t}^2 + 45 \mathfrak{t}^4 + 57 \mathfrak{t}^6 + 50 \mathfrak{t}^8 + 31 \mathfrak{t}^{10} + 9 \mathfrak{t}^{12})
+ \mathfrak{t}^{20 \Delta} (2 + 13 \mathfrak{t}^2 + 23 \mathfrak{t}^4 + 32 \mathfrak{t}^6 + 29 \mathfrak{t}^8 + 19 \mathfrak{t}^{10} + 10 \mathfrak{t}^{12})\nonumber \\
&\quad
+ \mathfrak{t}^{24 \Delta + 2} (1 + 4 \mathfrak{t}^2 + 7 \mathfrak{t}^4 + 9 \mathfrak{t}^6 + 6 \mathfrak{t}^8 + 3 \mathfrak{t}^{10})
+ \mathfrak{t}^{28 \Delta + 12}
),\\
&\langle W_{\text{sp}}W_{\text{sp}} \rangle^{Spin(3)^{(-)}+\text{(A/S)-}[2l](C)}(\mathfrak{t})
=
\frac{
3 - \mathfrak{t}^2 + \mathfrak{t}^{4 \Delta} (-1 + 3 \mathfrak{t}^2)
}{
(1 - \mathfrak{t}^4) (1 - \mathfrak{t}^{4 \Delta})},\\
&\langle W_{\text{sp}}W_{\text{sp}} \rangle^{Spin(5)^{(-)}+\text{(A/S)-}[2l](C)}(\mathfrak{t})
=
\frac{
5 - 4 \mathfrak{t}^2 + \mathfrak{t}^4
+ \mathfrak{t}^{4 \Delta} (-1 + 7 \mathfrak{t}^2 + 7 \mathfrak{t}^4 - \mathfrak{t}^6)
+ \mathfrak{t}^{8 \Delta + 2} (1 - 4 \mathfrak{t}^2 + 5 \mathfrak{t}^4)
}{(1 - \mathfrak{t}^2) (1 - \mathfrak{t}^8) (1 - \mathfrak{t}^{4 \Delta})^2},\\
&\langle W_{\text{sp}}W_{\text{sp}} \rangle^{Spin(7)^{(-)}+\text{(A/S)-}[2l](C)}(\mathfrak{t})\nonumber \\
&=
\frac{1}{(1 - \mathfrak{t}^2) (1 - \mathfrak{t}^8) (1 - \mathfrak{t}^{{12}}) (1 - \mathfrak{t}^{4 \Delta})^2 (1 - \mathfrak{t}^{12 \Delta})}
(
7 - 4 \mathfrak{t}^2 + 3 \mathfrak{t}^4 - 5 \mathfrak{t}^6 + 4 \mathfrak{t}^8 - \mathfrak{t}^{10}
+ \mathfrak{t}^{4 \Delta} (5 + 10 \mathfrak{t}^2 + 15 \mathfrak{t}^4 + 14 \mathfrak{t}^6\nonumber \\
&\quad\quad + 12 \mathfrak{t}^8 + 6 \mathfrak{t}^{10} + \mathfrak{t}^{12} + 2 \mathfrak{t}^{14} - \mathfrak{t}^{16})
+ \mathfrak{t}^{8 \Delta} (6 + 10 \mathfrak{t}^2 + 16 \mathfrak{t}^4 + 23 \mathfrak{t}^6 + 16 \mathfrak{t}^8 + 27 \mathfrak{t}^{10} + 13 \mathfrak{t}^{12} + 14 \mathfrak{t}^{14} - \mathfrak{t}^{16})\nonumber \\
&\quad
+ \mathfrak{t}^{12 \Delta} (-1 + 14 \mathfrak{t}^2 + 13 \mathfrak{t}^4 + 27 \mathfrak{t}^6 + 16 \mathfrak{t}^8 + 23 \mathfrak{t}^{10} + 16 \mathfrak{t}^{12} + 10 \mathfrak{t}^{14} + 6 \mathfrak{t}^{16})\nonumber \\
&\quad
+ \mathfrak{t}^{16 \Delta} (-1 + 2 \mathfrak{t}^2 + \mathfrak{t}^4 + 6 \mathfrak{t}^6 + 12 \mathfrak{t}^8 + 14 \mathfrak{t}^{10} + 15 \mathfrak{t}^{12} + 10 \mathfrak{t}^{14} + 5 \mathfrak{t}^{16})\nonumber \\
&\quad
+ \mathfrak{t}^{20 \Delta + 6} (-1 + 4 \mathfrak{t}^2 - 5 \mathfrak{t}^4 + 3 \mathfrak{t}^6 - 4 \mathfrak{t}^8 + 7 \mathfrak{t}^{10})),\\
&\langle W_{\text{sp}}W_{\text{sp}} \rangle^{Spin(9)^{(-)}+\text{(A/S)-}[2l](C)}(\mathfrak{t})\nonumber \\
&=
\frac{1}{(1 - \mathfrak{t}^2)^2 (1 - \mathfrak{t}^{12}) (1 - \mathfrak{t}^{16}) (1 - \mathfrak{t}^{4 \Delta})^2 (1 - \mathfrak{t}^{8 \Delta}) (1 - \mathfrak{t}^{12 \Delta})}
(
9 - 13 \mathfrak{t}^2 + 9 \mathfrak{t}^4 - 8 \mathfrak{t}^6 + 9 \mathfrak{t}^8                 - 5 \mathfrak{t}^{10}    + \mathfrak{t}^{12}\nonumber \\
&\quad
+ \mathfrak{t}^{4 \Delta} (11 + 11 \mathfrak{t}^2 + 2 \mathfrak{t}^4 + 33 \mathfrak{t}^8           - 13 \mathfrak{t}^{10} + 17 \mathfrak{t}^{12} - 15 \mathfrak{t}^{14} + 22 \mathfrak{t}^{16}  - 8 \mathfrak{t}^{18}    - \mathfrak{t}^{20}    + \mathfrak{t}^{22})\nonumber \\
&\quad
+ \mathfrak{t}^{8 \Delta} (14 + 17 \mathfrak{t}^2 + 28 \mathfrak{t}^4 + 26 \mathfrak{t}^6 + 34 \mathfrak{t}^8 + 35 \mathfrak{t}^{10} + 32 \mathfrak{t}^{12} + 29 \mathfrak{t}^{14} + 18 \mathfrak{t}^{16} + 18 \mathfrak{t}^{18}  + 4 \mathfrak{t}^{20}  + 3 \mathfrak{t}^{22}  - 2 \mathfrak{t}^{24})\nonumber \\
&\quad
+ \mathfrak{t}^{12 \Delta} (5 + 22 \mathfrak{t}^2 + 40 \mathfrak{t}^4 + 46 \mathfrak{t}^6 + 27 \mathfrak{t}^8  + 84 \mathfrak{t}^{10} + 46 \mathfrak{t}^{12} + 77 \mathfrak{t}^{14} + 17 \mathfrak{t}^{16} + 54 \mathfrak{t}^{18} + 14 \mathfrak{t}^{20} + 23 \mathfrak{t}^{22}  - 5 \mathfrak{t}^{24})\nonumber \\
&\quad
+ \mathfrak{t}^{16 \Delta} (-5 + 23 \mathfrak{t}^2 + 14 \mathfrak{t}^4 + 54 \mathfrak{t}^6 + 17 \mathfrak{t}^8 + 77 \mathfrak{t}^{10} + 46 \mathfrak{t}^{12} + 84 \mathfrak{t}^{14} + 27 \mathfrak{t}^{16} + 46 \mathfrak{t}^{18} + 40 \mathfrak{t}^{20} + 22 \mathfrak{t}^{22}  + 5 \mathfrak{t}^{24})\nonumber \\
&\quad
+ \mathfrak{t}^{20 \Delta} (-2 + 3 \mathfrak{t}^2 + 4 \mathfrak{t}^4 + 18 \mathfrak{t}^6 + 18 \mathfrak{t}^8  + 29 \mathfrak{t}^{10} + 32 \mathfrak{t}^{12} + 35 \mathfrak{t}^{14} + 34 \mathfrak{t}^{16} + 26 \mathfrak{t}^{18} + 28 \mathfrak{t}^{20} + 17 \mathfrak{t}^{22} + 14 \mathfrak{t}^{24})\nonumber \\
&\quad
+ \mathfrak{t}^{24 \Delta + 2} (1 - \mathfrak{t}^2 - 8 \mathfrak{t}^4 + 22 \mathfrak{t}^6 - 15 \mathfrak{t}^8 + 17 \mathfrak{t}^{10} - 13 \mathfrak{t}^{12} + 33 \mathfrak{t}^{14}              + 2 \mathfrak{t}^{18} + 11 \mathfrak{t}^{20} + 11 \mathfrak{t}^{22})\nonumber \\
&\quad
+ \mathfrak{t}^{28 \Delta + 12} (1 - 5 \mathfrak{t}^2 + 9 \mathfrak{t}^4 - 8 \mathfrak{t}^6 + 9 \mathfrak{t}^8- 13 \mathfrak{t}^{10}  + 9 \mathfrak{t}^{12})
).
\end{align}
\end{subequations}
}
In particular, we find in the limit $\mathfrak{t}\rightarrow 1$
\begin{align}
\lim_{\mathfrak{t}\rightarrow 1}
\langle {\cal W}_{\text{sp}} {\cal W}_{\text{sp}}\rangle^{Spin(2N+1)^{(\chi)}+\text{(A/S)-}[2l](C)}(\mathfrak{t})
=2^N,
\end{align}
which is consistent with our conjecture \eqref{Witten_indC_conj_int}.

\subsubsection{More general representations in $SO(2N+1)^{(\chi\zeta)}$}
\label{app_SOoddC_moregeneral}
Let us consider the Wilson line defect indices of $SO(2N+1)$ theory in more general representations, whose characters are given by (see e.g.~\cite{Hatsuda:2025jze})
\begin{align}
\chi^{\mathfrak{so}(2N+1)}_\lambda(s,\chi)=
\frac{
\text{det}[
s_j^{\lambda_i+N-i+\frac{1}{2}}
-\chi s_j^{-(\lambda_i+N-i+\frac{1}{2})}
]
}{
\text{det}[
s_j^{N-i+\frac{1}{2}}
-\chi s_j^{-(N-i+\frac{1}{2})}
]
}.
\label{SOoddcharacter}
\end{align}
Here we have introduced the $\mathbb{Z}^{{\cal C}}_2$ fugacity $\chi=\pm 1$ so that the coefficients of the monomials whose total power of $s_i$ is $\nu$ are multiplied with $\chi^{|\lambda|-\nu}$.

For example, for $\lambda=\ydiagram{2}$, we find
\begin{align}
\chi^{\mathfrak{so}(2N+1)}_{\ydiagram{2}}(s,\chi)=\sum_{i\le j}\Bigl(s_i+\frac{1}{s_i}\Bigr)\Bigl(s_j+\frac{1}{s_j}\Bigr)+\chi\sum_{i=1}^N\Bigl(s_i+\frac{1}{s_i}\Bigr)-N,
\end{align}
and hence the one- and two-point function of the Wilson lines $W_{\ydiagram{2}}$ can be calculated by choosing the single variable function $f(s)=\frac{1}{1-a(s_i+s_i^{-1})}\frac{1}{1-b(s_i+s_i^{-1})}$ as
\begin{subequations}
\label{SOodd_rank2symcalculation}
\begin{align}
&\langle W_{\ydiagram{2}}\rangle^{SO(2N+1)^{(\chi\zeta)}+\text{(A/S)-}[2l](C)}(\mathfrak{t})\nonumber \\
&=
\biggl\langle\prod_i f(s_i)\biggr\rangle^{SO(2N+1)^{(\chi\zeta)}+\text{(A/S)-}[2l](C)}(\mathfrak{t})\Bigr|_{a^2b^0}
+\chi\biggl\langle\prod_i f(s_i)\biggr\rangle^{SO(2N+1)^{(\chi\zeta)}+\text{(A/S)-}[2l](C)}(\mathfrak{t})\Bigr|_{a^1b^0}\nonumber \\
&\quad
-N\biggl\langle\prod_i f(s_i)\biggr\rangle^{SO(2N+1)^{(\chi\zeta)}+\text{(A/S)-}[2l](C)}(\mathfrak{t})\Bigr|_{a^0b^0},\\
&\langle W_{\ydiagram{2}}W_{\ydiagram{2}}\rangle^{SO(2N+1)^{(\chi\zeta)}+\text{(A/S)-}[2l](C)}(\mathfrak{t})\nonumber \\
&=
\biggl\langle\prod_i f(s_i)\biggr\rangle^{SO(2N+1)^{(\chi\zeta)}+\text{(A/S)-}[2l](C)}(\mathfrak{t})\Bigr|_{a^2b^2}
+\biggl\langle\prod_i f(s_i)\biggr\rangle^{SO(2N+1)^{(\chi\zeta)}+\text{(A/S)-}[2l](C)}(\mathfrak{t})\Bigr|_{a^1b^1}\nonumber \\
&\quad
+N^2\biggl\langle\prod_i f(s_i)\biggr\rangle^{SO(2N+1)^{(\chi\zeta)}+\text{(A/S)-}[2l](C)}(\mathfrak{t})\Bigr|_{a^0b^0}
+2\chi\biggl\langle\prod_i f(s_i)\biggr\rangle^{SO(2N+1)^{(\chi\zeta)}+\text{(A/S)-}[2l](C)}(\mathfrak{t})\Bigr|_{a^2b^1}\nonumber \\
&\quad
-2\chi N\biggl\langle\prod_i f(s_i)\biggr\rangle^{SO(2N+1)^{(\chi\zeta)}+\text{(A/S)-}[2l](C)}(\mathfrak{t})\Bigr|_{a^1b^0}
-2N\biggl\langle\prod_i f(s_i)\biggr\rangle^{SO(2N+1)^{(\chi\zeta)}+\text{(A/S)-}[2l](C)}(\mathfrak{t})\Bigr|_{a^2b^0}.
\end{align}
\end{subequations}
Calculating the building blocks $\tilde{\cal I}^{(0)}_{B,1,\chi}[f;\nu]$ and $\tilde{\cal I}^{(1)}[f;\nu]$ for this choice of $f(s)$ and plugging them into \eqref{SOoddCfiniteN} and then \eqref{SOodd_rank2symcalculation}, we obtain
{\fontsize{9pt}{1pt}\selectfont
\begin{subequations}
\begin{align}
&\langle W_{\ydiagram{2}}\rangle^{SO(3)^{(\chi\zeta)}+\text{(A/S)-}[2l](C)}(\mathfrak{t})=\frac{\mathfrak{t}^4 + \zeta \mathfrak{t}^{2\Delta} (1 + \mathfrak{t}^2 - \mathfrak{t}^4)}{(1 - \mathfrak{t}^4) (1 - \mathfrak{t}^{2\Delta} \zeta)},\\
&\langle W_{\ydiagram{2}}\rangle^{SO(5)^{(\chi\zeta)}+\text{(A/S)-}[2l](C)}(\mathfrak{t})=\frac{\mathfrak{t}^4 + \zeta \mathfrak{t}^{2\Delta} (1 + \mathfrak{t}^2 - \mathfrak{t}^4 + \mathfrak{t}^6) + \mathfrak{t}^{4 \Delta + 2} (1 + \mathfrak{t}^2 - \mathfrak{t}^4)}{(1 - \mathfrak{t}^4)^2 (1 - \zeta \mathfrak{t}^{2\Delta})^2},\\
&\langle W_{\ydiagram{2}}\rangle^{SO(7)^{(\chi\zeta)}+\text{(A/S)-}[2l](C)}(\mathfrak{t})\nonumber \\
&=
\frac{
1}{(1 - \mathfrak{t}^4)^2 (1 - \mathfrak{t}^8) (1 - \zeta \mathfrak{t}^{2\Delta})^2 (1 - \mathfrak{t}^{4\Delta})}
(\mathfrak{t}^4
+ \zeta \mathfrak{t}^{2\Delta} (1 + \mathfrak{t}^2 - \mathfrak{t}^4 + \mathfrak{t}^6 + \mathfrak{t}^8 + \mathfrak{t}^{10})
+ \mathfrak{t}^{4 \Delta + 2} (1 + 2 \mathfrak{t}^2 + 2 \mathfrak{t}^4 + \mathfrak{t}^6)\nonumber \\
&\quad
+ \zeta \mathfrak{t}^{6 \Delta + 2} (1 + 2 \mathfrak{t}^2 + \mathfrak{t}^4 - \mathfrak{t}^8 + \mathfrak{t}^{10})
+ \mathfrak{t}^{8 \Delta + 8} (1 + \mathfrak{t}^2 - \mathfrak{t}^4)),\\
&\langle W_{\ydiagram{2}}W_{\ydiagram{2}}\rangle^{SO(3)^{(\chi\zeta)}+\text{(A/S)-}[2l](C)}(\mathfrak{t})=\frac{1 + \mathfrak{t}^2 + \mathfrak{t}^4 + \mathfrak{t}^6 + \mathfrak{t}^8 + \zeta \mathfrak{t}^{2\Delta} (4 + 4 \mathfrak{t}^2 - \mathfrak{t}^4 - \mathfrak{t}^6 - \mathfrak{t}^8)}{(1 - \mathfrak{t}^4) (1 - \zeta \mathfrak{t}^{2\Delta})},\\
&\langle W_{\ydiagram{2}}W_{\ydiagram{2}}\rangle^{SO(5)+\text{(A/S)-}[2l](C)}(\mathfrak{t})\nonumber \\
&=\frac{1}{(1 - \mathfrak{t}^4)^2 (1 - \zeta \mathfrak{t}^{2\Delta}) (1 - \mathfrak{t}^{4\Delta})} (1 + \mathfrak{t}^2 + \mathfrak{t}^4 + \mathfrak{t}^6 + 2 \mathfrak{t}^8 + \mathfrak{t}^{10} + \mathfrak{t}^{12}
+ \zeta \mathfrak{t}^{2\Delta} (5 + 8 \mathfrak{t}^2 + 7 \mathfrak{t}^4 + 5 \mathfrak{t}^6 - \mathfrak{t}^{12})\nonumber \\
&\quad + \mathfrak{t}^{4\Delta} (5 + 12 \mathfrak{t}^2 + 9 \mathfrak{t}^4 + 2 \mathfrak{t}^6 - 2 \mathfrak{t}^8 - \mathfrak{t}^{10} - \mathfrak{t}^{12}) + \zeta \mathfrak{t}^{6\Delta} (5 + 11 \mathfrak{t}^2 - \mathfrak{t}^4 - 8 \mathfrak{t}^6 + \mathfrak{t}^{12})),\\
&\langle W_{\ydiagram{2}}W_{\ydiagram{2}}\rangle^{SO(7)+\text{(A/S)-}[2l](C)}(\mathfrak{t})\nonumber \\
&=
\frac{
1}{(1 - \mathfrak{t}^4)^2 (1 - \mathfrak{t}^8) (1 - \zeta \mathfrak{t}^{2\Delta})^2 (1 - \mathfrak{t}^{4\Delta})}
(
                     1 + \mathfrak{t}^2 + \mathfrak{t}^4 + \mathfrak{t}^6 + 2 \mathfrak{t}^8            + \mathfrak{t}^{10} + 2 \mathfrak{t}^{12}   + \mathfrak{t}^{14}   + \mathfrak{t}^{16}\nonumber \\
&\quad
+ \zeta \mathfrak{t}^{2\Delta} (    4 + 7 \mathfrak{t}^2 + 7 \mathfrak{t}^4 + 8 \mathfrak{t}^6 + 7 \mathfrak{t}^8    + 7 \mathfrak{t}^{10}   + \mathfrak{t}^{12} + 2 \mathfrak{t}^{14}            + \mathfrak{t}^{18})\nonumber \\
&\quad
+ \mathfrak{t}^{4\Delta} (        1 + 8 \mathfrak{t}^2 + 18 \mathfrak{t}^4 + 22 \mathfrak{t}^6 + 15 \mathfrak{t}^8 + 7 \mathfrak{t}^{10}   + \mathfrak{t}^{12} - 3 \mathfrak{t}^{14} - 2 \mathfrak{t}^{16} - \mathfrak{t}^{18})\nonumber \\
&\quad
+ \zeta \mathfrak{t}^{6\Delta} (5 + 12 \mathfrak{t}^2 + 11 \mathfrak{t}^4 + 13 \mathfrak{t}^6 + 5 \mathfrak{t}^8 - 4 \mathfrak{t}^{10} + 3 \mathfrak{t}^{12}   - \mathfrak{t}^{14}   + \mathfrak{t}^{16} - \mathfrak{t}^{18})
+ \mathfrak{t}^{8\Delta+2} (         5 + 7 \mathfrak{t}^2 + 4 \mathfrak{t}^6                           - 7 \mathfrak{t}^{10}   + \mathfrak{t}^{12}            + \mathfrak{t}^{16})
).
\end{align}
\end{subequations}
}
Interestingly, we observe that the one-point function and the two-point function satisfy the following linear recursion relations which are analogous respectively to the recursion relation for the two-point function of the fundamental Wilson lines \eqref{R_uN_C_Wasym} and that for the two-point function of the Wilson lines in the rank $2$ symmetric in the $U(N)$ ADHM theory \eqref{R_uN_C_Wsym2}:
\begin{subequations}
\label{SOoddCrank2symrecreln}
\begin{align}
&\langle W_{\ydiagram{2}}\rangle^{SO(2N+1)^{(\chi\zeta)}+\text{(A/S)-}[2l](C)}(\mathfrak{t})
=
{\cal I}^{SO(2N-1)+\text{(A/S)-}[2l](C)}(\mathfrak{t})
\langle W_{\ydiagram{2}}\rangle^{SO(3)^{(\chi\zeta)}+\text{(A/S)-}[2l](C)}(\mathfrak{t}),\\
&\langle W_{\ydiagram{2}}W_{\ydiagram{2}}\rangle^{SO(2N+1)^{(\chi\zeta)}+\text{(A/S)-}[2l](C)}(\mathfrak{t})\nonumber \\
&
=
c_1(\mathfrak{t}){\cal I}^{SO(2N-1)^{(\chi\zeta)}+\text{(A/S)-}[2l](C)}(\mathfrak{t})
+c_2(\mathfrak{t}){\cal I}^{SO(2N-3)^{(\chi\zeta)}+\text{(A/S)-}[2l](C)}(\mathfrak{t}),
\end{align}
\end{subequations}
with
\begin{subequations}
\begin{align}
&c_1=
\langle W_{\ydiagram{2}}W_{\ydiagram{2}}\rangle^{SO(3)^{(\chi\zeta)}+\text{(A/S)-}[2l](C)}(\mathfrak{t})
=
\frac{
1 + \mathfrak{t}^2 + \mathfrak{t}^4 + \mathfrak{t}^6 + \mathfrak{t}^8 + \zeta \mathfrak{t}^{2\Delta} (4 + 4 \mathfrak{t}^2 - \mathfrak{t}^4 - \mathfrak{t}^6 - \mathfrak{t}^8)
}{
(1 - \mathfrak{t}^4) (1 - \zeta \mathfrak{t}^{2\Delta})},\\
&c_2=
\langle W_{\ydiagram{2}}W_{\ydiagram{2}}\rangle^{SO(5)+\text{(A/S)-}[2l](C)}(\mathfrak{t})
-{\cal I}^{SO(3)+\text{(A/S)-}[2l](C)}(\mathfrak{t})\langle W_{\ydiagram{2}}W_{\ydiagram{2}}\rangle^{SO(3)+\text{(A/S)-}[2l](C)}(\mathfrak{t})\nonumber \\
&
=
\frac{1}{(1 - \mathfrak{t}^4)^2 (1 - \zeta \mathfrak{t}^{2\Delta}) (1 - \mathfrak{t}^{4\Delta})}
(\mathfrak{t}^8 + \mathfrak{t}^{10} + \mathfrak{t}^{12}
+ \zeta \mathfrak{t}^{2\Delta + 2} (2 + 6 \mathfrak{t}^2 + 4 \mathfrak{t}^4 - \mathfrak{t}^6 - \mathfrak{t}^8 - \mathfrak{t}^{10})\nonumber \\
&\quad
+ \mathfrak{t}^{4\Delta} (1 + 3 \mathfrak{t}^2 + 5 \mathfrak{t}^4 + 3 \mathfrak{t}^6 - \mathfrak{t}^8 - \mathfrak{t}^{10} - \mathfrak{t}^{12})
+ \zeta \mathfrak{t}^{6\Delta} (5 + 7 \mathfrak{t}^2 - 5 \mathfrak{t}^4 - 7 \mathfrak{t}^6 + \mathfrak{t}^8 + \mathfrak{t}^{10} + \mathfrak{t}^{12})
),\\
&{\cal I}^{SO(1)^{\chi\zeta}+\text{(A/S)-}[2l](C)}(\mathfrak{t})\equiv 1,\quad
{\cal I}^{SO(-1)^{\chi\zeta}+\text{(A/S)-}[2l](C)}(\mathfrak{t})\equiv 0.
\end{align}
\end{subequations}
Note that the Coulomb indices without line defect ${\cal I}^{SO(2N+1)^{(\chi\zeta)}+\text{(A/S)-}[2l](C)}(\mathfrak{t})$ can be generated from the grand canonical sum \eqref{ISOoddGC}.
Assuming these recursion relations, we find the limit $\mathfrak{t}\rightarrow 1$ of the normalized line defect indices as
\begin{align}
&\lim_{\mathfrak{t}\rightarrow 1}
\langle {\cal W}_{\ydiagram{2}}\rangle^{SO(2N+1)^{(\chi+)}+\text{(A/S)-}[2l](C)}(\mathfrak{t})
=
\lim_{\mathfrak{t}\rightarrow 1}
\langle {\cal W}_{\ydiagram{2}}\rangle^{Spin(2N+1)^{(\chi)}+\text{(A/S)-}[2l](C)}(\mathfrak{t})\nonumber \\
&
=N=\oint \prod_{i=1}^N\frac{ds_i}{2\pi is_i}\chi^{\mathfrak{so}(2N+1)}_{\ydiagram{2}}(s,\chi),\\
&\lim_{\mathfrak{t}\rightarrow 1}
\langle {\cal W}_{\ydiagram{2}}{\cal W}_{\ydiagram{2}}\rangle^{SO(2N+1)^{(\chi+)}+\text{(A/S)-}[2l](C)}(\mathfrak{t})
=
\lim_{\mathfrak{t}\rightarrow 1}
\langle {\cal W}_{\ydiagram{2}}{\cal W}_{\ydiagram{2}}\rangle^{Spin(2N+1)^{(\chi)}+\text{(A/S)-}[2l](C)}(\mathfrak{t})\nonumber \\
&
=N(3N+2)
=\oint \prod_{i=1}^N\frac{ds_i}{2\pi is_i}\chi^{\mathfrak{so}(2N+1)}_{\ydiagram{2}}(s,\chi)^2,
\end{align}
which are consistent with our conjecture \eqref{Witten_indC_conj_int}.
On the other hand, for $\mathbb{Z}^{{\cal M}}_2$ fugacity $\zeta=-1$, we find
\begin{subequations}
\begin{align}
&\lim_{\mathfrak{t}\rightarrow 1}
\langle {\cal W}_{\ydiagram{2}}\rangle^{SO(2N+1)^{(\chi-)}+\text{(A/S)-}[2l](C)}(\mathfrak{t})
=\begin{cases}
\displaystyle 0&\quad (N\text{: even})\vspace{0.2cm} \\
\displaystyle \frac{\Delta-3}{\Delta+1}&\quad (N\text{: odd})\\
\end{cases},\\
&\lim_{\mathfrak{t}\rightarrow 1}
\langle {\cal W}_{\ydiagram{2}}{\cal W}_{\ydiagram{2}}\rangle^{SO(2N+1)^{(\chi-)}+\text{(A/S)-}[2l](C)}(\mathfrak{t})
=\begin{cases}
\displaystyle -5N&\quad (N\text{: even})\vspace{0.2cm} \\
-5\Bigl(N-\frac{2(\Delta-1)}{\Delta+1}\Bigr)&\quad (N\text{: odd})\\
\end{cases}.
\end{align}
\end{subequations}

For $\lambda=\ydiagram{1,1}$, the character \eqref{SOoddcharacter} is
\begin{align}
\chi^{\mathfrak{so}(2N+1)}_{\ydiagram{1,1}}(s,\chi)=\sum_{i<j}\Bigl(s_i+\frac{1}{s_i}\Bigr)\Bigl(s_j+\frac{1}{s_j}\Bigr)+\chi\sum_i\Bigl(s_i+\frac{1}{s_i}\Bigr)+N.
\end{align}
Hence the one-point function and the two-point function of Wilson lines in the antisymmetric representations can be calculated by choosing the single variable function $f(s)=(1+a(s+s^{-1}))(1+b(s+s^{-1}))$ as
\begin{subequations}
\begin{align}
\label{SOodd_rank2antisymcalculation}
&\langle W_{\ydiagram{1,1}}\rangle^{SO(2N+1)^{(\chi\zeta)}+\text{(A/S)-}[2l](C)}(\mathfrak{t})\nonumber \\
&
=
\biggl\langle\prod_{i=1}^Nf(s_i)\biggr\rangle^{SO(2N+1)^{(\chi\zeta)}+\text{(A/S)-}[2l](C)}(\mathfrak{t})\Bigr|_{a^2b^0}
+\chi\biggl\langle\prod_{i=1}^Nf(s_i)\biggr\rangle^{SO(2N+1)^{(\chi\zeta)}+\text{(A/S)-}[2l](C)}(\mathfrak{t})\Bigr|_{a^1b^0}\nonumber \\
&\quad
+N\biggl\langle\prod_{i=1}^Nf(s_i)\biggr\rangle^{SO(2N+1)^{(\chi\zeta)}+\text{(A/S)-}[2l](C)}(\mathfrak{t})\Bigr|_{a^0b^0},\\
&\langle W_{\ydiagram{1,1}}W_{\ydiagram{1,1}}\rangle^{SO(2N+1)^{(\chi\zeta)}+\text{(A/S)-}[2l](C)}(\mathfrak{t})\nonumber \\
&
=
\biggl\langle\prod_{i=1}^Nf(s_i)\biggr\rangle^{SO(2N+1)^{(\chi\zeta)}+\text{(A/S)-}[2l](C)}(\mathfrak{t})\Bigr|_{a^2b^2}
+\biggl\langle\prod_{i=1}^Nf(s_i)\biggr\rangle^{SO(2N+1)^{(\chi\zeta)}+\text{(A/S)-}[2l](C)}(\mathfrak{t})\Bigr|_{a^1b^1}\nonumber \\
&\quad
+N^2\biggl\langle\prod_{i=1}^Nf(s_i)\biggr\rangle^{SO(2N+1)^{(\chi\zeta)}+\text{(A/S)-}[2l](C)}(\mathfrak{t})\Bigr|_{a^0b^0}
+2\chi\biggl\langle\prod_{i=1}^Nf(s_i)\biggr\rangle^{SO(2N+1)^{(\chi\zeta)}+\text{(A/S)-}[2l](C)}(\mathfrak{t})\Bigr|_{a^2b^1}\nonumber \\
&\quad
+2\chi N\biggl\langle\prod_{i=1}^Nf(s_i)\biggr\rangle^{SO(2N+1)^{(\chi\zeta)}+\text{(A/S)-}[2l](C)}(\mathfrak{t})\Bigr|_{a^1b^0}
+2N\biggl\langle\prod_{i=1}^Nf(s_i)\biggr\rangle^{SO(2N+1)^{(\chi\zeta)}+\text{(A/S)-}[2l](C)}(\mathfrak{t})\Bigr|_{a^2b^{0}}.
\end{align}
\end{subequations}
Calculating the building blocks $\tilde{\cal I}^{(0)}_{B,1,\chi}[f;\nu]$ and $\tilde{\cal I}^{(1)}[f;\nu]$ for this choice of $f(s)$ and plugging them into \eqref{SOoddCfiniteN} and then \eqref{SOodd_rank2antisymcalculation}, we obtain
{\fontsize{9pt}{1pt}\selectfont
\begin{subequations}
\begin{align}
&\langle W_{\ydiagram{1,1}}\rangle^{SO(3)^{(\chi\zeta)}+\text{(A/S)-}[2l](C)}(\mathfrak{t})
=
\frac{
\mathfrak{t}^2 + \mathfrak{t}^{2\Delta} \zeta
}{(1 - \mathfrak{t}^4) (1 - \zeta \mathfrak{t}^{2\Delta})},\\
&\langle W_{\ydiagram{1,1}}\rangle^{SO(5)^{(\chi\zeta)}+\text{(A/S)-}[2l](C)}(\mathfrak{t})
=
\frac{
(\mathfrak{t}^2 + \zeta \mathfrak{t}^{2\Delta}) (1 + \zeta \mathfrak{t}^{2\Delta+2})
}{(1 - \mathfrak{t}^4)^2 (1 - \zeta \mathfrak{t}^{2\Delta})^2},\\
&\langle W_{\ydiagram{1,1}}\rangle^{SO(7)^{(\chi\zeta)}+\text{(A/S)-}[2l](C)}(\mathfrak{t})\nonumber \\
&
=
\frac{
\mathfrak{t}^2 + \zeta \mathfrak{t}^{2\Delta}}{
(1 - \mathfrak{t}^4)^2 (1 - \mathfrak{t}^8) (1 - \zeta \mathfrak{t}^{2\Delta})^2 (1 - \mathfrak{t}^{4\Delta})}
(1 + \zeta \mathfrak{t}^{2\Delta + 2} (1 + \mathfrak{t}^2 + \mathfrak{t}^4)
+ \mathfrak{t}^{4\Delta + 2} (1 + \mathfrak{t}^2 + \mathfrak{t}^4) + \zeta \mathfrak{t}^{6\Delta + 8}),\\
&\langle W_{\ydiagram{1,1}} W_{\ydiagram{1,1}}\rangle^{SO(3)^{(\chi\zeta)}+\text{(A/S)-}[2l](C)}(\mathfrak{t})
=
\frac{
1 + \mathfrak{t}^2 + \mathfrak{t}^4 + \zeta \mathfrak{t}^{2\Delta} (2 + 2 \mathfrak{t}^2 - \mathfrak{t}^4)
}{
(1 - \mathfrak{t}^4) (1 - \zeta \mathfrak{t}^{2\Delta})},\\
&\langle W_{\ydiagram{1,1}} W_{\ydiagram{1,1}}\rangle^{SO(5)^{(\chi\zeta)}+\text{(A/S)-}[2l](C)}(\mathfrak{t})\nonumber \\
&
=
\frac{
1}{(1 - \mathfrak{t}^4)^2 (1 - \zeta \mathfrak{t}^{2\Delta}) (1 - \mathfrak{t}^{4\Delta})}
(1 + \mathfrak{t}^2 + 2 \mathfrak{t}^4 + \mathfrak{t}^6 + \mathfrak{t}^8
+ \zeta \mathfrak{t}^{2\Delta} (3 + 8 \mathfrak{t}^2 + 6 \mathfrak{t}^4 + 2 \mathfrak{t}^6 - \mathfrak{t}^8)
+ \mathfrak{t}^{4\Delta} (5 + 10 \mathfrak{t}^2 + 4 \mathfrak{t}^4 - \mathfrak{t}^8)\nonumber \\
&\quad
+ \zeta \mathfrak{t}^{6\Delta} (3 + 5 \mathfrak{t}^2 - 3 \mathfrak{t}^6 + \mathfrak{t}^8)),\\
&\langle W_{\ydiagram{1,1}} W_{\ydiagram{1,1}}\rangle^{SO(7)^{(\chi\zeta)}+\text{(A/S)-}[2l](C)}(\mathfrak{t})\nonumber \\
&
=
\frac{
1}{(1 - \mathfrak{t}^4)^2 (1 - \mathfrak{t}^8) (1 - \zeta \mathfrak{t}^{2\Delta})^2 (1 - \mathfrak{t}^{4\Delta})}
(
1 + \mathfrak{t}^2 + 2 \mathfrak{t}^4 + \mathfrak{t}^6 + 2 \mathfrak{t}^8 + \mathfrak{t}^{10} + \mathfrak{t}^{12}\nonumber \\
&\quad
+ \zeta \mathfrak{t}^{2\Delta} (2 + 7 \mathfrak{t}^2 + 6 \mathfrak{t}^4 + 9 \mathfrak{t}^6 + 7 \mathfrak{t}^8 + 4 \mathfrak{t}^{10} + \mathfrak{t}^{14})
+ \mathfrak{t}^{4\Delta} (3 + 8 \mathfrak{t}^2 + 15 \mathfrak{t}^4 + 16 \mathfrak{t}^6 + 9 \mathfrak{t}^8 + 4 \mathfrak{t}^{10} - \mathfrak{t}^{14})\nonumber \\
&\quad
+ \zeta \mathfrak{t}^{6\Delta} (3 + 8 \mathfrak{t}^2 + 10 \mathfrak{t}^4 + 9 \mathfrak{t}^6 + 6 \mathfrak{t}^8 - \mathfrak{t}^{10} + 2 \mathfrak{t}^{12} - \mathfrak{t}^{14})
+ \mathfrak{t}^{8\Delta + 2} (3 + 3 \mathfrak{t}^2 + \mathfrak{t}^4 + 3 \mathfrak{t}^6 + \mathfrak{t}^8 - 3 \mathfrak{t}^{10} + \mathfrak{t}^{12})).
\end{align}
\end{subequations}
}
Interestingly, we observe the following linear recursion relation for the line defect indices with respect to the rank $N$ of the gauge group:
\begin{subequations}
\label{SOoddCrank2antisymrecreln}
\begin{align}
&\langle
W_{\ydiagram{1,1}}
\rangle^{SO(2N+1)^{(\chi\zeta)}+\text{(A/S)-}[2l](C)}(\mathfrak{t})
=
{\cal I}^{SO(2N-1)^{(\chi\zeta)}+\text{(A/S)-}[2l](C)}(\mathfrak{t})
\langle
W_{\ydiagram{1,1}}
\rangle^{SO(3)^{(\chi\zeta)}+\text{(A/S)-}[2l](C)}(\mathfrak{t}),\\
&\langle
W_{\ydiagram{1,1}}
W_{\ydiagram{1,1}}
\rangle^{SO(2N+1)^{(\chi\zeta)}+\text{(A/S)-}[2l](C)}(\mathfrak{t})
=
c_1(\mathfrak{t})
{\cal I}^{SO(2N-1)^{(\chi\zeta)}+\text{(A/S)-}[2l](C)}(\mathfrak{t})
+c_2(\mathfrak{t})
{\cal I}^{SO(2N-3)^{(\chi\zeta)}+\text{(A/S)-}[2l](C)}(\mathfrak{t}),
\end{align}
\end{subequations}
with
\begin{subequations}
\begin{align}
&c_1(\mathfrak{t})=
\langle
W_{\ydiagram{1,1}}
W_{\ydiagram{1,1}}
\rangle^{SO(3)^{(\chi\zeta)}+\text{(A/S)-}[2l](C)}(\mathfrak{t})
=
\frac{1 + \mathfrak{t}^2 + \mathfrak{t}^4 + \zeta \mathfrak{t}^{2\Delta} (2 + 2 \mathfrak{t}^2 - \mathfrak{t}^4)}{(1 - \mathfrak{t}^4) (1 - \zeta \mathfrak{t}^{2\Delta})},\\
&c_2(\mathfrak{t})\nonumber \\
&=
\langle
W_{\ydiagram{1,1}}
W_{\ydiagram{1,1}}
\rangle^{SO(5)^{(\chi\zeta)}+\text{(A/S)-}[2l](C)}(\mathfrak{t})
-
{\cal I}^{SO(3)^{(\chi\zeta)}+\text{(A/S)-}[2l](C)}(\mathfrak{t})
\langle
W_{\ydiagram{1,1}}
W_{\ydiagram{1,1}}
\rangle^{SO(3)^{(\chi\zeta)}+\text{(A/S)-}[2l](C)}(\mathfrak{t})\nonumber \\
&
=
\frac{
1}{(1 - \mathfrak{t}^4)^2 (1 - \zeta \mathfrak{t}^{2\Delta}) (1 - \mathfrak{t}^{4\Delta})}
(\mathfrak{t}^4 + \mathfrak{t}^6 + \mathfrak{t}^8
+ \zeta \mathfrak{t}^{2\Delta + 2} (4 + 5 \mathfrak{t}^2 + \mathfrak{t}^4 - \mathfrak{t}^6)\nonumber \\
&\quad
+ \mathfrak{t}^{4\Delta} (3 + 5 \mathfrak{t}^2 + 2 \mathfrak{t}^4 - \mathfrak{t}^8)
+ \zeta \mathfrak{t}^{6\Delta} (3 + 3 \mathfrak{t}^2 - 2 \mathfrak{t}^4 - 2 \mathfrak{t}^6 + \mathfrak{t}^8)
),\\
&{\cal I}^{SO(1)^{(\chi\zeta)}+\text{(A/S)-}[2l](C)}(\mathfrak{t})\equiv 1,\quad
{\cal I}^{SO(-1)^{(\chi\zeta)}+\text{(A/S)-}[2l](C)}(\mathfrak{t})\equiv 0.
\end{align}
\end{subequations}
Assuming these recursion relation to be valid for higher $N$, we find the $\mathfrak{t}\rightarrow 1$ limit of the line defect indices as
\begin{align}
&\lim_{\mathfrak{t}\rightarrow 1} \langle{\cal W}_{\ydiagram{1,1}}\rangle^{SO(2N+1)^{(\chi+)}+\text{(A/S)-}[2l](C)}(\mathfrak{t})
=\lim_{\mathfrak{t}\rightarrow 1}\langle{\cal W}_{\ydiagram{1,1}}\rangle^{Spin(2N+1)^{(\chi)}+\text{(A/S)-}[2l](C)}(\mathfrak{t})\nonumber \\
&=N=\oint\prod_{i=1}^N\frac{ds_i}{2\pi is_i}\chi^{\mathfrak{so}(2N+1)}_{\ydiagram{1,1}}(s,\chi),\\
&\lim_{\mathfrak{t}\rightarrow 1}\langle{\cal W}_{\ydiagram{1,1}}{\cal W}_{\ydiagram{1,1}}\rangle^{SO(2N+1)^{(\chi+)}+\text{(A/S)-}[2l](C)}(\mathfrak{t})
=\lim_{\mathfrak{t}\rightarrow 1}\langle{\cal W}_{\ydiagram{1,1}}{\cal W}_{\ydiagram{1,1}}\rangle^{Spin(2N+1)^{(\chi)}+\text{(A/S)-}[2l](C)}(\mathfrak{t})\nonumber \\
&
=3N^2=\oint\prod_{i=1}^N\frac{ds_i}{2\pi is_i}\chi^{\mathfrak{so}(2N+1)}_{\ydiagram{1,1}}(s,\chi)^2,
\end{align}
which are consistent with our conjecture \eqref{Witten_indC_conj_int}.
For $\zeta=-1$, on the other hand, we find
\begin{subequations}
\begin{align}
&\lim_{\mathfrak{t}\rightarrow 1}
\langle{\cal W}_{\ydiagram{1,1}}\rangle^{SO(2N+1)^{(\chi-)}+\text{(A/S)-}[2l](C)}(\mathfrak{t})
=
\begin{cases}
\displaystyle 0&\quad (N\text{: even})\vspace{0.2cm} \\
\displaystyle \frac{\Delta-1}{\Delta+1}&\quad (N\text{: odd})
\end{cases},\\
&\lim_{\mathfrak{t}\rightarrow 1}
\langle {\cal W}_{\ydiagram{1,1}}{\cal W}_{\ydiagram{1,1}}\rangle^{SO(2N+1)^{(\chi-)}+\text{(A/S)-}[2l](C)}(\mathfrak{t})
=
\begin{cases}
\displaystyle -N&\quad (N\text{: even})\vspace{0.2cm} \\
\displaystyle -N+\frac{2(2\Delta-1)}{\Delta+1}&\quad (N\text{: odd})
\end{cases}.
\end{align}
\end{subequations}

For $\lambda=\ydiagram{4}$, whose character is
\begin{align}
&\chi^{\mathfrak{so}(2N+1)}_{\ydiagram{4}}(s,\chi)\nonumber \\
&=
\sum_{i\le j\le k\le k'}
\Bigl(s_i+\frac{1}{s_i}\Bigr)
\Bigl(s_j+\frac{1}{s_j}\Bigr)
\Bigl(s_k+\frac{1}{s_k}\Bigr)
\Bigl(s_{k'}+\frac{1}{s_{k'}}\Bigr)
+
\chi\sum_{i\le j\le k}
\Bigl(s_i+\frac{1}{s_i}\Bigr)
\Bigl(s_j+\frac{1}{s_j}\Bigr)
\Bigl(s_k+\frac{1}{s_k}\Bigr)\nonumber \\
&\quad
-(N+2)
\sum_{i\le j}
\Bigl(s_i+\frac{1}{s_i}\Bigr)
\Bigl(s_j+\frac{1}{s_j}\Bigr)
-\chi(N+1)
\sum_{i}
\Bigl(s_i+\frac{1}{s_i}\Bigr)
+\frac{N(N+1)}{2},
\end{align}
we observe the two-term recursion relation for the one-point function at least for $N=1$, $2$ and $3$
\begin{align}
\langle W_{\ydiagram{4}}\rangle^{SO(2N+1)^{(\chi\zeta)}+\text{(A/S)-}[2l](C)}(\mathfrak{t})
&=
c_1(\mathfrak{t})
{\cal I}^{SO(2N-1)^{(\chi\zeta)}+\text{(A/S)-}[2l](C)}(\mathfrak{t})\nonumber \\
&\quad
+c_2(\mathfrak{t})
{\cal I}^{SO(2N-3)^{(\chi\zeta)}+\text{(A/S)-}[2l](C)}(\mathfrak{t}),
\end{align}
with
\begin{subequations}
\begin{align}
&c_1(\mathfrak{t})=
\langle W_{\ydiagram{4}}\rangle^{SO(3)^{(\chi\zeta)}+\text{(A/S)-}[2l](C)}(\mathfrak{t})
=
\frac{
\mathfrak{t}^8 + \zeta \mathfrak{t}^{2\Delta} (1 + \mathfrak{t}^2 - \mathfrak{t}^8)
}{
(1 - \mathfrak{t}^4) (1 - \zeta \mathfrak{t}^{2\Delta})
},\\
&c_2(\mathfrak{t})=
\langle W_{\ydiagram{4}}\rangle^{SO(5)^{(\chi\zeta)}+\text{(A/S)-}[2l](C)}(\mathfrak{t})
-{\cal I}^{SO(3)^{(\chi\zeta)}+\text{(A/S)-}[2l](C)}(\mathfrak{t})
\langle W_{\ydiagram{4}}\rangle^{SO(3)^{(\chi\zeta)}+\text{(A/S)-}[2l](C)}(\mathfrak{t})\nonumber \\
&
=
\frac{
1
}{
(1 - \mathfrak{t}^4) (1 - \mathfrak{t}^8) (1 - \zeta \mathfrak{t}^{2\Delta}) (1 - \mathfrak{t}^{4 \Delta})
}
(\mathfrak{t}^{16} + \zeta \mathfrak{t}^{2\Delta + 4} (1 + \mathfrak{t}^2 + \mathfrak{t}^4 + \mathfrak{t}^6 - \mathfrak{t}^{12})\nonumber \\
&\quad + \mathfrak{t}^{4\Delta + 4} (1 + \mathfrak{t}^2 + \mathfrak{t}^4 + \mathfrak{t}^6 - \mathfrak{t}^{12}) + \zeta \mathfrak{t}^{6\Delta} (1 + 2 \mathfrak{t}^2 - \mathfrak{t}^8 - 2 \mathfrak{t}^{10} + \mathfrak{t}^{16})).
\end{align}
\end{subequations}
By using these recursion relation we find
\begin{subequations}
\begin{align}
&\lim_{\mathfrak{t}\rightarrow 1}
\langle {\cal W}_{\ydiagram{4}}\rangle^{SO(2N+1)^{(\chi+)}+\text{(A/S)-}[2l](C)}(\mathfrak{t})
=\lim_{\mathfrak{t}\rightarrow 1}
\langle {\cal W}_{\ydiagram{4}}\rangle^{Spin(2N+1)^{(\chi)}+\text{(A/S)-}[2l](C)}(\mathfrak{t})\nonumber \\
&
=\frac{N(N+1)}{2}=\oint \prod_{i=1}^N\frac{ds_i}{2\pi is_i}\chi^{\mathfrak{so}(2N+1)}_{\ydiagram{4}}(s,\chi),\\
&\lim_{\mathfrak{t}\rightarrow 1}
\langle {\cal W}_{\ydiagram{4}}\rangle^{SO(2N+1)^{(\chi-)}+\text{(A/S)-}[2l](C)}(\mathfrak{t})
=
\begin{cases}
\displaystyle -\frac{3N}{2}&\quad (N\text{: even})\vspace{0.2cm} \\
\displaystyle -\frac{3N}{2}+\frac{5\Delta-11}{2(\Delta+1)}&\quad (N\text{: odd})
\end{cases}.
\end{align}
\end{subequations}
The first line is consistent with our conjecture \eqref{Witten_indC_conj_int}.

\subsubsection{Fundamental Wilson lines in $SO(2N)^{(+\zeta)}$}
\label{app_SO2Nchi+fund}

In the theory with $SO(2N)$ gauge group, the one- and two-point function of the fundamental Wilson line are calculated by inserting $\chi^{\mathfrak{so}(2N)}_{\ydiagram{1}}(s)=\sum_{i=1}^N(s_i+s_i^{-1})$ and $\chi^{\mathfrak{so}(2N)}_{\ydiagram{1}}(s)^2$.
These can be extracted by inserting $\prod_{i=1}^N f(s_i)$ with $f(s)=(1+a(s+s^{-1}))(1+b(s+s^{-1}))$ as
\begin{subequations}
\begin{align}
&\langle W_{\ydiagram{1}}\rangle^{SO(2N)^{(+\zeta)}+\text{(A/S)-}[2l](C)}(\mathfrak{t})
=
\biggl\langle \prod_if(s_i)\biggr\rangle^{SO(2N)^{(+\zeta)}+\text{(A/S)-}[2l](C)}(\mathfrak{t})\Bigr|_{a^1b^0}=0,\\
&\langle W_{\ydiagram{1}} W_{\ydiagram{1}}\rangle^{SO(2N)^{(+\zeta)}+\text{(A/S)-}[2l](C)}(\mathfrak{t})
=
\biggl\langle \prod_if(s_i)\biggr\rangle^{SO(2N)^{(+\zeta)}+\text{(A/S)-}[2l](C)}(\mathfrak{t})\Bigr|_{a^1b^1}.
\end{align}
\end{subequations}
Note that due to the invariance of the integrand \eqref{SOoddC1} with $\gamma=0$ and $\chi=1$ except the insertion $\prod_if(s_i)$ under $s_i\rightarrow -s_i$, the one-point function vanishes trivially.

For this choice of $f(s)$, we propose the following expressions for $\tilde{\cal I}^{(0)}_{B,\gamma=0,\chi=1}[f;\nu]$ \eqref{calI0SOodd}
\begin{align}
&\tilde{\cal I}^{(0)}_{B,0,1}[(1+a(s+s^{-1}))(1+b(s+s^{-1}));\nu]\nonumber \\
&=\frac{1+\mathfrak{t}^{2\nu}}{2(\mathfrak{t}^4;\mathfrak{t}^4)_\nu}\Bigl(1+\frac{ab(1+\mathfrak{t}^{2\nu-2})(1-\mathfrak{t}^{2\nu})}{1-\mathfrak{t}^2}\Bigr)+{\cal O}(a^2,b^2),
\label{tildecalI(0)SOevenchi1fundguess}
\end{align}
which we have confirmed for $\nu\le 4$, while $\tilde{\cal I}^{(1)}[f;\nu]$ is the same as \eqref{calI1fundgeneralnu}.

We can evaluate the grand canonical sum of the indices \eqref{XiBfactorized} by plugging these $\tilde{\cal I}^{(0)}_{B,0,1}[f;\nu]$ and $\tilde{\cal I}^{(1)}[f;\nu]$ into \eqref{Xitilde(1)}, as
\begin{align}
&\Xi_{B,0,1}[(1+a(s+s^{-1}))(1+b(s+s^{-1}));u]\nonumber \\
&=
\Xi_{B,0,1}[1;u]
\biggl[1+\frac{abu(1+\zeta\mathfrak{t}^{2\Delta})}{(1-\mathfrak{t}^2)(1-\zeta\mathfrak{t}^{2\Delta})}
+{\cal O}(a^2,b^2)\biggr].
\label{SO2Nchi1fundGC}
\end{align}
Here
\begin{align}
&\Xi_{B,0,1}[1;u]=
\Bigl(\frac{1}{(u;\mathfrak{t}^4)_\infty}+\frac{1}{(u\mathfrak{t}^2;\mathfrak{t}^4)_\infty}\Bigr)
\prod_{m=1}^\infty \frac{1}{(\zeta^m \mathfrak{t}^{2\Delta m}u;\mathfrak{t}^2)_\infty}
\label{GCofICSO2N+1}
\end{align}
is the grand canonical sum of the Coulomb indices without Wilson lines.
From the grand canonical sum \eqref{SO2Nchi1fundGC}, we obtain the recursion relation for the two-point function
\begin{align}
&\langle W_{\ydiagram{1}}W_{\ydiagram{1}}\rangle^{SO(2N)^{(+\zeta)}+\text{(A/S)-}[2l](C)}(\mathfrak{t})\nonumber \\
&=
{\cal I}^{SO(2N-2)^{(+\zeta)}+\text{(A/S)-}[2l](C)}(\mathfrak{t})
\langle W_{\ydiagram{1}}W_{\ydiagram{1}}\rangle^{SO(2)^{(+\zeta)}+\text{(A/S)-}[2l](C)}(\mathfrak{t}),
\label{SO2Nchi+recrelnappendix}
\end{align}
for $N\ge 1$ (we have defined ${\cal I}^{SO(0)^{(+\zeta)}+\text{(A/S)-}[2l](C)}(\mathfrak{t})=2$), as we have claimed in section \eqref{R_so2N_C_W1}, with
\begin{align}
\langle W_{\ydiagram{1}}W_{\ydiagram{1}}\rangle^{SO(2)^{(+\zeta)}+\text{(A/S)-}[2l](C)}(\mathfrak{t})=\frac{1+\zeta\mathfrak{t}^{2\Delta}}{(1-\mathfrak{t}^2)(1-\zeta\mathfrak{t}^{2\Delta})}.
\label{WWSO2chi1}
\end{align}

In particular, for $\zeta=1$ or the theory with $\mathbb{Z}^{{\cal M}}_2$ gauged, we find
\begin{align}
&\lim_{\mathfrak{t}\rightarrow 1}
\langle {\cal W}_{\ydiagram{1}} {\cal W}_{\ydiagram{1}}\rangle^{SO(2N)^{(++)}+\text{(A/S)-}[2l](C)}(\mathfrak{t})
=
\lim_{\mathfrak{t}\rightarrow 1}
\langle {\cal W}_{\ydiagram{1}} {\cal W}_{\ydiagram{1}}\rangle^{Spin(2N)^{(+)}+\text{(A/S)-}[2l](C)}(\mathfrak{t})
=2N,
\end{align}
which are consistent with our conjecture \eqref{Witten_indC_conj_int}.
On the other hand, for $\zeta=-1$ we find
\begin{align}
&\lim_{\mathfrak{t}\rightarrow 1}
\langle {\cal W}_{\ydiagram{1}} {\cal W}_{\ydiagram{1}}\rangle^{SO(2N)^{(+-)}+\text{(A/S)-}[2l](C)}(\mathfrak{t})
=\begin{cases}
0&\quad (N\text{: even}),\\
2&\quad (N\text{: odd})
\end{cases}.
\label{SO2Nchi+fundzeta-1t1limit}
\end{align}

\subsubsection{(Anti)chiral spinor Wilson lines in $Spin(2N)^{(+)}$}
\label{SO+zeta_spinor}

Let us consider the two-point function of Wilson lines in the (anti)chiral spinor representations $\text{sp}$ (resp.~$\overline{\text{sp}}$) for $Spin(2N)^{(\chi)}$ gauge theories, whose characters are given by
\begin{subequations}
\begin{align}
&\chi^{\mathfrak{so}(2N)}_{\text{sp}}(s)=\frac{1}{2}\biggl(\prod_{i=1}^N(s_i^{\frac{1}{2}}+s_i^{-\frac{1}{2}})+\prod_{i=1}^N(s_i^{\frac{1}{2}}-s_i^{-\frac{1}{2}})\Bigr),\\
&\chi^{\mathfrak{so}(2N)}_{\overline{\text{sp}}}(s)=\frac{1}{2}\biggl(\prod_{i=1}^N(s_i^{\frac{1}{2}}+s_i^{-\frac{1}{2}})-\prod_{i=1}^N(s_i^{\frac{1}{2}}-s_i^{-\frac{1}{2}})\Bigr),
\end{align}
\end{subequations}
so that
\begin{subequations}
\begin{align}
&\langle \chi^{\mathfrak{so}(2N)}_{\text{sp}}(s)\chi^{\mathfrak{so}(2N)}_{\text{sp}}(s)\rangle^{SO(2N)^{(\chi\zeta)}+\text{(A/S)-}[2l]}(\mathfrak{t})\nonumber \\
&=
\frac{1}{4}
\biggl\langle
\prod_{i=1}^N(s_i+s_i^{-1}+2)
+2\prod_{i=1}^N(s_i-s_i^{-1})
+\prod_{i=1}^N(s_i+s_i^{-1}-2)
\biggr\rangle^{SO(2N)^{(\chi\zeta)}+\text{(A/S)-}[2l]}\nonumber \\
&=\frac{1+(-1)^N}{4}\biggl\langle
\prod_{i=1}^N(s_i+s_i^{-1}+2)
\biggr\rangle^{SO(2N)^{(\chi\zeta)}+\text{(A/S)-}[2l]},\\
&\langle \chi^{\mathfrak{so}(2N)}_{\text{sp}}(s)\chi^{\mathfrak{so}(2N)}_{\overline{\text{sp}}}(s)\rangle^{SO(2N)^{(\chi\zeta)}+\text{(A/S)-}[2l]}(\mathfrak{t})\nonumber \\
&=
\frac{1}{4}
\biggl\langle
\prod_{i=1}^N(s_i+s_i^{-1}+2)
-\prod_{i=1}^N(s_i+s_i^{-1}-2)
\biggr\rangle^{SO(2N)^{(\chi\zeta)}+\text{(A/S)-}[2l]}\nonumber \\
&=\frac{1-(-1)^N}{4}\biggl\langle
\prod_{i=1}^N(s_i+s_i^{-1}+2)
\biggr\rangle^{SO(2N)^{(\chi\zeta)}+\text{(A/S)-}[2l]}.
\end{align}
\end{subequations}
Here we have used the fact that the integrand of the $SO(2N)$ index is invariant under (i) $s_i\rightarrow s_i^{-1}$ (do for any single $i$) and (ii) $s_i\rightarrow -s_i$ (do for all $i$ at the same time).
Hence the line defect indices $\langle \chi_{\text{sp}}^{\mathfrak{so}(2N)}(s)\chi_{\text{sp}}^{\mathfrak{so}(2N)}(s)\rangle$ is non-vanishing only for even $N$, while $\langle \chi_{\text{sp}}^{\mathfrak{so}(2N)}(s)\chi_{\overline{\text{sp}}}^{\mathfrak{so}(2N)}(s)\rangle$ is non-vanishing only for odd $N$, both of which can be calculated
by choosing the single variable function as $f(s)=s+s^{-1}+2$.
For this choice of $f(s)$, we find
\begin{align}
&\tilde{\cal I}^{(0)}_{B,0,1}[s+s^{-1}+2;\nu]=\frac{1}{(\mathfrak{t}^2;\mathfrak{t}^2)_\nu},
\end{align}
which coincides with $\tilde{\cal I}^{(0)}_{B,1,1}[s+s^{-1}+2;\nu]$ for $SO(2N+1)^{(+\zeta)}$ gauge theory \eqref{tildecalISOoddspinor} and $\tilde{\cal I}^{(1)}(\nu)$ are also the same as those calculated for $SO(2N+1)^{(\chi\zeta)}$ (see \eqref{tildecalISOoddspinor}).
As a result, we obtain
\begin{subequations}
\begin{align}
&\langle W_{\text{sp}}W_{\overline{\text{sp}}}\rangle^{Spin(4N-2)^{(+)}+\text{(A/S)-}[2l](C)}(\mathfrak{t})=
\langle W_{\text{sp}}W_{\text{sp}}\rangle^{Spin(4N-1)^{(+)}+\text{(A/S)-}[2l](C)}(\mathfrak{t}),\\
&\langle W_{\text{sp}}W_{\text{sp}}\rangle^{Spin(4N)^{(+)}+\text{(A/S)-}[2l](C)}(\mathfrak{t})=
\langle W_{\text{sp}}W_{\text{sp}}\rangle^{Spin(4N+1)^{(+)}+\text{(A/S)-}[2l](C)}(\mathfrak{t}).
\end{align}
\end{subequations}

\subsection{$SO(2N)^{(-\zeta)}$}
\label{app_SOevenchim1}

Let us consider the theory with $SO(2N)$ gauge group with $\mathbb{Z}^{{\cal C}}_2$ fugacity $\chi=-1$.
In this case, one can set the $O(2N)$ holonomy to $\text{diag}(s_1$, $s_1^{-1}$, $\cdots$, $s_{N-1}$, $s_{N-1}^{-1}$, $1$, $-1)$ so that the gauge fugacity $s_{N}$ is simply replaced with $\pm 1$ and the magnetic flux $m_{N}$ is set to zero.
As a result, the supersymmetric index with the insertion of one variable function $\prod_{i=1}^{N-1}f(s_i)$ is given by \cite{Hayashi:2022ldo}
\begin{align}
\label{SO_index2}
&\biggl\langle \prod_{i=1}^{N-1}f(s_i)\biggr\rangle^{SO(2N)^{(-\zeta)}+\text{(A/S)-}[2l]}(t,x,y_{\alpha};q)\nonumber\\
&=
\frac{1}{2^{N-1} (N-1)!}
\frac{(-q^{\frac12}t^2;q)_{\infty}}
{(-q^{\frac12}t^{-2};q)_{\infty}}
\frac{(q^{\frac12}t^2;q)_{\infty}^{N-1}}
{(q^{\frac12}t^{-2};q)_{\infty}^{N-1}}
\sum_{m_1,\cdots, m_{N-1}\in \mathbb{Z}}
\oint \prod_{i=1}^{N-1}
\frac{ds_i}{2\pi is_i}f(s_i)
\nonumber\\
&\quad\times
\prod_{i=1}^{N-1}
\prod_\pm
(1-q^{|m_i|}s_i^{\pm2})
\prod_{i<j}^{N-1}
\prod_\pm
(1-q^{\frac{|m_i -m_j|}{2}} s_i^{\pm 1}s_j^{\mp 1})
(1-q^{\frac{|m_i +m_j|}{2}} s_i^{\pm 1}s_j^{\pm 1})
\nonumber\\
&\quad\times
\prod_{i=1}^{N-1}\prod_\pm
\frac{(q^{\frac12+\frac{|m_i|}{2}}t^2 s_i^{\mp 1};q)_{\infty}}
{(q^{\frac12+\frac{|m_i|}{2}}t^{-2}s_i^{\pm 1};q)_{\infty}}
\frac{(-q^{\frac12+\frac{|m_i|}{2}}t^2 s_i^{\mp 1};q)_{\infty}}
{(-q^{\frac12+\frac{|m_i|}{2}}t^{-2}s_i^{\pm 1};q)_{\infty}}
\nonumber\\
&\quad\times
\prod_{i<j}^{N-1}\prod_\pm
\frac{(q^{\frac12+\frac{|m_i-m_j|}{2}}t^2 s_i^{\mp 1}s_j^{\pm 1};q)_{\infty}}
{(q^{\frac12+\frac{|m_i-m_j|}{2}}t^{-2} s_i^{\pm 1}s_j^{\mp 1};q)_{\infty}}
\frac{(q^{\frac12+\frac{|m_i+m_j|}{2}}t^2 s_i^{\mp 1}s_j^{\mp 1};q)_{\infty}}
{(q^{\frac12+\frac{|m_i+m_j|}{2}}t^{-2} s_i^{\pm 1}s_j^{\pm 1};q)_{\infty}}
\nonumber\\
&\quad\times
\prod_\pm
\frac{(-q^{\frac34}t^{-1}x^{\mp 1};q)_{\infty}}
{(-q^{\frac14}tx^{\pm 1};q)_{\infty}}
\frac{(q^{\frac34}t^{-1}x^{\mp 1};q)_{\infty}^{N-1}}
{(q^{\frac14}tx^{\pm 1};q)_{\infty}^{N-1}}\nonumber \\
&\quad\times
\prod_{i=1}^{N-1}\prod_\pm
\frac{(q^{\frac34+\frac{|m_i|}{2}}t^{-1}s_i^{\mp 1}x^{\mp 1};q)_{\infty}}
{(q^{\frac14+\frac{|m_i|}{2}}ts_i^{\pm 1}x^{\pm 1};q)_{\infty}}
\frac{(q^{\frac34+\frac{|m_i|}{2}}t^{-1}s_i^{\mp 1}x^{\pm 1};q)_{\infty}}
{(q^{\frac14+\frac{|m_i|}{2}}ts_i^{\pm 1}x^{\mp 1};q)_{\infty}}
\nonumber\\
&\quad\times
\prod_{i=1}^{N-1}\prod_\pm
\frac{(-q^{\frac34+\frac{|m_i|}{2}}t^{-1}s_i^{\mp 1}x^{\mp 1};q)_{\infty}}
{(-q^{\frac14+\frac{|m_i|}{2}}ts_i^{\pm 1}x^{\pm 1};q)_{\infty}}
\frac{(-q^{\frac34+\frac{|m_i|}{2}}t^{-1}s_i^{\mp 1}x^{\pm 1};q)_{\infty}}
{(-q^{\frac14+\frac{|m_i|}{2}}ts_i^{\pm 1}x^{\mp 1};q)_{\infty}}
\nonumber\\
&\quad\times
\prod_{i<j}^{N-1}\prod_\pm
\frac{(q^{\frac34+\frac{|m_i-m_j|}{2}} t^{-1}s_i^{\mp 1} s_j^{\pm 1}x^{\mp 1};q)_{\infty}}
{(q^{\frac14+\frac{|m_i-m_j|}{2}} ts_i^{\pm 1} s_j^{\mp 1}x^{\pm 1};q)_{\infty}}
\frac{(q^{\frac34+\frac{|m_i-m_j|}{2}} t^{-1}s_i^{\mp 1} s_j^{\pm 1}x^{\pm 1};q)_{\infty}}
{(q^{\frac14+\frac{|m_i-m_j|}{2}} ts_i^{\pm 1} s_j^{\mp 1}x^{\mp 1};q)_{\infty}}
\nonumber\\
&\quad\times
\prod_{i<j}\prod_\pm
\frac{(q^{\frac34+\frac{|m_i+m_j|}{2}} t^{-1}s_i^{\mp 1} s_j^{\mp 1}x^{\mp 1};q)_{\infty}}
{(q^{\frac14+\frac{|m_i+m_j|}{2}} ts_i^{\pm 1} s_j^{\pm 1}x^{\pm 1};q)_{\infty}}
\frac{(q^{\frac34+\frac{|m_i+m_j|}{2}} t^{-1}s_i^{\mp 1} s_j^{\mp 1}x^{\pm 1};q)_{\infty}}
{(q^{\frac14+\frac{|m_i+m_j|}{2}} ts_i^{\pm 1} s_j^{\pm 1}x^{\mp 1};q)_{\infty}}
\nonumber\\
&\quad\times
\left[
\prod_{i=1}^{N-1}\prod_\pm
\frac{(q^{\frac34+|m_i|}t^{-1}s_i^{\mp 2}x^{\mp 1};q)_{\infty}}
{(q^{\frac14+|m_i|}t s_i^{\pm 2}x^{\pm 1};q)_{\infty}}
\frac{(q^{\frac34+|m_i|}t^{-1}s_i^{\mp 2}x^{\pm 1};q)_{\infty}}
{(q^{\frac14+|m_i|}t s_i^{\pm 2}x^{\mp 1};q)_{\infty}}
\prod_\pm
\frac{(q^{\frac34}t^{-1}x^{\mp 1};q)_{\infty}^2}
{(q^{\frac14}tx^{\pm 1};q)_{\infty}^2}
\right]^{\epsilon}
\nonumber\\
&\quad\times
\prod_{\alpha=1}^{2l}
\prod_{i=1}^{N-1}
\prod_\pm
\frac{(q^{\frac34+\frac{|m_i|}{2}} t^{-1}s_i^{\mp 1}y_{\alpha}^{-1};q)_{\infty}}
{(q^{\frac14+\frac{|m_i|}{2}} ts_i^{\pm 1}y_{\alpha};q)_{\infty}}
\prod_\pm
\frac{(\pm q^{\frac34}t^{-1}y_{\alpha}^{-1};q)_{\infty}}
{(\pm q^{\frac14}ty_{\alpha};q)_{\infty}}
\nonumber\\
&\quad\times
q^{\frac{\Delta}{2} \sum_{i=1}^{N-1}|m_i|}
t^{-2\Delta\sum_{i=1}^{N-1}|m_i|}
\zeta^{\sum_{i=1}^{N-1} m_i},
\end{align}
with $\Delta=l+2\epsilon$.

After restricting the summation over the monopole charges to $m_i\ge 0$ (see \eqref{Z2trick}) and taking the Coulomb limit, we obtain
\begin{align}
&\biggl\langle \prod_{i=1}^{N-1}f(s_i)\biggr\rangle^{SO(2N)^{(-\zeta)}+\text{(A/S)-}[2l](C)}(\mathfrak{t})\nonumber \\
&=\frac{1}{1+\mathfrak{t}^2}\frac{1}{2^{N-1}\cdot (N-1)!}\frac{1}{(1-\mathfrak{t}^2)^{N-1}}\sum_{m_1,\cdots,m_{N-1}\ge 0}\prod_{i=1}^{N-1} 2^{1-\delta_{m_i,0}} \int\prod_{i=1}^{N-1}\frac{ds_i}{2\pi is_i}f(s_i)\nonumber \\
&\quad\times
\prod_{\substack{i=1\\ (m_i=0)}}^{N-1}\prod_\pm\frac{1-s_i^{\pm 2}}{1-\mathfrak{t}^4 s_i^{\pm 2}}
\prod_{\substack{i<j\\ (m_i=m_j)}}^{N-1}\prod_\pm\frac{1-(\frac{s_i}{s_j})^{\pm 1}}{1-\mathfrak{t}^2(\frac{s_i}{s_j})^{\pm 1}}
\prod_{\substack{i<j\\ (m_i=-m_j)}}^{N-1}\prod_\pm\frac{1-(s_is_j)^{\pm 1}}{1-\mathfrak{t}^2(s_is_j)^{\pm 1}}
\mathfrak{t}^{2\Delta \sum_{i=1}^{N-1}|m_i|}\nonumber \\
&\quad\times \zeta^{\sum_{i=1}^{N-1}m_i}.
\end{align}
By classifying $N-1$ monopole charges by their multiplicities, we obtain
\begin{align}
&\biggl\langle \prod_{i=1}^{N-1}f(s_i)\biggr\rangle^{SO(2N)^{(-\zeta)}+\text{(A/S)-}[2l](C)}(\mathfrak{t})\nonumber \\
&=\frac{1}{1+\mathfrak{t}^2}\sum_{\substack{\nu_0,\nu_m=0\\ (\nu_0+\sum_{m=1}^\infty \nu_m=N-1)}}^\infty (\zeta \mathfrak{t}^{2\Delta})^{\sum_{m=1}^\infty m \nu_m}\tilde{\cal I}^{(0)}_\text{D}[f;\nu_0]\prod_{m=1}^\infty\tilde{\cal I}^{(1)}[f;\nu_m]\nonumber \\
&=
\frac{1}{1+\mathfrak{t}^2}
\sum_{\kappa=0}^{N-1}
\tilde{\cal I}_{\text{D}}^{(0)}[f;N-1-\kappa]
\sum_{\substack{\lambda\\ (|\lambda|=\kappa)}}
G_\lambda(\zeta,\mathfrak{t})
\prod_{i\ge 1}
\frac{
\tilde{\cal I}^{(1)}[f;i]^{m_i(\lambda)}
}{
m_i(\lambda)!},
\end{align}
where $\tilde{\cal I}^{(1)}[f;\nu]$ and $G_\lambda(\zeta,\mathfrak{t})$ are given by \eqref{calI1SOodd} and \eqref{formfactorG}, and
\begin{align}
&\tilde{\cal I}^{(0)}_\text{D}[f;\nu]\nonumber \\
&=\frac{1}{2^\nu(1-\mathfrak{t}^2)^\nu\nu!}\oint\prod_{i=1}^\nu\frac{d\sigma_i}{2\pi i\sigma_i}f(\sigma_i)\prod_\pm \frac{1-\sigma_i^{\pm 2}}{1-\mathfrak{t}^4\sigma_i^{\pm 2}}
\prod_{i<j}^\nu\prod_\pm \frac{
(1-\sigma_i^{\pm 1}\sigma_j^{\mp 1})(1-\sigma_i^{\pm 1}\sigma_j^{\pm 1})
}{
(1-\mathfrak{t}^2\sigma_i^{\pm 1}\sigma_j^{\mp 1})(1-\mathfrak{t}^2\sigma_i^{\pm 1}\sigma_j^{\pm 1})
}.
\end{align}

Again, the grand canonical sum
\begin{align}
\Xi_{\text{D}}[f;u]=\sum_{N=0}^\infty u^N\biggl\langle \prod_{i=1}^{N-1}f(s_i)\biggr\rangle^{SO(2N)^{(-\zeta)}+\text{(A/S)-}[2l](C)}(\mathfrak{t})
\end{align}
with ${\cal I}^{SO(0)^{(-\zeta)}+\text{(A/S)-}[2l](C)}(\mathfrak{t})\equiv 0$
factorizes as
\begin{align}
\Xi_{\text{D}}[f;u]=\frac{1}{1+\mathfrak{t}^2}\tilde{\Xi}^{(0)}_{\text{D}}[f;u]\prod_{m=1}^\infty\tilde{\Xi}^{(1)}[f;\zeta^m\mathfrak{t}^{2m\Delta} u],
\label{XiD}
\end{align}
with $\tilde{\Xi}^{(1)}(u)$ given by \eqref{Xitilde(1)} and
\begin{align}
\tilde{\Xi}_{\text{D}}^{(0)}[f;u]=\sum_{\nu=0}^\infty u^{\nu+1} \tilde{\cal I}^{(0)}_{\text{D}}[f;\nu].
\label{Xi(0)D}
\end{align}

\subsubsection{Fundamental Wilson lines}
\label{app_SOevenchim1fund}

We define the two-point function of fundamental Wilson lines in the $SO(2N)^{(-\zeta)}$ theory by the insertion of $\sum_{i,j=1}^{N-1}(s_i+s_i^{-1})(s_j+s_j^{-1})$, which can be calculated by choosing the single variable function $f(s)=(1+a(s+s^{-1}))(1+b(s+s^{-1}))$ as
\begin{align}
\langle W_{\ydiagram{1}}W_{\ydiagram{1}}\rangle^{SO(2N)^{(-\zeta)}+\text{(A/S)-}[2l](C)}(\mathfrak{t})= \biggl\langle \prod_{i=1}^{N-1} f(s_i)\biggr\rangle^{SO(2N)^{(-\zeta)}+\text{(A/S)-}[2l](C)}(\mathfrak{t})\Bigr|_{ab}.
\end{align}
For this choice of $f(s)$, we find
\begin{align}
&\tilde{\cal I}^{(0)}_D[(1+a(s+s^{-1}))(1+b(s+s^{-1}));\nu]\nonumber \\
&=\frac{1}{(1-\mathfrak{t}^2)(1+\mathfrak{t}^{2\nu+2})\prod_{j=1}^{\nu-1}(1-\mathfrak{t}^{4j+4})}\Bigl(1+ab\frac{(1-\mathfrak{t}^{2\nu})(1+\mathfrak{t}^{2\nu+2})}{1-\mathfrak{t}^2}+{\cal O}(a^2,b^2)\Bigr),
\end{align}
while $\tilde{\cal I}^{(1)}[f;\nu]$ are the same as those for $SO(2N+1)^{(\chi\zeta)}$ \eqref{calI1fundgeneralnu}.
Plugging these into the grand canonical sum \eqref{XiD} with \eqref{Xi(0)D}, we obtain
\begin{align}
\Xi_\text{D}[(1+a(s+s^{-1}))(1+b(s+s^{-1}));u]
=
\Xi_\text{D}[1;u]
\Bigl(1+\frac{(1+\zeta \mathfrak{t}^{2\Delta})abu}{(1-\mathfrak{t}^2)(1-\zeta\mathfrak{t}^{2\Delta})}+{\cal O}(a^2,b^2)\Bigr),
\end{align}
where
\begin{align}
\Xi_\text{D}[1;u]&=\sum_{N=0}^\infty u^N{\cal I}^{SO(2N)^{(-\zeta)}+\text{(A/S)-}[2l](C)}(\mathfrak{t})\nonumber \\
&=\Bigl(\frac{1}{(u;\mathfrak{t}^4)_\infty}-\frac{1}{(u\mathfrak{t}^2;\mathfrak{t}^4)_\infty}\Bigr)\prod_{m=1}^\infty\frac{1}{(\zeta^m\mathfrak{t}^{2m\Delta}u;\mathfrak{t}^2)_\infty}
\end{align}
is the generating function of the Coulomb index without Wilson line.
This implies that the two-point function satisfies the recursion relation as claimed in \eqref{R_so2N_C_W1}
\begin{align}
&\langle W_{\ydiagram{1}}W_{\ydiagram{1}}\rangle^{SO(2N)^{-\zeta}+\text{(A/S)-}[2l](C)}(\mathfrak{t})\nonumber \\
&
=
{\cal I}^{SO(2N-2)^{-\zeta}+\text{(A/S)-}[2l](C)}(\mathfrak{t})
\langle W_{\ydiagram{1}}W_{\ydiagram{1}}\rangle^{SO(2)^{+\zeta}+\text{(A/S)-}[2l](C)}(\mathfrak{t})
\label{SO2Nchi-recrelnappendix}
\end{align}
for $N\ge 1$, where $\langle W_{\ydiagram{1}}W_{\ydiagram{1}}\rangle^{SO(2)^{+\zeta}+\text{(A/S)-}[2l](C)}(\mathfrak{t})$ is given by \eqref{WWSO2chi1}.

In particular, for $\zeta=1$ and the theory with $\mathbb{Z}^{{\cal M}}_2$ gauged we find
\begin{align}
&\lim_{\mathfrak{t}\rightarrow 1}
\langle {\cal W}_{\ydiagram{1}}{\cal W}_{\ydiagram{1}}\rangle^{SO(2N)^{(-+)}+\text{(A/S)-}[2l](C)}(\mathfrak{t})
=\lim_{\mathfrak{t}\rightarrow 1}
\langle {\cal W}_{\ydiagram{1}}{\cal W}_{\ydiagram{1}}\rangle^{Spin(2N)^{(-)}+\text{(A/S)-}[2l](C)}(\mathfrak{t})
=2N-2,
\end{align}
which are consistent with our conjecture \eqref{Witten_indC_conj_int}, while for $\zeta=-1$ we find
\begin{align}
&\lim_{\mathfrak{t}\rightarrow 1}
\langle {\cal W}_{\ydiagram{1}}{\cal W}_{\ydiagram{1}}\rangle^{SO(2N)^{(--)}+\text{(A/S)-}[2l](C)}(\mathfrak{t})
=
\begin{cases}
\displaystyle \frac{2\Delta}{\Delta+2}&\quad (N\text{: even})\vspace{0.2cm} \\
\displaystyle 0&\quad (N\text{: odd})
\end{cases}.
\label{SO2Nchi-fundzeta-1t1limit}
\end{align}

We can also consider the (line defect) indices of the theory with $\mathbb{Z}^{{\cal C}}_2$ symmetry gauged
\begin{align}
\langle W_{{\cal R}_1}\cdots W_{{\cal R}_k}\rangle^{O(2N)^{(\zeta)}+\text{(A/S)-}[2l](C)}(\mathfrak{t})=
\frac{1}{2}\sum_{\chi=\pm 1}
\langle W_{{\cal R}_1}\cdots W_{{\cal R}_k}\rangle^{SO(2N)^{(\chi\zeta)}+\text{(A/S)-}[2l](C)}(\mathfrak{t}).
\end{align}
By combining the recursion relation \eqref{SO2Nchi-recrelnappendix} with that of $SO(2N)^{(+\zeta)}$ theory \eqref{SO2Nchi+recrelnappendix}, we find the recursion relation of $O(2N)^{(\zeta)}$ theory as
\begin{align}
\langle W_{\ydiagram{1}}W_{\ydiagram{1}}\rangle^{O(2N)^{(\zeta)}+\text{(A/S)-}[2l](C)}(\mathfrak{t})
={\cal I}^{O(2N-2)^{(\zeta)}+\text{(A/S)-}[2l](C)}(\mathfrak{t})
\langle W_{\ydiagram{1}}W_{\ydiagram{1}}\rangle^{SO(2)^{(+\zeta)}+\text{(A/S)-}[2l](C)}(\mathfrak{t}),
\end{align}
where ${\cal I}^{O(2N)^{(\zeta)}+\text{(A/S)-}[2l](C)}(\mathfrak{t})$ are obtained from the grand canonical sum
\begin{align}
\sum_{N=0}^\infty u^N {\cal I}^{O(2N)^{(\zeta)}+\text{(A/S)-}[2l](C)}(\mathfrak{t})=
\frac{
\Xi_{B,0,1}[1;u]+\Xi_{\text{D}}[1;u]
}{2}=
\frac{1}{(u;\mathfrak{t}^4)_\infty}\prod_{m=1}^\infty\frac{1}{(\zeta^m\mathfrak{t}^{2\Delta m}u;\mathfrak{t}^2)_\infty}.
\label{GCofICO2N}
\end{align}
Note that this coincides with the grand canonical sum of the supersymmetric index of $SO(2N+1)^{(\chi\zeta)}$ theory \eqref{GCofICSO2N+1}.
Namely, we have
\begin{align}
\label{ICO2N=ICO2N+1}
{\cal I}^{O(2N)^{(\zeta)}+\text{(A/S)-}[2l](C)}(\mathfrak{t})=
{\cal I}^{SO(2N+1)^{(\chi\zeta)}+\text{(A/S)-}[2l](C)}(\mathfrak{t})
={\cal I}^{O(2N+1)^{(\zeta)}+\text{(A/S)-}[2l](C)}(\mathfrak{t}),
\end{align}
where the last equality is simply due to the fact that the Coulomb index of $SO(2N+1)^{(\chi\zeta)}$ gauge theory is independent of $\chi=\pm 1$.
In particular, in the limit $\mathfrak{t}\rightarrow 1$ we find
\begin{subequations}
\begin{align}
&\lim_{\mathfrak{t}\rightarrow 1}
\langle {\cal W}_{\ydiagram{1}}{\cal W}_{\ydiagram{1}}\rangle^{O(2N)^{(+)}+\text{(A/S)-}[2l](C)}(\mathfrak{t})
=2N,\\
&\lim_{\mathfrak{t}\rightarrow 1}
\langle {\cal W}_{\ydiagram{1}}{\cal W}_{\ydiagram{1}}\rangle^{O(2N)^{(-)}+\text{(A/S)-}[2l](C)}(\mathfrak{t})
=
\begin{cases}
\displaystyle 0&\quad (N\text{: even})\vspace{0.2cm} \\
\displaystyle \frac{2\Delta}{\Delta+1}&\quad (N\text{: odd})
\end{cases}.
\end{align}
\end{subequations}
Again, the results for $\zeta=1$ are consistent with our conjecture \eqref{Witten_indC_conj_int}.

\subsubsection{(Anti)chiral spinor Wilson lines}
As explained in appendix \ref{SO+zeta_spinor}, the two-point function of the Wilson lines in (anti)chiral spinor representations can be calculated by choosing the single variable function $f(s)=s+s^{-1}+2$.
For this choice of $f(s)$, we find
\begin{align}
&{\cal I}^{(0)}_D[s+s^{-1}+2;1]=
\frac{2}{(1 - \mathfrak{t}^2) (1 + \mathfrak{t}^4)},\\
&{\cal I}^{(0)}_D[s+s^{-1}+2;2]=
\frac{
3 - 2 \mathfrak{t}^2 + \mathfrak{t}^4
}{(1 - \mathfrak{t}^2)^2 (1 + \mathfrak{t}^4) (1 + \mathfrak{t}^6)},\\
&{\cal I}^{(0)}_D[s+s^{-1}+2;3]=
\frac{
2 (2 - \mathfrak{t}^2 + \mathfrak{t}^4 - \mathfrak{t}^6 + \mathfrak{t}^8)
}{(1 - \mathfrak{t}^2)^2 (1 + \mathfrak{t}^4) (1 + \mathfrak{t}^8) (1 - \mathfrak{t}^{12})},
\end{align}
while ${\cal I}^{(1)}[s+s^{-1}+2;\nu]$ are given by \eqref{tildecalISOoddspinor}.
With these results, we find, after gauging $\mathbb{Z}^{{\cal M}}_2$ symmetry,
{\fontsize{9pt}{1pt}\selectfont
\begin{subequations}
\begin{align}
&\langle W_{\text{sp}} W_{\text{sp}}\rangle^{Spin(4)^{(-)}+\text{(A/S)-}[2l](C)}(\mathfrak{t})=
\frac{
2 + 2 \mathfrak{t}^{4 \Delta + 4}}{(1 - \mathfrak{t}^8) (1 - \mathfrak{t}^{4 \Delta})},\\
&\langle W_{\text{sp}} W_{\overline{\text{sp}}}\rangle^{Spin(6)^{(-)}+\text{(A/S)-}[2l](C)}(\mathfrak{t})\nonumber \\
&=
\frac{
1 + \mathfrak{t}^2}{(1 - \mathfrak{t}^4) (1 - \mathfrak{t}^8) (1 + \mathfrak{t}^6) (1 - \mathfrak{t}^{4 \Delta})^2}
(
3 - 2 \mathfrak{t}^2 + \mathfrak{t}^4
+ \mathfrak{t}^{4 \Delta} (1 + 2 \mathfrak{t}^2 + 3 \mathfrak{t}^4 + 3 \mathfrak{t}^6 + 2 \mathfrak{t}^8 + \mathfrak{t}^{10})
+ \mathfrak{t}^{8 \Delta + 6} (1 - 2 \mathfrak{t}^2 + 3 \mathfrak{t}^4)
),\\
&\langle W_{\text{sp}} W_{\text{sp}}\rangle^{Spin(8)^{(-)}+\text{(A/S)-}[2l](C)}(\mathfrak{t})\nonumber \\
&=
\frac{
1}{(1 - \mathfrak{t}^2) (1 - \mathfrak{t}^{12}) (1 - \mathfrak{t}^{16}) (1 - \mathfrak{t}^{4 \Delta})^2 (1 - \mathfrak{t}^{12 \Delta})}
(4 - 2 \mathfrak{t}^2 + 2 \mathfrak{t}^4 - 2 \mathfrak{t}^6 + 2 \mathfrak{t}^8\nonumber \\
&
+ \mathfrak{t}^{4 \Delta} (4 + 8 \mathfrak{t}^2 + 6 \mathfrak{t}^4 + 4 \mathfrak{t}^6 + 16 \mathfrak{t}^8    + 6 \mathfrak{t}^{10} + 10 \mathfrak{t}^{12}              + 8 \mathfrak{t}^{16}  + 2 \mathfrak{t}^{18})\nonumber \\
&
+ \mathfrak{t}^{8 \Delta} (6 + 4 \mathfrak{t}^2 + 12 \mathfrak{t}^4 + 10 \mathfrak{t}^6 + 16 \mathfrak{t}^8 + 16 \mathfrak{t}^{10} + 16 \mathfrak{t}^{12} + 14 \mathfrak{t}^{14} + 12 \mathfrak{t}^{16} + 10 \mathfrak{t}^{18} + 6 \mathfrak{t}^{20} + 2 \mathfrak{t}^{22})\nonumber \\
&
+ \mathfrak{t}^{12 \Delta} (2 + 6 \mathfrak{t}^2 + 10 \mathfrak{t}^4 + 12 \mathfrak{t}^6 + 14 \mathfrak{t}^8 + 16 \mathfrak{t}^{10} + 16 \mathfrak{t}^{12} + 16 \mathfrak{t}^{14} + 10 \mathfrak{t}^{16} + 12 \mathfrak{t}^{18} + 4 \mathfrak{t}^{20} + 6 \mathfrak{t}^{22})\nonumber \\
&
+ \mathfrak{t}^{16 \Delta + 4} (2 + 8 \mathfrak{t}^2 + 10 \mathfrak{t}^6 + 6 \mathfrak{t}^8       + 16 \mathfrak{t}^{10}  + 4 \mathfrak{t}^{12}  + 6 \mathfrak{t}^{14}  + 8 \mathfrak{t}^{16}  + 4 \mathfrak{t}^{18})
+ \mathfrak{t}^{20 \Delta + 14} (2 - 2 \mathfrak{t}^2 + 2 \mathfrak{t}^4 - 2 \mathfrak{t}^6 + 4 \mathfrak{t}^8)
),
\end{align}
\end{subequations}
}

In the limit $\mathfrak{t}\rightarrow 1$ we find
\begin{subequations}
\begin{align}
&\lim_{\mathfrak{t}\rightarrow 1}
\langle {\cal W}_{\text{sp}}{\cal W}_{\text{sp}}\rangle^{Spin(4)^{(-)}+\text{(A/S)-}[2l](C)}(\mathfrak{t})
=2,\\
&\lim_{\mathfrak{t}\rightarrow 1}
\langle {\cal W}_{\text{sp}}{\cal W}_{\overline{\text{sp}}}\rangle^{Spin(6)^{(-)}+\text{(A/S)-}[2l](C)}(\mathfrak{t})
=4,\\
&\lim_{\mathfrak{t}\rightarrow 1}
\langle {\cal W}_{\text{sp}}{\cal W}_{\text{sp}}\rangle^{Spin(8)^{(-)}+\text{(A/S)-}[2l](C)}(\mathfrak{t})
=8,
\end{align}
\end{subequations}
which are consistent with our conjecture \eqref{Witten_indC_conj_int}.

\subsection{$USp(2N)$}
\label{app_USp}

Next let us consider the theory with $USp(2N)$ gauge group with one hypermultiplet in symmetric or antisymmetric representation and $2l+\delta$ half-hypermultiplets in the fundamental representation ($\delta=0,1$).
Let us consider the supersymmetric index with the insertion of the product of single variable function $\prod_{i=1}^N f(s_i)$
\begin{align}
&\biggl\langle\prod_{i=1}^Nf(s_i)\biggr\rangle^{USp(2N)+\text{(A/S)-}[2l+\delta]}(t,x,y_\alpha;q)\nonumber \\
&
=\frac{1}{2^NN!}
\frac{
(q^{\frac{1}{2}}t^2;q)_\infty^N
}{
(q^{\frac{1}{2}}t^{-2};q)_\infty^N
}
\sum_{m_1,\cdots,m_N\in\mathbb{Z}}\oint\prod_{i=1}^N\frac{ds_i}{2\pi is_i}
f(s_i)
\prod_{i=1}^N\prod_\pm (1-q^{|m_i|}s_i^{\pm 2})\nonumber \\
&\quad\times
\prod_{i<j}^N\prod_\pm
\Bigl(1-q^{\frac{|m_i-m_j|}{2}}s_i^{\pm 1}s_j^{\mp 1}\Bigr)
\Bigl(1-q^{\frac{|m_i+m_j|}{2}}s_i^{\pm 1}s_j^{\pm 1}\Bigr)
\nonumber \\
&\quad\times
\prod_{i=1}^N\prod_\pm
\frac{
(q^{\frac{1}{2}+|m_i|}t^2s_i^{\mp 2};q)_\infty
}{
(q^{\frac{1}{2}+|m_i|}t^{-2}s_i^{\pm 2};q)_\infty
}\nonumber \\
&\quad\times
\prod_{i<j}^N\prod_\pm
\frac{
(q^{\frac{1}{2}+\frac{|m_i-m_j|}{2}}t^2s_i^{\mp 1}s_j^{\pm 1};q)_\infty
}{
(q^{\frac{1}{2}+\frac{|m_i-m_j|}{2}}t^{-2}s_i^{\pm 1}s_j^{\mp 1};q)_\infty
}
\frac{
(q^{\frac{1}{2}+\frac{|m_i+m_j|}{2}}t^2s_i^{\mp 1}s_j^{\mp 1};q)_\infty
}{
(q^{\frac{1}{2}+\frac{|m_i+m_j|}{2}}t^{-2}s_i^{\pm 1}s_j^{\pm 1};q)_\infty
}\nonumber \\
&\quad\times
\prod_\pm
\frac{
(q^{\frac{3}{4}}t^{-1}x^{\mp 1};q)_\infty^N
}{
(q^{\frac{1}{4}}t x^{\pm 1};q)_\infty^N
}\nonumber \\
&\quad\times
\prod_{i<j}^N\prod_\pm
\frac{
(q^{\frac{3}{4}+\frac{|m_i-m_j|}{2}}t^{-1}s_i^{\mp 1}s_j^{\pm 1}x^{\mp 1};q)_\infty
}{
(q^{\frac{1}{4}+\frac{|m_i-m_j|}{2}}t s_i^{\pm 1}s_j^{\mp 1}x^{\pm 1};q)_\infty
}
\frac{
(q^{\frac{3}{4}+\frac{|m_i-m_j|}{2}}t^{-1}s_i^{\mp 1}s_j^{\pm 1}x^{\pm 1};q)_\infty
}{
(q^{\frac{1}{4}+\frac{|m_i-m_j|}{2}}t s_i^{\pm 1}s_j^{\mp 1}x^{\mp 1};q)_\infty
}\nonumber \\
&\quad\times
\prod_{i<j}^N\prod_\pm
\frac{
(q^{\frac{3}{4}+\frac{|m_i+m_j|}{2}}t^{-1}s_i^{\mp 1}s_j^{\mp 1}x^{\mp 1};q)_\infty
}{
(q^{\frac{1}{4}+\frac{|m_i+m_j|}{2}}t s_i^{\pm 1}s_j^{\pm 1}x^{\pm 1};q)_\infty
}
\frac{
(q^{\frac{3}{4}+\frac{|m_i+m_j|}{2}}t^{-1}s_i^{\mp 1}s_j^{\mp 1}x^{\pm 1};q)_\infty
}{
(q^{\frac{1}{4}+\frac{|m_i+m_j|}{2}}t s_i^{\pm 1}s_j^{\pm 1}x^{\mp 1};q)_\infty
}\nonumber \\
&\quad\times
\prod_{i=1}^N\prod_{\alpha=1}^{2l+\delta}\prod_\pm
\frac{
(q^{\frac{3}{4}+\frac{|m_i|}{2}}t^{-1}s_i^{\mp 1}y_\alpha^{-1};q)_\infty
}{
(q^{\frac{1}{4}+\frac{|m_i|}{2}}t s_i^{\pm 1}y_\alpha;q)_\infty
}\nonumber \\
&\quad\times
\Biggl[
\prod_{i=1}^N\prod_\pm
\frac{
(q^{\frac{3}{4}+\frac{|m_i|}{2}}t^{-1}s_i^{\mp 2}x^{\mp 1};q)_\infty
}{
(q^{\frac{1}{4}+\frac{|m_i|}{2}}t s_i^{\pm 2}x^{\pm 1};q)_\infty
}
\frac{
(q^{\frac{3}{4}+\frac{|m_i|}{2}}t^{-1}s_i^{\mp 2}x^{\pm 1};q)_\infty
}{
(q^{\frac{1}{4}+\frac{|m_i|}{2}}t s_i^{\pm 2}x^{\mp 1};q)_\infty
}
\Biggr]^{\epsilon}\nonumber \\
&\quad\times q^{\frac{\Delta}{2}\sum_{i=1}^N|m_i|}t^{-2\Delta \sum_{i=1}^N|m_i|},
\label{USpfull}
\end{align}
with $\epsilon=0$ (resp.~$\epsilon=1$) for the hypermultiplet with fugacity $x$ being in antisymmetric (resp.~symmetric) representation, and the flavor fugacities $y_\alpha$ satisfying $y_{l+\alpha}=y_\alpha^{-1}$ and $y_{2l+1}=1$ when $\delta=1$.
Here we have also defined $\Delta$ as
\begin{align}
\Delta=l-2+2\epsilon+\frac{\delta}{2}.
\label{DeltaUSp}
\end{align}

After restricting the summation over the monopole charges to $m_i\ge 0$ (see \eqref{Z2trick}) and taking the Coulomb limit, we obtain
\begin{align}
&\biggl\langle\prod_{i=1}^Nf(s_i)\biggr\rangle^{USp(2N)+\text{(A/S)-}[2l+\delta](C)}(\mathfrak{t})\nonumber \\
&
=\frac{1}{2^NN!}
\frac{
1
}{
(1-\mathfrak{t}^2)^N
}
\sum_{m_1,\cdots,m_N\ge 0}\prod_{i=1}^N 2^{1-\delta_{m_i,0}}\oint\prod_{i=1}^N\frac{ds_i}{2\pi is_i}
f(s_i)
\prod_{\substack{i=1\\ (m_i=0)}}^N\prod_\pm
\frac{
1-s_i^{\pm 2}
}{
1-\mathfrak{t}^2s_i^{\pm 2}
}\nonumber \\
&\quad\times
\prod_{\substack{i<j\\ (m_i=m_j)}}^N\prod_\pm
\frac{
1-s_i^{\pm 1}s_j^{\mp 1}
}{
1-\mathfrak{t}^2s_i^{\pm 1}s_j^{\mp 1}
}
\prod_{\substack{i<j\\ (m_i=-m_j)}}^N\prod_\pm
\frac{
1-s_i^{\pm 1}s_j^{\pm 1}
}{
1-\mathfrak{t}^2s_i^{\pm 1}s_j^{\pm 1}
}
\mathfrak{t}^{2\Delta \sum_{i=1}^N|m_i|}\nonumber \\
&=
\sum_{\kappa=0}^N
\tilde{\cal I}_{\text{C}}^{(0)}[f;N-\kappa]
\sum_{\substack{\lambda\\ (|\lambda|=\kappa)}}
G_\lambda(\zeta=1,\mathfrak{t})
\prod_{i\ge 1}
\frac{
\tilde{\cal I}^{(1)}[f;i]^{m_i(\lambda)}
}{
m_i(\lambda)!},
\label{USpC}
\end{align}
where $\tilde{\cal I}^{(1)}[f;\nu]$ and $G_\lambda(\zeta,\mathfrak{t})$ are given by \eqref{calI1SOodd} and \eqref{formfactorG}, and
\begin{align}
&\tilde{\cal I}_\text{C}^{(0)}[f;\nu]\nonumber \\
&=\frac{1}{2^\nu\cdot \nu!}\frac{1}{(1-\mathfrak{t}^2)^{\nu}}\oint\prod_{i=1}^\nu\frac{d\sigma_i}{2\pi i\sigma_i}f(\sigma_i) \prod_\pm\frac{1-\sigma_i^{\pm 2}}{1-\mathfrak{t}^2\sigma_i^{\pm 2}}
\prod_{i<j}^\nu \prod_\pm
\frac{
(1-\sigma_i^{\pm 1}\sigma_j^{\mp 1})
(1-\sigma_i^{\pm 1}\sigma_j^{\pm 1})
}{
(1-\mathfrak{t}^2 \sigma_i^{\pm 1} \sigma_j^{\mp 1})(1-\mathfrak{t}^2 \sigma_i^{\pm 1}\sigma_j^{\pm 1})
}.\label{calI0forUSp}
\end{align}

The grand canonical sum
\begin{align}
\Xi_{\text{C}}[f;u]=\sum_{N=0}^\infty u^N\biggl\langle \prod_{i=1}^Nf(s_i)\biggr\rangle^{USp(2N)+\text{(A/S)-}[2l+\delta](C)}(\mathfrak{t})
\end{align}
factorizes as
\begin{align}
\Xi_{\text{C}}[f;u]=\tilde{\Xi}^{(0)}_{\text{C}}[f;u]\prod_{m=1}^\infty\tilde{\Xi}^{(1)}[f;\mathfrak{t}^{2m \Delta}u],
\label{XiC}
\end{align}
with $\tilde{\Xi}^{(1)}(u)$ given by \eqref{Xitilde(1)} and
\begin{align}
\tilde{\Xi}_{\text{C}}^{(0)}[f;u]=\sum_{\nu=0}^\infty u^\nu \tilde{\cal I}^{(0)}_{\text{C}}[f;\nu].
\label{XiC(0)}
\end{align}

\subsubsection{Fundamental Wilson lines}
\label{app_USp_fund}

Since the character of the fundamental representation of $USp(2N)$ is $\chi^{\mathfrak{usp}}_{\ydiagram{1}}(s)=\sum_{i=1}^N(s_i+s_i^{-1})$, we can calculate the one- and two-point function of fundamental Wilson lines by choosing the single variable function $f(s)=(1+a(s+s^{-1}))(1+b(s+s^{-1}))$ as
\begin{subequations}
\begin{align}
&\langle W_{\ydiagram{1}}\rangle^{USp(2N)+\text{(A/S)-}[2l+\delta](C)}(\mathfrak{t})=\biggl\langle \prod_if(s_i)\biggr\rangle^{USp(2N)+\text{(A/S)-}[2l+\delta](C)}(\mathfrak{t})\Bigr|_{a^1b^0}=0,\\
&\langle W_{\ydiagram{1}}W_{\ydiagram{1}}\rangle^{USp(2N)+\text{(A/S)-}[2l+\delta](C)}(\mathfrak{t})=\biggl\langle \prod_if(s_i)\biggr\rangle^{USp(2N)+\text{(A/S)-}[2l+\delta](C)}(\mathfrak{t})\Bigr|_{a^1b^1}.
\end{align}
\end{subequations}
where the one-point function vanishes identically because the integrand of \eqref{USpC} without insertion of $\prod_if(s_i)$ is invariant under $s_i\rightarrow -s_i$ for any single $i$.
For this choice of $f(s)$, we find
\begin{align}
\tilde{\cal I}^{(0)}_{\text{C}}[(1+a(s+s^{-1}))(1+b(s+s^{-1}));\nu]=\frac{1}{(\mathfrak{t}^4;\mathfrak{t}^4)_\nu}+\frac{ab}{(1-\mathfrak{t}^2)(\mathfrak{t}^4;\mathfrak{t}^4)_{\nu-1}}+{\cal O}(a^2,b^2),
\label{tildecalI(0)USpfundguess}
\end{align}
while $\tilde{\cal I}^{(1)}[f;\nu]$ are the same as those in $SO(2N+1)$ \eqref{calI1fundgeneralnu}.
Plugging these into the grand canonical sum \eqref{XiC} with \eqref{XiC(0)}, we find
\begin{align}
&\Xi_\text{C}[(1+a(s+s^{-1}))(1+b(s+s^{-1}));u]\nonumber \\
&=
\Xi_\text{C}[1;u]
\Bigl(1+\frac{1+\mathfrak{t}^{2\Delta}}{(1-\mathfrak{t}^2)(1-\mathfrak{t}^{2\Delta})}abu+{\cal O}(a^2,b^2)\Bigr),
\label{USpGCfinal}
\end{align}
where
\begin{align}
\Xi_\text{C}[1;u]
=
\frac{1}{(u;\mathfrak{t}^4)_\infty\prod_{m=1}^\infty (\mathfrak{t}^{2m\Delta}u;\mathfrak{t}^2)_\infty}
\end{align}
is the grand canonical sum of the Coulomb index without Wilson lines.

From the grand canonical sum \eqref{USpGCfinal} we obtain the recursion relations as claimed in \eqref{R_usp_C_W1}
\begin{align}
&\langle
W_{\ydiagram{1}}
W_{\ydiagram{1}}
\rangle^{USp(2N)+\text{(A/S)-}[2l+\delta](C)}(\mathfrak{t})\nonumber \\
&
={\cal I}^{USp(2N-2)+\text{(A/S)-}[2l+\delta](C)}(\mathfrak{t})
\langle
W_{\ydiagram{1}}
W_{\ydiagram{1}}
\rangle^{USp(2)+\text{(A/S)-}[2l+\delta](C)}(\mathfrak{t}),
\label{USprecrelnappendix}
\end{align}
with
\begin{align}
\langle
W_{\ydiagram{1}} W_{\ydiagram{1}}
\rangle^{USp(2)+\text{(A/S)-}[2l+\delta](C)}(\mathfrak{t})=\frac{1+\mathfrak{t}^{2\Delta}}{(1-\mathfrak{t}^2)(1-\mathfrak{t}^{2\Delta})}.
\end{align}
In particular, we find
\begin{align}
\lim_{\mathfrak{t}\rightarrow 1}
\frac{
\langle
W_{\ydiagram{1}}
W_{\ydiagram{1}}
\rangle^{USp(2N)+\text{(A/S)-}[2l+\delta](C)}(\mathfrak{t})
}{
{\cal I}^{USp(2N)+\text{(A/S)-}[2l+\delta](C)}(\mathfrak{t})
}
=2N,
\end{align}
which is consistent with our conjecture \eqref{Witten_indC_conj_int}.

Note that the recursion relation of the line defect indices and the grand canonical sum of the supersymmetric indices coincides with those of $O(2N)^{(+)}$ theory \eqref{GCofICO2N}.
Hence we have
\begin{subequations}
\begin{align}
&{\cal I}^{USp(2N)+\text{(A/S)-}[2l+\delta](C)}(\mathfrak{t})
={\cal I}^{O(2N)^{(+)}+\text{(A/S)}'\text{-}[2l'](C)}(\mathfrak{t}),\\
&\langle W_{\ydiagram{1}}W_{\ydiagram{1}}\rangle^{USp(2N)+\text{(A/S)-}[2l+\delta](C)}(\mathfrak{t})
=\langle W_{\ydiagram{1}}W_{\ydiagram{1}}\rangle^{O(2N)^{(+)}+\text{(A/S)}'\text{-}[2l'](C)}(\mathfrak{t}),
\end{align}
\end{subequations}
with $\text{(A/S)}'$ and $l'$ such that the value of $\Delta$ of the $USp(2N)$ theory \eqref{DeltaUSp} coincides with that of the $O(2N)$ theory \eqref{DeltaSO}.
In particular, the coincidence of the Coulomb indices of the $USp(2N)$ gauge theory $O(2N)^{(+)}$ gauge theory (and also $O(2N+1)^{(\chi\zeta)}$ gauge theory \eqref{ICO2N=ICO2N+1}) is consistent with the fact that these theories share the same Coulomb branch moduli space ${\cal M}_C=\text{Sym}^N(\mathbb{C}^2/\widehat{D}_\Delta)$.

\subsubsection{More general representations}
\label{app_USpC_moregeneral}

The line defect indices for more general representations are obtained by inserting character $\chi^{\mathfrak{usp}}_\lambda(s)$, which can be obtained by the determinant formula
\begin{align}
\chi_\lambda^{\mathfrak{usp}}(s)=
\frac{\text{det}(s_j^{\lambda_i+N-i+1}-s_j^{-\lambda_i-N+i-1})}{\text{det}(s_j^{N-i+1}-s_j^{-N+i-1})}.
\label{chiusplambda}
\end{align}

For example, for $\lambda=\ydiagram{2}$, the character is \eqref{chiusplambda}
\begin{align}
&\chi^{\mathfrak{usp}}_{\ydiagram{2}}(s)=\sum_{i\le j}
\Bigl(s_i+\frac{1}{s_i}\Bigr)
\Bigl(s_j+\frac{1}{s_j}\Bigr)
-N,
\end{align}
and hence the one- and two-point function of the Wilson lines $W_{\ydiagram{2}}$ can be calculated by choosing the single variable function $f(s)=\frac{1}{1-a(s_i+s_i^{-1})}\frac{1}{1-b(s_i+s_i^{-1})}$ as
\begin{subequations}
\label{USp_rank2symcalculation}
\begin{align}
&\langle W_{\ydiagram{2}}\rangle^{USp(2N)+\text{(A/S)-}[2l+\delta](C)}(\mathfrak{t})\nonumber \\
&=
\biggl\langle\prod_i f(s_i)\biggr\rangle^{USp(2N)+\text{(A/S)-}[2l+\delta](C)}(\mathfrak{t})\Bigr|_{a^2b^0}
-N\biggl\langle\prod_i f(s_i)\biggr\rangle^{USp(2N)+\text{(A/S)-}[2l+\delta](C)}(\mathfrak{t})\Bigr|_{a^0b^0},\\
&\langle W_{\ydiagram{2}}W_{\ydiagram{2}}\rangle^{USp(2N)+\text{(A/S)-}[2l+\delta](C)}(\mathfrak{t})\nonumber \\
&=
\biggl\langle\prod_i f(s_i)\biggr\rangle^{USp(2N)+\text{(A/S)-}[2l+\delta](C)}(\mathfrak{t})\Bigr|_{a^2b^2}
-2N\biggl\langle\prod_i f(s_i)\biggr\rangle^{USp(2N)+\text{(A/S)-}[2l+\delta](C)}(\mathfrak{t})\Bigr|_{a^2b^0}\nonumber \\
&\quad
+N^2\biggl\langle\prod_i f(s_i)\biggr\rangle^{USp(2N)+\text{(A/S)-}[2l+\delta](C)}(\mathfrak{t})\Bigr|_{a^0b^0}.
\end{align}
\end{subequations}
Calculating the building blocks $\tilde{\cal I}^{(0)}_{\text{C}}[f;\nu]$ and $\tilde{\cal I}^{(1)}[f;\nu]$ for this choice of $f(s)$ and plugging them into \eqref{USpC} and then \eqref{USp_rank2symcalculation}, we obtain
{\fontsize{9pt}{1pt}\selectfont
\begin{subequations}
\begin{align}
&\langle W_{\ydiagram{2}} \rangle^{USp(2)+\text{(A/S)-}[2l+\delta](C)}(\mathfrak{t})=
\frac{\mathfrak{t}^2 + \mathfrak{t}^{2\Delta}}{(1 - \mathfrak{t}^4) (1 - \mathfrak{t}^{2\Delta})},\\
&\langle W_{\ydiagram{2}} \rangle^{USp(4)+\text{(A/S)-}[2l+\delta](C)}(\mathfrak{t})=
\frac{\mathfrak{t}^2 + \mathfrak{t}^{2\Delta} (1 + \mathfrak{t}^4) + \mathfrak{t}^{4\Delta + 2}}{(1 - \mathfrak{t}^4)^2 (1 - \mathfrak{t}^{2\Delta})^2},\\
&\langle W_{\ydiagram{2}} \rangle^{USp(6)+\text{(A/S)-}[2l+\delta](C)}(\mathfrak{t})\nonumber \\
&=
\frac{
\mathfrak{t}^2
+ \mathfrak{t}^{2\Delta} (1 + \mathfrak{t}^4 + \mathfrak{t}^6 + \mathfrak{t}^8)
+ \mathfrak{t}^{4\Delta + 2} (1 + 2 \mathfrak{t}^2 + 2 \mathfrak{t}^4 + \mathfrak{t}^6)
+ \mathfrak{t}^{6\Delta + 2} (1 + \mathfrak{t}^2 + \mathfrak{t}^4 + \mathfrak{t}^8)
+ \mathfrak{t}^{8\Delta + 8}
}{(1 - \mathfrak{t}^4)^2 (1 - \mathfrak{t}^8) (1 - \mathfrak{t}^{2\Delta})^2 (1 - \mathfrak{t}^{4\Delta})},\\
&\langle W_{\ydiagram{2}}W_{\ydiagram{2}} \rangle^{USp(2)+\text{(A/S)-}[2l+\delta](C)}(\mathfrak{t})=
\frac{1 + \mathfrak{t}^2 + \mathfrak{t}^4 + \mathfrak{t}^{2\Delta} (2 + 2 \mathfrak{t}^2 - \mathfrak{t}^4)}{(1 - \mathfrak{t}^4) (1 - \mathfrak{t}^{2\Delta})},\\
&\langle W_{\ydiagram{2}}W_{\ydiagram{2}} \rangle^{USp(4)+\text{(A/S)-}[2l+\delta](C)}(\mathfrak{t})\nonumber \\
&=
\frac{1 + \mathfrak{t}^2 + 2 \mathfrak{t}^4 + \mathfrak{t}^6 + \mathfrak{t}^8 + \mathfrak{t}^{2\Delta} (5 + 10 \mathfrak{t}^2 + 4 \mathfrak{t}^4 - \mathfrak{t}^8) + \mathfrak{t}^{4\Delta} (3 + 8 \mathfrak{t}^2 + 6 \mathfrak{t}^4 + 2 \mathfrak{t}^6 - \mathfrak{t}^8) + \mathfrak{t}^{6\Delta} (3 + 5 \mathfrak{t}^2 - 3 \mathfrak{t}^6 + \mathfrak{t}^8)}{(1 - \mathfrak{t}^4)^2 (1 - \mathfrak{t}^{2\Delta}) (1 - \mathfrak{t}^{4\Delta})},\\
&\langle W_{\ydiagram{2}}W_{\ydiagram{2}} \rangle^{USp(6)+\text{(A/S)-}[2l+\delta](C)}(\mathfrak{t})\nonumber \\
&=\frac{1}{(1 - \mathfrak{t}^4)^2 (1 - \mathfrak{t}^8) (1 - \mathfrak{t}^{2\Delta})^2 (1 - \mathfrak{t}^{4\Delta})} (1 + \mathfrak{t}^2 + 2 \mathfrak{t}^4 + \mathfrak{t}^6 + 2 \mathfrak{t}^8 + \mathfrak{t}^{10} + \mathfrak{t}^{12}\nonumber \\
&\quad + \mathfrak{t}^{2\Delta} (4 + 9 \mathfrak{t}^2 + 6 \mathfrak{t}^4 + 9 \mathfrak{t}^6 + 5 \mathfrak{t}^8 + 2 \mathfrak{t}^{10} + \mathfrak{t}^{14}) + \mathfrak{t}^{4\Delta} (1 + 8 \mathfrak{t}^2 + 17 \mathfrak{t}^4 + 16 \mathfrak{t}^6 + 11 \mathfrak{t}^8 + 4 \mathfrak{t}^{10} - 2 \mathfrak{t}^{12} - \mathfrak{t}^{14})\nonumber \\
&\quad + \mathfrak{t}^{6\Delta} (3 + 6 \mathfrak{t}^2 + 8 \mathfrak{t}^4 + 9 \mathfrak{t}^6 + 6 \mathfrak{t}^8 + \mathfrak{t}^{10} + 4 \mathfrak{t}^{12} - \mathfrak{t}^{14}) + \mathfrak{t}^{8\Delta + 2} (3 + 3 \mathfrak{t}^2 + \mathfrak{t}^4 + 3 \mathfrak{t}^6 + \mathfrak{t}^8 - 3 \mathfrak{t}^{10} + \mathfrak{t}^{12})).
\end{align}
\end{subequations}
}
We observe the following recursion relations for the one-point function and the two-point function
\begin{subequations}
\label{USpCrank2symrecreln}
\begin{align}
&\langle W_{\ydiagram{2}}\rangle^{USp(2N)+\text{(A/S)-}[2l+\delta](C)}(\mathfrak{t})=
{\cal I}^{USp(2N-2)+\text{(A/S)-}[2l+\delta](C)}(\mathfrak{t})
\langle W_{\ydiagram{2}}\rangle^{USp(2)+\text{(A/S)-}[2l+\delta](C)}(\mathfrak{t}),\\
&\langle W_{\ydiagram{2}}W_{\ydiagram{2}}\rangle^{USp(2N)+\text{(A/S)-}[2l+\delta](C)}(\mathfrak{t})\nonumber \\
&=
c_1(\mathfrak{t})
{\cal I}^{USp(2N-2)+\text{(A/S)-}[2l+\delta](C)}(\mathfrak{t})
+
c_2(\mathfrak{t})
{\cal I}^{USp(2N-4)+\text{(A/S)-}[2l+\delta](C)}(\mathfrak{t})
\end{align}
\end{subequations}
with
\begin{subequations}
\begin{align}
&c_1(\mathfrak{t})=\langle W_{\ydiagram{2}}W_{\ydiagram{2}}\rangle^{USp(2)+\text{(A/S)-}[2l+\delta](C)}(\mathfrak{t})
=
\frac{1 + \mathfrak{t}^2 + \mathfrak{t}^4 + \mathfrak{t}^{2\Delta} (2 + 2 \mathfrak{t}^2 - \mathfrak{t}^4)}{(1 - \mathfrak{t}^4) (1 - \mathfrak{t}^{2\Delta})},\\
&c_2(\mathfrak{t})=
\langle W_{\ydiagram{2}}W_{\ydiagram{2}}\rangle^{USp(4)+\text{(A/S)-}[2l+\delta](C)}(\mathfrak{t})
-
{\cal I}^{USp(2)+\text{(A/S)-}[2l+\delta](C)}(\mathfrak{t})
\langle W_{\ydiagram{2}}W_{\ydiagram{2}}\rangle^{USp(2)+\text{(A/S)-}[2l+\delta](C)}(\mathfrak{t})\nonumber \\
&
=
\frac{
1}{(1 - \mathfrak{t}^4)^2 (1 - \mathfrak{t}^{2\Delta}) (1 - \mathfrak{t}^{4 \Delta})}
(\mathfrak{t}^4 + \mathfrak{t}^6 + \mathfrak{t}^8
+ \mathfrak{t}^{2 \Delta} (2 + 6 \mathfrak{t}^2 + 3 \mathfrak{t}^4 - \mathfrak{t}^6 - \mathfrak{t}^8)\nonumber \\
&\quad
+ \mathfrak{t}^{4 \Delta} (1 + 3 \mathfrak{t}^2 + 4 \mathfrak{t}^4 + 2 \mathfrak{t}^6 - \mathfrak{t}^8)
+ \mathfrak{t}^{6 \Delta} (3 + 3 \mathfrak{t}^2 - 2 \mathfrak{t}^4 - 2 \mathfrak{t}^6 + \mathfrak{t}^8)),\\
&{\cal I}^{USp(0)+\text{(A/S)-}[2l+\delta](C)}(\mathfrak{t})=1,\quad
{\cal I}^{USp(-2)+\text{(A/S)-}[2l+\delta](C)}(\mathfrak{t})=0.
\end{align}
\end{subequations}
Assuming the recursion relation, we find the limit $\mathfrak{t}\rightarrow 1$ of the line defect indices as
\begin{subequations}
\begin{align}
&\lim_{\mathfrak{t}\rightarrow 1}
\langle {\cal W}_{\ydiagram{2}}\rangle^{USp(2N)+\text{(A/S)-}[2l+\delta](C)}=N=\oint\prod_{i=1}^N\frac{ds_i}{2\pi is_i}\chi^{\mathfrak{usp}(2N)}_{\ydiagram{2}}(s),\\
&\lim_{\mathfrak{t}\rightarrow 1}
\langle {\cal W}_{\ydiagram{2}}{\cal W}_{\ydiagram{2}}\rangle^{USp(2N)+\text{(A/S)-}[2l+\delta](C)}=3N^2=\oint\prod_{i=1}^N\frac{ds_i}{2\pi is_i}\chi^{\mathfrak{usp}(2N)}_{\ydiagram{2}}(s)^2,
\end{align}
\end{subequations}
which are consistent with our conjecture \eqref{Witten_indC_conj_int}.

For $\lambda=\ydiagram{1,1}$ the character is \eqref{chiusplambda}
\begin{align}
&\chi^{\mathfrak{usp}}_{\ydiagram{1,1}}(s)=\sum_{i<j}\Bigl(s_i+\frac{1}{s_i}\Bigr)\Bigl(s_j+\frac{1}{s_j}\Bigr)+N-1,\quad (N\ge 2),
\end{align}
and hence the one- and two-point function of the Wilson lines $W_{\ydiagram{1,1}}$ can be calculated by choosing the single variable function $f(s)=(1+a(s_i+s_i^{-1}))(1+b(s_i+s_i^{-1}))$ as
\begin{subequations}
\label{USp_rank2antisymcalculation}
\begin{align}
&\langle
W_{\ydiagram{1,1}}
\rangle^{USp(2N)+\text{(A/S)-}[2l+\delta](C)}\nonumber \\
&
=
\biggl\langle \prod_i f(s_i)\biggr\rangle^{USp(2N)+\text{(A/S)-}[2l+\delta](C)}(\mathfrak{t})\Bigr|_{a^2b^0}
+(N-1)\biggl\langle \prod_i f(s_i)\biggr\rangle^{USp(2N)+\text{(A/S)-}[2l+\delta](C)}(\mathfrak{t})\Bigr|_{a^0b^0},\\
&\langle
W_{\ydiagram{1,1}}
W_{\ydiagram{1,1}}
\rangle^{USp(2N)+\text{(A/S)-}[2l+\delta](C)}\nonumber \\
&
=
\biggl\langle \prod_i f(s_i)\biggr\rangle^{USp(2N)+\text{(A/S)-}[2l+\delta](C)}(\mathfrak{t})\Bigr|_{a^2b^2}
+2(N-1)\biggl\langle \prod_i f(s_i)\rangle^{USp(2N)+\text{(A/S)-}[2l+\delta](C)}(\mathfrak{t})\biggr|_{a^2b^0}\nonumber \\
&\quad
+(N-1)^2\biggl\langle \prod_i f(s_i)\biggr\rangle^{USp(2N)+\text{(A/S)-}[2l+\delta](C)}(\mathfrak{t})\Bigr|_{a^0b^0}.
\end{align}
\end{subequations}
Calculating the building blocks $\tilde{\cal I}^{(0)}_{\text{C}}[f;\nu]$ and $\tilde{\cal I}^{(1)}[f;\nu]$ for this choice of $f(s)$ and plugging them into \eqref{USpC} and then \eqref{USp_rank2antisymcalculation}, we obtain
{\fontsize{9pt}{1pt}\selectfont
\begin{subequations}
\begin{align}
&\langle W_{\ydiagram{1,1}} \rangle^{USp(4)+\text{(A/S)-}[2l+\delta](C)}(\mathfrak{t})\nonumber \\
&=
\frac{\mathfrak{t}^4 + \mathfrak{t}^{2\Delta} (1 + \mathfrak{t}^2 + \mathfrak{t}^6) + \mathfrak{t}^{4\Delta+2} (1 + \mathfrak{t}^4 + \mathfrak{t}^6) + \mathfrak{t}^{6\Delta+4}}{(1 - \mathfrak{t}^4) (1 - \mathfrak{t}^8) (1 - \mathfrak{t}^{2\Delta}) (1 - \mathfrak{t}^{4\Delta})},\\
&\langle W_{\ydiagram{1,1}} \rangle^{USp(6)+\text{(A/S)-}[2l+\delta](C)}(\mathfrak{t})\nonumber \\
&=
\frac{
1
}{(1 - \mathfrak{t}^4)^2 (1 - \mathfrak{t}^{12}) (1 - \mathfrak{t}^{2\Delta})^2 (1 - \mathfrak{t}^{6\Delta})}
\bigl(
\mathfrak{t}^4
+ \mathfrak{t}^{2\Delta}(1 + \mathfrak{t}^2 - \mathfrak{t}^4 + \mathfrak{t}^6 + \mathfrak{t}^8 + \mathfrak{t}^{10})
+ \mathfrak{t}^{4\Delta+2} (1 + 2 \mathfrak{t}^2 + \mathfrak{t}^4 + \mathfrak{t}^6 + \mathfrak{t}^{8} + \mathfrak{t}^{10})\nonumber \\
&\quad + \mathfrak{t}^{6\Delta+2} (1 + \mathfrak{t}^2 + \mathfrak{t}^4 + \mathfrak{t}^6 + 2 \mathfrak{t}^{8} + \mathfrak{t}^{10})
+ \mathfrak{t}^{8\Delta+4} (1 + \mathfrak{t}^2 + \mathfrak{t}^4 - \mathfrak{t}^{6} + \mathfrak{t}^{8} + \mathfrak{t}^{10})
+ \mathfrak{t}^{10\Delta+10}
\bigr),\\
&\langle W_{\ydiagram{1,1}}W_{\ydiagram{1,1}} \rangle^{USp(4)+\text{(A/S)-}[2l+\delta](C)}(\mathfrak{t})\nonumber \\
&=
\frac{
1
}{(1 - \mathfrak{t}^4) (1 - \mathfrak{t}^8) (1 - \mathfrak{t}^{2\Delta}) (1 - \mathfrak{t}^{4\Delta})}(
1 + \mathfrak{t}^2 + \mathfrak{t}^4 + \mathfrak{t}^6 + \mathfrak{t}^8 + \mathfrak{t}^{2\Delta} (2 + 4 \mathfrak{t}^2 + 4 \mathfrak{t}^4 + 4 \mathfrak{t}^6 + \mathfrak{t}^8)\nonumber \\
&\quad + \mathfrak{t}^{4\Delta} (2 + 3 \mathfrak{t}^2 + 4 \mathfrak{t}^4 + 4 \mathfrak{t}^6 + \mathfrak{t}^8 + \mathfrak{t}^{10}) + \mathfrak{t}^{6\Delta+2} (2 + \mathfrak{t}^2 + \mathfrak{t}^4 + 2 \mathfrak{t}^6 - \mathfrak{t}^8)
),\\
&\langle W_{\ydiagram{1,1}}W_{\ydiagram{1,1}} \rangle^{USp(6)+\text{(A/S)-}[2l+\delta](C)}(\mathfrak{t})\nonumber \\
&=
\frac{
1}{(1 - \mathfrak{t}^4)^2 (1 - \mathfrak{t}^{12}) (1 - \mathfrak{t}^{2\Delta}) (1 - \mathfrak{t}^{4\Delta}) (1 - \mathfrak{t}^{6\Delta})}
\bigl(1 + \mathfrak{t}^{2} + \mathfrak{t}^{4} + \mathfrak{t}^{6} + 2 \mathfrak{t}^{8} + \mathfrak{t}^{10} + \mathfrak{t}^{12}\nonumber \\
&\quad + \mathfrak{t}^{2\Delta} (3 + 6 \mathfrak{t}^{2} + 7 \mathfrak{t}^{4} + 7 \mathfrak{t}^{6} + 6 \mathfrak{t}^{8} + 7 \mathfrak{t}^{10} + 3 \mathfrak{t}^{12} + \mathfrak{t}^{14})\nonumber \\
&\quad + \mathfrak{t}^{4\Delta} (5 + 12 \mathfrak{t}^{2} + 15 \mathfrak{t}^{4} + 15 \mathfrak{t}^{6} + 14 \mathfrak{t}^{8} + 15 \mathfrak{t}^{10} + 9 \mathfrak{t}^{12} + 3 \mathfrak{t}^{14})\nonumber \\
&\quad + \mathfrak{t}^{6\Delta} (5 + 15 \mathfrak{t}^{2} + 17 \mathfrak{t}^{4} + 19 \mathfrak{t}^{6} + 19 \mathfrak{t}^{8} + 18 \mathfrak{t}^{10} + 14 \mathfrak{t}^{12} + 4 \mathfrak{t}^{14} + \mathfrak{t}^{16})\nonumber \\
&\quad + \mathfrak{t}^{8\Delta} (2 + 10 \mathfrak{t}^{2} + 15 \mathfrak{t}^{4} + 14 \mathfrak{t}^{6} + 15 \mathfrak{t}^{8} + 15 \mathfrak{t}^{10} + 13 \mathfrak{t}^{12} + 4 \mathfrak{t}^{14})\nonumber \\
&\quad + \mathfrak{t}^{10\Delta+2} (4 + 7 \mathfrak{t}^{2} + 6 \mathfrak{t}^{4} + 7 \mathfrak{t}^{6} + 7 \mathfrak{t}^{8} + 7 \mathfrak{t}^{10} + 2 \mathfrak{t}^{12})
+ \mathfrak{t}^{12\Delta+4} (2 + 2 \mathfrak{t}^{2} + \mathfrak{t}^{4} + \mathfrak{t}^{6} + \mathfrak{t}^{8} + 2 \mathfrak{t}^{10} - \mathfrak{t}^{12})
\bigr).
\end{align}
\end{subequations}
}
We observe the following recursion relations for the one-point function and the two-point function
\begin{subequations}
\label{USpCrank2antisymrecreln}
\begin{align}
&\langle W_{\ydiagram{1,1}}\rangle^{USp(2N)+\text{(A/S)-}[2l+\delta](C)}(\mathfrak{t})
+{\cal I}^{USp(2N)+\text{(A/S)-}[2l+\delta](C)}(\mathfrak{t})\nonumber \\
&=
{\cal I}^{USp(2N-2)+\text{(A/S)-}[2l+\delta](C)}(\mathfrak{t})
\Bigl(
\langle W_{\ydiagram{1,1}}\rangle^{USp(2)+\text{(A/S)-}[2l+\delta](C)}(\mathfrak{t})
+{\cal I}^{USp(2)+\text{(A/S)-}[2l+\delta](C)}(\mathfrak{t})
\Bigr),\\
&\langle W_{\ydiagram{1,1}}W_{\ydiagram{1,1}}\rangle^{USp(2N)+\text{(A/S)-}[2l+\delta](C)}(\mathfrak{t})
-{\cal I}^{USp(2N)+\text{(A/S)-}[2l+\delta](C)}(\mathfrak{t})\nonumber \\
&=
c_1(\mathfrak{t})
{\cal I}^{USp(2N-2)+\text{(A/S)-}[2l+\delta](C)}(\mathfrak{t})
+
c_2(\mathfrak{t})
{\cal I}^{USp(2N-4)+\text{(A/S)-}[2l+\delta](C)}(\mathfrak{t})
\end{align}
\end{subequations}
with
\begin{subequations}
\begin{align}
&c_1(\mathfrak{t})=\langle W_{\ydiagram{1,1}}W_{\ydiagram{1,1}}\rangle^{USp(2)+\text{(A/S)-}[2l+\delta](C)}(\mathfrak{t})
-{\cal I}^{USp(2)+\text{(A/S)-}[2l+\delta](C)}(\mathfrak{t})\nonumber \\
&\quad =-\frac{1 + \mathfrak{t}^{2\Delta+2}}{(1 - \mathfrak{t}^4) (1 - \mathfrak{t}^{2\Delta})},\\
&c_2(\mathfrak{t})=\langle W_{\ydiagram{1,1}}W_{\ydiagram{1,1}}\rangle^{USp(4)+\text{(A/S)-}[2l+\delta](C)}(\mathfrak{t})
-{\cal I}^{USp(4)+\text{(A/S)-}[2l+\delta](C)}(\mathfrak{t})
-
c_1(\mathfrak{t}){\cal I}^{USp(2)+\text{(A/S)-}[2l+\delta](C)}(\mathfrak{t})\nonumber \\
&\quad =
\frac{
1 + \mathfrak{t}^2 + \mathfrak{t}^4 + \mathfrak{t}^{2\Delta} (3 + 5 \mathfrak{t}^2 + \mathfrak{t}^4) + \mathfrak{t}^{4 \Delta} (2 + 4 \mathfrak{t}^2 + 2 \mathfrak{t}^4 + \mathfrak{t}^6) + \mathfrak{t}^{6 \Delta + 2} (2 + 2 \mathfrak{t}^2 - \mathfrak{t}^4)
}{
(1 - \mathfrak{t}^4)^2 (1 - \mathfrak{t}^{2\Delta}) (1 - \mathfrak{t}^{4\Delta})
},\\
&\langle W_{\ydiagram{1,1}}\rangle^{USp(2)+\text{(A/S)-}[2l+\delta](C)}(\mathfrak{t})\equiv 0,\quad
\langle W_{\ydiagram{1,1}}W_{\ydiagram{1,1}}\rangle^{USp(2)+\text{(A/S)-}[2l+\delta](C)}(\mathfrak{t})\equiv 0.
\end{align}
\end{subequations}
In the limit $\mathfrak{t}\rightarrow 1$, we find
\begin{subequations}
\begin{align}
&\lim_{\mathfrak{t}\rightarrow 1}
\langle {\cal W}_{\ydiagram{1,1}}\rangle^{USp(2N)+\text{(A/S)-}[2l+\delta](C)}(\mathfrak{t})
=N-1
=\int\Bigl(\frac{ds}{2\pi is}\Bigr)^N\chi^{\mathfrak{usp}(2N)}_{\ydiagram{1,1}}(s),\\
&\lim_{\mathfrak{t}\rightarrow 1}
\langle {\cal W}_{\ydiagram{1,1}}{\cal W}_{\ydiagram{1,1}}\rangle^{USp(2N)+\text{(A/S)-}[2l+\delta](C)}(\mathfrak{t})
=(N-1)(3N-1)
=\int\Bigl(\frac{ds}{2\pi is}\Bigr)^N\chi^{\mathfrak{usp}(2N)}_{\ydiagram{1,1}}(s)^2,
\end{align}
\end{subequations}
which are consistent with our conjecture \eqref{Witten_indC_conj_int}.

For $\lambda=\ydiagram{4}$, whose character is
\begin{align}
\chi^{\mathfrak{usp}(2N)}_{\ydiagram{4}}(s)
&=
\sum_{i\le j\le k\le k'}
\Bigl(s_i+\frac{1}{s_i}\Bigr)\Bigl(s_j+\frac{1}{s_j}\Bigr)
\Bigl(s_k+\frac{1}{s_k}\Bigr)\Bigl(s_{k'}+\frac{1}{s_{k'}}\Bigr)
-(N+2)\sum_{i\le j}
\Bigl(s_i+\frac{1}{s_i}\Bigr)\Bigl(s_j+\frac{1}{s_j}\Bigr)\nonumber \\
&\quad
+\frac{N(N+1)}{2},
\end{align}
we find the two-term recursion relation for the one-point function at least for $N=1$, $2$ and $3$
\begin{align}
\langle W_{\ydiagram{4}}\rangle^{USp(2N)+\text{(A/S)-}[2l+\delta](C)}(\mathfrak{t})
=c_1(\mathfrak{t}){\cal I}^{USp(2N-2)+\text{(A/S)-}[2l+\delta](C)}(\mathfrak{t})
+c_2(\mathfrak{t}){\cal I}^{USp(2N-4)+\text{(A/S)-}[2l+\delta](C)}(\mathfrak{t})
\end{align}
with
\begin{subequations}
\begin{align}
&c_1(\mathfrak{t})=\langle W_{\ydiagram{4}}\rangle^{USp(2)+\text{(A/S)-}[2l+\delta](C)}(\mathfrak{t})
=\frac{\mathfrak{t}^4 + \mathfrak{t}^{2\Delta} (1 + \mathfrak{t}^2 - \mathfrak{t}^4)}{(1 - \mathfrak{t}^4) (1 - \mathfrak{t}^{2\Delta})},\\
&c_2(\mathfrak{t})=\langle W_{\ydiagram{4}}\rangle^{USp(4)+\text{(A/S)-}[2l+\delta](C)}(\mathfrak{t})
-{\cal I}^{USp(2)+\text{(A/S)-}[2l+\delta](C)}(\mathfrak{t})
\langle W_{\ydiagram{4}}\rangle^{USp(2)+\text{(A/S)-}[2l+\delta](C)}(\mathfrak{t})
\nonumber \\
&
=
\frac{
\mathfrak{t}^{12}
+ \mathfrak{t}^{2\Delta + 2} (1 + \mathfrak{t}^2 + \mathfrak{t}^4 + \mathfrak{t}^6 - \mathfrak{t}^{10})
+ \mathfrak{t}^{4\Delta + 4} (1 + \mathfrak{t}^2 + \mathfrak{t}^4 + \mathfrak{t}^6 - \mathfrak{t}^8)
+ \mathfrak{t}^{6\Delta} (1 + \mathfrak{t}^2 - \mathfrak{t}^8 - \mathfrak{t}^{10} + \mathfrak{t}^{12})
}{(1 - \mathfrak{t}^4) (1 - \mathfrak{t}^8) (1 - \mathfrak{t}^{2\Delta}) (1 - \mathfrak{t}^{4 \Delta})}.
\end{align}
\end{subequations}
In the $\mathfrak{t}\rightarrow 1$ limit we find \eqref{Witten_indC_conj_int}
\begin{align}
\lim_{\mathfrak{t}\rightarrow 1}
\frac{\langle W_{\ydiagram{4}}\rangle^{USp(2N)+\text{(A/S)-}[2l+\delta](C)}(\mathfrak{t})
}{
{\cal I}^{USp(2N)+\text{(A/S)-}[2l+\delta](C)}(\mathfrak{t})
}=\frac{N(N+1)}{2}=\int\prod_{i=1}^N \frac{ds_i}{2\pi is_i}\chi^{\mathfrak{usp}(2N)}_{\ydiagram{4}}(s).
\end{align}

\subsection{$U(N)_k\times U(N)_{-k}$ ABJM theory}
\label{ABJMalgorithm}

The supersymmetric index of $U(N)_k\times U(N)_{-k}$ ABJM theory with the insertion of the product of single variable functions $\prod_{i=1}^Nf(s^{(1)}_i)g(s^{(2)}_i)$ is given as \cite{Hayashi:2022ldo}
\begin{align}
&\biggl\langle\prod_{i=1}^Nf(s^{(1)}_i)g(s^{(2)}_i)\biggr\rangle^{U(N)_k\times U(N)_{-k}\text{ ABJM}}(t,x,z,y;q)\nonumber \\
&=\frac{1}{(N!)^2}
\sum_{m^{(1)}_1,\cdots,m^{(1)}_N, m^{(2)}_1,\cdots,m^{(2)}_N \in \mathbb{Z}}
\oint \prod_{i=1}^N \frac{ds^{(1)}_{i}}{2\pi is^{(1)}_i} f(s^{(1)}_i) (s_i^{(1)})^{km^{(1)}_i}
\oint \prod_{i=1}^N \frac{ds^{(2)}_{i}}{2\pi is^{(2)}_i} g(s^{(2)}_i) (s_i^{(2)})^{-km^{(2)}_i}
\nonumber\\
&\quad\times
\prod_{I=1}^2
\prod_{i<j}^N \prod_\pm
\Bigl(1-q^{\frac{|m^{(I)}_i-m^{(I)}_j|}{2}}(s^{(I)}_i)^{\pm 1}(s^{(I)}_j)^{\mp 1}\Bigr)
\nonumber\\
&\quad\times
\prod_{i,j=1}^N
\prod_\pm
\frac{
(q^{\frac34+\frac{|m_i^{(1)}-m_j^{(2)}|}{2}}t(s_i^{(1)})^{\mp 1} (s_j^{(2)})^{\pm 1} z^{\mp 1};q)_{\infty}
}{
(q^{\frac14+\frac{|m_i^{(1)}-m_j^{(2)}|}{2}}t^{-1} (s_i^{(1)})^{\pm 1} (s_j^{(2)})^{\mp 1}z^{\pm 1};q)_{\infty}
}
\frac{
(q^{\frac34+\frac{|m_i^{(2)} - m_j^{(1)}|}{2}}t^{-1}(s_i^{(2)})^{\mp 1} (s_j^{(1)})^{\pm 1}x^{\mp 1};q)_{\infty}
}{
(q^{\frac14+\frac{|m_i^{(2)} - m_j^{(1)}|}{2}}t (s_i^{(2)})^{\pm 1} (s_j^{(1)})^{\mp 1}x^{\pm 1};q)_{\infty}
}
\nonumber\\
&\quad\times
q^{-\frac12\sum_{I=1}^2\sum_{i<j}^N |m_{i}^{(I)}-m_{j}^{(I)}|+\frac12 \sum_{i,j=1}^N |m^{(1)}_i-m^{(2)}_j| }
y^{\frac12 \sum_{I=1}^2 \sum_{i=1}^N m_i^{(I)}}.
\label{ABJMindex}
\end{align}
In the Coulomb limit, the monopole charges $m^{(2)}_i$ for the second gauge node are restricted to those which coincide with $m^{(1)}_i$ up to an ${\cal S}_N$ permutation, which can be represented by $m^{(2)}_i=m^{(1)}_i$ with the overall multiplicity $\frac{N!}{\prod_{m=-\infty }^\infty (\#\{i|m^{(1)}_i=m\})!}$.
As a result, we obtain
\begin{align}
&\biggl\langle\prod_{i=1}^Nf(s^{(1)}_i)g(s^{(2)}_i)\biggr\rangle^{U(N)_k\times U(N)_{-k}\text{ ABJM}(C)}(z,y;\mathfrak{t})\nonumber \\
&=\frac{1}{N!}
\sum_{m^{(1)}_1,\cdots,m^{(1)}_N \in \mathbb{Z}}
\frac{1}{\prod_{m=-\infty}^\infty (\#\{i|m^{(1)}_i=m\})!}
\oint \prod_{i=1}^N \frac{ds^{(1)}_{i}}{2\pi is^{(1)}_i} f(s^{(1)}_i) (s_i^{(1)})^{km^{(1)}_i}\nonumber \\
&\quad\times
\oint \prod_{i=1}^N \frac{ds^{(2)}_{i}}{2\pi is^{(2)}_i} g(s^{(2)}_i) (s_i^{(2)})^{-km^{(1)}_i}
\prod_{I=1}^2
\prod_{\substack{i<j\\ (m^{(1)}_i=m^{(1)}_j)}}^N \prod_\pm
(1-(s^{(I)}_i)^{\pm 1}(s^{(I)}_j)^{\mp 1})
\nonumber\\
&\quad\times
\prod_{\substack{i,j=1\\ (m^{(1)}_i=m^{(1)}_j)}}^N
\prod_\pm
\frac{
1
}{
1-\mathfrak{t}(s^{(1)}_i)^{\pm 1}(s^{(2)}_j)^{\mp 1}z^{\pm 1}
}
y^{\sum_{i=1}^N m_i^{(1)}}.
\label{ABJMCstep1}
\end{align}
By classifying the choice of the monopole charges $\{m^{(1)}_i\}$ by the multiplicities $\nu_m=\#\{i|m^{(1)}_i=m\}$, we can rewrite \eqref{ABJMCstep1} as
\begin{align}
&\biggl\langle\prod_{i=1}^Nf(s^{(1)}_i)g(s^{(2)}_i)\biggr\rangle^{U(N)_k\times U(N)_{-k}\text{ ABJM}(C)}(z,y;\mathfrak{t})\nonumber \\
&=\sum_{\substack{\nu_m=0\\ (\sum_{m=-\infty}^\infty \nu_m=N)}}
y^{\sum_{m=-\infty}^\infty m\nu_m}
\prod_{m=-\infty}^\infty
\tilde{\cal I}_{\text{ABJM},km}[f,g;\nu_m],
\label{ABJMCstep2}
\end{align}
where
\begin{align}
\tilde{\cal I}_{\text{ABJM},k}[f,g;\nu]
&=\frac{1}{(\nu!)^2}
\oint \prod_{i=1}^{\nu} \frac{d\sigma^{(1)}_{i}}{2\pi i\sigma^{(1)}_i} f(\sigma^{(1)}_i) (\sigma_i^{(1)})^{k}
\oint \prod_{i=1}^{\nu} \frac{d\sigma^{(2)}_{i}}{2\pi i\sigma^{(2)}_i} g(\sigma^{(2)}_i) (\sigma_i^{(2)})^{-k}\nonumber \\
&\quad\times
\frac{
\prod_{I=1}^2
\prod_{i<j}^{\nu} \prod_\pm (1-(\sigma^{(I)}_i)^{\pm 1}(\sigma^{(I)}_j)^{\mp 1})
}{
\prod_{i,j=1}^{\nu}
\prod_\pm
(1-\mathfrak{t}(\sigma^{(1)}_i)^{\pm 1}(\sigma^{(2)}_j)^{\mp 1}z^{\pm 1})
}.
\label{ItildeABJM}
\end{align}

The grand canonical sum
\begin{align}
\Xi_{\text{ABJM}}[f,g;u]=\sum_{N=0}^\infty
u^N
\biggl\langle\prod_{i=1}^Nf(s^{(1)}_i)g(s^{(2)}_i)\biggr\rangle^{U(N)_k\times U(N)_{-k}\text{ ABJM}(C)}(z,y;\mathfrak{t})
\end{align}
factorizes as
\begin{align}
\Xi_{\text{ABJM}}[f,g;u]=\prod_{m=-\infty}^\infty \tilde{\Xi}_{\text{ABJM},km}[f,g;y^mu],
\label{XiABJM}
\end{align}
with
\begin{align}
\tilde{\Xi}_{\text{ABJM},k}[f,g;u]=\sum_{\nu=0}^\infty u^{\nu}\tilde{\cal I}_{\text{ABJM},k}[f,g;\nu]
\label{XitildeABJM}
\end{align}

As is the case in the $U(N)$ ADHM theory \cite{Hayashi:2024jof}, these $\tilde{\cal I}_{\text{ABJM},k}[f,g;\nu]$ can be calculated efficiently through the Fermi gas formalism.
For this purpose, first we rewrite $\tilde{\cal I}_{\text{ABJM},k}[f,g;\nu]$ by using the Cauchy determinant formula
\begin{align}
\frac{\prod_{i<j}^\nu (x_i-x_j)(y_i-y_j)}{\prod_{i,j}^\nu (x_i+y_j)}=\det\Bigl[\frac{1}{x_i+y_j}\Bigr]
\label{Cauchy}
\end{align}
and the Andr\'eief identity
\begin{align}
\frac{1}{\nu!}\int d^\nu x\det[f_i(x_j)]\det[g_j(x_i)]=\det\biggl[\int dxf_i(x)g_j(x)\biggr]
\label{Andreief}
\end{align}
as
\begin{align}
\tilde{\cal I}_{\text{ABJM},k}[f,g;\nu]
=\frac{\mathfrak{t}^{-\nu^2+\nu}}{\nu!}
\prod_{i=1}^\nu \oint\frac{d\sigma^{(1)}_i}{2\pi i\sigma^{(1)}_i}
\det
\rho(\sigma^{(1)}_i,\sigma^{(1)}_j)
\label{calIABJMnew}
\end{align}
with
\begin{align}
\rho(\sigma,\sigma')=
\oint\frac{d\sigma''}{2\pi i\sigma''}
\frac{
f(\sigma)\sigma^k
(\sigma'')^{-k}
g(\sigma'')
}{(1-\mathfrak{t}z^{-1}\sigma^{-1}\sigma'')
(1-\mathfrak{t}z (\sigma'')^{-1} \sigma')
},
\end{align}
This new expression \eqref{calIABJMnew} implies that $\tilde{\cal I}_{\text{ABJM},k}[f,g;\nu]$ can be obtained from the functional traces through the following generating function
\begin{align}
\sum_{\nu=0}^\infty\mathfrak{t}^{\nu^2-\nu} \kappa^\nu \tilde{\cal I}_{\text{ABJM},k}[f,g;\nu]=\text{exp}[\text{Tr}\log(1+\kappa\rho)]=\text{exp}\biggl[\sum_{n=1}^\infty\frac{(-1)^{n-1}}{n}\kappa^n\text{Tr}\rho^n\biggr],
\end{align}
with
\begin{align}
\text{Tr}\rho^n=
\prod_{i=1}^n
\oint\frac{d\sigma_i}{2\pi i\sigma_i}
\rho(\sigma_1,\sigma_2)
\rho(\sigma_2,\sigma_3)
\cdots
\rho(\sigma_n,\sigma_1).
\label{Trrhon}
\end{align}
By using the notation of 1d quantum mechanics on a circle adopted in \cite{Hayashi:2024jof}:
\begin{subequations}
\label{1dQMnotation}
\begin{align}
&[{\hat \alpha},{\hat p}]=\frac{i}{2\pi},\\
&|\alpha\rangle\text{: position eigenstate},\quad
\langle \alpha|\alpha'\rangle=\delta(\alpha-\alpha'),\quad
1=\int_0^1d\alpha |\alpha\rangle\langle\alpha|,\\
&|p\rangle\!\rangle\text{: momentum eigenstate},\quad
\langle\!\langle p|p'\rangle\!\rangle=\delta_{p,p'},\quad
1=\sum_{p=-\infty}^\infty |p\rangle\!\rangle\langle\!\langle p|,\\
&\langle\alpha|p\rangle\!\rangle=e^{2\pi i\alpha p},\quad
\langle\!\langle p|\alpha\rangle=e^{-2\pi i\alpha p},
\end{align}
\end{subequations}
we can write the functional trace $\text{Tr}\rho^n$ \eqref{Trrhon} as a trace $\text{tr}\hat{\rho}^n$ of quantum mechanical operator $\hat{\rho}$ given by
\begin{align}
\hat{\rho}=f(e^{2\pi i\hat{\alpha}})
e^{2\pi ik\hat{\alpha}}
\theta_{\mathbb{Z}}(-\hat{p})
(\mathfrak{t}z^{-1})^{-\hat{p}}
g(e^{2\pi i\hat{\alpha}})
e^{-2\pi ik\hat{\alpha}}\theta_{\mathbb{Z}}(-\hat{p})(\mathfrak{t}z)^{-\hat{p}}
\label{rhohat}
\end{align}
with\footnote{
Note that we have changed the definition of $\theta_{\mathbb{Z}}(x)$ from the one in \cite{Hayashi:2024jof} as $\theta_{\mathbb{Z}}(x)=(\theta_{\mathbb{Z}}(-x))^{\text{\cite{Hayashi:2024jof}}}$, so that it is a more natural discrete analogue of the step function.
}
\begin{align}
\theta_\mathbb{Z}(p)=\begin{cases}
1&\quad (p=0,1,\cdots,\infty)\\
0&\quad (p=-1,-2,\cdots,-\infty)
\end{cases}.
\label{thetaZ}
\end{align}
The 1d quantum mechanical notation allows us to calculate the traces $\text{Tr}\rho^n=\text{tr}\hat{\rho}^n$ for any $f(s)$ and $g(s)$ which can be expanded in the Laurent series
\begin{align}
s^k f(s)=\sum_{\ell=-\infty}^\infty A_\ell s^\ell,\quad
s^{-k} g(s)=\sum_{\ell=-\infty}^\infty B_\ell s^\ell
\end{align}
efficiently by taking the trace in the momentum basis as (see \cite{Hayashi:2024jof})
\begin{align}
\text{Tr}\rho^n=
\text{tr}\hat{\rho}^n=
\frac{1}{1-\mathfrak{t}^{2n}}
\sum_{\substack{\ell_1,\cdots,\ell_n,\ell'_1,\cdots,\ell'_n=-\infty\\ (\sum_{i=1}^n(\ell_i+\ell'_i)=0)}}^\infty
\prod_{i=1}^n A_{\ell_i}B_{\ell'_i}
\mathfrak{t}^{\sum_{i=1}^{2n}(L_i-L_\text{min})}z^{\sum_{i=1}^n(-L_{2i-1}+L_{2i})}
\end{align}
with
\begin{align}
L_{2i-1}=\sum_{j=1}^i\ell_j+\sum_{j=1}^{i-1}\ell'_j,\quad
L_{2i}=\sum_{j=1}^i(\ell_j+\ell'_j),\quad
L_\text{min}=\text{min}\{L_1,L_2,\cdots,L_{2n-1},0\}.
\end{align}

Plugging $\tilde{\cal I}_{\text{ABJM},k}[f,g;\nu]$ with $\nu\le N$ into \eqref{ABJMCstep2}, or equivalently into the grand canonical sum \eqref{XiABJM} with \eqref{XitildeABJM} and extracting the coefficient of $u^N$, we obtain the supersymmetric index $\langle\prod_if(s^{(1)}_i)g(s^{(2)}_i)\rangle^{U(N)_k\times U(N)_{-k}\text{ABJ}(C)}(y,z;\mathfrak{t})$ with the insertion of the single-variable functions.

\subsubsection{Wilson line only in one gauge node}
The Coulomb line defect index of $U(N)_k\times U(N)_{-k}$ ABJM theory without any insertion in the gauge node $U(N)_{-k}$ can be calculated by setting the second single variable function as $g(s)=1$.
In this case the Fermi gas density matrix $\hat{\rho}$ \eqref{rhohat} simplifies, up to a similarity transformation (which does not affect the traces $\text{tr}\hat{\rho}^n$) as
\begin{align}
\hat{\rho}=f(e^{2\pi i\hat{\alpha}})\mathfrak{t}^{-2\hat{p}}\theta_{\mathbb{Z}}(-\hat{p})\mathfrak{t}^{|k|}z^{-k}.
\end{align}
Note that this density matrix is equivalent to the one used for the $U(N)$ ADHM theory in \cite{Hayashi:2024jof}, up to the overall factor $\mathfrak{t}^{|k|}z^{-k}$ which we can remove by replacing $u$ in the right-hand side of $\tilde{\Xi}_{\text{ABJM},k}[f,1;u]$ \eqref{XitildeABJM} with $\mathfrak{t}^{|k|}z^{-k} u$.
This implies that the grand canonical sum \eqref{XiABJM} is nothing but the grand canonical sum of the supersymmetric indices of $U(N)$ ADHM theory with $|k|$ flavors with the insertion of $\prod_if(s_i)$, where the monopole fugacity of the ADHM theory is given by $z_{\text{ADHM}}=zy^{-\frac{1}{k}}$.
Namely, we conclude
\begin{align}
\langle W_{\lambda;\mathbf{1}}\rangle^{U(N)_k\times U(N)_{-k}\text{ ABJM}(C)}(z,y;\mathfrak{t})
=\langle W_{\lambda}\rangle^{U(N)_k\text{ ADHM-}[|k|](C)}(z_\text{ADHM}=zy^{-\frac{1}{k}};\mathfrak{t})
\end{align}
for any representation $\lambda$.

\subsubsection{Antisymmetric Wilson lines}

Let us consider the line defect index of the $U(N)_k\times U(N)_{-k}$ ABJM theory with the Wilson line in the antisymmetric representations in the two gauge nodes
\begin{align}
\langle W_{(1^{\ell});\overline{(1^{\ell'})}}\rangle^{U(N)_k\times U(N)_{-k}\text{ ABJM}(C)}(z,y;\mathfrak{t}),
\end{align}
which can be calculated by choosing the one-variable functions $f(s)$ and $g(s)$ as $f(s)=1+as$ and $g(s)=1+bs^{-1}$
\begin{align}
&\biggl\langle\prod_{i=1}^N (1+as_i^{(1)})(1+b(s_i^{(2)})^{-1})\biggr\rangle^{U(N)_k\times U(N)_{-k}\text{ ABJM}(C)}(z,y;\mathfrak{t})\nonumber \\
&=\sum_{\ell,\ell'=0}^\infty a^\ell b^{\ell'} \langle W_{(1^{\ell});\overline{(1^{\ell'})}}\rangle^{U(N)_k\times U(N)_{-k}\text{ ABJM}(C)}(z,y;\mathfrak{t}).
\end{align}
For this choice of $f(s)$ and $g(s)$, we find that $\tilde{\cal I}_{\text{ABJM},k}[f,g;\nu]$ \eqref{ItildeABJM} is given by
\begin{align}
\tilde{\cal I}_{\text{ABJM},k}[1+as,1+bs^{-1};\nu]=
\begin{cases}
\displaystyle
\frac{\mathfrak{t}^{|k|\nu}}{z^{k\nu}}\sum_{j=0}^\nu
\Bigl(
\frac{ab\mathfrak{t}}{z}\Bigr)^j
\frac{1
}{
(\mathfrak{t}^2;\mathfrak{t}^2)_j
(\mathfrak{t}^2;\mathfrak{t}^2)_{\nu-j}
}&\quad (k\ge 0)\vspace{0.2cm}\\
\displaystyle
\frac{\mathfrak{t}^{|k|\nu}}{z^{k\nu}}\sum_{j=0}^\nu
\Bigl(
\frac{ab\mathfrak{t}^{-1}}{z}\Bigr)^j
\frac{1
}{
(\mathfrak{t}^2;\mathfrak{t}^2)_j
(\mathfrak{t}^2;\mathfrak{t}^2)_{\nu-j}
}&\quad (k<0)
\end{cases}.
\label{ABJMantisym1antisymbar2Omegaknu}
\end{align}
Plugging this into the grand canonical sum \eqref{XiABJM} with \eqref{XitildeABJM}, we finally obtain (assuming $k>0$)
\begin{align}
\Xi_{\text{ABJM}}[1+as,1+bs^{-1};u]
=
\frac{
\Xi_\text{ABJM}[1,1;u]
}{
(\mathfrak{t}z^{-1}abu;\mathfrak{t}^2)_\infty}
\prod_{m=1}^\infty\prod_\pm
\frac{1}{
(\mathfrak{t}^{km\pm 1}z^{\mp km-1}y^{\pm m}abu;\mathfrak{t}^2)_\infty
},
\label{XiABJMantisymantisymbar}
\end{align}
where
\begin{align}
\Xi_{\text{ABJM}}[1,1;u]
=
\sum_{N=0}^\infty u^N {\cal I}^{U(N)_k\times U(N)_{-k}\text{ ABJM}(C)}(y,z;\mathfrak{t})=\prod_{m=-\infty}^\infty \frac{1}{(\mathfrak{t}^{k|m|}z^{-km}y^mu;\mathfrak{t}^2)_\infty}
\end{align}
is the grand canonical sum of the Coulomb indices without Wilson lines.

From \eqref{XiABJMantisymantisymbar}, we find that the line defect index $\langle W_{(1^{\ell});\overline{(1^{\ell'})}}\rangle^{U(N)_k\times U(N)_{-k}\text{ ABJM}(C)}$ is non-vanishing only when $\ell=\ell'$ and satisfies the recursion relation
\begin{align}
&\langle W_{(1^{\ell});\overline{(1^{\ell})}}\rangle^{U(N)_k\times U(N)_{-k}\text{ ABJM}(C)}(y,z;\mathfrak{t})\nonumber \\
&
=
{\cal I}^{U(N-\ell)_k\times U(N-\ell)_{-k}\text{ ABJM}(C)}(y,z;\mathfrak{t})
\langle W_{(1^{\ell});\overline{(1^{\ell})}}\rangle^{U(\ell)_k\times U(\ell)_{-k}\text{ ABJM}(C)}(y,z;\mathfrak{t}),
\end{align}
as we have claimed in \eqref{R_uNuNABJM_W}.
Here the line defect index for $\ell=N$ on the right-hand side is obtained from the generating function
\begin{align}
&\sum_{N=0}^\infty u^N\langle W_{(1^N);\overline{(1^N)}}\rangle^{U(N)_k\times U(N)_{-k}\text{ ABJM}(C)}(y,z;\mathfrak{t})\nonumber \\
&
=
\frac{1}{
(\mathfrak{t}z^{-1}u;\mathfrak{t}^2)_\infty}
\prod_{m=1}^\infty\prod_\pm
\frac{1}{
(\mathfrak{t}^{km\pm 1}z^{\mp km-1}y^{\pm m}u;\mathfrak{t}^2)_\infty
}.
\end{align}

\subsubsection{More general representations}
Wilson lines in a more general representation can be decomposed into the multi point function of the Wilson lines in (anti)symmetric representations, which can be calculated by inserting the single variable functions of the form
\begin{align}
f(s)=\frac{\prod_{I=1}^{m}(1+a_I s)}{\prod_{I=1}^n(1-b_Is)},\quad
g(s)=\frac{\prod_{I=1}^{m'}(1+a_I' s)}{\prod_{I=1}^{n'}(1-b_I' s)}.
\end{align}
For example, for $k=1$ and $z=y=1$ we find
\begin{subequations}
\begin{align}
&\langle W_{\ydiagram{2};\overline{\ydiagram{2}}}\rangle^{U(3)_1\times U(3)_{-1}\text{ ABJM}(C)}(\mathfrak{t})=
\frac{1 + 2 \mathfrak{t} + 2 \mathfrak{t}^3 - \mathfrak{t}^4}{(1 - \mathfrak{t})^4 (1 - \mathfrak{t}^2)^2},\\
&\langle W_{\ydiagram{3};\overline{\ydiagram{3}}}\rangle^{U(3)_1\times U(3)_{-1}\text{ ABJM}(C)}(\mathfrak{t})\nonumber \\
&=
\frac{2 + 4 \mathfrak{t} + 13 \mathfrak{t}^2 + 14 \mathfrak{t}^3 + 17 \mathfrak{t}^4 + 10 \mathfrak{t}^5 + 5 \mathfrak{t}^6 - 2 \mathfrak{t}^7 - 2 \mathfrak{t}^8 - 2 \mathfrak{t}^{11} + \mathfrak{t}^{12}}{(1 - \mathfrak{t})^2 (1 - \mathfrak{t}^2)^2 (1 - \mathfrak{t}^3)^2},\\
&\langle W_{\ydiagram{2,1};\overline{\ydiagram{2,1}}}\rangle^{U(3)_1\times U(3)_{-1}\text{ ABJM}(C)}(\mathfrak{t})=
\frac{1 + 6 \mathfrak{t} + 12 \mathfrak{t}^2 + 20 \mathfrak{t}^3 + 13 \mathfrak{t}^4 + 8 \mathfrak{t}^5 + \mathfrak{t}^6 - \mathfrak{t}^8}{(1 - \mathfrak{t})^2 (1 - \mathfrak{t}^2)^2 (1 - \mathfrak{t}^3)^2},\\
&\langle W_{\ydiagram{2,1};\overline{\ydiagram{2,1}}}\rangle^{U(4)_1\times U(4)_{-1}\text{ ABJM}(C)}(\mathfrak{t})=
\frac{
1 + 4 \mathfrak{t} + 8 \mathfrak{t}^2 + 14 \mathfrak{t}^3 + 9 \mathfrak{t}^4 + 6 \mathfrak{t}^5 + \mathfrak{t}^6 - \mathfrak{t}^8}{(1 - \mathfrak{t})^4 (1 - \mathfrak{t}^2)^2 (1 - \mathfrak{t}^3)^2},\\
&\langle W_{\ydiagram{3,1};\overline{\ydiagram{3,1}}}\rangle^{U(3)_1\times U(3)_{-1}\text{ ABJM}(C)}(\mathfrak{t})=
\frac{
2 + 8 \mathfrak{t} + 14 \mathfrak{t}^2 + 11 \mathfrak{t}^3 + 16 \mathfrak{t}^4 - 7 \mathfrak{t}^5 - \mathfrak{t}^7 - 2 \mathfrak{t}^8 + \mathfrak{t}^9}{(1 - \mathfrak{t})^3 (1 - \mathfrak{t}^2)^2 (1 - \mathfrak{t}^3)},\\
&\langle W_{\ydiagram{3,1};\overline{\ydiagram{3,1}}}\rangle^{U(4)_1\times U(4)_{-1}\text{ ABJM}(C)}(\mathfrak{t})\nonumber \\
&=
\frac{
1}{(1 - \mathfrak{t})^3 (1 - \mathfrak{t}^2)^2 (1 - \mathfrak{t}^3) (1 - \mathfrak{t}^4)^2}(2 + 10 \mathfrak{t} + 27 \mathfrak{t}^2 + 60 \mathfrak{t}^3 + 103 \mathfrak{t}^4 + 126 \mathfrak{t}^5 + 146 \mathfrak{t}^6\nonumber \\
&\quad + 81 \mathfrak{t}^7 + 75 \mathfrak{t}^8 - 3 \mathfrak{t}^9
+ \mathfrak{t}^{10} - 21 \mathfrak{t}^{11} - 6 \mathfrak{t}^{12} - 2 \mathfrak{t}^{13} - \mathfrak{t}^{14} + 2 \mathfrak{t}^{15} - 2 \mathfrak{t}^{16} + 3 \mathfrak{t}^{17} - \mathfrak{t}^{18}).
\end{align}
\end{subequations}
In the limit $\mathfrak{t}\rightarrow 1$, we find
\begin{align}
\lim_{\mathfrak{t}\rightarrow 1}\langle {\cal W}_{\mu;\overline{\lambda}}\rangle^{U(N)_1\times U(N)_{-1}\text{ ABJM}(C)}(\mathfrak{t})
=
\oint\prod_{i=1}^N\frac{ds_i}{2\pi is_i}\chi_\mu(s)\chi_{\overline{\lambda}}(s),
\end{align}
which is consistent with our conjecture \eqref{Witten_indC_conj_int}.

\subsection{$U(N)_k\times U(N+M)_{-k}$ ABJ theory}
\label{app_ABJrank}

Solving the vacuum equations of the rank deformed ABJ theories, as we did in section \ref{sec_rankABJ}, gives the stabilizer \eqref{H0ABJrank},
\begin{align}
\label{H0ABJ}
H_0=U(1)^N_{\text{diag}}\times U(M)\;\subset\; U(N)_k\times U(N+M)_{-k}\,.
\end{align}

The supersymmetric index is obtained from \eqref{ABJMindex} by simply replacing the two ranks by $N$ and $N+M$. In the Coulomb limit of the index itself, with no line defect inserted, only the configurations for which the total power of $q$, which is the dimension of the corresponding bare monopole operator,
\begin{align}
\label{ABJEexp}
E=-\frac12\sum_{I=1}^2\sum_{i<j}\bigl|m^{(I)}_i-m^{(I)}_j\bigr|+\frac12\sum_{i,j}\bigl|m^{(1)}_i-m^{(2)}_j\bigr|\,,
\end{align}
vanishes contribute, exactly as in \eqref{ABJMCstep1}.
This is no longer true once a Wilson line is inserted, since the inserted character can compensate the gauge fugacity dependence of a configuration with $E<0$ and let it survive the holonomy integrals, as we discuss below.
Which configurations these are is conveniently described by the counting functions
$\alpha_t=\#\{i\,|\,m^{(1)}_i\le t\}$ and $\beta_t=\#\{i\,|\,m^{(2)}_i\le t\}$, in terms of which
\begin{align}
\label{ABJEexpcount}
E=\frac12\sum_{t\in\mathbb{Z}}d_t\,(d_t-M)\,,\qquad d_t=\beta_t-\alpha_t\,,
\end{align}
where $d_t$ goes from $d_{-\infty}=0$ to $d_{+\infty}=M$ so that the sum is finite.
For $M=0$ this gives $E=\frac12\sum_td_t^2$, which vanishes only for $\{m^{(2)}_{i}\}=\{m^{(1)}_{i}\}$, while for $M=1$ every term $d_t(d_t-1)$ is non-negative and $E=0$ amounts to the interlacing condition $d_t\in\{0,1\}$, so that a larger set of configurations survives.
A new phenomenon appears for $M\ge2$, namely that the terms with $0<d_t<M$ are negative, so that $E$ itself can be negative.
A configuration with $E<0$ carries an overall $q^{E}$, which has to be compensated by terms of order $q^{-E}$ in the expansion of the rest of the integrand, that is by its subleading terms rather than by the leading ones that survive the naive $q\rightarrow0$ limit, and whether it contributes depends on the inserted characters. We leave the line defect indices of the theories with $M\ge2$ for future work.
Below we set $z=y=1$ and take $N=1$, so that the relevant group theory is that of $U(1)\times U(M)\subset U(1)_k\times U(1+M)_{-k}$.

We first note that the Coulomb index itself is insensitive to the rank deformation,
\begin{align}
{\cal I}^{U(1)_k\times U(1+M)_{-k}\text{ ABJ}(C)}(\mathfrak{t})
={\cal I}^{U(1)_k\times U(1)_{-k}\text{ ABJM}(C)}(\mathfrak{t})
=\frac{1+\mathfrak{t}^k}{(1-\mathfrak{t}^2)(1-\mathfrak{t}^k)}\,,
\end{align}
which we have checked for $M=1,2,3$ and $k=1,\cdots,4$, in agreement with the fact that the moduli space is $\mathbb{C}^4/\mathbb{Z}_k$ for any $M$.
For $M=1$, inserting Wilson lines on the second node in the representations $(m)=\mathrm{Sym}^m$ and its conjugate we find

\begin{align}
\lim_{\mathfrak{t}\rightarrow1}
\bigl\langle {\cal W}_{\mathbf{1};(m)}{\cal W}_{\mathbf{1};\overline{(m)}}\bigr\rangle^{U(1)_k\times U(2)_{-k}\text{ ABJ}(C)}(\mathfrak{t})=m+1\,,
\end{align}
for $k=1,2,3$ and $m\le3$.

Let us see how the same numbers come out of the character average \eqref{Witten_ind_unified}.
With $H_0$ as in \eqref{H0ABJ} and $N=1$, a group element of $H_0$ is
\begin{align}
\label{H0ABJelement}
h=\bigl(\,s\;;\;\mathrm{diag}(s)\oplus g\,\bigr),
\qquad s\in U(1),\quad g\in U(M),
\end{align}
the first entry being the holonomy of the first gauge node and the second that of the second gauge node.
Under $H_0$ the fundamental representation of the second node decomposes as $\mathbb{C}^{1+M}=\mathbb{C}_{s}\oplus\mathbb{C}^{M}$, so that
$\mathrm{Sym}^m(\mathbb{C}^{1+M})=\bigoplus_{p=0}^{m}\mathbb{C}_{s^p}\otimes\mathrm{Sym}^{m-p}(\mathbb{C}^M)$ and
\begin{align}
\label{ABJcharrestriction}
\chi_{(m)}(h)=\sum_{p=0}^{m}s^{p}\,\chi^{\mathfrak{u}(M)}_{(m-p)}(g)\,.
\end{align}

Since the Wilson lines sit on the second node only, $\prod_j\chi_{\mathcal{R}_j}(h)=\chi_{(m)}(h)\chi_{\overline{(m)}}(h)$ and \eqref{Witten_ind_unified} reads
\begin{align}
\label{ABJunifiedeval}
&\int_{H_0}dh\,\bigl|\chi_{(m)}(h)\bigr|^2
=\oint\frac{ds}{2\pi is}\int_{U(M)}\!\!dg\,
\sum_{p,p'=0}^{m}s^{\,p-p'}\,\chi^{\mathfrak{u}(M)}_{(m-p)}(g)\,\chi^{\mathfrak{u}(M)}_{\overline{(m-p')}}(g)\nonumber\\
&\qquad=\sum_{p=0}^{m}\int_{U(M)}\!\!dg\,\bigl|\chi^{\mathfrak{u}(M)}_{(m-p)}(g)\bigr|^2=\sum_{p=0}^{m}1=m+1\,,
\end{align}
where the $U(1)$ integral sets $p=p'$ and the $U(M)$ integral is unity by the orthonormality of the irreducible characters of $U(M)$.
Equivalently, in the form \eqref{Witten_ind_unified_inv}, the $m+1$ summands in \eqref{ABJcharrestriction} are pairwise inequivalent irreducible representations of $U(1)\times U(M)$, so that $(m)\otimes \overline{(m)}$ contains exactly $m+1$ invariants.

The computation \eqref{ABJunifiedeval} holds for any $M$, and it is for $M\ge2$ that it would test the non-Abelian factor, since replacing $U(M)$ by its maximal torus, or dropping it altogether, would give different numbers. Whether the Coulomb limit agrees with it there is the question we leave for future work.

\subsection{$SO(2N+\gamma)_{2k}^{(\chi\zeta)} \times USp(2N)_{-k}$ ABJ theory with $(\gamma,\chi)\neq (0,-1)$}
\label{app_OSpgammachinot0-1}

Lastly let us consider the line defect indices in $SO(2N+\gamma)_{2k}\times USp(2N)_{-k}$ ABJ theory, with $\gamma=0,1$.
As in the theory with single gauge node considered in appendices \ref{app_SO2N+1algorithm}-\ref{app_USp}, we can turn on two discrete fugacities $(\chi,\zeta)$ for $\mathbb{Z}^{{\cal C}}_2\times \mathbb{Z}^{{\cal M}}_2$.
In this appendix we consider the case with $(\gamma,\chi)\neq (0,-1)$, where the supersymmetric index with the insertion of $\prod_{i=1}^Nf(s_i^{(1)})g(s_i^{(2)})$ is given as
\begin{align}
&\biggl\langle\prod_{i=1}^Nf(s^{(1)}_i)g(s^{(2)}_i)\biggr\rangle^{SO(2N+\gamma)^{(\chi\zeta)}_{2k}\times USp(2N)_{-k}\text{ABJ}}(t;q)
\nonumber\\
&=\frac{1}{2^{2N+\gamma-1}(N!)^2}
\sum_{m^{(1)}_1,\cdots,m^{(1)}_N, m^{(2)}_1,\cdots,m^{(2)}_N \in \mathbb{Z}}
\oint \prod_{i=1}^N \frac{ds^{(1)}_{i}}{2\pi is^{(1)}_i}f(s_i^{(1)}) (s_i^{(1)})^{2km^{(1)}_i}\nonumber \\
&\quad\times
\oint \prod_{i=1}^N \frac{ds^{(2)}_{i}}{2\pi is^{(2)}_i}g(s_i^{(2)}) (s_i^{(2)})^{-2km^{(2)}_i}
\prod_{i=1}^{N}\prod_\pm \Bigl(1-\chi q^{\frac{|m_i^{(1)}|}{2}} (s_i^{(1)})^{\pm 1}\Bigr)^{\gamma}
\Bigl(1-q^{|m_i^{(2)}|} (s_i^{(2)})^{\pm 2}\Bigr)\nonumber \\
&\quad\times
\prod_{I=1}^2 \prod_{i<j}^N \prod_\pm \Bigl(1-q^{\frac{|m^{(I)}_i-m^{(I)}_j|}{2}}(s_i^{(I)})^{\pm 1}(s_j^{(I)})^{\mp 1}\Bigr)
\Bigl(1-q^{\frac{|m^{(I)}_i+m^{(I)}_j|}{2}}(s_i^{(I)})^{\pm 1} (s_j^{(I)})^{\pm 1}\Bigr)
\nonumber\\
&\quad \times
\prod_{i,j=1}^N
\prod_\pm
\frac{(q^{\frac34+\frac{|m_i^{(1)}-m_j^{(2)}|}{2}}t^{-1}(s_i^{(1)})^{\mp 1} (s_j^{(2)})^{\pm 1};q)_{\infty}}
{(q^{\frac14+\frac{|m_i^{(1)}-m_j^{(2)}|}{2}}t (s_i^{(1)})^{\pm 1} (s_j^{(2)})^{\mp 1};q)_{\infty}}
\frac{(q^{\frac34+\frac{|m_i^{(1)}+m_j^{(2)}|}{2}}t^{-1}(s_i^{(1)})^{\mp 1} (s_j^{(2)})^{\mp 1};q)_{\infty}}
{(q^{\frac14+\frac{|m_i^{(1)}+m_j^{(2)}|}{2}}t (s_i^{(1)})^{\pm 1} (s_j^{(2)})^{\pm 1};q)_{\infty}}
\nonumber\\
&\quad\quad \times
\frac{(q^{\frac34+\frac{|m_i^{(2)} - m_j^{(1)}|}{2}}t(s_i^{(2)})^{\mp 1} (s_j^{(1)})^{\pm 1};q)_{\infty}}
{(q^{\frac14+\frac{|m_i^{(2)} - m_j^{(1)}|}{2}}t^{-1} (s_i^{(2)})^{\pm 1} (s_j^{(1)})^{\mp 1};q)_{\infty}}
\frac{(q^{\frac34+\frac{|m_i^{(2)} + m_j^{(1)}|}{2}}t (s_i^{(2)})^{\mp 1} (s_j^{(1)})^{\mp 1};q)_{\infty}}
{(q^{\frac14+\frac{|m_i^{(2)} + m_j^{(1)}|}{2}}t^{-1} (s_i^{(2)})^{\pm 1} (s_j^{(1)})^{\pm 1};q)_{\infty}}
\nonumber\\
&\quad \times
\prod_{i=1}^N\prod_\pm
\left[
\frac{(\chi q^{\frac34+\frac{|m_i^{(2)}|}{2}}t^{-1}(s_i^{(2)})^{\mp 1};q)_{\infty}}
{(\chi q^{\frac14+\frac{|m_i^{(2)}|}{2}}t          (s_i^{(2)})^{\pm 1};q)_{\infty}}
\frac{(\chi q^{\frac34+\frac{|m_i^{(2)}|}{2}}t     (s_i^{(2)})^{\mp 1};q)_{\infty}}
{(\chi q^{\frac14+\frac{|m_i^{(2)}|}{2}}t^{-1}     (s_i^{(2)})^{\pm 1};q)_{\infty}}
\right]^{\gamma}
\nonumber\\
&\quad \times
q^{-\frac{\gamma}{2}\sum_{i=1}^N|m_i^{(1)}| +(\frac{\gamma}{2}-1)\sum_{i=1}^N|m_i^{(2)}|
-\frac12\sum_{I=1}^2\sum_{i<j}^N\sum_\pm
|m_{i}^{(I)}\pm m_{j}^{(I)}|
+\frac12 \sum_{i,j=1}^N\sum_\pm
|m^{(1)}_i\pm m^{(2)}_j|
}\nonumber \\
&\quad\times
\zeta^{\sum_{i=1}^N m_i^{(1)}},
\label{abj_index1}
\end{align}
while the case with $(\gamma,\chi)=(0,-1)$ is treated in the next appendix.

Let us assume that $f(s)$ and $g(s)$ satisfy $f(s)=f(s^{-1})$ and $g(s)=g(s^{-1})$ so that the summand of the summation over the monopole charges in \eqref{abj_index1} is invariant under $m^{(I)}_i\rightarrow -m^{(I)}_i$ for any one of $m^{(I)}_i$, which allows us to restrict the summation to $m^{(I)}_i\ge 0$ (see \eqref{Z2trick}).
By further taking the Coulomb limit, we obtain
\begin{align}
&\biggl\langle\prod_{i=1}^Nf(s^{(1)}_i)g(s^{(2)}_i)\biggr\rangle^{SO(2N+\gamma)^{(\chi\zeta)}_{2k}\times USp(2N)_{-k}\text{ABJ}(C)}(\mathfrak{t})\nonumber \\
&=\frac{1}{2^{2N+\gamma-1}(N!)^2}\sum_{\substack{m^{(1)}_1,\cdots,m^{(1)}_N,m^{(2)}_1,\cdots,m^{(2)}_N\ge 0\\ (\{m^{(2)}_i\}=\{m^{(1)}_i\})}}\prod_{I=1}^2\prod_{i=1}^N2^{1-\delta_{m^{(I)}_i,0}}
\oint\prod_{i=1}^N\frac{ds^{(1)}_i}{2\pi is^{(1)}_i}
f(s^{(1)}_i)(s^{(1)}_i)^{2km^{(1)}_i}\nonumber \\
&\quad\times
\oint\prod_{i=1}^N\frac{ds^{(2)}_i}{2\pi is^{(2)}_i}
g(s^{(2)}_i)(s^{(2)}_i)^{-2km^{(2)}_i}
\prod_{\substack{i=1\\ (m^{(1)}_i=0)}}^N\prod_\pm (1-\chi(s^{(1)}_i)^{\pm 1})^\gamma\nonumber \\
&\quad\times
\prod_{\substack{i=1\\ (m^{(2)}_i=0)}}^N \prod_\pm \frac{1-(s^{(2)}_i)^{\pm 2}}{(1-\chi\mathfrak{t}(s^{(2)}_i)^{\pm 1})^\gamma}
\prod_{I=1}^2
\prod_{\substack{i<j\\ (m^{(I)}_i=m^{(I)}_j)}}^N \prod_\pm (1-(s^{(I)}_i)^{\pm 1}(s^{(I)}_j)^{\mp 1})\nonumber \\
&\quad\times \prod_{I=1}^2
\prod_{\substack{i<j\\ (m^{(I)}_i=-m^{(I)}_j)}}^N\prod_\pm (1-(s^{(I)}_i)^{\pm 1}(s^{(I)}_j)^{\pm 1})
\prod_{\substack{i,j=1\\ (m^{(2)}_i=m^{(1)}_j)}}^N\prod_\pm
\frac{1}{1-\mathfrak{t}(s^{(2)}_i)^{\pm 1}(s^{(1)}_j)^{\mp 1}}\nonumber \\
&\quad\times
\prod_{\substack{i,j=1\\ (m^{(2)}_i=-m^{(1)}_j)}}^N\prod_\pm
\frac{1}{1-\mathfrak{t}(s^{(2)}_i)^{\pm 1}(s^{(1)}_j)^{\pm 1}}
\zeta^{\sum_{i=1}^Nm^{(1)}_i},
\end{align}
where the monopole charges $m^{(2)}_i$ are restricted in the Coulomb limit to those which coincide with $m^{(1)}_i$ up to a permutation.
By characterizing $\{m^{(1)}_i\}$ with the multiplicities $\nu_m=\#\{i|m^{(1)}_i=m\}$, we obtain
\begin{align}
&\biggl\langle\prod_{i=1}^Nf(s^{(1)}_i)g(s^{(2)}_i)\biggr\rangle^{SO(2N+\gamma)^{\chi\zeta}_{2k}\times USp(2N)_{-k}\text{ABJ}(C)}(\mathfrak{t})\nonumber \\
&=\frac{1}{2^{\gamma-1}}\sum_{\substack{\nu_0,\nu_m=0\\ (\nu_0+\sum_{m=1}^\infty \nu_m=N)}}\zeta^{\sum_{m=1}^\infty m\nu_m}\tilde{\cal I}^{(0)}_{\text{OSp},\gamma,\chi}[f,g;\nu_0]\prod_{m=1}^\infty\tilde{\cal I}^{(1)}_{\text{OSp},km}[f,g;\nu_m],
\label{OSpABJ_index2}
\end{align}
with
\begin{subequations}
\begin{align}
&\tilde{\cal I}^{(0)}_{\text{OSp},\gamma,\chi}[f,g;\nu]\nonumber \\
&=\frac{1}{4^\nu(\nu!)^2}
\oint\prod_{i=1}^\nu\frac{d\sigma^{(1)}_i}{2\pi i\sigma^{(1)}_i}
f(\sigma^{(1)}_i)
\oint\prod_{i=1}^\nu\frac{d\sigma^{(2)}_i}{2\pi i\sigma^{(2)}_i}
g(\sigma^{(2)}_i)\nonumber \\
&\quad\times
\prod_{i=1}^\nu \prod_\pm\biggl(\frac{1-\chi(\sigma_i^{(1)})^{\pm 1}}{1-\chi\mathfrak{t}(\sigma^{(2)}_i)^{\pm 1}}\biggr)^\gamma
\frac{1-(\sigma^{(2)}_i)^{\pm 2}
}{
(1-\mathfrak{t}(\sigma^{(2)}_i)^{\pm 1}(\sigma^{(1)}_i)^{\mp 1})(1-\mathfrak{t}(\sigma^{(2)}_i)^{\pm 1}(\sigma^{(1)}_i)^{\pm 1})
}\nonumber \\
&\quad\times
\prod_{i<j}^\nu
\prod_\pm
\prod_{\pm'}
\frac{
(1-(\sigma^{(1)}_i)^{\pm 1}(\sigma^{(1)}_j)^{\pm' 1})
(1-(\sigma^{(2)}_i)^{\pm 1}(\sigma^{(2)}_j)^{\pm' 1})
}{
(1-\mathfrak{t}(\sigma^{(2)}_i)^{\pm 1}(\sigma^{(1)}_j)^{\pm' 1})
(1-\mathfrak{t}(\sigma^{(2)}_j)^{\pm 1}(\sigma^{(1)}_i)^{\pm' 1})
},
\label{calI0OSp}\\
&\tilde{\cal I}^{(1)}_{\text{OSp},k}[f,g;\nu]\nonumber \\
&=\frac{1}{(\nu!)^2}
\oint\prod_{i=1}^\nu\frac{d\sigma^{(1)}_i}{2\pi i\sigma^{(1)}_i}
f(\sigma^{(1)}_i)(\sigma^{(1)}_i)^{2k}
\oint\prod_{i=1}^\nu\frac{d\sigma^{(2)}_i}{2\pi i\sigma^{(2)}_i}
g(\sigma^{(2)}_i)(\sigma^{(2)}_i)^{-2k}\nonumber \\
&\quad\times
\prod_{i=1}^\nu\prod_\pm
\frac{1
}{
1-\mathfrak{t}(\sigma^{(2)}_i)^{\pm 1}(\sigma^{(1)}_i)^{\mp 1}
}
\prod_{i<j}^\nu
\prod_\pm
\frac{
(1-(\sigma^{(1)}_i)^{\pm 1}(\sigma^{(1)}_j)^{\mp 1})
(1-(\sigma^{(2)}_i)^{\pm 1}(\sigma^{(2)}_j)^{\mp 1})
}{
(1-\mathfrak{t}(\sigma^{(2)}_i)^{\pm 1}(\sigma^{(1)}_j)^{\mp 1})
(1-\mathfrak{t}(\sigma^{(2)}_j)^{\pm 1}(\sigma^{(1)}_i)^{\mp 1})
}\nonumber \\
&=\tilde{\cal I}_{\text{ABJM},2k}[f,g;\nu]\Bigr|_{z=1}.
\label{calI1OSp}
\end{align}
\end{subequations}

The powers of two follow from the counting of the monopole charges.
For a configuration with multiplicities $\nu_m$ the overall factor in \eqref{abj_index1} is
$\frac{1}{2^{2N+\gamma-1}(N!)^2}\prod_{I,i}2^{1-\delta_{m^{(I)}_i,0}}$ times the number $\frac{(N!)^2}{\prod_m(\nu_m!)^2}$ of ways of distributing the charges over the two nodes, which equals
$\frac{1}{2^{\gamma-1}}\frac{1}{4^{\nu_0}(\nu_0!)^2}\prod_{m\ge1}\frac{1}{(\nu_m!)^2}$.
We have checked \eqref{OSpABJ_index2} against a direct evaluation of the Coulomb limit for $N\le2$, $\gamma=0,1$ and $k=2,3$.

The grand canonical sum ($(\gamma,\chi)\neq(0,-1)$)
\begin{align}
\Xi_{\text{OSp},\gamma,\chi}[f,g;u]=\sum_{N=0}^\infty
u^N
\biggl\langle\prod_{i=1}^Nf(s^{(1)}_i)g(s^{(2)}_i)\biggr\rangle^{SO(2N+\gamma)^{(\chi\zeta)}_{2k}\times USp(2N)_{-k}\text{ABJ}(C)}(\mathfrak{t}),
\label{XiOSpgammachinot0-1}
\end{align}
factorizes as
\begin{subequations}
\begin{align}
&\Xi_{\text{OSp},1,\chi}[f,g;u]=
\tilde{\Xi}_{\text{OSp},1,\chi}^{(0)}[f,g;u]
\prod_{m=1}^\infty \tilde{\Xi}_{\text{OSp},m}^{(1)}[f,g;\zeta^m u],\\
&\Xi_{\text{OSp},0,1}[f,g;u]=
2
\tilde{\Xi}_{\text{OSp},0,1}^{(0)}[f,g;u]
\prod_{m=1}^\infty \tilde{\Xi}_{\text{OSp},m}^{(1)}[f,g;\zeta^m u],
\end{align}
\end{subequations}
with
\begin{subequations}
\begin{align}
&\tilde{\Xi}_{\text{OSp},1,\chi}^{(0)}[f,g;u]=\sum_{\nu=0}^\infty u^{\nu}\tilde{\cal I}_{\text{OSp},1,\chi}^{(0)}[f,g;\nu],\\
&\tilde{\Xi}_{\text{OSp},0,1}^{(0)}[f,g;u]=\sum_{\nu=0}^\infty u^{\nu}\tilde{\cal I}_{\text{OSp},0,1}^{(0)}[f,g;\nu],\\
&\tilde{\Xi}_{\text{OSp},m}^{(1)}[f,g;u]=\sum_{\nu=0}^\infty u^{\nu}\tilde{\cal I}_{\text{OSp},km}^{(1)}[f,g;\nu]
\label{XitildeOSp(1)}
\end{align}
\end{subequations}

Since the building blocks $\tilde{\cal I}^{(0)}_{\text{OSp},\gamma,\chi}[f,g;\nu]$ \eqref{calI0OSp} and $\tilde{\cal I}^{(1)}_{\text{OSp},k}[f,g;\nu]$ \eqref{calI1OSp} are $2\nu$ dimensional integrals, it is impractical to evaluate them directly for $\nu\ge 2$.
However, the second building block $\tilde{\cal I}^{(1)}_{\text{OSp},k}[f,g;\nu]$ coincides with the building block $\tilde{\cal I}_{\text{ABJM},2k}[f,g;\nu]$ \eqref{ItildeABJM} for the $U(N)_{2k}\times U(N)_{-2k}$ ABJM theory with $z=1$, and hence we can evaluate it efficiently by the Fermi gas technique as we have explained in appendix \ref{ABJMalgorithm}.
In order to apply the same technique to the other building block $\tilde{\cal I}^{(0)}_{\text{OSp},\gamma,\chi}[f,g;\nu]$, first we notice that the Cauchy determinant formula \eqref{Cauchy} can be generalized for the $BC$ type factors as\footnote{
Determinant formulas of this $BC$ type, written as a $\nu\times\nu$ determinant with kernel $[(u_i+v_j)(u_i v_j+1)]^{-1}$, have appeared in the study of matrix models with orthogonal and symplectic gauge groups \cite{Mezei:2013gqa} and, before that, in combinatorics \cite{MR1954236}.
}
\begin{align}
&\frac{
\prod_{i<j}^\nu\prod_{\pm,\pm'}
(1-(\sigma^{(1)}_i)^{\pm 1}(\sigma^{(1)}_j)^{\pm' 1})
(1-(\sigma^{(2)}_i)^{\pm 1}(\sigma^{(2)}_j)^{\pm' 1})
\prod_i\prod_\pm(1-(\sigma^{(2)}_i)^{\pm 2})
}{
\prod_{i,j=1}^{\nu}\prod_{\pm,\pm'}(1-\mathfrak{t}(\sigma^{(1)}_i)^{\pm 1}(\sigma^{(2)}_j)^{\pm' 1})
}\nonumber \\
&
=\mathfrak{t}^{-\nu(2\nu-1)}\prod_{i=1}^\nu\frac{
\sigma^{(2)}_i-(\sigma^{(2)}_i)^{-1}
}{
\sigma^{(1)}_i-(\sigma^{(1)}_i)^{-1}
}
\det
\begin{pmatrix}
\Bigl[\frac{1}{1-\mathfrak{t}(\sigma^{(1)}_i)^{-1}\sigma^{(2)}_j}\Bigr]_{i,j}^{\nu\times \nu}&
\Bigl[\frac{1}{1-\mathfrak{t}(\sigma^{(1)}_i\sigma^{(2)}_j)^{-1}}\Bigr]_{i,j}^{\nu\times \nu}\vspace{0.2cm} \\
\Bigl[\frac{1}{1-\mathfrak{t}\sigma^{(1)}_i\sigma^{(2)}_j}\Bigr]_{i,j}^{\nu\times \nu}&
\Bigl[\frac{1}{1-\mathfrak{t}\sigma^{(1)}_i(\sigma^{(2)}_j)^{-1}}\Bigr]_{i,j}^{\nu\times \nu}
\end{pmatrix},
\label{CauchyBC}
\end{align}
which is obtained by doubling the number of variables in the Cauchy determinant formula \eqref{Cauchy} and specifying them as $\{x_I\}_{I=1}^{2\nu}=\{\sigma_i^{(1)},(\sigma^{(1)}_i)^{-1}\}$ and $\{y_I\}_{I=1}^{2\nu}=\{-\mathfrak{t}\sigma^{(2)}_i,-\mathfrak{t}(\sigma^{(2)}_i)^{-1}\}$.
The $BC$ structure of the numerator arises because the doubled set is closed under inversion, so that its Vandermonde determinant splits into the factors $\sigma^{(a)}_i-(\sigma^{(a)}_i)^{-1}$ and $\bigl(1-(\sigma^{(a)}_i)^{\pm1}(\sigma^{(a)}_j)^{\pm1}\bigr)$ with $a=1,2$, the latter being the Weyl denominator of $D_\nu$ for the orthogonal node and of the corresponding part of $C_\nu$ for the symplectic one.
The difference between the two nodes does not enter the Cauchy kernel but the prefactor: the long roots of $USp(2\nu)$ give $\prod_\pm(1-(\sigma^{(2)}_i)^{\pm2})=-\bigl(\sigma^{(2)}_i-(\sigma^{(2)}_i)^{-1}\bigr)^2$, one power of which is supplied by the doubled Vandermonde determinant and the other by the prefactor, whereas on the orthogonal node the factor $\sigma^{(1)}_i-(\sigma^{(1)}_i)^{-1}$ produced by the doubling is absent from the integrand and is divided out.
For $\gamma=1$ the short roots of $B_\nu$ appear instead as $\prod_i\prod_\pm(1-\chi (\sigma^{(1)}_i)^{\pm1})$, which together with $\prod_i\prod_\pm(1-\chi\mathfrak{t}(\sigma^{(2)}_i)^{\pm1})$ multiplies both sides and plays no role in the identity.

By applying the Cauchy determinant formula \eqref{CauchyBC} to the $BC$ type product in $\tilde{\cal I}^{(0)}_{\text{OSp},\gamma,\chi}[f,g;\nu]$ \eqref{calI0OSp}, we obtain
\begin{align}
&\tilde{\cal I}^{(0)}_{\text{OSp},\gamma,\chi}[f,g;\nu]\nonumber \\
&
=\frac{\mathfrak{t}^{-\nu(2\nu-1)}}{(\nu!)^2}\oint\prod_{i=1}^\nu
\frac{d\sigma^{(1)}_i}{2\pi i\sigma^{(1)}_i}
\frac{d\sigma^{(2)}_i}{2\pi i\sigma^{(2)}_i}
\det\begin{pmatrix}
\Bigl[\frac{f'(\sigma^{(1)}_i)g'(\sigma^{(2)}_j)}{2(1-\mathfrak{t}(\sigma^{(1)}_i)^{-1}\sigma^{(2)}_j)}\Bigr]_{i,j}^{\nu\times \nu}
&\Bigl[\frac{f'(\sigma^{(1)}_i)}{2(1-\mathfrak{t}(\sigma^{(1)}_i\sigma^{(2)}_j)^{-1})}\Bigr]_{i,j}^{\nu\times \nu}\vspace{0.2cm} \\
\Bigl[\frac{g'(\sigma^{(2)}_j)}{2(1-\mathfrak{t}\sigma^{(1)}_i\sigma^{(2)}_j)}\Bigr]_{i,j}^{\nu\times \nu}
&\Bigl[\frac{1}{2(1-\mathfrak{t}\sigma^{(1)}_i(\sigma^{(2)}_j)^{-1})}\Bigr]_{i,j}^{\nu\times \nu}
\end{pmatrix}\nonumber \\
&
=\frac{\mathfrak{t}^{-\nu(2\nu-1)}}{(\nu!)^2}\int_0^1\prod_{i=1}^\nu
d\alpha^{(1)}_i
d\alpha^{(2)}_i\nonumber \\
&\quad\times
\det\begin{pmatrix}
 [\frac{1}{2}\langle \alpha^{(1)}_i|\hat{f}'\mathfrak{t}^{-\hat{p}}\theta_{\mathbb{Z}}(-\hat{p})\hat{g}'|\alpha^{(2)}_j\rangle]_{i,j}^{\nu\times \nu}
&[\frac{1}{2}\langle \alpha^{(1)}_i|\hat{f}'(e^{2\pi i\hat{\alpha}})\mathfrak{t}^{-\hat{p}}\theta_{\mathbb{Z}}(-\hat{p})\hat{R}|\alpha^{(2)}_j\rangle]_{i,j}^{\nu\times \nu}\vspace{0.2cm} \\
 [\frac{1}{2}\langle \alpha^{(1)}_i|\hat{R}\mathfrak{t}^{-\hat{p}}\theta_{\mathbb{Z}}(-\hat{p})\hat{g}'|\alpha^{(2)}_j\rangle]_{i,j}^{\nu\times \nu}
&[\frac{1}{2}\langle \alpha^{(1)}_i|\hat{R}\mathfrak{t}^{-\hat{p}}\theta_{\mathbb{Z}}(-\hat{p})\hat{R}|\alpha^{(2)}_j\rangle]_{i,j}^{\nu\times \nu}
\end{pmatrix},
\label{calI0OSpinFGstep1}
\end{align}
where we have defined $f'(\sigma)$ and $g'(\sigma)$ as
\begin{align}
f'(\sigma)=\frac{f(\sigma)\prod_\pm(1-\chi\sigma^{\pm 1})^\gamma}{\sigma-\sigma^{-1}},\quad
g'(\sigma)=\frac{g(\sigma)(\sigma-\sigma^{-1})}{\prod_\pm (1-\chi\mathfrak{t}\sigma^{\pm 1})^\gamma},
\end{align}
and in the third line of \eqref{calI0OSpinFGstep1} we have rewritten the matrix elements by using the notation of 1d quantum mechanics \eqref{1dQMnotation} and denoting $f'(e^{2\pi i\hat{\alpha}})$ and $g'(e^{2\pi i\hat{\alpha}})$ simply as $\hat{f}'$ and $\hat{g}'$.
Here we have also defined the reflection operator $\hat{R}$ as
\begin{align}
\hat{R}|\alpha\rangle=|-\alpha\rangle,\quad
\langle \alpha|\hat{R}=\langle -\alpha|,\quad
\hat{R}^2=1,\quad
\hat{R}\hat{\alpha}\hat{R}=-\hat{\alpha},\quad
\hat{R}\hat{p}\hat{R}=-\hat{p}.
\end{align}
We can further perform the $\alpha^{(2)}_i$ integrations by using the following generalization of the Andr\'eief identity \eqref{Andreief}
\begin{align}
&\frac{1}{\nu!}\int d^\nu x\det\Biggl(
\Bigl[X_m(x_n)\Bigr]_{m,n}^{2\nu\times \nu}
\quad
\Bigl[Y_m(x_n)\Bigr]_{m,n}^{2\nu\times \nu}
\Biggr)\nonumber \\
&
=(-1)^{\frac{\nu(\nu-1)}{2}}\operatorname{pf}\Biggl(\int dx(X_m(x)Y_n(x)-Y_m(x)X_n(x))\Biggr),
\end{align}
which is the special case of de Bruijn's formula (see e.g.~\cite{Babinet:2022nxg}), as
\begin{align}
\tilde{\cal I}^{(0)}_{\text{OSp},\gamma,\chi}[f,g;\nu]
&
=\frac{\mathfrak{t}^{-\nu(2\nu-1)}(-1)^{\frac{\nu(\nu-1)}{2}}}{\nu!}\int_0^1\prod_{i=1}^\nu d\alpha^{(1)}_i\nonumber \\
&\quad\times \operatorname{pf}\begin{pmatrix}
 [\langle\alpha^{(1)}_i|(-\hat{f}'\hat{R}\hat{V}\hat{f}'\hat{R})|\alpha^{(1)}_j\rangle]_{i,j}^{\nu\times \nu}
&[\langle\alpha^{(1)}_i|\hat{f}'\hat{R}\hat{V}|\alpha^{(1)}_j\rangle]_{i,j}^{\nu\times \nu}\vspace{0.2cm} \\
 [\langle\alpha^{(1)}_i|(-\hat{V}\hat{f}'\hat{R})|\alpha^{(1)}_j\rangle]_{i,j}^{\nu\times \nu}
&[\langle\alpha^{(1)}_i|\hat{V}|\alpha^{(1)}_j\rangle]_{i,j}^{\nu\times \nu}
\end{pmatrix},
\end{align}
with
\begin{align}
\hat{V}=\frac{1}{2}\hat{R}\mathfrak{t}^{-\hat{p}}\theta_\mathbb{Z}(-\hat{p})\hat{g}'\mathfrak{t}^{-\hat{p}}\theta_{\mathbb{Z}}(-\hat{p}).
\end{align}
Note that we have used the symmetry properties $f'(\sigma^{-1})=-f'(\sigma)$ and $g'(\sigma^{-1})=-g'(\sigma)$ of $f'$ and $g'$.

In analogy to the rearrangement of $\tilde{\cal I}_{\text{ABJM},k}[f,g;\nu]$ in section \ref{ABJMalgorithm}, the generating function of $\tilde{\cal I}^{(0)}_{\text{OSp},\gamma,\chi}[f,g;\nu]$ with the overall power of $\mathfrak{t}$ removed can be written by the Fredholm determinant by using the following Fredholm Pfaffian formula (see e.g.~\cite{Kubo:2024raz})
\begin{align}
\sum_{\nu=0}^\infty (-1)^{\frac{\nu(\nu-1)}{2}}\frac{u^\nu}{\nu!}\int d^\nu x\operatorname{pf}
\begin{pmatrix}
[X(x_i,x_j)]_{i,j}^{\nu\times \nu}&[Y(x_i,x_j)]_{i,j}^{\nu\times \nu}\vspace{0.2cm} \\
[Z(x_i,x_j)]_{i,j}^{\nu\times \nu}&[W(x_i,x_j)]_{i,j}^{\nu\times \nu}
\end{pmatrix}
=\sqrt{\operatorname{Det}\begin{pmatrix}
uX&1+uY\\
-1+uZ&uW
\end{pmatrix}
},
\end{align}
as
\begin{align}
\sum_{\nu=0}^\infty u^\nu\mathfrak{t}^{\nu(2\nu-1)}\tilde{\cal I}^{(0)}_{\text{OSp},\gamma,\chi}[f,g;\nu]
&=\sqrt{\operatorname{Det}\begin{pmatrix}
-u\hat{f}'\hat{R}\hat{V}\hat{f}'\hat{R}
&1+u\hat{f}'\hat{R}\hat{V}\\
-1-u\hat{V}\hat{f}'\hat{R}
&u\hat{V}
\end{pmatrix}
}\nonumber \\
&=\sqrt{\operatorname{Det}\begin{pmatrix}
1+u\hat{\rho}_{\text{OSp},\gamma,\chi}
\end{pmatrix}
},
\label{calI0OSpinFGstep3}
\end{align}
where the second line is obtained by performing the elementary row transformations to the $2\times 2$ determinant, with
\begin{align}
\hat{\rho}_{\text{OSp},\gamma,\chi}=2\hat{f}'\hat{R}\hat{V}=
\hat{f}'\mathfrak{t}^{-\hat{p}}\theta_{\mathbb{Z}}(-\hat{p})
\hat{g}'\mathfrak{t}^{-\hat{p}}\theta_{\mathbb{Z}}(-\hat{p}).
\end{align}

The final expression \eqref{calI0OSpinFGstep3} implies that we can calculate $\tilde{\cal I}^{(0)}_{\text{OSp},\gamma,\chi}[f,g;\nu]$ by calculating instead $\operatorname{tr}\hat{\rho}_{\text{OSp},\gamma,\chi}^n$ with $n=1,\cdots,\nu$.
Note that due to the factor $(\sigma-\sigma^{-1})^{-1}$ in $f'(\sigma)$, the calculations of $\operatorname{tr}\hat{\rho}_{\text{OSp},\gamma,\chi}^n$ are not as straightforward as those in the $U(N)_k\times U(N)_{-k}$ ABJM theory in section \ref{ABJMalgorithm} using the momentum basis.
One may instead calculate the traces by re-expressing the matrix elements in $\sigma=e^{2\pi i\alpha}$ and performing the circle integrations explicitly
\begin{align}
\operatorname{tr}\hat{\rho}_{\text{OSp},\gamma,\chi}^n=\oint\prod_{i=1}^n\frac{d\sigma_i}{2\pi i\sigma_i}\rho_{\text{OSp},\gamma,\chi}(\sigma_i,\sigma_{i+1}),\quad (\sigma_{n+1}=\sigma_1)
\label{trrhoOSp^nindsigma}
\end{align}
with ($\sigma_i=e^{2\pi i\alpha_i}$ and $\sigma=e^{2\pi i\alpha}$)
\begin{align}
&\rho_{\text{OSp},\gamma,\chi}(\sigma_1,\sigma_2)\nonumber \\
&=f'(\sigma_1)\int_0^1 d\alpha\langle \alpha_1|\mathfrak{t}^{-\hat{p}}\theta_{\mathbb{Z}}(-\hat{p})|\alpha\rangle g'(e^{2\pi i\alpha})\langle \alpha|\mathfrak{t}^{-\hat{p}}\theta_{\mathbb{Z}}(-\hat{p})|\alpha_2\rangle\nonumber \\
&=\frac{f(\sigma_1)\prod_\pm (1-\chi\sigma_1^{\pm 1})^\gamma}{\sigma_1-\sigma_1^{-1}}\oint\frac{d\sigma}{2\pi i\sigma}\frac{g(\sigma)(\sigma-\sigma^{-1})}{(1-\mathfrak{t}\sigma_1^{-1}\sigma)\prod_\pm (1-\chi\mathfrak{t}\sigma^{\pm 1})^\gamma (1-\mathfrak{t}\sigma^{-1}\sigma_2)}.
\label{rhoOSp(s1,s2)}
\end{align}
Here in \eqref{trrhoOSp^nindsigma} we include the residues at $\sigma_i=\pm 1$ with weight $\frac{1}{2}$.
When the second single variable function $g(\sigma)$ is not so complicated, one can obtain the closed form expression of $\rho_{\text{OSp},\gamma,\chi}(\sigma_1,\sigma_2)$ by performing the one dimensional integration \eqref{rhoOSp(s1,s2)}.
Then the only non-trivial task is to perform the $n$-dimensional integration with $n\le \nu$.
Hence the Fermi gas formalism indeed simplifies the calculation compared with the original fully mixed $2\nu$ dimensional integration \eqref{calI0OSp}.

\subsubsection{Fundamental Wilson lines}
\label{app_OSp2Nchi1}

The two-point function of fundamental Wilson lines in $SO(2N+\gamma)^{(\chi\zeta)}_{2k}\times USp(2N)_{-k}$ ABJ theory with $(\gamma,\chi)\neq (0,-1)$ can be calculated by choosing the single variable functions $f(s)$ and $g(s)$ as $f(s)=1+a(s+s^{-1})$ and $g(s)=1+b(s+s^{-1})$.
Namely, for $SO(2N+1)^{(\chi\zeta)}_{2k}\times USp(2N)_{-k}$ ABJ theory, since $\chi^{\mathfrak{so}(2N+1)}_{\ydiagram{1}}(s,\chi)=\chi+\sum_{i=1}^N(s_i+s_i^{-1})$ and $\chi^{\mathfrak{usp}(2N)}_{\ydiagram{1}}(s)=\sum_{i=1}^N(s_i+s_i^{-1})$, we can calculate the two-point function as
\begin{align}
&\langle W_{\ydiagram{1};\ydiagram{1}}\rangle^{SO(2N+1)_{2k}^{(\chi\zeta)}\times USp(2N)_{-k}\text{ABJ}(C)}(\mathfrak{t})\nonumber \\
&
=
\chi \langle \prod_i(1+a(s_i^{(1)}+(s_i^{(1)})^{-1}))(1+b(s_i^{(2)}+(s_i^{(2)})^{-1}))\rangle^{SO(2N+1)_{2k}^{(\chi\zeta)}\times USp(2N)_{-k}\text{ABJ}(C)}(\mathfrak{t})\biggr|_{a^0b}\nonumber \\
&\quad
+\langle \prod_i(1+a(s_i^{(1)}+(s_i^{(1)})^{-1}))(1+b(s_i^{(2)}+(s_i^{(2)})^{-1}))\rangle^{SO(2N+1)^{(\chi\zeta)}_{2k}\times USp(2N)_{-k}\text{ABJ}(C)}(\mathfrak{t})\biggr|_{ab},
\end{align}
while for $SO(2N)^{(+\zeta)}_{2k}\times USp(2N)_{-k}$ ABJ theory, since $\chi^{\mathfrak{so}(2N)}_{\ydiagram{1}}(s)=\sum_{i=1}^N(s_i+s_i^{-1})$, we can calculate the two-point function as
\begin{align}
&\langle W_{\ydiagram{1};\ydiagram{1}}\rangle^{SO(2N)^{(+\zeta)}_{2k}\times USp(2N)_{-k}\text{ABJ}(C)}(\mathfrak{t})\nonumber \\
&
=
\langle \prod_i(1+a(s_i^{(1)}+(s_i^{(1)})^{-1}))(1+b(s_i^{(2)}+(s_i^{(2)})^{-1}))\rangle^{SO(2N)^{(+\zeta)}_{2k}\times USp(2N)_{-k}\text{ABJ}(C)}(\mathfrak{t})\biggr|_{ab}.
\end{align}

For this choice of $f(s)$ and $g(s)$, we find
\begin{subequations}
\label{calIOSp(1)}
\begin{align}
&\tilde{\cal I}^{(0)}_{\text{OSp},1,\chi}[1+a(s+s^{-1});1+b(s+s^{-1});\nu]\nonumber \\
&=
\frac{
1
}{
(\mathfrak{t}^4;\mathfrak{t}^4)_\nu}\Bigl(1+
\chi a(-1+\mathfrak{t}^{2\nu})+\frac{ab\mathfrak{t}(1-\mathfrak{t}^{4\nu})}{1-\mathfrak{t}^2}\Bigr)
+{\cal O}(a^2,b^2),\\
&\tilde{\cal I}^{(0)}_{\text{OSp},0,1}[1+a(s+s^{-1});1+b(s+s^{-1});\nu]\nonumber \\
&=
\frac{
1 + \mathfrak{t}^{2\nu}
}{
(\mathfrak{t}^4;\mathfrak{t}^4)_\nu}\Bigl(1+
\frac{ab\mathfrak{t}(1+\mathfrak{t}^{2\nu-2})(1-\mathfrak{t}^{2\nu})}{1-\mathfrak{t}^2}\Bigr)
+{\cal O}(a^2,b^2),\\
&\tilde{\cal I}^{(1)}_{\text{OSp},m}[1+a(s+s^{-1});1+b(s+s^{-1});\nu]\nonumber \\
&=\frac{\mathfrak{t}^{2\nu|km|}}{(\mathfrak{t}^2;\mathfrak{t}^2)_\nu}\Bigl(1+ab\Bigl(\mathfrak{t}^{-1}+\mathfrak{t}^{2\nu-1}+\frac{2\mathfrak{t}(1-\mathfrak{t}^{2\nu-2})}{1-\mathfrak{t}^2}\Bigr)\Bigr)+{\cal O}(a^2,b^2),
\end{align}
\end{subequations}
where $\tilde{\cal I}^{(0)}_{\text{OSp},1,\chi}[f,g;\nu]$ and $\tilde{\cal I}^{(0)}_{\text{OSp},0,1}[f,g;\nu]$ for general $\nu$ are proposed based on the closed form expressions for $\nu=1,2$ and $3$.
With these results, we can calculate the grand canonical sum of the supersymmetric indices with the insertion of $\prod_if(s^{(1)}_i)g(s^{(2)}_i)$ \eqref{XiOSpgammachinot0-1} as
\begin{subequations}
\begin{align}
&\Xi_{\text{OSp},1,\chi}[1+a(s+s^{-1});1+b(s+s^{-1});u]\nonumber \\
&=\Xi_{\text{OSp},1,\chi}[1;1;u]\biggl[1+\chi a\Bigl(-1+\frac{(u;\mathfrak{t}^4)_\infty}{(u\mathfrak{t}^2;\mathfrak{t}^4)_\infty}\Bigr)+
\frac{abu(\mathfrak{t}+\zeta\mathfrak{t}^{2|k|-1})}{(1-\mathfrak{t}^2)(1-\zeta\mathfrak{t}^{2|k|})}
\biggr]+{\cal O}(a^2,b^2),\\
&\Xi_{\text{OSp},0,1}[1+a(s+s^{-1});1+b(s+s^{-1});u]\nonumber \\
&=\Xi_{\text{OSp},0,1}[1;1;u]\biggl[1+\frac{abu(\mathfrak{t}+\zeta\mathfrak{t}^{2|k|-1})}{(1-\mathfrak{t}^2)(1-\zeta\mathfrak{t}^{2|k|})}\biggr]+{\cal O}(a^2,b^2),
\end{align}
\end{subequations}
where
\begin{subequations}
\begin{align}
&\Xi_{\text{OSp},1,\chi}[1;1;u]=\frac{1}{(u;\mathfrak{t}^4)_\infty}\prod_{m=1}^\infty\frac{1}{(\mathfrak{t}^{2|km|}\zeta^mu;\mathfrak{t}^2)_\infty},\\
&\Xi_{\text{OSp},0,1}[1;1;u]=\Bigl(\frac{1}{(u;\mathfrak{t}^4)_\infty}+\frac{1}{(u\mathfrak{t}^2;\mathfrak{t}^4)_\infty}\Bigr)\prod_{m=1}^\infty\frac{1}{(\mathfrak{t}^{2|km|}\zeta^mu;\mathfrak{t}^2)_\infty}.
\end{align}
\end{subequations}
are the grand canonical sum of the Coulomb indices without line defect.
By extracting the expansion coefficients in $a$ and $b$, we find the following recursion relations as claimed in \eqref{SO2Nchi+recreln}
\begin{subequations}
\begin{align}
&\langle W_{\ydiagram{1};\ydiagram{1}}\rangle^{SO(2N+1)^{(\chi\zeta)}_{2k}\times USp(2N)_{-k}\text{ABJ}(C)}(\mathfrak{t})\nonumber \\
&={\cal I}^{SO(2N-1)^{(\chi\zeta)}_{2k}\times USp(2N-2)_{-k}\text{ABJ}(C)}(\mathfrak{t})
\langle W_{\ydiagram{1};\ydiagram{1}}\rangle^{SO(3)^{(\chi\zeta)}_{2k}\times USp(2)_{-k}\text{ABJ}(C)}(\mathfrak{t}),\\
&\langle W_{\ydiagram{1};\ydiagram{1}}\rangle^{SO(2N)^{(+\zeta)}_{2k}\times USp(2N)_{-k}\text{ABJ}(C)}(\mathfrak{t})\nonumber \\
&={\cal I}^{SO(2N-2)^{(+\zeta)}_{2k}\times USp(2N-2)_{-k}\text{ABJ}(C)}(\mathfrak{t})
\langle W_{\ydiagram{1};\ydiagram{1}}\rangle^{SO(3)^{(\chi\zeta)}_{2k}\times USp(2)_{-k}\text{ABJ}(C)}(\mathfrak{t})
\label{OSp2Nchi1recreln}
\end{align}
\end{subequations}
with
\begin{align}
\langle W_{\ydiagram{1};\ydiagram{1}}\rangle^{SO(3)^{(\chi\zeta)}_{2k}\times USp(2)_{-k}\text{ABJ}(C)}(\mathfrak{t})
=\frac{\mathfrak{t}+\mathfrak{t}^{2k-1}\zeta}{(1-\mathfrak{t}^2)(1-\zeta\mathfrak{t}^{2k})}.
\label{WWOSp32}
\end{align}
Here we have also defined
\begin{subequations}
\begin{align}
&{\cal I}^{SO(1)^{(\chi\zeta)}_{2k}\times USp(0)_{{-k}}\text{ABJ}(C)}(\mathfrak{t})=1,\\
&{\cal I}^{SO(0)^{(+\zeta)}_{2k}\times USp(0)_{{-k}}\text{ABJ}(C)}(\mathfrak{t})=2.
\end{align}
\end{subequations}

In particular, we find
\begin{subequations}
\label{OSpABJt1limits}
\begin{align}
&\lim_{\mathfrak{t}\rightarrow 1}
\langle {\cal W}_{\ydiagram{1};\ydiagram{1}}\rangle^{SO(2N+1)^{(\chi+)}_{2k}\times USp(2N)_{-k}\text{ABJ}(C)}(\mathfrak{t})
=
\lim_{\mathfrak{t}\rightarrow 1}
\langle {\cal W}_{\ydiagram{1};\ydiagram{1}}\rangle^{Spin(2N+1)^{(\chi)}_{2k}\times USp(2N)_{-k}\text{ABJ}(C)}(\mathfrak{t})
\nonumber \\
&=2N,\\
&\lim_{\mathfrak{t}\rightarrow 1}
\langle {\cal W}_{\ydiagram{1};\ydiagram{1}}\rangle^{SO(2N+1)^{(\chi-)}_{2k}\times USp(2N)_{-k}\text{ABJ}(C)}(\mathfrak{t})
=
\begin{cases}
\displaystyle 0&\quad (N\text{: even})\vspace{0.2cm}\\
\displaystyle \frac{2(k-1)}{k+1}&\quad (N\text{: odd})
\end{cases},\\
&\lim_{\mathfrak{t}\rightarrow 1}
\langle {\cal W}_{\ydiagram{1};\ydiagram{1}}\rangle^{SO(2N)^{(++)}_{2k}\times USp(2N)_{-k}\text{ABJ}(C)}(\mathfrak{t})
=
\lim_{\mathfrak{t}\rightarrow 1}
\langle {\cal W}_{\ydiagram{1};\ydiagram{1}}\rangle^{Spin(2N)^{(+)}_{2k}\times USp(2N)_{-k}\text{ABJ}(C)}(\mathfrak{t})
\nonumber \\
&=2N,\\
&\lim_{\mathfrak{t}\rightarrow 1}
\langle {\cal W}_{\ydiagram{1};\ydiagram{1}}\rangle^{SO(2N)^{(+-)}_{2k}\times USp(2N)_{-k}\text{ABJ}(C)}(\mathfrak{t})
=
\begin{cases}
\displaystyle 0&\quad (N\text{: even})\vspace{0.2cm}\\
\displaystyle \frac{2(k-1)}{k}&\quad (N\text{: odd})
\end{cases}.
\end{align}
\end{subequations}
For $\zeta=1$ and the theory with $\mathbb{Z}^{{\cal M}}_2$ gauged, the results coincide with the dimension of the fundamental representation in $USp(2N)$.
We also find
\begin{subequations}
\begin{align}
&\langle {\cal W}_{\ydiagram{1};\ydiagram{1}}\rangle^{SO(2N+1)^{(\chi\zeta)}_{2k}\times USp(2N)_{-k}\text{ABJ}(C)}(\mathfrak{t}^{-1})
=\langle {\cal W}_{\ydiagram{1};\ydiagram{1}}\rangle^{SO(2N+1)^{(\chi\zeta)}_{2k}\times USp(2N)_{-k}\text{ABJ}(C)}(\mathfrak{t}),\\
&\langle {\cal W}_{\ydiagram{1};\ydiagram{1}}\rangle^{SO(2N)^{(+\zeta)}_{2k}\times USp(2N)_{-k}\text{ABJ}(C)}(\mathfrak{t}^{-1})
=\langle {\cal W}_{\ydiagram{1};\ydiagram{1}}\rangle^{SO(2N)^{(+\zeta)}_{2k}\times USp(2N)_{-k}\text{ABJ}(C)}(\mathfrak{t}).
\end{align}
\end{subequations}

\subsubsection{Other representations}
\label{app_OSp_otherreps}

The Coulomb limit of \eqref{abj_index1} can be evaluated for an arbitrary insertion, and this provides a further test of the general formula \eqref{Witten_ind_unified}.
Recall from section \ref{sec_CSmatter} that a generic point of the moduli space of these theories leaves the diagonal torus $H_0=U(1)^N_{\text{diag}}$ unbroken, so that the prediction for a set of Wilson lines in the representations $\mathcal{R}_j$ of $SO(2N+\gamma)\times USp(2N)$ is
\begin{align}
\label{OSpABJprediction}
\lim_{\mathfrak{t}\rightarrow1}\langle \mathcal{W}_{\mathcal{R}_1}\cdots\mathcal{W}_{\mathcal{R}_k}\rangle^{SO(2N+\gamma)^{(\chi+)}_{2k}\times USp(2N)_{-k}\text{ABJ}(C)}(\mathfrak{t})
=\oint\prod_{i=1}^{N}\frac{ds_i}{2\pi is_i}\prod_{j=1}^{k}\chi_{\mathcal{R}_j}(s)\,,
\end{align}
where the characters are restricted to $H_0$, that is the characters of $SO(2N+\gamma)$ are evaluated on the holonomy with eigenvalues $\{s_i^{\pm1}\}\cup\{1\}^{\gamma}$ and those of $USp(2N)$ on $\{s_i^{\pm1}\}$, with the same $s_i$ on both nodes.
Table \ref{tab_OSpABJreps} collects the results, mostly for $N=1$, that is for the $SO(3)_{2k}\times USp(2)_{-k}$ and $SO(2)_{2k}\times USp(2)_{-k}$ theories, with $k=2,3$ and $\zeta=1$.
The last line is a rank two example.
Here $(m)$ denotes the traceless symmetric power of the vector representation on the orthogonal node and the symmetric power of the fundamental representation on the symplectic node, and $\mathbf{1}$ denotes the trivial representation.
In every case \eqref{OSpABJprediction} is reproduced.

\begin{table}[htbp]
\centering
\begin{tabular}{l|cc|cc}
\hline
&\multicolumn{2}{c|}{$SO(3)_{2k}\times USp(2)_{-k}$}&\multicolumn{2}{c}{$SO(2)_{2k}\times USp(2)_{-k}$}\\
Wilson lines & $\lim_{\mathfrak{t}\rightarrow1}\langle\mathcal{W}\cdots\rangle^{(C)}$ & $\oint\prod_{i=1}^{N}\frac{ds_i}{2\pi is_i}\prod_j\chi_{\mathcal{R}_j}$ & $\lim_{\mathfrak{t}\rightarrow1}\langle\mathcal{W}\cdots\rangle^{(C)}$ & $\oint\prod_{i=1}^{N}\frac{ds_i}{2\pi is_i}\prod_j\chi_{\mathcal{R}_j}$ \\ \hline
$\mathcal{W}_{\ydiagram{1};\ydiagram{1}}$ & $2$ & $2$ & $2$ & $2$\\
$\mathcal{W}_{\ydiagram{1};\mathbf{1}}\mathcal{W}_{\ydiagram{1};\mathbf{1}}$ & $3$ & $3$ & $2$ & $2$\\
$\mathcal{W}_{\mathbf{1};\ydiagram{1}}\mathcal{W}_{\mathbf{1};\ydiagram{1}}$ & $2$ & $2$ & $2$ & $2$\\
$\mathcal{W}_{\ydiagram{2};\mathbf{1}}$ & $1$ & $1$ & $0$ & $0$\\
$\mathcal{W}_{\mathbf{1};\ydiagram{2}}$ & $1$ & $1$ & $1$ & $1$\\
$\mathcal{W}_{\ydiagram{2};\ydiagram{2}}$ & $3$ & $3$ & $2$ & $2$\\
$\mathcal{W}_{\ydiagram{3};\ydiagram{3}}$ & $4$ & $4$ & $2$ & $2$\\
$\mathcal{W}_{\ydiagram{3};\ydiagram{1}}$ & $2$ & $2$ & $0$ & $0$\\
$\mathcal{W}_{\ydiagram{1};\ydiagram{1}}\mathcal{W}_{\ydiagram{1};\ydiagram{1}}$ & $8$ & $8$ & $6$ & $6$\\ \hline
\multicolumn{5}{l}{$N=2$, that is $SO(5)_{2k}\times USp(4)_{-k}$ and $SO(4)_{2k}\times USp(4)_{-k}$}\\ \hline
$\mathcal{W}_{\mathbf{1};\ydiagram{2}}$ & $2$ & $2$ & --- & ---\\ \hline
\end{tabular}
\caption{The $\mathfrak{t}\rightarrow1$ limit of the normalized Coulomb line defect indices of the orthosymplectic ABJ theories at $\zeta=1$, compared with the character integral \eqref{OSpABJprediction} over the unbroken diagonal torus $H_0=U(1)^N_{\text{diag}}$. The upper block is $N=1$ and the last line is a $\gamma=1$ example at $N=2$. The values are independent of $k$, and were obtained for $k=2$ and $k=3$.}
\label{tab_OSpABJreps}
\end{table}

These entries are sensitive to the precise identification of $H_0$.
Had the whole $USp(2)$ survived, that is $H_0=SU(2)$ embedded diagonally, the character average would give $0$ for $\mathcal{W}_{\mathbf{1};\ydiagram{2}}$ and $1$ for $\mathcal{W}_{\ydiagram{1};\mathbf{1}}\mathcal{W}_{\ydiagram{1};\mathbf{1}}$ instead of $1$ and $3$.
Had the two nodes retained independent tori, $H_0=U(1)\times U(1)$, one would find $0$ instead of $2$ for the bifundamental Wilson line.
Both alternatives are therefore excluded, and the diagonal identification of the two Cartan subalgebras, which is what the $N$ M2-branes impose, is essential.

\subsection{$SO(2N)_{2k}^{(-\zeta)}\times USp(2N)_{-k}$ ABJ theory}
\label{app_OSp_SO2Nchi-1}
Let us consider the $SO(2N)_{2k}\times USp(2N)_{-k}$ ABJ theory with $\chi=-1$.
In this case, we can set the last two components of the $SO(2N)$ holonomy to $(1,-1)$ and the last component of the magnetic flux to $m_{N}=0$.
As a result, the supersymmetric index with the insertion of one variable function $\prod_{i=1}^{N-1}f(s_i^{(1)})\prod_{i=1}^Ng(s^{(2)}_i)$ is given by \cite{Hayashi:2022ldo}
\begin{align}
&\biggl\langle\prod_{i=1}^{N-1}f(s^{(1)}_i)\prod_{i=1}^Ng(s^{(2)}_i)\biggr\rangle^{SO(2N)_{2k}^{(-\zeta)}\times USp(2N)_{-k}\text{ABJ}}(t;q)
\nonumber\\
&
=\frac{1}{2^{2N-1}(N-1)!N!}
\sum_{m^{(1)}_1,\cdots,m^{(1)}_{N-1}, m^{(2)}_1,\cdots,m^{(2)}_N \in \mathbb{Z}}
\oint \prod_{i=1}^{N-1} \frac{ds^{(1)}_{i}}{2\pi is^{(1)}_i}(s_i^{(1)})^{2km^{(1)}_i}f(s^{(1)}_i)\nonumber \\
&\quad\times
\oint \prod_{i=1}^N \frac{ds^{(2)}_{i}}{2\pi is^{(2)}_i}(s_i^{(2)})^{-2km^{(2)}_i}g(s^{(2)}_i)
\nonumber\\
&\quad\times
\prod_{i=1}^{N-1} \prod_\pm \Bigl(1-q^{|m_i^{(1)}|} (s_i^{(1)})^{\pm 2}\Bigr)
\prod_{i=1}^{N} \prod_\pm \Bigl(1-q^{|m_i^{(2)}|} (s_i^{(2)})^{\pm 2}\Bigr)\nonumber \\
&\quad\times
 \prod_{i<j}^{N-1}\prod_\pm
\Bigl(1-q^{\frac{|m^{(1)}_i-m^{(1)}_j|}{2}}(s_i^{(1)})^{\pm 1}(s_j^{(1)})^{\mp 1}\Bigr)
\Bigl(1-q^{\frac{|m^{(1)}_i+m^{(1)}_j|}{2}}(s_i^{(1)})^{\pm 1} (s_j^{(1)})^{\pm 1}\Bigr)\nonumber \\
&\quad\times
\prod_{i<j}^{N}\prod_\pm
\Bigl(1-q^{\frac{|m^{(2)}_i-m^{(2)}_j|}{2}}(s_i^{(2)})^{\pm 1}(s_j^{(2)})^{\mp 1}\Bigr)
\Bigl(1-q^{\frac{|m^{(2)}_i+m^{(2)}_j|}{2}}(s_i^{(2)})^{\pm 1} (s_j^{(2)})^{\pm 1}\Bigr)\nonumber \\
&\quad\times
\prod_{i=1}^{N-1}\prod_{j=1}^N\prod_\pm
\frac{
(q^{\frac34+\frac{|m_i^{(1)}-m_j^{(2)}|}{2}}t^{-1}(s_i^{(1)})^{\mp 1} (s_j^{(2)})^{\pm 1};q)_{\infty}
}{
(q^{\frac14+\frac{|m_i^{(1)}-m_j^{(2)}|}{2}}t (s_i^{(1)})^{\pm 1} (s_j^{(2)})^{\mp 1};q)_{\infty}
}\nonumber \\
&\quad\times
\prod_{i=1}^{N-1}\prod_{j=1}^N\prod_\pm
\frac{
(q^{\frac34+\frac{|m_i^{(1)}+m_j^{(2)}|}{2}}t^{-1}(s_i^{(1)})^{\mp 1} (s_j^{(2)})^{\mp 1};q)_{\infty}
}{
(q^{\frac14+\frac{|m_i^{(1)}+m_j^{(2)}|}{2}}t     (s_i^{(1)})^{\pm 1} (s_j^{(2)})^{\pm 1};q)_{\infty}
}\nonumber\\
&\quad\times
\prod_{i=1}^{N}\prod_{j=1}^{N-1}\prod_\pm
\frac{(q^{\frac34+\frac{|m_i^{(2)} - m_j^{(1)}|}{2}}t (s_i^{(2)})^{\mp 1} (s_j^{(1)})^{\pm 1};q)_{\infty}}
{(q^{\frac14+\frac{|m_i^{(2)} - m_j^{(1)}|}{2}}t^{-1} (s_i^{(2)})^{\pm 1} (s_j^{(1)})^{\mp 1};q)_{\infty}}\nonumber \\
&\quad\times
\prod_{i=1}^{N}\prod_{j=1}^{N-1}\prod_\pm
\frac{(q^{\frac34+\frac{|m_i^{(2)} + m_j^{(1)}|}{2}}t (s_i^{(2)})^{\mp 1} (s_j^{(1)})^{\mp 1};q)_{\infty}}
{(q^{\frac14+\frac{|m_i^{(2)} + m_j^{(1)}|}{2}}t^{-1} (s_i^{(2)})^{\pm 1} (s_j^{(1)})^{\pm 1};q)_{\infty}}
\nonumber\\
&\quad\times
\prod_{i=1}^N\prod_\pm
\frac{(\mp q^{\frac34+\frac{|m_i^{(2)}|}{2}}t^{-1}(s_i^{(2)})^{\pm 1};q)_{\infty}}
{(\pm q^{\frac14+\frac{|m_i^{(2)}|}{2}}t          (s_i^{(2)})^{\mp 1};q)_{\infty}}
\frac{(\pm q^{\frac34+\frac{|m_i^{(2)}|}{2}}t^{-1}(s_i^{(2)})^{\pm 1};q)_{\infty}}
{(\mp q^{\frac14+\frac{|m_i^{(2)}|}{2}}t          (s_i^{(2)})^{\mp 1};q)_{\infty}}
\nonumber\\
&\quad\times
\prod_{i=1}^N\prod_\pm
\frac{(\pm q^{\frac34+\frac{|m_i^{(2)}|}{2}}t (s_i^{(2)})^{\mp 1};q)_{\infty}}
{(\mp q^{\frac14+\frac{|m_i^{(2)}|}{2}}t^{-1} (s_i^{(2)})^{\pm 1};q)_{\infty}}
\frac{(\mp q^{\frac34+\frac{|m_i^{(2)}|}{2}}t (s_i^{(2)})^{\mp 1};q)_{\infty}}
{(\pm q^{\frac14+\frac{|m_i^{(2)}|}{2}}t^{-1} (s_i^{(2)})^{\pm 1};q)_{\infty}}
\nonumber\\
&\quad\times
q^{-\sum_{i=1}^{N-1}|m_i^{(1)}|
-\frac12\sum_{i<j}^{N-1}\sum_\pm
|m_{i}^{(1)}\pm m_{j}^{(1)}|
-\frac12\sum_{i<j}^N\sum_\pm
|m_{i}^{(2)}\pm m_{j}^{(2)}|
+\frac12 \sum_{i=1}^{N-1}\sum_{j=1}^N\sum_\pm
|m^{(1)}_i\pm m^{(2)}_j|
}
\nonumber\\
&\quad\times
\zeta^{\sum_{i=1}^{N-1} m_i^{(1)}}.
\label{abj_index1p}
\end{align}
If we assume $f(s)=f(s^{-1})$ and $g(s)=g(s^{-1})$, we can restrict the monopole charges to $\mathbb{Z}_{\ge 0}$ rather than $\mathbb{Z}$, with weight $1$ for $m^{(I)}_i=0$ and weight $2$ for $m^{(I)}_i> 0$, which one may write $2^{1-\delta_{m^{(I)}_i,0}}$.
Hence we have
\begin{align}
&\biggl\langle\prod_{i=1}^{N-1}f(s^{(1)}_i)\prod_{i=1}^Ng(s^{(2)}_i)\biggr\rangle^{SO(2N)_{2k}^{(-\zeta)}\times USp(2N)_{-k}\text{ABJ}}(t;q)
\nonumber\\
&
=\frac{1}{2^{2N-1}(N-1)!N!}
\sum_{m^{(1)}_1,\cdots,m^{(1)}_{N-1}, m^{(2)}_1,\cdots,m^{(2)}_N \in \mathbb{Z}_{\ge 0}}
\prod_{i=1}^{N-1}2^{1-\delta_{m^{(1)}_i,0}}
\prod_{i=1}^{N}2^{1-\delta_{m^{(2)}_i,0}}\nonumber \\
&\quad\times
\oint \prod_{i=1}^{N-1} \frac{ds^{(1)}_{i}}{2\pi is^{(1)}_i}(s_i^{(1)})^{2km^{(1)}_i}f(s^{(1)}_i)
\oint \prod_{i=1}^N \frac{ds^{(2)}_{i}}{2\pi is^{(2)}_i}(s_i^{(2)})^{-2km^{(2)}_i}   g(s^{(2)}_i)
\nonumber\\
&\quad\times
\prod_{i=1}^{N-1} \prod_\pm \Bigl(1-q^{|m_i^{(1)}|} (s_i^{(1)})^{\pm 2}\Bigr)
\prod_{i=1}^{N} \prod_\pm \Bigl(1-q^{|m_i^{(2)}|} (s_i^{(2)})^{\pm 2}\Bigr)\nonumber \\
&\quad\times
 \prod_{i<j}^{N-1}\prod_\pm
\Bigl(1-q^{\frac{|m^{(1)}_i-m^{(1)}_j|}{2}}(s_i^{(1)})^{\pm 1}(s_j^{(1)})^{\mp 1}\Bigr)
\Bigl(1-q^{\frac{|m^{(1)}_i+m^{(1)}_j|}{2}}(s_i^{(1)})^{\pm 1} (s_j^{(1)})^{\pm 1}\Bigr)\nonumber \\
&\quad\times
\prod_{i<j}^{N}\prod_\pm
\Bigl(1-q^{\frac{|m^{(2)}_i-m^{(2)}_j|}{2}}(s_i^{(2)})^{\pm 1}(s_j^{(2)})^{\mp 1}\Bigr)
\Bigl(1-q^{\frac{|m^{(2)}_i+m^{(2)}_j|}{2}}(s_i^{(2)})^{\pm 1} (s_j^{(2)})^{\pm 1}\Bigr)\nonumber \\
&\quad\times
\prod_{i=1}^{N-1}\prod_{j=1}^N\prod_\pm
\frac{
(q^{\frac34+\frac{|m_i^{(1)}-m_j^{(2)}|}{2}}t^{-1}(s_i^{(1)})^{\mp 1} (s_j^{(2)})^{\pm 1};q)_{\infty}
}{
(q^{\frac14+\frac{|m_i^{(1)}-m_j^{(2)}|}{2}}t (s_i^{(1)})^{\pm 1} (s_j^{(2)})^{\mp 1};q)_{\infty}
}\nonumber \\
&\quad\times
\prod_{i=1}^{N-1}\prod_{j=1}^N\prod_\pm
\frac{
(q^{\frac34+\frac{|m_i^{(1)}+m_j^{(2)}|}{2}}t^{-1}(s_i^{(1)})^{\mp 1} (s_j^{(2)})^{\mp 1};q)_{\infty}
}{
(q^{\frac14+\frac{|m_i^{(1)}+m_j^{(2)}|}{2}}t     (s_i^{(1)})^{\pm 1} (s_j^{(2)})^{\pm 1};q)_{\infty}
}\nonumber\\
&\quad\times
\prod_{i=1}^{N}\prod_{j=1}^{N-1}\prod_\pm
\frac{(q^{\frac34+\frac{|m_i^{(2)} - m_j^{(1)}|}{2}}t (s_i^{(2)})^{\mp 1} (s_j^{(1)})^{\pm 1};q)_{\infty}}
{(q^{\frac14+\frac{|m_i^{(2)} - m_j^{(1)}|}{2}}t^{-1} (s_i^{(2)})^{\pm 1} (s_j^{(1)})^{\mp 1};q)_{\infty}}\nonumber \\
&\quad\times
\prod_{i=1}^{N}\prod_{j=1}^{N-1}\prod_\pm
\frac{(q^{\frac34+\frac{|m_i^{(2)} + m_j^{(1)}|}{2}}t (s_i^{(2)})^{\mp 1} (s_j^{(1)})^{\mp 1};q)_{\infty}}
{(q^{\frac14+\frac{|m_i^{(2)} + m_j^{(1)}|}{2}}t^{-1} (s_i^{(2)})^{\pm 1} (s_j^{(1)})^{\pm 1};q)_{\infty}}
\nonumber\\
&\quad\times
\prod_{i=1}^N\prod_\pm
\frac{(\mp q^{\frac34+\frac{|m_i^{(2)}|}{2}}t^{-1}(s_i^{(2)})^{\pm 1};q)_{\infty}}
{(\pm q^{\frac14+\frac{|m_i^{(2)}|}{2}}t          (s_i^{(2)})^{\mp 1};q)_{\infty}}
\frac{(\pm q^{\frac34+\frac{|m_i^{(2)}|}{2}}t^{-1}(s_i^{(2)})^{\pm 1};q)_{\infty}}
{(\mp q^{\frac14+\frac{|m_i^{(2)}|}{2}}t          (s_i^{(2)})^{\mp 1};q)_{\infty}}
\nonumber\\
&\quad\times
\prod_{i=1}^N\prod_\pm
\frac{(\pm q^{\frac34+\frac{|m_i^{(2)}|}{2}}t (s_i^{(2)})^{\mp 1};q)_{\infty}}
{(\mp q^{\frac14+\frac{|m_i^{(2)}|}{2}}t^{-1} (s_i^{(2)})^{\pm 1};q)_{\infty}}
\frac{(\mp q^{\frac34+\frac{|m_i^{(2)}|}{2}}t (s_i^{(2)})^{\mp 1};q)_{\infty}}
{(\pm q^{\frac14+\frac{|m_i^{(2)}|}{2}}t^{-1} (s_i^{(2)})^{\pm 1};q)_{\infty}}
\nonumber\\
&\quad\times
q^{-\sum_{i=1}^{N-1}|m_i^{(1)}|
-\frac12\sum_{i<j}^{N-1}\sum_\pm
|m_{i}^{(1)}\pm m_{j}^{(1)}|
-\frac12\sum_{i<j}^N\sum_\pm
|m_{i}^{(2)}\pm m_{j}^{(2)}|
+\frac12 \sum_{i=1}^{N-1}\sum_{j=1}^N\sum_\pm
|m^{(1)}_i\pm m^{(2)}_j|
}
\nonumber\\
&\quad\times
\zeta^{\sum_{i=1}^{N-1} m_i^{(1)}}.
\label{abj_index2}
\end{align}

Note that in this model the number of monopole charges are different for $SO(2N)^{(-\zeta)}_{2k}$ and $USp(2N)_{-k}$, which makes it difficult to generalize systematic classification of the monopole charges contributing in the Coulomb limit in the $U(N)_k\times U(N)_{-k}$ ABJM theory and $SO(2N+\gamma)^{(\chi\zeta)}_{2k}\times USp(2N)_{-k}$ ABJ theory with $(\gamma,\chi)\neq (0,-1)$.
Nevertheless, we observe that the power of $q$ in the overall factor in the line defect index \eqref{abj_index2}
\begin{align}
&-\sum_{i=1}^{N-1}|m_i^{(1)}|
-\frac12\sum_{i<j}^{N-1}\sum_\pm
|m_{i}^{(1)}\pm m_{j}^{(1)}|
-\frac12\sum_{i<j}^N\sum_\pm
|m_{i}^{(2)}\pm m_{j}^{(2)}|\nonumber \\
&\quad
+\frac12 \sum_{i=1}^{N-1}\sum_{j=1}^N\sum_\pm
|m^{(1)}_i\pm m^{(2)}_j|
\end{align}
is positive semi-definite, which vanishes if and only if
\begin{align}
0\le m^{(2)}_{1}\le m^{(1)}_{1}\le m^{(2)}_{2}\le m^{(1)}_{2}\le \cdots\le m^{(1)}_{N-1}\le m^{(2)}_{N}
\label{OSp2Nchi-1Cmonopolerestriction}
\end{align}
after the appropriate permutations of $\{m^{(1)}_i\}_{i=1}^{N-1}$ and $\{m^{(2)}_i\}_{i=1}^N$.
Therefore, the line defect index \eqref{abj_index2} reduces in the Coulomb limit as
\begin{align}
&\biggl\langle\prod_{i=1}^{N-1}f(s^{(1)}_i)\prod_{i=1}^Ng(s^{(2)}_i)\biggr\rangle^{SO(2N)_{2k}^{(-\zeta)}\times USp(2N)_{-k}\text{ABJ}(C)}(\mathfrak{t})
\nonumber\\
&
=\frac{1}{2^{2N-1}(N-1)!N!}
\sum_{\substack{m^{(1)}_1,\cdots,m^{(1)}_{N-1}, m^{(2)}_1,\cdots,m^{(2)}_N \in \mathbb{Z}_{\ge 0}\\ (\text{satisfying }\eqref{OSp2Nchi-1Cmonopolerestriction})}}
\prod_{i=1}^{N-1}2^{1-\delta_{m^{(1)}_i,0}}
\prod_{i=1}^{N}2^{1-\delta_{m^{(2)}_i,0}}\nonumber \\
&\quad\times
\oint \prod_{i=1}^{N-1} \frac{ds^{(1)}_{i}}{2\pi is^{(1)}_i}(s_i^{(1)})^{2km^{(1)}_i} f(s^{(1)}_i)
\oint \prod_{i=1}^N \frac{ds^{(2)}_{i}}{2\pi is^{(2)}_i}(s_i^{(2)})^{-2km^{(2)}_i}    g(s^{(2)}_i)
\nonumber\\
&\quad\times
\prod_{\substack{i=1\\ (m^{(1)}_i=0)}}^{N-1} \prod_\pm (1-(s_i^{(1)})^{\pm 2})
\prod_{\substack{i=1\\ (m^{(2)}_i=0)}}^{N} \prod_\pm
\frac{
1-(s_i^{(2)})^{\pm 2}
}{
1- \mathfrak{t}^2 (s_i^{(2)})^{\pm 2}
}\nonumber\\
&\quad\times \prod_{\substack{i<j\\ (m^{(1)}_i=m^{(1)}_j)}}^{N-1}\prod_\pm \Bigl(1-(s_i^{(1)})^{\pm 1}(s_j^{(1)})^{\mp 1}\Bigr)
\prod_{\substack{i<j\\ (m^{(1)}_i=m^{(1)}_j=0)}}^{N-1}\prod_\pm \Bigl(1-(s_i^{(1)})^{\pm 1} (s_j^{(1)})^{\pm 1}\Bigr)\nonumber \\
&\quad\times \prod_{\substack{i<j\\ (m^{(2)}_i=m^{(2)}_j)}}^{N}\prod_\pm \Bigl(1-(s_i^{(2)})^{\pm 1}(s_j^{(2)})^{\mp 1}\Bigr)
\prod_{\substack{i<j\\ (m^{(2)}_i=m^{(2)}_j=0)}}^{N}\prod_\pm \Bigl(1-(s_i^{(2)})^{\pm 1} (s_j^{(2)})^{\pm 1}\Bigr)\nonumber \\
&\quad\times\mathop{\prod_{i=1}^{N}\prod_{j=1}^{N-1}}_{(m^{(1)}_i=m^{(2)}_j)}\prod_\pm \frac{1}{1-\mathfrak{t} (s_i^{(2)})^{\pm 1} (s_j^{(1)})^{\mp 1}}
\mathop{\prod_{i=1}^{N}\prod_{j=1}^{N-1}}_{(m^{(1)}_i=m^{(2)}_j=0)}\prod_\pm \frac{1}{1-\mathfrak{t} (s_i^{(2)})^{\pm 1} (s_j^{(1)})^{\pm 1}}\nonumber\\
&\quad\times
\zeta^{\sum_{i=1}^{N-1} m_i^{(1)}}.
\end{align}

Note that the solutions of the condition \eqref{OSp2Nchi-1Cmonopolerestriction} can be classified by the locations where the inequalities ($\le$) are saturated ($=$), which we shall call ``clustering pattern''.
For example, for $N=2$ we have the following clustering patterns:
\begin{align}
\begin{array}{|c|c|}
\hline
\text{inequality}&\text{clustering pattern }((\nu_0,\nu'_0);(\nu_{m_1},\nu'_{m_1}),(\nu_{m_2},\nu'_{m_2}),\cdots)\\ \hline
0=m^{(2)}_1=m^{(1)}_1=m^{(2)}_2&((1,2);\emptyset)\\ \hline
0=m^{(2)}_1=m^{(1)}_1<m^{(2)}_2&((1,1);(0,1))\\ \hline
0=m^{(2)}_1<m^{(1)}_1=m^{(2)}_2&((0,1);(1,1))\\ \hline
0=m^{(2)}_1<m^{(1)}_1<m^{(2)}_2&((0,1);(1,0),(0,1))\\ \hline
0<m^{(2)}_1=m^{(1)}_1=m^{(2)}_2&((0,0);(1,2))\\ \hline
0<m^{(2)}_1<m^{(1)}_1=m^{(2)}_2&((0,0);(0,1),(1,1))\\ \hline
0<m^{(2)}_1=m^{(1)}_1<m^{(2)}_2&((0,0);(1,1),(0,1))\\ \hline
0<m^{(2)}_1<m^{(1)}_1<m^{(2)}_2&((0,0);(0,1),(1,0),(0,1))\\ \hline
\end{array}
\end{align}
where $m_1,m_2,\cdots>0$, and $\nu_0$ or $\nu_m$ (resp.~$\nu_0'$ or $\nu_m'$) are the number of monopole charges satisfying $m^{(1)}_i=0$ or $m^{(1)}_i=m>0$ (resp.~$m^{(2)}_i=0$ or $m^{(2)}_i=m>0$), and hence belonging to the same cluster delimited by the strict inequality ($<$).
Note that by definition $\nu_0'$ is either $\nu_0$ or $\nu_0+1$, and each $\nu_m'$ is either $\nu_m$, $\nu_m+1$ or $\nu_m-1$.
By reorganizing the summation over the monopole charges in the Coulomb line defect index \eqref{abj_index2} into the summation over the clustering patterns and the extra degrees of freedom $m_1,m_2,\cdots$ ($0<m_1<m_2<\cdots$), we obtain
\begin{align}
&\biggl\langle\prod_{i=1}^{N-1}f(s^{(1)}_i)\prod_{i=1}^Ng(s^{(2)}_i)\biggr\rangle^{SO(2N)_{2k}^{(-\zeta)}\times USp(2N)_{-k}\text{ABJ}(C)}(\mathfrak{t})
\nonumber\\
&
=\sum_{\text{clustering pattern}}
{\cal I}^{(0)}_{\text{OSp},0,-1}[f,g;\nu_0,\nu_0']
\sum_{0<m_1<m_2<\cdots}
\zeta^{\sum_{i\ge 1}m_i\nu_{m_i}}\prod_{i\ge 1}{\cal I}_{\text{OSp},km_i}^{(1)}[f,g;\nu_{m_i},\nu'_{m_i}],
\label{OSp2Nchi-1intildecalI}
\end{align}
where
\begin{subequations}
\begin{align}
&{\cal I}^{(0)}_{\text{OSp},0,-1}[f,g;\nu,\nu']\nonumber \\
&=\frac{1}{2^{\nu+\nu'}\cdot \nu!\nu'!}
\prod_{i=1}^{\nu}\oint\frac{d\sigma^{(1)}_i}{2\pi i\sigma^{(1)}_i}f(\sigma^{(1)}_i)
\prod_{i=1}^{\nu'}\oint\frac{d\sigma^{(2)}_i}{2\pi i\sigma^{(2)}_i}g(\sigma^{(2)}_i)
\prod_{i=1}^\nu\prod_\pm(1-(\sigma^{(1)}_i)^{\pm 2})\nonumber \\
&\quad\times
\prod_{i=1}^{\nu'}\prod_\pm\frac{1-(\sigma^{(2)}_i)^{\pm 2}}{1-\mathfrak{t}^2(\sigma^{(2)}_i)^{\pm 2}}
\prod_{i<j}^{\nu}\prod_\pm
(1-(\sigma^{(1)}_i)^{\pm 1}(\sigma^{(1)}_j)^{\mp 1})
(1-(\sigma^{(1)}_i)^{\pm 1}(\sigma^{(1)}_j)^{\pm 1})\nonumber \\
&\quad\times
\prod_{i<j}^{\nu'}\prod_\pm
(1-(\sigma^{(2)}_i)^{\pm 1}(\sigma^{(2)}_j)^{\mp 1})
(1-(\sigma^{(2)}_i)^{\pm 1}(\sigma^{(2)}_j)^{\pm 1})\nonumber \\
&\quad\times
\frac{1}{\prod_{i=1}^\nu\prod_{j=1}^{\nu'}
(1-\mathfrak{t}(\sigma^{(1)}_i)^{\mp 1}(\sigma^{(2)}_j)^{\pm 1})
(1-\mathfrak{t}(\sigma^{(1)}_i)^{\pm 1}(\sigma^{(2)}_j)^{\pm 1})
},\\
&{\cal I}^{(1)}_{\text{OSp},k}[f,g;\nu,\nu']\nonumber \\
&=
\frac{1}{\nu!\nu'!}
\prod_{i=1}^\nu\oint\frac{d\sigma^{(1)}_i}{2\pi i\sigma^{(1)}_i}\prod_{i=1}^\nu(\sigma^{(1)}_i)^{2k}f(\sigma^{(1)}_i)
\prod_{i=1}^{\nu'}\oint\frac{d\sigma^{(2)}_i}{2\pi i\sigma^{(2)}_i}\prod_{i=1}^{{\nu'}}(\sigma^{(2)}_i)^{-2k}g(\sigma^{(2)}_i)\nonumber \\
&\quad\times
\frac{
\prod_{i<j}^\nu \prod_\pm(1-(\sigma^{(1)}_i)^{\pm 1}(\sigma^{(1)}_j)^{\mp 1})
\prod_{i<j}^{\nu'} \prod_\pm(1-(\sigma^{(2)}_i)^{\pm 1}(\sigma^{(2)}_j)^{\mp 1})
}{
\prod_{i=1}^\nu\prod_{j=1}^{\nu'} \prod_\pm(1-(\sigma^{(1)}_i)^{\mp 1}(\sigma^{(2)}_j)^{\pm 1})
}.
\label{calI1OSpnunu'}
\end{align}
\end{subequations}
Note that ${\cal I}^{(1)}_{\text{OSp},k}[f,g;\nu,\nu]$ coincides with ${\cal I}^{(1)}_{\text{OSp},k}[f,g;\nu]$ \eqref{calI1OSp} we have introduced in section \ref{app_OSpgammachinot0-1} for $SO(2N+\gamma)^{(\chi\zeta)}_{2k}\times USp(2N)_{-k}$ ABJ theory with $(\gamma,\chi)\neq (0,-1)$.

In particular, for the choices of the single variable functions $f(s)$ and $g(s)$ such that ${\cal I}^{(1)}_{\text{OSp},k}[f,g;\nu,\nu+1]={\cal I}^{(1)}_{\text{OSp},k}[f,g;\nu+1,\nu]=0$ for all $\nu\ge 0$ and $k\ge 0$, only the clustering patterns of the form
\begin{align}
((\nu_0,\nu_0+1);(\nu_{m_1},\nu_{m_1}),(\nu_{m_2},\nu_{m_2}),\cdots)
\end{align}
contribute to the line defect index \eqref{OSp2Nchi-1intildecalI}.
As a result, the summation \eqref{OSp2Nchi-1intildecalI} simplifies as
\begin{align}
&\biggl\langle\prod_{i=1}^{N-1}f(s^{(1)}_i)\prod_{i=1}^Ng(s^{(2)}_i)\biggr\rangle^{SO(2N)_{2k}^{(-\zeta)}\times USp(2N)_{-k}\text{ABJ}(C)}(\mathfrak{t})
\nonumber\\
&
=\sum_{\substack{\nu_0,\nu_m=0\\ (\nu_0+\sum_{m=1}^\infty \nu_m=N)}}^\infty \zeta^{\sum_{m=1}^\infty m\nu_m}\tilde{\cal I}^{(0)}_{\text{OSp},0,-1}[f,g;\nu_0,\nu_0+1]
\prod_{m=1}^\infty\tilde{\cal I}^{(1)}_{\text{OSp},km}[f,g;\nu_m],
\label{OSp2Nchi-1simplified}
\end{align}
which has the same structure as in $SO(2N+\gamma)^{(\chi\zeta)}_{2k}\times USp(2N)_{-k}$ ABJ theory with $(\gamma,\chi)\neq (0,-1)$ \eqref{OSpABJ_index2}.
In this case, the grand canonical sum factorizes in the same way as $(\gamma,\chi)\neq (0,-1)$, as
\begin{align}
\Xi_{\text{OSp},0,-1}[f,g;u]&=\sum_{N=0}^\infty u^N\biggl\langle\prod_{i=1}^{N-1}f(s^{(1)}_i)\prod_{i=1}^Ng(s^{(2)}_i)\biggr\rangle^{SO(2N)_{2k}^{(-\zeta)}\times USp(2N)_{-k}\text{ABJ}(C)}(\mathfrak{t})\nonumber \\
&=\tilde{\Xi}^{(0)}_{\text{OSp},0,-1}[f,g;u]\prod_{m=1}^\infty\tilde{\Xi}^{(1)}_{\text{OSp},m}[f,g;\zeta^mu],
\label{OSp2Nchi-1simplified_GC}
\end{align}
with
\begin{align}
&\tilde{\Xi}^{(0)}_{\text{OSp},0,-1}[f,g;u]=\sum_{\nu=0}^\infty u^{\nu+1} \tilde{\cal I}^{(0)}_{\text{OSp},0,-1}[f,g;\nu,\nu+1],
\label{Xi0OSp2Nchi-1}
\end{align}
while $\tilde{\Xi}^{(1)}_{\text{OSp},m}[f,g;u]$ is given by \eqref{XitildeOSp(1)}.
Here we have defined ${\cal I}^{SO(0)^{(-\zeta)}_{2k}\times USp(0)_{-k}\text{ABJ}(C)}(\mathfrak{t})=0$.

We can calculate $\tilde{\cal I}^{(1)}_{\text{OSp},k}[f,g;\nu,\nu']$ systematically by generalizing the Fermi gas formalism in appendix \ref{ABJMalgorithm} by including the rank deformation.
For concreteness let us consider $\tilde{\cal I}^{(1)}_{\text{OSp},k}[f,g;\nu,\nu+\mu]$ with $\nu,\mu\ge 0$.\footnote{
For our purpose of calculating the line defect index \eqref{OSp2Nchi-1intildecalI}, only the formulas for $\mu=0$ and $\mu=1$ are relevant.
}
The formula for $\tilde{\cal I}^{(1)}_{\text{OSp},k}[f,g;\nu+\mu,\nu]$ can be obtained just by replacing $k\rightarrow -k$ and $f(s)\leftrightarrow g(s)$.
By using the Cauchy-Vandermonde determinant formula (see e.g.~\cite{Matsumoto:2013nya})
\begin{align}
\frac{\prod_{i<j}^N(x_i-x_j)\prod_{i<j}^{N+M}(y_i-y_j)}{\prod_{i=1}^N\prod_{j=1}^{N+M}(x_i+y_j)}=(-1)^{MN}
\text{det}\begin{pmatrix}
\Bigl[\frac{1}{x_i+y_j}\Bigr]_{i,j}^{N\times (N+M)}\vspace{0.2cm} \\
[y_j^{M-\mathsf{r}}]_{\mathsf{r},j}^{M\times (N+M)}
\end{pmatrix},
\end{align}
we can rewrite the integrand of $\tilde{\cal I}^{(1)}_{\text{OSp},k}[f,g;\nu,\nu+\mu]$ \eqref{calI1OSpnunu'} as
\begin{align}
&\tilde{\cal I}^{(1)}_{\text{OSp},k}[f,g;\nu,\nu+\mu]\nonumber \\
&
=\mathfrak{t}^{-\nu^2-2\mu\nu+\nu}\frac{1}{\nu!(\nu+\mu)!}
\prod_{i=1}^\nu
\oint\frac{d\sigma^{(1)}_i}{2\pi i\sigma^{(1)}_i}(\sigma^{(1)}_i)^{2k}f(\sigma^{(1)}_i)
\prod_{i=1}^{\nu+\mu}
\oint\frac{d\sigma^{(2)}_i}{2\pi i\sigma^{(2)}_i}(\sigma^{(2)}_i)^{-2k}g(\sigma^{(2)}_i)\nonumber \\
&\quad\times
\text{det}\begin{pmatrix}
\Biggl[\frac{1}{1-\mathfrak{t}(\sigma^{(1)}_i)^{-1}\sigma^{(2)}_j}\Biggr]_{i,j}^{\nu\times (\nu+\mu)}\vspace{0.2cm} \\
[(\sigma^{(2)}_j)^{\mu-\mathsf{r}}]_{\mathsf{r},j}^{\mu\times (\nu+\mu)}
\end{pmatrix}
\text{det}\Biggl(
\Biggl[\frac{1}{1-\mathfrak{t}(\sigma^{(2)}_i)^{-1}\sigma^{(1)}_j}\Biggr]_{i,j}^{(\nu+\mu)\times \nu}\,\, [(\sigma^{(2)}_i)^{\mathsf{s}-\mu}]_{i,\mathsf{s}}^{(\nu+\mu)\times \mu}
\Biggr)\nonumber \\
&=\mathfrak{t}^{-\nu^2-2\mu\nu+\nu}\frac{1}{\nu!(\nu+\mu)!}
\prod_{i=1}^\nu \int_0^1d\alpha^{(1)}_i
\prod_{i=1}^{\nu+\mu}
\int_0^1 d\alpha^{(2)}_i\nonumber \\
&\quad\times \det\begin{pmatrix}
\Bigl[\langle \alpha^{(1)}_i|e^{4\pi ik\hat{\alpha}}f(e^{2\pi i\hat{\alpha}})\mathfrak{t}^{-\hat{p}}\theta_{\mathbb{Z}}(-\hat{p})|\alpha^{(2)}_j\rangle\Bigr]_{i,j}^{\nu\times (\nu+\mu)}\vspace{0.2cm} \\
[\langle\!\langle \mathsf{r}-\mu|\alpha^{(2)}_j\rangle]_{\mathsf{r},j}^{\mu\times (\nu+\mu)}
\end{pmatrix}\nonumber \\
&\quad\times \text{det}\Bigl(
\Bigl[\langle\alpha^{(2)}_i|e^{-4\pi ik\hat{\alpha}}g(e^{2\pi i\hat{\alpha}})\mathfrak{t}^{-\hat{p}}\theta_{\mathbb{Z}}(-\hat{p})|\alpha^{(1)}_j\rangle\Bigr]_{i,j}^{(\nu+\mu)\times \nu}
\,\,
 [\langle \alpha^{(2)}_i|e^{-4\pi ik\hat{\alpha}}g(e^{2\pi i\hat{\alpha}})|\mathsf{s}-\mu\rangle\!\rangle]_{i,\mathsf{s}}^{(\nu+\mu)\times \mu}
\Bigr).
\end{align}
Here in the second line we have rewritten the integration measures and the matrix elements by using the notation of 1d quantum mechanics \eqref{1dQMnotation}.
We can combine the two determinants and $\alpha^{(2)}_i$-integration by using the Andr\'eief identity \eqref{Andreief}
as
\begin{align}
&\tilde{\cal I}^{(1)}_{\text{OSp},k}[f,g;\nu,\nu+\mu]\nonumber \\
&
=\mathfrak{t}^{-\nu^2-2\mu\nu+\nu}\frac{1}{\nu!}
\prod_{i=1}^\nu \int_0^1d\alpha^{(1)}_i
\text{det}\begin{pmatrix}
\Bigl[\langle \alpha^{(1)}_i|\hat{H}_{11}|\alpha^{(1)}_j\rangle\Bigr]_{i,j}^{\nu\times \nu}
&\Bigl[\langle \alpha^{(1)}_i|\hat{H}_{12}|\mathsf{s}-\mu\rangle\!\rangle\Bigr]_{i,\mathsf{s}}^{\nu\times \mu}\vspace{0.2cm} \\
\Bigl[\langle\!\langle \mathsf{r}-\mu|\hat{H}_{21}|\alpha^{(1)}_j\rangle\Bigr]_{\mathsf{r},j}^{\mu\times \nu}
&\Bigl[\langle\!\langle \mathsf{r}-\mu|\hat{H}_{22}|\mathsf{s}-\mu\rangle\!\rangle\Bigr]_{\mathsf{r},\mathsf{s}}^{\mu\times \mu}
\end{pmatrix},
\end{align}
with
\begin{subequations}
\begin{align}
&\hat{H}_{11}=
e^{4\pi ik\hat{\alpha}}
f(e^{2\pi i\hat{\alpha}})
\mathfrak{t}^{-\hat{p}}\theta_{\mathbb{Z}}(-\hat{p})
e^{-4\pi ik\hat{\alpha}}
g(e^{2\pi i\hat{\alpha}})
\mathfrak{t}^{-\hat{p}}\theta_{\mathbb{Z}}(-\hat{p}),\\
&\hat{H}_{12}=
e^{4\pi ik\hat{\alpha}}
f(e^{2\pi i\hat{\alpha}})
\mathfrak{t}^{-\hat{p}}\theta_{\mathbb{Z}}(-\hat{p})
e^{-4\pi ik\hat{\alpha}}
g(e^{2\pi i\hat{\alpha}}),\\
&\hat{H}_{21}=
e^{-4\pi ik\hat{\alpha}}
g(e^{2\pi i\hat{\alpha}})
\mathfrak{t}^{-\hat{p}}\theta_{\mathbb{Z}}(-\hat{p}),\\
&\hat{H}_{22}=
e^{-4\pi ik\hat{\alpha}}
g(e^{2\pi i\hat{\alpha}}).
\end{align}
\end{subequations}
As in the case without rank deformation, in order to calculate $\tilde{\cal I}^{(1)}_k[f,g;\nu,\nu+\mu]$ it is convenient to consider the generating function
\begin{align}
\sum_{\nu=0}^\infty \kappa^\nu\mathfrak{t}^{\nu^2+2\mu\nu-\nu}\tilde{\cal I}^{(1)}_k[f,g;\nu,\nu+\mu].
\end{align}
By using the Fredholm determinant formula (see e.g.~\cite{Matsumoto:2013nya}),
we can write the generating function as
\begin{align}
&\sum_{\nu=0}^\infty \kappa^\nu\mathfrak{t}^{\nu^2+2\mu\nu-\nu}\tilde{\cal I}^{(1)}_k[f,g;\nu,\nu+\mu]\nonumber \\
&=
\text{Det}\begin{pmatrix}
1+\kappa\hat{H}_{11}& [\kappa \hat{H}_{12}|\mathsf{s}-\mu\rangle\!\rangle]_{\mathsf{s}=1}^\mu\vspace{0.2cm}  \\
[\langle\!\langle \mathsf{r}-\mu|\hat{H}_{21}]_{\mathsf{r=1}}^\mu &[\langle\!\langle \mathsf{r}-\mu|\hat{H}_{22}|\mathsf{s}-\mu\rangle\!\rangle]_{\mathsf{r},\mathsf{s}}^{\mu\times \mu}
\end{pmatrix}\nonumber \\
&=\text{Det}(1+\kappa\hat{H}_{11})
\det_{\mu\times \mu}\Bigl(\langle\!\langle \mathsf{r}-\mu|\frac{1}{1+\kappa\hat{\rho}_{\text{OSp}}}e^{-4\pi ik\hat{\alpha}}g(e^{2\pi i\hat{\alpha}})|\mathsf{s}-\mu\rangle\!\rangle\Bigr),
\label{openstringformalism}
\end{align}
(the second expression is obtained by performing elementary row transformation) with
\begin{align}
\hat{\rho}_{\text{OSp}}=
e^{-4\pi ik\hat{\alpha}}g(e^{2\pi i\hat{\alpha}})
\mathfrak{t}^{-\hat{p}}\theta_{\mathbb{Z}}(-\hat{p})
e^{4\pi ik\hat{\alpha}}f(e^{2\pi i\hat{\alpha}})
\mathfrak{t}^{-\hat{p}}\theta_{\mathbb{Z}}(-\hat{p}).
\end{align}
Note that $\hat{H}_{11}$ and $\hat{\rho}_{\text{OSp}}$ are equivalent up to the similarity transformation, which does not affect the traces $\text{tr}\hat{H}_{11}^n$.
Our final expression \eqref{openstringformalism} for the generating function implies that we can obtain $\tilde{\cal I}^{(1)}_{\text{OSp},k}[f,g;\nu,\nu+\mu]$ by calculating
$\text{tr}(\hat{H}_{11}^n)$ with $n=1,\cdots,\nu$ and $\langle\!\langle \mathsf{r}-\mu|\hat{\rho}_{\text{OSp}}^n e^{-4\pi ik\hat{\alpha}}g(e^{2\pi i\hat{\alpha}})|\mathsf{s}-\mu\rangle\!\rangle$ with $n=0,\cdots,\nu$ and $\mathsf{r},\mathsf{s}=1,\cdots,\mu$.

As we have done in appendix \ref{ABJMalgorithm} for the calculation of $\text{tr}\hat{\rho}^n$, the latter quantities $\langle\!\langle \mathsf{r}-\mu|\hat{\rho}_{\text{OSp}}^n e^{-4\pi ik\hat{\alpha}}g(e^{2\pi i\hat{\alpha}})|\mathsf{s}-\mu\rangle\!\rangle$ can be calculated by using the coefficients of the Laurent expansions of $f(s)$ and $g(s)$
\begin{align}
s^{2k}f(s)=\sum_{\ell=-\infty}^\infty A_\ell s^\ell,\quad
s^{-2k}g(s)=\sum_{\ell=-\infty}^\infty B_\ell s^\ell
\end{align}
as
\begin{align}
\langle\!\langle \mathsf{r}-\mu|e^{-4\pi ik\hat{\alpha}}g(e^{2\pi i\hat{\alpha}})|\mathsf{s}-\mu\rangle\!\rangle
=B_{\mathsf{r}-\mathsf{s}}
\end{align}
for $n=0$, and
\begin{align}
&\langle\!\langle \mathsf{r}-\mu|\hat{\rho}_{\text{OSp}}^ne^{-4\pi ik\hat{\alpha}}g(e^{2\pi i\hat{\alpha}})|\mathsf{s}-\mu\rangle\!\rangle\nonumber \\
&
=\sum_{\substack{
\ell_1,\cdots,\ell_{n+1}\\
\ell'_1,\cdots,\ell'_n\\
(L_{2n}+\ell_{n+1}=\mathsf{r}-\mathsf{s})
}}
\prod_{i=1}^{n+1}B_{\ell_i}
\prod_{i=1}^{n}A_{\ell'_i}
\prod_{i=1}^{2n}\mathfrak{t}^{-r+\mu+L_i}
\theta_{\mathbb{Z}}(-r+\mu+L_i)
\end{align}
for $n\ge 1$, with
\begin{align}
L_{2i-1}=\sum_{j=1}^i\ell_j+\sum_{j=1}^{i-1}\ell'_j,\quad
L_{2i}=\sum_{j=1}^i(\ell_j+\ell'_j).
\end{align}

\subsubsection{Fundamental Wilson lines}
\label{app_OSp2Nchim1}
Let us consider the supersymmetric index with insertion of a bifundamental Wilson line $W_{\ydiagram{1};\ydiagram{1}}$, which can be calculated by choosing the single variable functions $f(s)=1+a(s+s^{-1})$ and $g(s)=1+b(s+s^{-1})$ as
\begin{align}
&\langle W_{\ydiagram{1};\ydiagram{1}}\rangle^{SO(2N)^{(-\zeta)}_{2k}\times USp(2N)_{-k}\text{ABJ}(C)}(\mathfrak{t})\nonumber \\
&=\biggl\langle \prod_{i=1}^{N-1}f(\sigma^{(1)}_i)\prod_{j=1}^Ng(\sigma^{(2)}_j)\biggr\rangle^{SO(2N)^{(-\zeta)}_{2k}\times USp(2N)_{-k}\text{ABJ}(C)}(\mathfrak{t})\Bigr|_{ab}.
\end{align}
For this choice of $f(s)$ and $g(s)$, we can easily show that the matrix elements $\langle\!\langle 0|\hat{\rho}^n_{\text{OSp}}e^{-4\pi ik\hat{\alpha}}g(e^{2\pi i\hat{\alpha}})|0\rangle\!\rangle$ vanish for any $n\ge 0$ and $k\ge 1$, which implies $\tilde{\cal I}^{(1)}_{\text{OSp},k}[f,g;\nu,\nu+1]$ vanishes identically.
Similarly, it also follows that $\tilde{\cal I}^{(1)}_{\text{OSp},k}[f,g;\nu+1,\nu]$ vanishes identically.
Hence the calculation of the line defect indices simplifies as \eqref{OSp2Nchi-1simplified} or \eqref{OSp2Nchi-1simplified_GC}, which consists only of $\tilde{\cal I}^{(0)}_{\text{OSp},0,-1}[f,g;\nu,\nu+1]$ and $\tilde{\cal I}^{(1)}_{\text{OSp},k}[f,g;\nu]$.
For $\tilde{\cal I}^{(0)}_{\text{OSp},0,-1}[f,g;\nu,\nu+1]$, we propose
\begin{align}
&{\cal I}^{(0)}_{\text{OSp},0,-1}[1+a(s+s^{-1}),1+b(s+s^{-1});\nu,\nu+1]\nonumber \\
&
=
\frac{1}{(\mathfrak{t}^4;\mathfrak{t}^4)_{\nu+1}}\biggl[1-\mathfrak{t}^{2\nu+2}+ab\frac{\mathfrak{t}(1-\mathfrak{t}^{2\nu})(1-\mathfrak{t}^{4\nu+4})}{1-\mathfrak{t}^2}\biggr]+{\cal O}(a^2,b^2)
\end{align}
which we have confirmed for $\nu=0,1,2$.
From this expression, we can calculate $\tilde{\Xi}^{(0)}_{\text{OSp},0,-1}[f,g;u]$ \eqref{Xi0OSp2Nchi-1} as
\begin{align}
\tilde{\Xi}^{(0)}_{\text{OSp},0,-1}[f,g;u]=\Bigl[\frac{1}{(u;\mathfrak{t}^4)_\infty}-\frac{1}{(u\mathfrak{t}^2;\mathfrak{t}^4)_\infty}\Bigr]\Bigl(1+\frac{abu\mathfrak{t}}{1-\mathfrak{t}^2}\Bigr)+{\cal O}(a^2,b^2).
\end{align}
Plugging this and $\tilde{\Xi}^{(1)}_{\text{OSp},m}[f,g;u]$ \eqref{XitildeOSp(1)} with $\tilde{\cal I}^{(1)}_{\text{OSp},k}[f,g;\nu,\nu]=\tilde{\cal I}^{(1)}_{\text{OSp},k}[f,g;\nu]$ \eqref{calIOSp(1)} into the full grand canonical sum \eqref{OSp2Nchi-1simplified_GC}, we obtain
\begin{align}
\Xi_{\text{OSp},0,-1}[f,g;u]=\Xi_{\text{OSp},0,-1}[1,1;u]\Bigl(1+\frac{abu(\mathfrak{t}+\zeta\mathfrak{t}^{2k-1})}{(1-\mathfrak{t}^2)(1-\zeta\mathfrak{t}^{2k})}\Bigr)+{\cal O}(a^2,b^2),
\end{align}
with
\begin{align}
\Xi_{\text{OSp},0,-1}[1,1;u]
&=
\sum_{N=0}^\infty u^N {\cal I}^{SO(2N)^{(-\zeta)}_{2k}\times USp(2N)_{-k}\text{ABJ}(C)}(\mathfrak{t})\nonumber \\
&=\Bigl[\frac{1}{(u;\mathfrak{t}^4)_\infty}-\frac{1}{(u\mathfrak{t}^2;\mathfrak{t}^4)_\infty}\Bigr]
\prod_{m=1}^\infty\frac{1}{(\zeta^m \mathfrak{t}^{2km}u;\mathfrak{t}^2)_\infty}.
\end{align}
By extracting the coefficient of $ab$, we find the following recursion relation as claimed in \eqref{SO2Nchi+recreln}
\begin{align}
&\langle W_{\ydiagram{1};\ydiagram{1}}\rangle^{SO(2N)^{(-\zeta)}_{2k}\times USp(2N)_{-k}\text{ABJ}(C)}(\mathfrak{t})\nonumber \\
&={\cal I}^{SO(2N-2)^{(-\zeta)}_{2k}\times USp(2N-2)_{-k}\text{ABJ}(C)}(\mathfrak{t})
\langle W_{\ydiagram{1};\ydiagram{1}}\rangle^{SO(3)^{(\chi\zeta)}_{2k}\times USp(2)_{-k}\text{ABJ}(C)}(\mathfrak{t}),
\label{OSp2Nchim1recreln}
\end{align}
for $N\ge 1$, with ${\cal I}^{SO(0)^{(-\zeta)}_{2k}\times USp(0)_{-k}\text{ABJ}(C)}(\mathfrak{t})=0$ and $\langle W_{\ydiagram{1};\ydiagram{1}}\rangle^{SO(3)^{(\chi\zeta)}_{2k}\times USp(2)_{-k}\text{ABJ}(C)}(\mathfrak{t})$ given by \eqref{WWOSp32}.

In particular, we find in the limit $\mathfrak{t}\rightarrow 1$
\begin{align}
&\lim_{\mathfrak{t}\rightarrow 1}\langle {\cal W}_{\ydiagram{1};\ydiagram{1}}\rangle^{SO(2N)^{(-+)}_{2k}\times USp(2N)_{-k}\text{ABJ}(C)}(\mathfrak{t})
=\lim_{\mathfrak{t}\rightarrow 1}\langle {\cal W}_{\ydiagram{1};\ydiagram{1}}\rangle^{Spin(2N)^{(-)}_{2k}\times USp(2N)_{-k}\text{ABJ}(C)}(\mathfrak{t})\nonumber \\
&
=2(N-1),\\
&\lim_{\mathfrak{t}\rightarrow 1}\langle {\cal W}_{\ydiagram{1};\ydiagram{1}}\rangle^{SO(2N)^{(--)}_{2k}\times USp(2N)_{-k}\text{ABJ}(C)}(\mathfrak{t})
=
\begin{cases}
\displaystyle \frac{2(k-1)}{k+2}&\quad (N\text{: even})\vspace{0.2cm} \\
\displaystyle 0&\quad (N\text{: odd})\vspace{0.2cm}
\end{cases}.
\end{align}
For $\zeta=1$ and the theory with $\mathbb{Z}^{{\cal M}}_2$ gauged, the results coincide with the dimension of the fundamental representation in $SO(2N)^{(-\zeta)}$.
We also find
\begin{align}
\langle {\cal W}_{\ydiagram{1};\ydiagram{1}}\rangle^{SO(2N)^{(-\zeta)}_{2k}\times USp(2N)_{-k}\text{ABJ}(C)}(\mathfrak{t}^{-1})
=\langle {\cal W}_{\ydiagram{1};\ydiagram{1}}\rangle^{SO(2N)^{(-\zeta)}_{2k}\times USp(2N)_{-k}\text{ABJ}(C)}(\mathfrak{t}).
\end{align}

By combining the recursion relation \eqref{OSp2Nchim1recreln} with that of $SO(2N)^{(+\zeta)}_{2k}\times USp(2N)_{-k}$ ABJ theory \eqref{OSp2Nchi1recreln}, we find the recursion relation of $O(2N)^{(\zeta)}_{2k}\times USp(2N)_{-k}$ ABJ theory with $\mathbb{Z}^{{\cal C}}_2$ gauged as
\begin{align}
&\langle W_{\ydiagram{1};\ydiagram{1}}\rangle^{O(2N)^{(\zeta)}_{2k}\times USp(2N)_{-k}\text{ABJ}(C)}(\mathfrak{t})\nonumber \\
&
={\cal I}^{O(2N-2)^{(\zeta)}_{2k}\times USp(2N-2)_{-k}\text{ABJ}(C)}(\mathfrak{t})
\langle W_{\ydiagram{1};\ydiagram{1}}\rangle^{SO(3)^{(\chi\zeta)}_{2k}\times USp(2)_{-k}\text{ABJ}(C)}(\mathfrak{t}),
\end{align}
with ${\cal I}^{O(2N)^{(\zeta)}_{2k}\times USp(2N)_{-k}\text{ABJ}(C)}(\mathfrak{t})$ obtained by the grand canonical sum
\begin{align}
\sum_{N=0}^\infty u^N {\cal I}^{O(2N)^{(\zeta)}_{2k}\times USp(2N)_{-k}\text{ABJ}(C)}(\mathfrak{t})=\frac{1}{(u;\mathfrak{t}^4)_\infty}\prod_{m=1}^\infty \frac{1}{(\zeta^m\mathfrak{t}^{2km}u;\mathfrak{t}^2)_\infty}.
\end{align}
Here we have defined ${\cal I}^{O(0)^{(\zeta)}_{2k}\times USp(0)_{-k}\text{ABJ}(C)}(\mathfrak{t})\equiv 1$.
Remarkably, both the Coulomb indices and the bifundamental line defect indices of $O(2N)^{(\zeta)}_{2k}\times USp(2N)_{-k}$ ABJ theory precisely coincide with those of $SO(2N+1)^{(\chi\zeta)}_{2k}\times USp(2N)_{-k}$ ABJ theory (which is independent of $\chi=\pm 1$)
\begin{align}
&{\cal I}^{O(2N)^{(\zeta)}_{2k}\times USp(2N)_{-k}\text{ABJ}(C)}(\mathfrak{t})={\cal I}^{O(2N+1)^{(\zeta)}_{2k}\times USp(2N)_{-k}\text{ABJ}(C)}(\mathfrak{t})\nonumber \\
&\quad ={\cal I}^{SO(2N+1)^{(\chi \zeta)}_{2k}\times USp(2N)_{-k}\text{ABJ}(C)}(\mathfrak{t}),\\
&\langle W_{\ydiagram{1};\ydiagram{1}}\rangle^{O(2N)^{(\zeta)}_{2k}\times USp(2N)_{-k}\text{ABJ}(C)}(\mathfrak{t})
=\langle W_{\ydiagram{1};\ydiagram{1}}\rangle^{O(2N+1)^{(\zeta)}_{2k}\times USp(2N)_{-k}\text{ABJ}(C)}(\mathfrak{t})\nonumber \\
&\quad =\langle W_{\ydiagram{1};\ydiagram{1}}\rangle^{SO(2N+1)^{(\chi \zeta)}_{2k}\times USp(2N)_{-k}\text{ABJ}(C)}(\mathfrak{t}).
\end{align}

\subsection{$\mathcal{N}=4$ SQCD}
\label{app_C_SQCD}

All the theories treated so far contain a hypermultiplet in the adjoint or in the (anti)symmetric representation.
To check that the conjecture \eqref{Witten_indC_conj_int} is not restricted to this class, we collect here the Coulomb line defect Witten indices of $\mathcal{N}=4$ SQCD, whose only matter consists of $N_f$ hypermultiplets in the fundamental (vector) representation.
In the Coulomb limit the hypermultiplets drop out of the integrand and survive only through the dimension of the bare monopole operators, so that the Coulomb line defect index reduces to the monopole sum
\begin{align}
\label{monopoleformula}
&\langle W_{\mathcal{R}_1}\cdots W_{\mathcal{R}_k}\rangle^{G\text{-}[N_f](C)}(\mathfrak{t})\nonumber\\
&=\sum_{\mathbf{m}}\frac{\mathfrak{t}^{2\Delta(\mathbf{m})}}{|\mathscr{W}_{H_{\mathbf{m}}}|}
\oint\prod_{i=1}^{r}\frac{ds_i}{2\pi is_i}\,\prod_{j=1}^{k}\chi_{\mathcal{R}_j}(s)\,
\frac{1}{(1-\mathfrak{t}^2)^{r}}\prod_{\substack{\alpha\in\Delta_{\mathfrak{g}}\\ \alpha(\mathbf{m})=0}}\frac{1-s^{\alpha}}{1-\mathfrak{t}^2s^{\alpha}}\,,
\end{align}
where $H_{\mathbf{m}}\subseteq G$ is the subgroup left unbroken by the flux $\mathbf{m}$, $\mathscr{W}_{H_{\mathbf{m}}}$ its Weyl group, and
$\Delta(\mathbf{m})=\frac12\sum_{\rho}|\rho(\mathbf{m})|-\sum_{\alpha>0}|\alpha(\mathbf{m})|$
is the bare monopole dimension, the first sum running over the weights of the hypermultiplet representation.
Without the Wilson lines the contour integral is the Molien series of $H_{\mathbf m}$ and \eqref{monopoleformula} reduces to the usual monopole formula for the Coulomb branch Hilbert series.
We have checked that \eqref{monopoleformula} reproduces the exact expressions
${\cal I}^{U(1)\text{ ADHM-}[l](C)}$ of \eqref{u1u1_W1-1} and
${\cal I}^{USp(2)+\text{(A/S)-}[2l](C)}$ together with $\langle \mathcal{W}_{\ydiagram{1}}\mathcal{W}_{\ydiagram{1}}\rangle^{USp(2)+\text{(A/S)-}[2l](C)}$ of \eqref{usp2_C_W1}.
Note also that for $USp(2)$ the antisymmetric hypermultiplet is a singlet and for $SO(3)$ it is again a vector, so that at rank one SQCD coincides with theories already treated in this paper. The genuinely new cases are $U(N)$ for any $N$, and $USp(4)$ and $SO(5)$.
We restrict ourselves to good theories, $N_f\ge 2N$ for $U(N)$, $N_f\ge 2N+1$ for $USp(2N)$ and $N_f\ge N_c-1$ for $SO(N_c)$, and set the topological fugacities to unity. Here the bracket counts hypermultiplets in the fundamental representation, which is the standard flavor counting for SQCD. For the ADHM-type theories with orthogonal and symplectic gauge groups the bracket instead counts fundamental half-hypermultiplets, as in $USp(2N)+$(A/S)-$[2l]$, since there the number of half-hypermultiplets can be odd and the flavor symmetry acts on them.

For the indices without line defects we find, for example,
\begin{subequations}
\begin{align}
{\cal I}^{U(2)\text{-}[4](C)}&=\frac{(1+\mathfrak{t}^4)(1+\mathfrak{t}^2+\mathfrak{t}^4)}{(1-\mathfrak{t}^2)^4(1+\mathfrak{t}^2)^2},\\
{\cal I}^{U(3)\text{-}[6](C)}&=\frac{(1+\mathfrak{t}^4)(1-\mathfrak{t}^2+\mathfrak{t}^4)(1+\mathfrak{t}^2+\mathfrak{t}^4+\mathfrak{t}^6+\mathfrak{t}^8)}{(1-\mathfrak{t}^2)^6(1+\mathfrak{t}^2)(1+\mathfrak{t}^2+\mathfrak{t}^4)^2},\\
{\cal I}^{USp(4)\text{-}[5](C)}={\cal I}^{SO(5)\text{-}[4](C)}&=\frac{(1+\mathfrak{t}^8)(1-\mathfrak{t}^2+\mathfrak{t}^4)}{(1-\mathfrak{t}^2)^4(1+\mathfrak{t}^2)^2(1+\mathfrak{t}^4)},
\end{align}
\end{subequations}
the order of the pole at $\mathfrak{t}=1$ being twice the quaternionic dimension of the Coulomb branch in each case.
Line defect indices are equally compact, for instance
\begin{subequations}
\begin{align}
\langle \mathcal{W}_{\ydiagram{1}}\mathcal{W}_{\overline{\ydiagram{1}}}\rangle^{U(2)\text{-}[4](C)}
&=\frac{(1+\mathfrak{t}^2)^2(1+2\mathfrak{t}^2+2\mathfrak{t}^6+\mathfrak{t}^8)}{(1+\mathfrak{t}^4)^2(1+\mathfrak{t}^2+\mathfrak{t}^4)},\\
\langle \mathcal{W}_{\ydiagram{1}}\mathcal{W}_{\ydiagram{1}}\rangle^{USp(4)\text{-}[5](C)}
&=\frac{(1+\mathfrak{t}^2)^2}{1-\mathfrak{t}^2+\mathfrak{t}^4}.
\end{align}
\end{subequations}

Taking $\mathfrak{t}\rightarrow1$ we find in every case we examined
\begin{align}
\lim_{\mathfrak{t}\rightarrow1}\langle \mathcal{W}_{\mathcal{R}_1}\cdots\mathcal{W}_{\mathcal{R}_k}\rangle^{G\text{-}[N_f](C)}(\mathfrak{t})
=\oint\prod_{i=1}^{r}\frac{ds_i}{2\pi is_i}\prod_{j=1}^{k}\chi_{\mathcal{R}_j}(s)\,,
\end{align}
in agreement with \eqref{Witten_indC_conj_int}. The results are listed in table \ref{tab_SQCD_Coulomb}.

\begin{table}[htbp]
\centering
\begin{tabular}{l|l|c|c}
\hline
theory & Wilson lines & $\lim_{\mathfrak{t}\rightarrow1}\langle\mathcal{W}\cdots\rangle^{(C)}$ & $\oint\prod_{i=1}^r\frac{ds_i}{2\pi is_i}\prod_j\chi_{\mathcal{R}_j}$ \\ \hline
$U(2)$-$[4]$, $U(2)$-$[5]$ & $\mathcal{W}_{\ydiagram{1}}\mathcal{W}_{\overline{\ydiagram{1}}}$ & $2$ & $2$ \\
 & $\mathcal{W}_{\ydiagram{2}}\mathcal{W}_{\overline{\ydiagram{2}}}$ & $3$ & $3$ \\
 & $\mathcal{W}_{\ydiagram{3}}\mathcal{W}_{\overline{\ydiagram{3}}}$ & $4$ & $4$ \\
 & $\mathcal{W}_{\ydiagram{1,1}}\mathcal{W}_{\overline{\ydiagram{1,1}}}$ & $1$ & $1$ \\
 & $\mathcal{W}_{\ydiagram{1}}\mathcal{W}_{\ydiagram{1}}\mathcal{W}_{\overline{\ydiagram{1,1}}}$ & $2$ & $2$ \\ \hline
$U(3)$-$[6]$ & $\mathcal{W}_{\ydiagram{1}}\mathcal{W}_{\overline{\ydiagram{1}}}$ & $3$ & $3$ \\ \hline
$USp(2)$-$[3]$, $USp(2)$-$[4]$ & $\mathcal{W}_{\ydiagram{1}}\mathcal{W}_{\ydiagram{1}}$ & $2$ & $2$ \\
 & $\mathcal{W}_{\ydiagram{2}}$, $\mathcal{W}_{\ydiagram{2}}\mathcal{W}_{\ydiagram{2}}$ & $1$, $3$ & $1$, $3$ \\
 & $\mathcal{W}_{\ydiagram{3}}\mathcal{W}_{\ydiagram{1}}$ & $2$ & $2$ \\ \hline
$USp(4)$-$[5]$ & $\mathcal{W}_{\ydiagram{1}}\mathcal{W}_{\ydiagram{1}}$, $\mathcal{W}_{\ydiagram{1}}\mathcal{W}_{\ydiagram{1,1}}$ & $4$, $0$ & $4$, $0$ \\
 & $\mathcal{W}_{\ydiagram{1,1}}$, $\mathcal{W}_{\ydiagram{1,1}}\mathcal{W}_{\ydiagram{1,1}}$ & $1$, $5$ & $1$, $5$ \\
 & $\mathcal{W}_{\ydiagram{2}}$, $\mathcal{W}_{\ydiagram{2}}\mathcal{W}_{\ydiagram{2}}$ & $2$, $12$ & $2$, $12$ \\ \hline
$SO(5)$-$[4]$ & $\mathcal{W}_{\ydiagram{1}}$, $\mathcal{W}_{\ydiagram{1}}\mathcal{W}_{\ydiagram{1}}$ & $1$, $5$ & $1$, $5$ \\
 & $\mathcal{W}_{\ydiagram{1,1}}$, $\mathcal{W}_{\ydiagram{1,1}}\mathcal{W}_{\ydiagram{1,1}}$ & $2$, $12$ & $2$, $12$ \\
 & $\mathcal{W}_{\ydiagram{2}}$, $\mathcal{W}_{\ydiagram{2}}\mathcal{W}_{\ydiagram{2}}$ & $2$, $16$ & $2$, $16$ \\ \hline
\end{tabular}
\caption{The $\mathfrak{t}\rightarrow1$ limit of the normalized Coulomb line defect indices of $\mathcal{N}=4$ SQCD, compared with the character integral \eqref{Witten_indC_conj_int}. Here $G$-$[N_f]$ denotes the gauge group $G$ with $N_f$ fundamental hypermultiplets.}
\label{tab_SQCD_Coulomb}
\end{table}

It is worth stressing that the agreement with \eqref{Witten_indC_conj_int} persists even though the factorized recursion relations of section \ref{sec_recursion} do \textit{not} extend to SQCD.
For the theories with an (anti)symmetric hypermultiplet the bare monopole dimension is linear in the flux with a rank independent coefficient, which is what makes the recursion relations such as \eqref{R_usp_C_W1} possible.
In SQCD the coefficient depends on the rank, and indeed the right hand side of \eqref{R_usp_C_W1} evaluated for $USp(4)$-$[5]$ gives $\frac{(1+\mathfrak{t}^2)^2(1+\mathfrak{t}^4)}{(1+\mathfrak{t}^2+\mathfrak{t}^4)^2}$, which differs from the correct answer and tends to $\frac{8}{9}$ rather than to $4$ as $\mathfrak{t}\rightarrow1$.
The Witten index formula \eqref{Witten_indC_conj_int} is therefore insensitive to the factorization structure that organizes the ADHM-type theories.

\section{Higgs line defect indices}
\label{app_Higgs}

In this appendix, we consider Higgs line defect indices for various 3d $\mathcal{N}\geq 4$ theories. We also compute the $\bt\to1$ limit of those indices and compare the results with \eqref{Witten_indH_char} or \eqref{Witten_indC_conj_int}. The comparison with \eqref{Witten_indC_conj_int} may appear when there are residual $U(1)$ symmetries on the branch as discussed in section \ref{sec_globalandCS}. Both comparisons are instances of the character average \eqref{Witten_ind_unified}, with $H_0$ trivial in the first case and a torus in the second.
All the Higgs line defect indices below are obtained from the general zero flux expression \eqref{Higgs_general_int}, specialized to the matter content at hand.
Appendices \ref{app_Higgs_UN}--\ref{app_Higgs_SON} deal with the ADHM-type theories, which always contain a hypermultiplet in the adjoint or (anti)symmetric representation, while appendix \ref{app_Higgs_SQCD} treats $\mathcal{N}=4$ SQCD, for which no such matter is present.

\subsection{$U(N)$ ADHM theory}
\label{app_Higgs_UN}
For the $U(N)$ ADHM theory with $l$ flavors, let us consider the multi-point function of Wilson lines in symmetric representations and their conjugates, which form a complete basis of Wilson lines in arbitrary representations.
In the Higgs limit, the generating function of $L$ or smaller number of symmetric Wilson lines and their conjugates is given as
\begin{align}
&\sum_{k_1,\cdots,k_{2L}\ge 0}\prod_{I=1}^La_I^{k_I}b_I^{k_{L+I}}
\biggl\langle
\prod_{I=1}^LW_{(k_I)}W_{\overline{(k_{L+I})}}
\biggr\rangle^{U(N)\text{ ADHM-}[l](H)}(x,y_\alpha;\mathfrak{t})\nonumber \\
&=
\biggl\langle
\prod_{I=1}^L\prod_{i=1}^N\frac{1}{(1-a_Is_i)(1-b_Is_i^{-1})}
\biggr\rangle^{U(N)\text{ ADHM-}[l](H)}(x,y_\alpha;\mathfrak{t})\nonumber \\
&
=
\frac{(1-\mathfrak{t}^{2})^N}{N!\prod_\pm(1-x^{\pm 1}\mathfrak{t})^N}\oint\prod_{i=1}^N
\frac{ds_i}{2\pi is_i}
\frac{
\prod_{i<j}\prod_\pm(1-s_i^{\pm 1}s_j^{\mp 1})(1-\mathfrak{t}^2s_i^{\pm 1}s_j^{\mp 1})
}{
\prod_{i<j}\prod_{\pm,\pm'}(1-\mathfrak{t}(s_is_j^{-1}x^{\pm' 1})^{\pm 1})}\nonumber \\
&\quad\times
\frac{1}{\prod_{i=1}^N\prod_{\alpha=1}^{l+L}\prod_\pm(1-\mathfrak{t}s_i^{\pm 1}y_\alpha^{(\pm)})},
\label{IHU(N)ADHM}
\end{align}
with
\begin{subequations}
\begin{align}
&y_\alpha^{(\pm)}=y_\alpha^{\pm 1},\quad (\alpha=1,\cdots,l),\\
&y_{l+I}^{(+)}=\mathfrak{t}^{-1}a_I,\quad (I=1,\cdots,L),\\
&y_{l+I}^{(-)}=\mathfrak{t}^{-1}b_I,\quad (I=1,\cdots,L).
\end{align}
\end{subequations}
As studied in \cite{Hayashi:2024jof}, section 4.2, the integral \eqref{IHU(N)ADHM} can be evaluated exactly by a finite residue sum.
For example, for $l=1$ and $L=1$ we have, after setting $x=1$
\begin{align}
&\sum_{k_1,k_2\ge 0}
a^{k_1}b^{k_2}
\langle
W_{(k_1)}W_{\overline{(k_2)}}
\rangle^{U(2)\text{ ADHM-}[1](H)}(\mathfrak{t})\nonumber \\
&
=
\frac{1}{
(1 - a b) (1 - \mathfrak{t})^2 (1 - \mathfrak{t}^2)^2 (1 - a \mathfrak{t}) (1 - b \mathfrak{t}) (1 - a b \mathfrak{t})^2 (1 - a \mathfrak{t}^2)^2 (1 - b \mathfrak{t}^2)^2
}
(
1 + \mathfrak{t}^2\nonumber \\
&\quad
- 2 \mathfrak{t}^4 (a + b)
- 4 \mathfrak{t}^3 (1 + \mathfrak{t} - \mathfrak{t}^3) a b
- \mathfrak{t}^3 (1 - 4 \mathfrak{t}^2 - 2 \mathfrak{t}^3 - \mathfrak{t}^4) (a^2 b + a b^2)\nonumber \\
&\quad
+ \mathfrak{t}^4 (1 + 4 \mathfrak{t} + \mathfrak{t}^2 - \mathfrak{t}^4 - 4 \mathfrak{t}^5 - \mathfrak{t}^6) a^2 b^2
- \mathfrak{t}^7 (1 + 2 \mathfrak{t} + 4 \mathfrak{t}^2 - \mathfrak{t}^4) (a^3 b^2 + a^2 b^3)\nonumber \\
&\quad
- 4 \mathfrak{t}^8 (1 - \mathfrak{t}^2 - \mathfrak{t}^3) a^3 b^3
+ 2 \mathfrak{t}^{10} (a^4 b^3 + a^3 b^4)
- \mathfrak{t}^{12} (1 + \mathfrak{t}^2) a^4 b^4
).
\end{align}
Dividing both sides by the coefficient of $a^0b^0$ and taking the limit $\mathfrak{t}\rightarrow 1$, we find
\begin{align}
\sum_{k_1,k_2\ge 0}
a^{k_1}b^{k_2}
\lim_{\mathfrak{t}\rightarrow 1}
\langle
{\cal W}_{(k_1)}{\cal W}_{\overline{(k_2)}}
\rangle^{U(2)\text{ ADHM-}[1](H)}(\mathfrak{t})
&=
\frac{1}{(1-a)^2(1-b)^2}\nonumber \\
&
=\sum_{k_1,k_2\ge 0}
a^{k_1}b^{k_2}
\chi_{(k_1)}(1) \chi_{(k_2)}(1).
\end{align}
Namely, we have $
\lim_{\mathfrak{t}\rightarrow 1}
\langle
{\cal W}_{(k_1)}{\cal W}_{\overline{(k_2)}}
\rangle^{U(2)\text{ ADHM-}[1](H)}(\mathfrak{t})
=
\chi_{(k_1)}(1) \chi_{(k_2)}(1)$.
We have checked the same relation hold also for other values of $l$ ($l\ge 1$) and $L$ as
\begin{align}
\lim_{\mathfrak{t}\rightarrow 1}
\biggl\langle
\prod_{I=1}^L
{\cal W}_{(k_I)}{\cal W}_{\overline{(k_{L+I})}}
\biggr\rangle^{U(N)\text{ ADHM-}[l](H)}(\mathfrak{t})
=
\prod_{I=1}^L\chi_{(k_I)}(1) \chi_{(k_{L+I})}(1),
\end{align}
which verifies our conjecture \eqref{Witten_indH_char} for the Wilson lines in any representations.

The residue sum formula in \cite{Hayashi:2024jof} is also applicable to the case with $l=0$, which corresponds to the 3d ${\cal N}=8$ super Yang-Mills theory, and $L\ge 1$.
In this case we find
\begin{align}
&\sum_{k_1,\cdots,k_{2L}\ge 0}
\prod_{I=1}^La_I^{k_I}b_I^{k_{L+I}}
\lim_{\mathfrak{t}\rightarrow 1}
\biggl\langle
\prod_{I=1}^L{\cal W}_{(k_I)}{\cal W}_{\overline{(k_{L+I})}}
\biggr\rangle^{{\cal N}=8\,U(N)(H)}(\mathfrak{t})\nonumber \\
&
=
\biggl(\oint\frac{ds}{2\pi is}\prod_{I=1}^L\frac{1}{(1-a_Is)(1-b_Is^{-1})}\biggr)^N.
\end{align}
Namely, we have
\begin{align}
\lim_{\mathfrak{t}\rightarrow 1}
\biggl\langle
\prod_{I=1}^L{\cal W}_{(k_I)}{\cal W}_{\overline{(k_{L+I})}}
\biggr\rangle^{{\cal N}=8\,U(N)(H)}(\mathfrak{t})
=
\oint\prod_{i=1}^N\frac{ds_i}{2\pi is_i}\prod_{I=1}^L\chi_{(k_I)}(s)\chi_{(k_{L+I})}(s),
\end{align}
which coincides with our conjecture for the {\it Coulomb} line defect indices \eqref{Witten_indC_conj_int}
rather than that of the Higgs line defect indices \eqref{Witten_indH_char}.
This is because in the current theory the $U(1)^N$ gauge group always remains since there are no fundamental hypermultiplets, so that $H_0=U(1)^N$ rather than the trivial group, as explained in section \ref{sec_globalandCS}.

\subsection{$USp(2N)$}
In the theory with $USp(2N)$ gauge group with one (anti)symmetric hypermultiplet and $2l+\delta$ fundamental half-hypermultiplets, the Higgs line defect indices are given as
\begin{align}
&\langle W_{{\cal R}_1}\cdots W_{{\cal R}_k} \rangle^{USp(2N)+\text{(A/S)-}[2l+\delta](H)}(x,y_\alpha;\mathfrak{t})\nonumber \\
&=\frac{1}{2^NN!}\frac{(1-\mathfrak{t}^2)^N}{\prod_\pm (1-\mathfrak{t}x^{\pm 1})^N}\oint\prod_{i=1}^N\frac{ds_i}{2\pi is_i}
\prod_i\prod_\pm \frac{(1-s_i^{\pm 2})(1-\mathfrak{t}^2s^{\mp 2}_i)}{\prod_{\alpha=1}^{2l+\delta}(1-\mathfrak{t}s_i^{\pm 1}y_\alpha)\prod_{\pm'} (1-\mathfrak{t}s_i^{\pm 2}x^{\pm'1})^\epsilon}\nonumber \\
&\quad\times
\prod_{i<j}\prod_\pm \frac{(1-(\frac{s_i}{s_j})^{\pm 1})(1-(s_is_j)^{\pm 1})(1-\mathfrak{t}^2(\frac{s_i}{s_j})^{\mp 1})(1-\mathfrak{t}^2(s_is_j)^{\mp 1})
}{
\prod_{\pm'}
(1-\mathfrak{t}(\frac{s_i}{s_j})^{\pm 1}x^{\pm' 1})
(1-\mathfrak{t}(s_is_j)^{\pm 1}x^{\pm' 1})
}\prod_{j=1}^k\chi_{{\cal R}_j}(s_i).
\end{align}

For $x=y_\alpha=1$, we find, for example,
\begin{subequations}
\begin{align}
&{\cal I}_H^{USp(2)+\text{A-}[6](H)}= \frac{1 + 9 \mathfrak{t}^2 + 9 \mathfrak{t}^4 + \mathfrak{t}^6}{(1 - \mathfrak{t})^2 (1 - \mathfrak{t}^2)^6},\\
&{\cal I}_H^{USp(4)+\text{A-}[6](H)}\nonumber \\
&=
\frac{1}{(1 - \mathfrak{t}) (1 - \mathfrak{t}^2)^8 (1 - \mathfrak{t}^3)^7} (1 + \mathfrak{t} + 11 \mathfrak{t}^2 + 34 \mathfrak{t}^3 + 88 \mathfrak{t}^4 + 216 \mathfrak{t}^5 + 473 \mathfrak{t}^6 + 797 \mathfrak{t}^7 + 1243 \mathfrak{t}^8\nonumber \\
&\quad  + 1738 \mathfrak{t}^9\nonumber + 2080 \mathfrak{t}^{10} + 2152 \mathfrak{t}^{11}
+\cdots
+ \mathfrak{t}^{22}),
\end{align}
\end{subequations}
and
\begin{subequations}
\begin{align}
&\langle W_{\ydiagram{1}}\rangle^{USp(2)+\text{A-}[6](H)}=
\frac{6 \mathfrak{t} + 28 \mathfrak{t}^3 + 6 \mathfrak{t}^5}{(1 - \mathfrak{t})^2 (1 - \mathfrak{t}^2)^6},\\
&\langle W_{\ydiagram{2}}\rangle^{USp(2)+\text{A-}[6](H)}=
\frac{20 \mathfrak{t}^2 + 55 \mathfrak{t}^4 - 21 \mathfrak{t}^6 + 7 \mathfrak{t}^8 - \mathfrak{t}^{10}}{(1 - \mathfrak{t})^2 (1 - \mathfrak{t}^2)^6},\\
&\langle W_{\ydiagram{1}}W_{\ydiagram{1}}\rangle^{USp(2)+\text{A-}[6](H)}=
\frac{
1 + 29 \mathfrak{t}^2 + 64 \mathfrak{t}^4 - 20 \mathfrak{t}^6 + 7 \mathfrak{t}^8 - \mathfrak{t}^{10}}{(1 - \mathfrak{t})^2 (1 - \mathfrak{t}^2)^6},\\
&\langle W_{\ydiagram{1}}\rangle^{USp(4)+\text{A-}[6](H)}\nonumber \\
&=
\frac{1}{(1 - \mathfrak{t}) (1 - \mathfrak{t}^2)^8 (1 - \mathfrak{t}^3)^7} (6 \mathfrak{t} + 18 \mathfrak{t}^2 + 78 \mathfrak{t}^3 + 272 \mathfrak{t}^4 + 692 \mathfrak{t}^5 + 1532 \mathfrak{t}^6 + 3034 \mathfrak{t}^7 + 5002 \mathfrak{t}^8\nonumber \\
&\quad + 6994 \mathfrak{t}^9 + 8692 \mathfrak{t}^{10} + 9424 \mathfrak{t}^{11}
+\cdots
+ 6 \mathfrak{t}^{21}),\\
&\langle W_{\ydiagram{1,1}}\rangle^{USp(4)+\text{A-}[6](H)}\nonumber \\
&=
\frac{1}{(1 - \mathfrak{t}) (1 - \mathfrak{t}^2)^8 (1 - \mathfrak{t}^3)^7} (2 \mathfrak{t} + 17 \mathfrak{t}^2 + 77 \mathfrak{t}^3 + 242 \mathfrak{t}^4 + 743 \mathfrak{t}^5 + 1760 \mathfrak{t}^6 + 3507 \mathfrak{t}^7 + 6039 \mathfrak{t}^8\nonumber \\
&\quad + 9066 \mathfrak{t}^9 + 11286 \mathfrak{t}^{10} + 12102 \mathfrak{t}^{11}
+\cdots
+ 2 \mathfrak{t}^{21}).
\end{align}
\end{subequations}
Here we have abbreviated higher order terms in the numerator if it is a palindromic polynomial.
From these results, we find
\begin{subequations}
\begin{align}
&\lim_{\mathfrak{t}\rightarrow 1}
\langle {\cal W}_{\ydiagram{1}}\rangle^{USp(2)+\text{A-}[6](H)}
=2=
\chi^{\mathfrak{usp}(2)}_{\ydiagram{1}}(1),\quad
\lim_{\mathfrak{t}\rightarrow 1}
\langle {\cal W}_{\ydiagram{2}}\rangle^{USp(2)+\text{A-}[6](H)}
=3=
\chi^{\mathfrak{usp}(2)}_{\ydiagram{2}}(1),\\
&\lim_{\mathfrak{t}\rightarrow 1}
\langle {\cal W}_{\ydiagram{1}}{\cal W}_{\ydiagram{1}}\rangle^{USp(2)+\text{A-}[6](H)}
=4=
(\chi^{\mathfrak{usp}(2)}_{\ydiagram{1}}(1))^2,\quad
\lim_{\mathfrak{t}\rightarrow 1}
\langle {\cal W}_{\ydiagram{1}}\rangle^{USp(4)+\text{A-}[6](H)}
=4=
\chi^{\mathfrak{usp}(4)}_{\ydiagram{1}}(1),\\
&\lim_{\mathfrak{t}\rightarrow 1}
\langle {\cal W}_{\ydiagram{1,1}}\rangle^{USp(4)+\text{A-}[6](H)}
=5=
\chi^{\mathfrak{usp}(4)}_{\ydiagram{1,1}}(1),
\end{align}
\end{subequations}
which are consistent with our conjecture \eqref{Witten_indH_char}.

\subsection{$SO(N)$}
\label{app_Higgs_SON}
We consider the Higgs line defect index of the $SO(N)$ or $O(N)$ gauge theory with one hypermultiplet in the (anti)symmetric representation and $l$ hypermultiplets in the fundamental representation. The Higgs line defect index is given by applying the limit \eqref{Higgs} to \eqref{SOoddandevenwithchi1fullindex} or \eqref{SO_index2}. In the cases where we have the $SO(2N+\gamma)$ gauge group and the discrete fugacity $\chi$ with $\left(\gamma, \chi\right)=(0, 1), (1, \pm 1)$, the Higgs line defect index becomes
\begin{equation}
\begin{split}
&\Braket{W_{{\cal R}_1}\cdots W_{{\cal R}_k} }^{SO(2N+\gamma)^{(\chi)}+\text{(A/S)-}[2l](H)}(x,y_\alpha;\mathfrak{t})\cr
&=\frac{1}{2^{N+\gamma-1}N!}\oint\prod_{i=1}^N\frac{ds_i}{2\pi i s_i}\prod_{i=1}^k\chi_{{\cal R}_i}(s)\prod_{i=1}^N\prod_{\pm}\left(1-\chi s_i^{\pm 1}\right)^{\gamma}\prod_{i<j}\prod_{\pm}\left(1-s_i^{\pm 1}s_j^{\mp 1}\right)\left(1-s_i^{\pm 1}s_j^{\pm 1}\right)\cr
&\frac{\left(1-\mathfrak{t}^2\right)^N\prod_{i=1}^N\prod_{\pm}\left(1-\chi\mathfrak{t}^2s_i^{\pm 1}\right)^{\gamma}\prod_{i<j}\prod_{\pm}\left(1-\mathfrak{t}^2s_i^{\mp 1}s_j^{\pm 1}\right)\left(1 - \mathfrak{t}^2s_i^{\mp 1}s_j^{\mp 1}\right)}{\prod_{i<j}\prod_{\pm}\left(1-\mathfrak{t}s_i^{\pm 1}s_j^{\mp 1}x^{\pm 1}\right)\left(1-\mathfrak{t}s_i^{\pm 1}s_j^{\mp 1}x^{\mp 1}\right)\left(1-\mathfrak{t}s_i^{\pm 1}s_j^{\pm 1}x^{\pm 1}\right)\left(1-\mathfrak{t}s_i^{\pm 1}s_j^{\pm 1}x^{\mp 1}\right)}\cr
&\frac{1}{\left(1-\mathfrak{t}x^{\pm 1}\right)^{N+\gamma\epsilon}\prod_{i=1}^N\prod_{\pm}\left(1-\chi \mathfrak{t}s_i^{\pm 1}x^{\pm 1}\right)^{\gamma}\left(1-\chi\mathfrak{t}s_i^{\pm 1}x^{\mp 1}\right)^{\gamma}\left(1- \mathfrak{t}s_i^{\pm 2}x^{\pm 1}\right)^{\epsilon}\left(1-\mathfrak{t}s_i^{\pm 2}x^{\mp 1}\right)^{\epsilon}}\cr
&\prod_{\alpha=1}^{2l}\frac{1}{\left(1-\chi \mathfrak{t} y_{\alpha}\right)^{\gamma}}\prod_{i=1}^N\prod_{\alpha=1}^{2l}\prod_{\pm}\frac{1}{\left(1-\mathfrak{t}s_i^{\pm 1}y_{\alpha}\right)},
\end{split}
\end{equation}
where $\epsilon = 1$ or $\epsilon = 0$ means the theory possesses a symmetric hypermultiplet or an antisymmetric hypermultiplet respectively. Also, the flavor fugacity $y_{\alpha}$ satisfies $y_{l+\alpha} = y_{\alpha}^{-1}$ for $\alpha=1, 2, \cdots, l$. On the other hand, when the gauge group is $SO(2N)$ with the discrete fugacity $\chi = -1$, the Higgs line defect index is given by
\begin{equation}
\begin{split}
&\Braket{W_{{\cal R}_1}\cdots W_{{\cal R}_k} }^{SO(2N)^{(-)}+\text{(A/S)-}[2l](H)}(x,y_\alpha;\mathfrak{t})\cr
&=\frac{1}{2^{N-1}(N-1)!}\oint\prod_{i=1}^{N-1}\frac{ds_i}{2\pi i s_i}\prod_{i=1}^k\chi_{{\cal R}_i}(s)\prod_{i=1}^{N-1}\prod_{\pm}\left(1-s_i^{\pm 2}\right)\prod_{i<j}\prod_{\pm}\left(1-s_i^{\pm 1}s_j^{\mp 1}\right)\left(1-s_i^{\pm 1}s_j^{\pm 1}\right)\cr
&\frac{\left(1+\mathfrak{t}^2\right)\left(1-\mathfrak{t}^2\right)^{N-1}\prod_{i=1}^{N-1}\prod_{\pm}\left(1-\mathfrak{t}^2s_i^{\pm 1}\right)\left(1+\mathfrak{t}^2s_i^{\pm 1}\right)\prod_{i<j}^{N-1}\prod_{\pm}\left(1-\mathfrak{t}^2s_i^{\mp 1}s_j^{\pm 1}\right)\left(1 - \mathfrak{t}^2s_i^{\mp 1}s_j^{\mp 1}\right)}{\prod_{i<j}^{N-1}\prod_{\pm}\left(1-\mathfrak{t}s_i^{\pm 1}s_j^{\mp 1}x^{\pm 1}\right)\left(1-\mathfrak{t}s_i^{\pm 1}s_j^{\mp 1}x^{\mp 1}\right)\left(1-\mathfrak{t}s_i^{\pm 1}s_j^{\pm 1}x^{\pm 1}\right)\left(1-\mathfrak{t}s_i^{\pm 1}s_j^{\pm 1}x^{\mp 1}\right)}\cr
&\frac{1}{\left(1+\mathfrak{t}x^{\pm 1}\right)\left(1-\mathfrak{t}x^{\pm 1}\right)^{N-1}\prod_{i=1}^{N-1}\prod_{\pm}\left(1- \mathfrak{t}s_i^{\pm 1}x^{\pm 1}\right)\left(1-\mathfrak{t}s_i^{\pm 1}x^{\mp 1}\right)\left(1+ \mathfrak{t}s_i^{\pm 1}x^{\pm 1}\right)\left(1+ \mathfrak{t}s_i^{\pm 1}x^{\mp 1}\right)}\cr
&\prod_{i=1}^{N-1}\prod_{\pm}\frac{1}{\left(1-\mathfrak{t}s_i^{\pm 2}x^{\pm 1}\right)^{\epsilon}\left(1-\mathfrak{t}s_i^{\pm 2}x^{\mp 1}\right)^{\epsilon}}\prod_\pm \frac{1}{\left(1-\mathfrak{t}x^{\pm 1}\right)^{2\epsilon}}\cr
&\prod_{\alpha=1}^{2l}\frac{1}{\left(1- \mathfrak{t} y_{\alpha}\right)\left(1+ \mathfrak{t} y_{\alpha}\right)}\prod_{i=1}^{N-1}\prod_{\alpha=1}^{2l}\prod_{\pm}\frac{1}{\left(1-\mathfrak{t}s_i^{\pm 1}y_{\alpha}\right)}.
\end{split}
\end{equation}
Since the Higgs line defect index consists of the contribution only from the zero magnetic flux, it does not depend on $\zeta$ and the value of $\zeta$ was suppressed to denote the indices. For simplicity, we focus on rank $1$ gauge groups with two fundamental hypermultiplets for the explicit computations. In the cases without Wilson lines, we find, for example,
\begin{align}
&\mathcal{I}^{SO(2)^{(+)}+S+[4](H)}\nonumber\\
&=\frac{1+\mathfrak{t}+14\mathfrak{t}^2+49\mathfrak{t}^3+103\mathfrak{t}^4+198\mathfrak{t}^5+352\mathfrak{t}^6 + 442\mathfrak{t}^7 + 460\mathfrak{t}^8 + 442\mathfrak{t}^{9} + \cdots + \mathfrak{t}^{16}}{(1-\mathfrak{t})^{12}(1+\mathfrak{t})^6(1+\mathfrak{t}+\mathfrak{t}^2)^5},\\
&\mathcal{I}^{O(2)+S+[4](H)}\nonumber\\
&=\frac{1+\mathfrak{t}+8\mathfrak{t}^2+23\mathfrak{t}^3+50\mathfrak{t}^4+95\mathfrak{t}^5+177\mathfrak{t}^6 + 222\mathfrak{t}^7 + 236\mathfrak{t}^8 + 222\mathfrak{t}^{9} + \cdots + \mathfrak{t}^{16}}{(1-\mathfrak{t})^{12}(1+\mathfrak{t})^6(1+\mathfrak{t}+\mathfrak{t}^2)^5},\label{IO2S2FH}\\
&\mathcal{I}^{SO(2)^{(+)}+A+[4](H)} =\frac{1+9\mathfrak{t}^2+9\mathfrak{t}^4+\mathfrak{t}^6}{(1-\mathfrak{t})^8(1+\mathfrak{t})^6},\\
&\mathcal{I}^{O(2)+A+[4](H)} =\frac{1-2\mathfrak{t}+8\mathfrak{t}^2-4\mathfrak{t}^3+8\mathfrak{t}^4-2\mathfrak{t}^5 +\mathfrak{t}^6}{(1-\mathfrak{t})^8(1+\mathfrak{t})^6},
\end{align}
for the $SO(2)$ or $O(2)$ gauge group and
\begin{align}
\mathcal{I}^{SO(3)^{(+)}+S+[4](H)} &=\frac{P_{SO(3)+S+[4](H)}(\mathfrak{t})}{(1-\mathfrak{t})^{18}(1+\mathfrak{t})^{10}(1+\mathfrak{t}+\mathfrak{t}^2)^6},\\
\mathcal{I}^{O(3)+S+[4](H)} &=\frac{P_{O(3)+S+[4](H)}(\mathfrak{t})}{(1-\mathfrak{t})^{18}(1+\mathfrak{t})^{10}(1+\mathfrak{t}^2)^5(1+\mathfrak{t}+\mathfrak{t}^2)^6},\label{IO3S2FH}\\
\mathcal{I}^{SO(3)^{(+)}+A+[4](H)} &=\frac{1-2\mathfrak{t}+12\mathfrak{t}^2-8\mathfrak{t}^3+34\mathfrak{t}^4-12\mathfrak{t}^5+34\mathfrak{t}^6 + \cdots +\mathfrak{t}^{10}}{(1-\mathfrak{t})^{12}(1+\mathfrak{t})^{10}},\\
\mathcal{I}^{O(3)+A+[4](H)} &=\frac{P_{O(3)+A+[4](H)}(\mathfrak{t})}{(1-\mathfrak{t})^{12}(1+\mathfrak{t})^{10}(1+\mathfrak{t}^2)^{5}},
\end{align}
for the $SO(3)$ or $O(3)$ gauge group, where
\begin{align}
P_{SO(3)+S+[4](H)}(\mathfrak{t})&=1+3\mathfrak{t}^2+22\mathfrak{t}^3+76\mathfrak{t}^4+140\mathfrak{t}^5+320\mathfrak{t}^6+636\mathfrak{t}^7+1008\mathfrak{t}^8+1292\mathfrak{t}^9\cr
&\qquad+1647\mathfrak{t}^{10}+1804\mathfrak{t}^{11}+1647\mathfrak{t}^{12}+ \cdots +\mathfrak{t}^{22},\\
P_{O(3)+S+[4](H)}(\mathfrak{t})&=1+8\mathfrak{t}^2+18\mathfrak{t}^3+61\mathfrak{t}^4+142\mathfrak{t}^5+388\mathfrak{t}^6+792\mathfrak{t}^7+1691\mathfrak{t}^8+2996\mathfrak{t}^9\cr
&\qquad+5255\mathfrak{t}^{10}+7994\mathfrak{t}^{11}+11713\mathfrak{t}^{12}+15134\mathfrak{t}^{13}+18773\mathfrak{t}^{14}\cr
&\qquad+20796\mathfrak{t}^{15}+21980\mathfrak{t}^{16} + 20796\mathfrak{t}^{17} + \cdots +\mathfrak{t}^{32}\\
P_{O(3)+A+[4](H)}(\mathfrak{t})&=1-2\mathfrak{t}+9\mathfrak{t}^2-6\mathfrak{t}^3+48\mathfrak{t}^4-40\mathfrak{t}^5+174\mathfrak{t}^6-80\mathfrak{t}^7+311\mathfrak{t}^8-128\mathfrak{t}^9\cr
&\qquad+418\mathfrak{t}^{10}-128\mathfrak{t}^{11} + \cdots  + \mathfrak{t}^{20}
\end{align}
and we set $x=y_{\alpha}=1$. The numerator of each index is a palindromic polynomial. Note that \eqref{IO2S2FH} and \eqref{IO3S2FH} agree with the Hilbert series of the moduli spaces of the two  and three $USp(4)$ instantons respectively \cite{Hanany:2012dm, Cremonesi:2014xha} since the Higgs branch moduli spaces of the 3d theories realize the corresponding instanton moduli spaces. With Wilson lines, we find, for example,
\begin{align}
&\Braket{W_{\ydiagram{1}}W_{\ydiagram{1}}}^{SO(2)^{(+)}+S+[4](H)}=2\Braket{W_{\ydiagram{1}}W_{\ydiagram{1}}}^{O(2)+S+[4](H)}\nonumber\\
&=\frac{2\left(1+3\mathfrak{t}+26\mathfrak{t}^2+85\mathfrak{t}^3+209\mathfrak{t}^4+408\mathfrak{t}^5+682\mathfrak{t}^6+884\mathfrak{t}^7+964\mathfrak{t}^8+884\mathfrak{t}^9 + \cdots + \mathfrak{t}^{16}\right)}{(1-\mathfrak{t})^{12}(1+\mathfrak{t})^6(1+\mathfrak{t}+\mathfrak{t}^2)^5},\\
&\Braket{W_{\ydiagram{1}}W_{\ydiagram{1}}}^{SO(2)^{(+)}+A+[4](H)}=2\Braket{W_{\ydiagram{1}}W_{\ydiagram{1}}}^{O(2)+A+[4](H)}=\frac{2\left(1+19\mathfrak{t}^2+19\mathfrak{t}^4+\mathfrak{t}^6\right)}{(1-\mathfrak{t})^{8}(1+\mathfrak{t})^6},
\end{align}
for the $SO(2)$ or $O(2)$ gauge group and
\begin{align}
\Braket{W_{\ydiagram{1}}W_{\ydiagram{1}}}^{SO(3)^{(+)}+S+[4](H)}&=\frac{\mathcal{P}_{SO(3)+S+[4](H)}(\mathfrak{t})}{(1-\mathfrak{t})^{18}(1+\mathfrak{t})^{10}(1+\mathfrak{t}+\mathfrak{t}^2)^4},\\
\Braket{W_{\ydiagram{2}}}^{SO(3)^{(+)}+S+[4](H)}&=\frac{\mathcal{P}'_{SO(3)+S+[4](H)}(\mathfrak{t})}{(1-\mathfrak{t})^{18}(1+\mathfrak{t})^{10}(1+\mathfrak{t}+\mathfrak{t}^2)^6},\\
\Braket{W_{\ydiagram{1}}W_{\ydiagram{1}}}^{O(3)+S+[4](H)}&=\frac{\mathcal{P}_{O(3)+S+[4](H)}(\mathfrak{t})}{(1-\bt)^{18}(1+\bt)^{10}(1+\bt+2\bt^2+\bt^3+\bt^4)^5},\\
\Braket{W_{\ydiagram{1}}W_{\ydiagram{1}}}^{SO(3)^{(+)}+A+[4](H)}&=\frac{1+4\mathfrak{t}+35\mathfrak{t}^2+52\mathfrak{t}^{3}+135\mathfrak{t}^4+104\mathfrak{t}^5+135\mathfrak{t}^6 + \cdots + \mathfrak{t}^{10}}{(1-\mathfrak{t})^{12}(1+\mathfrak{t})^{10}},\\
\Braket{W_{\ydiagram{2}}}^{SO(3)^{(+)}+A+[4](H)}&=\frac{\mathfrak{t}^2\left(21+22\mathfrak{t}+88\mathfrak{t}^2+48\mathfrak{t}^3+88\mathfrak{t}^4+22\mathfrak{t}^5+21\mathfrak{t}^6\right)}{(1-\mathfrak{t})^{12}(1+\mathfrak{t})^{10}},\\
\Braket{W_{\ydiagram{1}}W_{\ydiagram{1}}}^{O(3)+A+[4](H)}&=\frac{\mathcal{P}_{O(3)+A+[4](H)}(\mathfrak{t})}{(1-\bt)^{12}(1+\bt)^{10}(1+\bt^2)^5},
\end{align}
for the $SO(3)$ or $O(3)$ gauge group, where
\begin{align}
\mathcal{P}_{SO(3)+S+[4](H)}(\mathfrak{t})&=1+4\mathfrak{t}+27\mathfrak{t}^2+84\mathfrak{t}^3+243\mathfrak{t}^4+514\mathfrak{t}^5+952\mathfrak{t}^6+1410\mathfrak{t}^7\cr
&\qquad +1832\mathfrak{t}^8+1960\mathfrak{t}^9 + 1832\mathfrak{t}^{10}+ \cdots + \mathfrak{t}^{18},\\
\mathcal{P}'_{SO(3)+S+[4](H)}(\mathfrak{t})&=\mathfrak{t}\left(2+21\mathfrak{t}+74\mathfrak{t}^2+253\mathfrak{t}^3+708\mathfrak{t}^4+1605\mathfrak{t}^5+2962\mathfrak{t}^6+4852\mathfrak{t}^7\right.\cr
&\qquad \left.+6880\mathfrak{t}^8+8418\mathfrak{t}^9+8920\mathfrak{t}^{10}+8418\mathfrak{t}^{11}+\cdots + 2\mathfrak{t}^{20}\right),\\
\mathcal{P}_{O(3)+S+[4](H)}(\mathfrak{t})&=1+\mathfrak{t}+25\mathfrak{t}^2+72\mathfrak{t}^3+260\mathfrak{t}^4+754\mathfrak{t}^5+1905\mathfrak{t}^6\cr
&\qquad +4133\mathfrak{t}^7+8304\mathfrak{t}^8+14531\mathfrak{t}^9+23665\mathfrak{t}^{10}+34494\mathfrak{t}^{11}\cr
&\qquad+46518\mathfrak{t}^{12}+57010\mathfrak{t}^{13}+64954\mathfrak{t}^{14}+67258\mathfrak{t}^{15}\cr
&\qquad+64954\mathfrak{t}^{16}+ \cdots +\mathfrak{t}^{30},\\
\mathcal{P}_{O(3)+A+[4](H)}&=1+24\mathfrak{t}^2+36\mathfrak{t}^3+152\mathfrak{t}^4+208\mathfrak{t}^5+586\mathfrak{t}^6+568\mathfrak{t}^7\cr
&\qquad+1199\mathfrak{t}^8+916\mathfrak{t}^9+1548\mathfrak{t}^{10}+916\mathfrak{t}^{11}+\cdots+\mathfrak{t}^{20}.
\end{align}
Again, the numerator of each index is a palindromic polynomial. Then we can see that the normalized Higgs line defect indices in the limit $\mathfrak{t} \to 1$ satisfy
\begin{align}
&\lim_{\mathfrak{t}\to1}\Braket{\mathcal{W}_{\ydiagram{1}}\mathcal{W}_{\ydiagram{1}}}^{SO(2)^{(+)}+S+[4](H)} = \lim_{\mathfrak{t}\to1}\Braket{\mathcal{W}_{\ydiagram{1}}\mathcal{W}_{\ydiagram{1}}}^{SO(2)^{(+)}+A+[4](H)}\cr
&=\lim_{\mathfrak{t}\to1}\Braket{\mathcal{W}_{\ydiagram{1}}\mathcal{W}_{\ydiagram{1}}}^{O(2)+S+[4](H)}=\lim_{\mathfrak{t}\to1}\Braket{\mathcal{W}_{\ydiagram{1}}\mathcal{W}_{\ydiagram{1}}}^{O(2)+A+[4](H)}  = 4 = \left(\chi_{\ydiagram{1}}^{\mathfrak{so}(2)}(1)\right)^2,\\
&\lim_{\mathfrak{t}\to1}\Braket{\mathcal{W}_{\ydiagram{1}}\mathcal{W}_{\ydiagram{1}}}^{SO(3)^{(+)}+S+[4](H)} = \lim_{\mathfrak{t}\to1}\Braket{\mathcal{W}_{\ydiagram{1}}\mathcal{W}_{\ydiagram{1}}}^{SO(3)^{(+)}+A+[4](H)}\cr
&=\lim_{\mathfrak{t}\to1}\Braket{\mathcal{W}_{\ydiagram{1}}\mathcal{W}_{\ydiagram{1}}}^{O(3)+S+[4](H)} = \lim_{\mathfrak{t}\to1}\Braket{\mathcal{W}_{\ydiagram{1}}\mathcal{W}_{\ydiagram{1}}}^{O(3)+A+[4](H)} = 9 = \left(\chi_{\ydiagram{1}}^{\mathfrak{so}(3)}(1)\right)^2,
\end{align}
and
\begin{equation}
\lim_{\mathfrak{t}\to1}\Braket{\mathcal{W}_{\ydiagram{2}}}^{SO(3)^{(+)}+S+[4](H)} =\lim_{\mathfrak{t}\to1}\Braket{\mathcal{W}_{\ydiagram{2}}}^{SO(3)^{(+)}+A+[4](H)} =5 = \chi_{\ydiagram{2}}^{\mathfrak{so}(3)}(1),
\end{equation}
which are consistent with \eqref{Witten_indH_char}.

\subsection{$\mathcal{N}=4$ SQCD}
\label{app_Higgs_SQCD}

Every theory considered so far contains a hypermultiplet in the adjoint or in the (anti)symmetric representation, so that all of them are of ADHM type.
In order to confirm that the conjecture \eqref{Witten_indH_char} is a property of the Higgs limit itself rather than an artifact of this class of theories,
we collect here the results for $\mathcal{N}=4$ SQCD, namely the gauge theory whose only matter consists of $N_f$ hypermultiplets in the fundamental (vector) representation.
We restrict ourselves to good theories, i.e.~$N_f\ge 2N$ for $U(N)$, $N_f\ge 2N+1$ for $USp(2N)$ and $N_f\ge N_c-1$ for $SO(N_c)$ or $O(N_c)$,
Here the bracket counts hypermultiplets in the fundamental representation, which is the standard flavor counting for SQCD. For the ADHM-type theories with orthogonal and symplectic gauge groups the bracket instead counts fundamental half-hypermultiplets, as in $USp(2N)+$(A/S)-$[2l]$, since there the number of half-hypermultiplets can be odd and the flavor symmetry acts on them.
so that only the zero magnetic flux sector survives the Higgs limit and \eqref{Higgs_general_int} applies.
For $G=U(N)$ it reads
\begin{align}
\label{HiggsUNSQCD}
&\langle W_{\mathcal{R}_1}\cdots W_{\mathcal{R}_k}\rangle^{U(N)\text{-}[N_f](H)}(y_\alpha;\mathfrak{t})\nonumber\\
&=\frac{(1-\mathfrak{t}^2)^N}{N!}\oint\prod_{i=1}^N\frac{ds_i}{2\pi is_i}
\prod_{j=1}^k\chi^{\mathfrak{u}(N)}_{\mathcal{R}_j}(s)
\frac{\prod_{i\neq j}\bigl(1-\frac{s_i}{s_j}\bigr)\bigl(1-\mathfrak{t}^2\frac{s_i}{s_j}\bigr)}
{\prod_{i=1}^N\prod_{\alpha=1}^{N_f}\prod_\pm\bigl(1-\mathfrak{t}(s_iy_\alpha)^{\pm 1}\bigr)}\,,
\end{align}
and the symplectic and orthogonal cases are obtained from \eqref{Higgs_general_int} in the same way,
by replacing the $\mathfrak{u}(N)$ root system $\{e_i-e_j\}$ with that of $\mathfrak{usp}(2N)$ or $\mathfrak{so}(N_c)$
and the weights $\pm e_i$ of the hypermultiplets accordingly.
In the following we set $y_\alpha=1$.
For the indices without Wilson lines we find, for example,
\begin{subequations}
\begin{align}
\mathcal{I}^{U(1)\text{-}[2](H)}&=\frac{1+\mathfrak{t}^2}{(1-\mathfrak{t}^2)^2},\qquad
\mathcal{I}^{U(1)\text{-}[3](H)}=\frac{1+4\mathfrak{t}^2+\mathfrak{t}^4}{(1-\mathfrak{t}^2)^4},\\
\mathcal{I}^{U(2)\text{-}[4](H)}&=\frac{(1+\mathfrak{t}^2)^2(1+5\mathfrak{t}^2+\mathfrak{t}^4)}{(1-\mathfrak{t}^2)^8},\\
\mathcal{I}^{U(3)\text{-}[6](H)}&=\frac{(1+\mathfrak{t}^2)^3\left(1+14\mathfrak{t}^2+72\mathfrak{t}^4+133\mathfrak{t}^6+\cdots+\mathfrak{t}^{12}\right)}{(1-\mathfrak{t}^2)^{18}},\\
\mathcal{I}^{USp(2)\text{-}[3](H)}&=\frac{(1+\mathfrak{t}^2)(1+8\mathfrak{t}^2+\mathfrak{t}^4)}{(1-\mathfrak{t}^2)^6},\\
\mathcal{I}^{USp(4)\text{-}[5](H)}&=\frac{(1+\mathfrak{t}^2)^2\left(1+23\mathfrak{t}^2+223\mathfrak{t}^4+925\mathfrak{t}^6+1532\mathfrak{t}^8+\cdots+\mathfrak{t}^{16}\right)}{(1-\mathfrak{t}^2)^{20}},\\
\mathcal{I}^{SO(3)^{(+)}\text{-}[3](H)}&=\frac{(1+\mathfrak{t}^2)\left(1-2\mathfrak{t}+11\mathfrak{t}^2-6\mathfrak{t}^3+23\mathfrak{t}^4-6\mathfrak{t}^5+11\mathfrak{t}^6-2\mathfrak{t}^7+\mathfrak{t}^8\right)}{(1-\mathfrak{t})^{12}(1+\mathfrak{t})^{10}},\\
\mathcal{I}^{SO(5)^{(+)}\text{-}[4](H)}&=\frac{(1+\mathfrak{t}^2)^2\left(1+14\mathfrak{t}^2+79\mathfrak{t}^4+223\mathfrak{t}^6+317\mathfrak{t}^8+\cdots+\mathfrak{t}^{16}\right)}{(1-\mathfrak{t}^2)^{20}}.
\end{align}
\end{subequations}

In each case the order of the pole at $\mathfrak{t}=1$ is $2\dim_{\mathbb{H}}\mathcal{M}_H=2(\dim_{\mathbb{C}}\mathcal{H}-\dim G)$, in agreement with the general scaling argument given in section \ref{sec_Higgsindices}.
With Wilson lines inserted we find, for example,
\begin{subequations}
\begin{align}
\langle\mathcal{W}_{\ydiagram{1}}\rangle^{U(2)\text{-}[4](H)}&=\frac{4\mathfrak{t}}{1+\mathfrak{t}^2},\qquad
\langle\mathcal{W}_{\ydiagram{1}}\rangle^{U(3)\text{-}[6](H)}=\frac{6\mathfrak{t}}{1+\mathfrak{t}^2},\\
\langle\mathcal{W}_{\ydiagram{1}}\mathcal{W}_{\overline{\ydiagram{1}}}\rangle^{U(2)\text{-}[4](H)}
&=\frac{1+22\mathfrak{t}^2+66\mathfrak{t}^4+22\mathfrak{t}^6+\mathfrak{t}^8}{(1+\mathfrak{t}^2)^2(1+5\mathfrak{t}^2+\mathfrak{t}^4)},\\
\langle\mathcal{W}_{\ydiagram{1}}\rangle^{USp(2)\text{-}[3](H)}
&=\frac{2\mathfrak{t}(3+14\mathfrak{t}^2+3\mathfrak{t}^4)}{(1+\mathfrak{t}^2)(1+8\mathfrak{t}^2+\mathfrak{t}^4)},\\
\langle\mathcal{W}_{\ydiagram{1}}\mathcal{W}_{\ydiagram{1}}\rangle^{USp(2)\text{-}[3](H)}
&=\frac{1+29\mathfrak{t}^2+64\mathfrak{t}^4-20\mathfrak{t}^6+7\mathfrak{t}^8-\mathfrak{t}^{10}}{(1+\mathfrak{t}^2)(1+8\mathfrak{t}^2+\mathfrak{t}^4)}.
\end{align}
\end{subequations}

Taking the limit $\mathfrak{t}\rightarrow 1$ we obtain
\begin{align}
\lim_{\mathfrak{t}\rightarrow 1}\langle\mathcal{W}_{\mathcal{R}_1}\cdots\mathcal{W}_{\mathcal{R}_k}\rangle^{G\text{-}[N_f](H)}(\mathfrak{t})
=\prod_{j=1}^k\dim\mathcal{R}_j
\end{align}
for all the cases we have examined, which are listed in table \ref{tab_SQCD_Higgs}.
Together with the ADHM-type results of appendices \ref{app_Higgs_UN}--\ref{app_Higgs_SON}, this supports \eqref{Witten_indH_conj} as a general statement about the Higgs limit,
valid for any good 3d $\mathcal{N}=4$ gauge theory whose Higgs branch completely breaks the gauge group.
\begin{table}[htbp]
\centering
\footnotesize
\begin{tabular}{l|l|c|c}
\hline
theory & Wilson lines & $\lim_{\mathfrak{t}\rightarrow1}\langle\mathcal{W}\cdots\rangle^{(H)}$ & $\prod_j\dim\mathcal{R}_j$ \\ \hline
$U(1)$-$[2]$, $U(1)$-$[3]$ & $\mathcal{W}_{(n)}$, $\mathcal{W}_{(n)}\mathcal{W}_{\overline{(n')}}$, $n,n'\le 3$ & $1$ & $1$ \\ \hline
$U(2)$-$[4]$ & $\mathcal{W}_{\ydiagram{1}}$ & $2$ & $2$ \\
 & $\mathcal{W}_{\ydiagram{1}}\mathcal{W}_{\overline{\ydiagram{1}}}$ & $4$ & $4$ \\
 & $\mathcal{W}_{\ydiagram{2}}\mathcal{W}_{\overline{\ydiagram{2}}}$ & $9$ & $9$ \\
 & $\mathcal{W}_{\ydiagram{1,1}}\mathcal{W}_{\overline{\ydiagram{1,1}}}$ & $1$ & $1$ \\
 & $\mathcal{W}_{\ydiagram{2,1}}$ & $2$ & $2$ \\
 & $\mathcal{W}_{\ydiagram{1}}\mathcal{W}_{\ydiagram{1}}\mathcal{W}_{\overline{\ydiagram{1,1}}}$ & $4$ & $4$ \\ \hline
$U(2)$-$[5]$ & $\mathcal{W}_{\ydiagram{2}}\mathcal{W}_{\overline{\ydiagram{1,1}}}$ & $3$ & $3$ \\
 & $\mathcal{W}_{\ydiagram{3}}\mathcal{W}_{\overline{\ydiagram{3}}}$ & $16$ & $16$ \\ \hline
$U(3)$-$[6]$ & $\mathcal{W}_{\ydiagram{1}}$ & $3$ & $3$ \\
 & $\mathcal{W}_{\ydiagram{1}}\mathcal{W}_{\overline{\ydiagram{1}}}$, $\mathcal{W}_{\ydiagram{1,1}}\mathcal{W}_{\overline{\ydiagram{1,1}}}$ & $9$ & $9$ \\
 & $\mathcal{W}_{\ydiagram{2}}\mathcal{W}_{\overline{\ydiagram{1,1}}}$ & $18$ & $18$ \\
 & $\mathcal{W}_{\ydiagram{2,1}}\mathcal{W}_{\overline{\ydiagram{2,1}}}$ & $64$ & $64$ \\
 & $\mathcal{W}_{\ydiagram{1,1,1}}$ & $1$ & $1$ \\ \hline
$USp(2)$-$[3]$ & $\mathcal{W}_{\ydiagram{1}}$, $\mathcal{W}_{\ydiagram{1}}\mathcal{W}_{\ydiagram{1}}$ & $2$, $4$ & $2$, $4$ \\
 & $\mathcal{W}_{\ydiagram{2}}$, $\mathcal{W}_{\ydiagram{2}}\mathcal{W}_{\ydiagram{2}}$ & $3$, $9$ & $3$, $9$ \\
 & $\mathcal{W}_{\ydiagram{3}}\mathcal{W}_{\ydiagram{1}}$ & $8$ & $8$ \\ \hline
$USp(4)$-$[5]$ & $\mathcal{W}_{\ydiagram{1}}$, $\mathcal{W}_{\ydiagram{1}}\mathcal{W}_{\ydiagram{1}}$ & $4$, $16$ & $4$, $16$ \\
 & $\mathcal{W}_{\ydiagram{1,1}}$, $\mathcal{W}_{\ydiagram{1,1}}\mathcal{W}_{\ydiagram{1,1}}$ & $5$, $25$ & $5$, $25$ \\
 & $\mathcal{W}_{\ydiagram{2}}$, $\mathcal{W}_{\ydiagram{2}}\mathcal{W}_{\ydiagram{2}}$ & $10$, $100$ & $10$, $100$ \\
 & $\mathcal{W}_{\ydiagram{1}}\mathcal{W}_{\ydiagram{1,1}}$ & $20$ & $20$ \\ \hline
$SO(3)^{(+)}$-$[2]$, $SO(3)^{(+)}$-$[3]$ & $\mathcal{W}_{\ydiagram{1}}$, $\mathcal{W}_{\ydiagram{1}}\mathcal{W}_{\ydiagram{1}}$ & $3$, $9$ & $3$, $9$ \\
 & $\mathcal{W}_{\ydiagram{2}}$, $\mathcal{W}_{\ydiagram{2}}\mathcal{W}_{\ydiagram{2}}$ & $5$, $25$ & $5$, $25$ \\ \hline
$SO(5)^{(+)}$-$[4]$ & $\mathcal{W}_{\ydiagram{1}}$, $\mathcal{W}_{\ydiagram{1}}\mathcal{W}_{\ydiagram{1}}$ & $5$, $25$ & $5$, $25$ \\
 & $\mathcal{W}_{\ydiagram{1,1}}$, $\mathcal{W}_{\ydiagram{1,1}}\mathcal{W}_{\ydiagram{1,1}}$ & $10$, $100$ & $10$, $100$ \\
 & $\mathcal{W}_{\ydiagram{2}}$, $\mathcal{W}_{\ydiagram{2}}\mathcal{W}_{\ydiagram{2}}$ & $14$, $196$ & $14$, $196$ \\
 & $\mathcal{W}_{\ydiagram{3}}$, $\mathcal{W}_{\ydiagram{3}}\mathcal{W}_{\ydiagram{3}}$ & $30$, $900$ & $30$, $900$ \\
$SO(5)^{(+)}$-$[5]$ & $\mathcal{W}_{\ydiagram{1}}$, $\mathcal{W}_{\ydiagram{1,1}}$ & $5$, $10$ & $5$, $10$ \\ \hline
\end{tabular}
\caption{The $\mathfrak{t}\rightarrow1$ limit of the normalized Higgs line defect indices of $\mathcal{N}=4$ SQCD, compared with the product of the dimensions of the representations. Here $G$-$[N_f]$ denotes the gauge group $G$ with $N_f$ fundamental hypermultiplets.}
\label{tab_SQCD_Higgs}
\end{table}

Finally we turn to the orthogonal theories with the charge conjugation $\mathbb{Z}^{\mathcal{C}}_2$ gauged, that is with gauge group $O(N_c)=SO(N_c)\rtimes\mathbb{Z}^{\mathcal{C}}_2$, for which the line defect indices are obtained by averaging the $\chi=\pm1$ sectors,
$\langle\cdots\rangle^{O(N_c)}=\frac{1}{2}\bigl(\langle\cdots\rangle^{SO(N_c)^{(+)}}+\langle\cdots\rangle^{SO(N_c)^{(-)}}\bigr)$,
with the $O(N_c)$ holonomy taken to be $\mathrm{diag}(s_1,s_1^{-1},\cdots,s_N,s_N^{-1},\chi)$ for $N_c=2N+1$.
Here $\chi=\pm1$ is the determinant of the holonomy, so that the two values label the two connected components of $O(N_c)$, and the vector representation has the $N_c$ eigenvalues $\{s_i^{\pm1}\}\cup\{\chi\}$. Since the representations labeled by Young diagrams are built from tensor powers of the vector representation, their characters are the corresponding symmetric functions of these eigenvalues,
\begin{align}
\label{chiOdd}
\chi_{(1^m)}^{\mathfrak{so}(N_c)}(s,\chi)=e_m\bigl(\{s_i^{\pm1}\},\chi\bigr),\qquad
\chi_{(m)}^{\mathfrak{so}(N_c)}(s,\chi)=h_m\bigl(\{s_i^{\pm1}\},\chi\bigr)-h_{m-2}\bigl(\{s_i^{\pm1}\},\chi\bigr),
\end{align}
where the super/subscript distinguishes the characters from the discrete fugacity $\chi$.
Here $(m)=\text{Sym}_0^m$ denotes the traceless part.
At $\chi=+1$ these are the ordinary $SO(N_c)$ characters, while at $\chi=-1$ they are evaluated on the component with determinant $-1$, and setting $s_i=1$ there gives the values at the reflection $g_0$ listed below. For $O(3)$, for instance,
\begin{align}
\label{chiO3}
\chi_{\ydiagram{1}}^{\mathfrak{so}(3)}(s,\chi)=s+s^{-1}+\chi,\qquad
\chi_{\ydiagram{2}}^{\mathfrak{so}(3)}(s,\chi)=s^2+1+s^{-2}+\chi(s+s^{-1}),
\end{align}
which give $3$ and $5$ at $s=\chi=1$, and $1$ and $1$ at $s=1$, $\chi=-1$, while for $O(5)$
\begin{align}
\label{chiO5}
\chi_{\ydiagram{1}}^{\mathfrak{so}(5)}(s,\chi)&=\sum_{i=1,2}(s_i+s_i^{-1})+\chi,\nonumber\\
\chi_{\ydiagram{1,1}}^{\mathfrak{so}(5)}(s,\chi)&=(s_1+s_1^{-1})(s_2+s_2^{-1})+2+\chi\sum_{i=1,2}(s_i+s_i^{-1}),\nonumber\\
\chi_{\ydiagram{2}}^{\mathfrak{so}(5)}(s,\chi)&=\chi_{\ydiagram{1,1}}^{\mathfrak{so}(5)}(s,\chi)+\sum_{i=1,2}(s_i^2+s_i^{-2}),
\end{align}
which give $5$, $10$ and $14$ at $s_i=\chi=1$, and $3$, $2$ and $6$ at $s_i=1$, $\chi=-1$.
The terms linear in $\chi$ are the ones that distinguish the two sectors, and for $\ydiagram{1,1}$ and $\ydiagram{2}$ they enter with the sign shown here.
Exactly as anticipated in section \ref{sec_orthSQCD}, the balanced case $N_f=N_c-1$ is special, since there the two sectors carry the same leading divergence. In fact,
\begin{align}
\mathcal{I}^{SO(3)^{(-)}\text{-}[2](H)}=\mathcal{I}^{SO(3)^{(+)}\text{-}[2](H)},
\qquad
\mathcal{I}^{SO(5)^{(-)}\text{-}[4](H)}=\mathcal{I}^{SO(5)^{(+)}\text{-}[4](H)},
\end{align}
while for $N_f\ge N_c$ the $\chi=-1$ sector has a strictly weaker divergence, e.g.~$\mathcal{I}^{SO(3)^{(\pm)}\text{-}[3](H)}$ have poles of order $12$ and $10$ at $\mathfrak{t}=1$ respectively,
and $\mathcal{I}^{SO(5)^{(\pm)}\text{-}[5](H)}$ have poles of order $30$ and $28$.
The resulting Witten indices are collected in table \ref{tab_ON_Higgs}.
They are reproduced neither by $\prod_j\dim\mathcal{R}_j$ alone nor by a torus integral, but precisely by \eqref{Witten_indH_O}, i.e.~by the average of the characters over the unbroken $\mathbb{Z}_2=\{e,g_0\}$ with $g_0=\mathrm{diag}(1,\cdots,1,-1)$, in agreement with the character average \eqref{Witten_ind_unified}.

Let us make the evaluation of \eqref{Witten_ind_unified} explicit here as well.
Since $H_0=\{e,g_0\}$ has two elements its normalized Haar measure is $\int_{H_0}dh\,f(h)=\frac12(f(e)+f(g_0))$, so that the only new ingredient is the value of the characters at the reflection $g_0$.
For the antisymmetric and the traceless symmetric powers of the vector representation these follow from the elementary and complete homogeneous polynomials evaluated at the eigenvalues $(\underbrace{1,\cdots,1}_{N_c-1},-1)$ of $g_0$,
\begin{subequations}
\label{chiatg0}
\begin{align}
\chi_{(1^m)}^{\mathfrak{so}(N_c)}(g_0)&=e_m(1^{N_c-1},-1)=\binom{N_c-1}{m}-\binom{N_c-1}{m-1},\\
\chi_{(m)}^{\mathfrak{so}(N_c)}(g_0)&=h_m(1^{N_c-1},-1)-h_{m-2}(1^{N_c-1},-1).
\end{align}
\end{subequations}
For the representations appearing in table \ref{tab_ON_Higgs} this gives
\begin{align}
\begin{array}{c|ccc}
\hline
&\chi_{\ydiagram{1}}^{\mathfrak{so}(N_c)}(g_0)&\chi_{\ydiagram{1,1}}^{\mathfrak{so}(N_c)}(g_0)&\chi_{\ydiagram{2}}^{\mathfrak{so}(N_c)}(g_0)\\ \hline
O(3)&1&-&1\\
O(5)&3&2&6\\ \hline
\end{array}
\end{align}
to be compared with the dimensions $3,-,5$ for $O(3)$ and $5,10,14$ for $O(5)$.
Substituting into \eqref{Witten_indH_O} one obtains, for instance,
\begin{subequations}
\begin{align}
\lim_{\mathfrak{t}\rightarrow1}\langle\mathcal{W}_{\ydiagram{1}}\rangle^{O(3)\text{-}[2](H)}
&=\tfrac12(3+1)=2,\qquad
\lim_{\mathfrak{t}\rightarrow1}\langle\mathcal{W}_{\ydiagram{1}}\mathcal{W}_{\ydiagram{1}}\rangle^{O(3)\text{-}[2](H)}=\tfrac12(3^2+1^2)=5,\\
\lim_{\mathfrak{t}\rightarrow1}\langle\mathcal{W}_{\ydiagram{1}}\mathcal{W}_{\ydiagram{2}}\rangle^{O(3)\text{-}[2](H)}
&=\tfrac12(3\cdot5+1\cdot1)=8,\\
\lim_{\mathfrak{t}\rightarrow1}\langle\mathcal{W}_{\ydiagram{2}}\mathcal{W}_{\ydiagram{2}}\rangle^{O(5)\text{-}[4](H)}
&=\tfrac12(14^2+6^2)=116,
\end{align}
\end{subequations}
which are exactly the values found from the index.
Note that the answers are integers even though \eqref{Witten_indH_O} is a half sum, because $\frac12(\dim\mathcal{R}+\chi_{\mathcal{R}}^{\mathfrak{so}(N_c)}(g_0))=\dim\mathcal{R}^{+}$ is the dimension of the $g_0$ even subspace, in agreement with \eqref{Witten_ind_unified_inv}.

\begin{table}[htbp]
\centering
\small
\begin{tabular}{l|l|c|c|c}
\hline
theory & Wilson lines & $\lim_{\mathfrak{t}\rightarrow1}\langle\mathcal{W}\cdots\rangle^{(H)}$ & $\prod_j\dim\mathcal{R}_j$ & \eqref{Witten_indH_O}\\ \hline
$O(3)$-$[2]$ & $\mathcal{W}_{\ydiagram{1}}$ & $2$ & $3$ & $2$ \\
 & $\mathcal{W}_{\ydiagram{1}}\mathcal{W}_{\ydiagram{1}}$ & $5$ & $9$ & $5$ \\
 & $\mathcal{W}_{\ydiagram{1}}\mathcal{W}_{\ydiagram{1}}\mathcal{W}_{\ydiagram{1}}$ & $14$ & $27$ & $14$ \\
 & $\mathcal{W}_{\ydiagram{2}}$ & $3$ & $5$ & $3$ \\
 & $\mathcal{W}_{\ydiagram{2}}\mathcal{W}_{\ydiagram{2}}$ & $13$ & $25$ & $13$ \\
 & $\mathcal{W}_{\ydiagram{1}}\mathcal{W}_{\ydiagram{2}}$ & $8$ & $15$ & $8$ \\ \hline
$O(5)$-$[4]$ & $\mathcal{W}_{\ydiagram{1}}$, $\mathcal{W}_{\ydiagram{1}}\mathcal{W}_{\ydiagram{1}}$ & $4$, $17$ & $5$, $25$ & $4$, $17$ \\
 & $\mathcal{W}_{\ydiagram{1,1}}$, $\mathcal{W}_{\ydiagram{1,1}}\mathcal{W}_{\ydiagram{1,1}}$ & $6$, $52$ & $10$, $100$ & $6$, $52$ \\
 & $\mathcal{W}_{\ydiagram{2}}$, $\mathcal{W}_{\ydiagram{2}}\mathcal{W}_{\ydiagram{2}}$ & $10$, $116$ & $14$, $196$ & $10$, $116$ \\ \hline
$O(3)$-$[3]$, $O(3)$-$[4]$ & $\mathcal{W}_{\ydiagram{1}}$, $\mathcal{W}_{\ydiagram{1}}\mathcal{W}_{\ydiagram{1}}$ & $3$, $9$ & $3$, $9$ & --- \\
 & $\mathcal{W}_{\ydiagram{2}}$, $\mathcal{W}_{\ydiagram{2}}\mathcal{W}_{\ydiagram{2}}$ & $5$, $25$ & $5$, $25$ & --- \\ \hline
$O(5)$-$[5]$ & $\mathcal{W}_{\ydiagram{1}}$, $\mathcal{W}_{\ydiagram{1,1}}$ & $5$, $10$ & $5$, $10$ & --- \\ \hline
\end{tabular}
\caption{The $\mathfrak{t}\rightarrow1$ limit of the normalized Higgs line defect indices of $O(N_c)$ SQCD. For the balanced case $N_f=N_c-1$ the results deviate from $\prod_j\dim\mathcal{R}_j$ and are instead reproduced by the average over the unbroken $\mathbb{Z}_2$, \eqref{Witten_indH_O}. For $N_f\ge N_c$ the gauge group is completely broken and \eqref{Witten_indH_conj} is recovered.}
\label{tab_ON_Higgs}
\end{table}

\bibliographystyle{utphys}
\bibliography{ref}

\end{document}